\documentclass[10pt]{article}
\pdfoutput=1
\usepackage{setspace}
\newcounter{iequation}
\newcounter{cequation}
\usepackage{jheplikemod}
\usepackage{xcolor}
\usepackage{ifpdf,tocloft,rotating,textcomp,mathrsfs,amssymb,amsmath,amsthm,mathtools,float,caption,tikz-cd,subcaption,epsfig,dcolumn,upgreek,enumitem,array,multirow,bigdelim,arydshln,appendix,xparse,microtype,bm,braket,dsfont,nccmath,scalerel}
\usetikzlibrary{cd}
\usepackage{bold-extra}

\usepackage{hyperref}
\newcommand{\dbar}{d\hspace*{-0.08em}\bar{}\hspace*{0.1em}}
\newcommand{\ieq}[1]{\begin{equation} #1 \tag{$\mathcal{I}$.\arabic{iequation}}\end{equation}\stepcounter{iequation}}
\newcommand{\ceq}[1]{\begin{equation} #1 \tag{$\mathcal{C}$.\arabic{cequation}}\end{equation}\stepcounter{cequation}}
\usetikzlibrary{shapes.geometric,arrows,svg.path}
\newcommand{\fract}[2]{\frac{\mathtt{#1}}{\mathtt{#2}}}
\usepackage[export]{adjustbox}
\usepackage[T1]{fontenc}
\usepackage[utf8]{inputenc}
\usepackage{blkarray}
\definecolor{lapis}{rgb}{0.0.0470,0.2941,0.5568}
\newcommand{\msymb}[1]{\normalfont{\textit{\varc{#1\_}}}}
\definecolor{burgundy}{rgb}{0.5, 0.0, 0.13}
\definecolor{functioncolor}{rgb}{0.08,0.28,0.6}
\definecolor{wine}{rgb}{0.44, 0.18, 0.21}
\definecolor{refgreen}{rgb}{0.0,0.4,0.4}
\definecolor{paper_blue}{rgb}{0.3,0.2,0.75}
\definecolor{hblue}{rgb}{0,0,0.575}
\definecolor{varcolor}{rgb}{0.08,0.44,0.2}
\definecolor{functioncolor}{rgb}{0.08,0.28,0.6}
\newcommand{\mathematica}[3]{\vspace{0.35cm}\noindent\begin{minipage}{#1\textwidth}\begin{tabular}{lp{11cm}}{\color{lapis}{\scriptsize{\tt In:}}\raisebox{-0.65pt}{{\scriptsize{\tt=}}}}&{\tt #2}\\{\color{lapis}{\scriptsize {\tt Out:}}\raisebox{-0.65pt}{{\scriptsize{\tt=}}}}&{\tt #3}\end{tabular}\end{minipage}\vspace{0.35cm}}
\newcommand{\bl}[1]{\textcolor{hblue}{#1}}
\newcommand{\var}[1]{\textcolor{functioncolor}{#1}}

\newcommand{\varc}[1]{\textcolor{varcolor}{#1}}
\newcommand{\fun}[1]{\textcolor{functioncolor}{#1}}
\newcommand{\bur}[1]{\textcolor{burgundy}{#1}}

\newcommand{\lab}[1]{\langle{#1\rangle}}

\makeatletter
\newcommand{\bigs}[1]{{\hbox{$\left#1\vbox to24\p@{}\right.\n@space$}}}
\makeatother
\hypersetup{colorlinks=true,urlcolor=lapis,linkcolor=lapis,citecolor=lapis,pdfauthor={Nikhil Kalyanapuram}, pdftitle={},	pdfdisplaydoctitle=true, pdfstartview=FitH}
\newtheorem*{acknowledgements}{Acknowledgements}
\newtheorem*{outline}{Outline}

\allowdisplaybreaks
\graphicspath{{figures/}}
\renewcommand{\thefigure}{\thesection.\arabic{figure}}

\title{Prescriptive Unitarity and Rigidity at Two Loops}
\author{\vspace{-0.5cm}Nikhil Kalyanapuram}
\affiliation{Department of Physics and Institute for Gravitation and the Cosmos, Pennsylvania State University, University Park PA 16802, USA}
\emailAdd{nkalyanapuram@psu.edu}
\abstract{We elaborate upon and consolidate various recent developments focusing on the triality of questions offered by issues of basis building, unitarity and non-polylogarithmicity in quantum field theory, specifically for planar two loops. The interplay between the dual questions of setting up bases of integrands and accurately preparing a complete set of cuts to secure correct \emph{ans\"atze} of loop integrands expanded thereby is enriched by the appearance of non-polylogarithmic structures, first seen in planar two loops in the form of elliptic polylogarithms. We strengthen this by presenting an extended discussion of a new method of building bases, classifying loop integrands by power counting, or their behaviour in the ultraviolet and studying a convenient, albeit manifestly non-canonical set of cuts of full rank. By studying cut equations derived from poorly chosen contours in loop momentum space, the question of finding morally good sets of cuts to accommodate ellipticity at two loops is forced upon us. We discuss a generalization of the notion of a leading singularity in this case---something we call an elliptic leading singularity---a concept which only makes reference to the underlying geometry of the elliptic curve. We also expand upon the task of constructing master integrand bases that neatly distinguish between elliptic and ordinary polylogs. This stratification of the basis---where each master is either pure elliptic or polylog---is carried out by drawing on an expanded basis at two loops, the so-called triangle power counting basis. In the course of developing such a master integrand basis, we emphasize the importance of choosing, intelligently, spanning sets of cuts, and writing down integrand numerators dual to these cuts that are diagonal---or prescriptive---with regard to these choices, to highlight the conceptual and technical simplifications arising therefrom.}

\begin{document}
\maketitle
\setcounter{page}{1}
\clearpage
\phantomsection
\label{intro}
\addcontentsline{toc}{section}{\emph{Introduction}}
\section*{\emph{Introduction}}
{\vspace{-15pt}\color{lapis}\rule{\textwidth}{.6pt}}

\vspace{10pt}

Unitarity-based methods \cite{Bern:1987tw,Bern:1990cu,Bern:1990ux,Bern:1990qr,Bern:1991aq,Bern:1992ee,Bern:1992em,Bern:1993mq,Bern:1993qk,Bern:1994ju,Bern:1994zx,Bern:1994fz,Bern:1994cg,Bern:1995ix,Bern:1995db,Bern:1996ka,Bern:1997nh,Bern:1997sc,Ossola:2006us,Britto:2007tt,Forde:2007mi,Badger:2008cm,Arkani-Hamed:2010pyv,Bourjaily:2011hi,Bourjaily:2013mma,Bourjaily:2015jna,Bourjaily:2016evz,Bourjaily:2019iqr,Bourjaily:2019gqu} have proven extremely valuable in guiding the computation of scattering amplitudes in a variety of quantum field theories for high loop order and multiplicity, domains where the Feynman expansion becomes famously intractable. Computational methods that make use of this become especially interesting and rich when we want to deal with calculations beyond tree level. At tree level, the use of unitarity to rapidly evaluate scattering amplitudes leads one to the study of what are now called recursion relations and on-shell diagrams \cite{Britto:2004ap,Britto:2005fq,ArkaniHamed:2010kv,Arkani-Hamed:2013jha,Trnka:2013eth,Franco:2013nwa,Arkani-Hamed:2014bca,Franco:2015rma,Frassek:2015rka,Jin:2015pua,Benincasa:2015zna,Heslop:2016plj,Herrmann:2016qea,Bourjaily:2016mnp,Bork:2016xfn,Benincasa:2016awv,Boels:2016jmi,Herderschee:2019ofc,Armstrong:2020ljm,Bartsch:2022pyi}, which recast the evaluation of tree amplitudes as sums of canonical diagrams evaluated on the solutions of cut equations. Downstream, the study of these structures has often bled into the study of mathematical forms for their own sake, including those of Grassmannians \cite{ArkaniHamed:2008yf,ArkaniHamed:2010gh,ArkaniHamed:2012nw,Arkani-Hamed:2013jha}, scattering equations and moduli spaces of Riemann surfaces \cite{Cachazo:2013hca,Cachazo:2013iaa,Cachazo:2014fwa,Cachazo:2014nsa,Cachazo:2014xea,Mizera:2017rqa,Mizera:2019gea,Mizera:2019blq}.

The nomenclature of \emph{generalized} unitarity refers to the broad class of such methods applied to the study of scattering amplitudes at one loop order and beyond. Most techniques of generalized unitarity exploit the fact that it is often better to---despite in a broad sense essentially equivalent to---consider as fundamental a loop \emph{integrand}, as opposed to a loop integral. The paradigmatic case of the scalar triangle exemplifies this. For general external kinematics and massless internal legs, the scalar triangle, expressed as a Feynman \emph{integral}, takes the form

\ieq{
   F_{\text{triangle}} = \int\frac{\dbar^4\ell}{(\ell|\bur{a_1})(\ell|\bur{a_2})(\ell|\bur{a_3})(\ell|\bl{X})}
}where the $|\bur{a_i})$ have been used to denote coordinates in dual momentum space and $(\ell|\bur{a_{i}})$ have been used to indicate the inner product in embedding space\footnote{An expanded review and summary of the attendant notation may be found in appendix \ref{app:A}.}. $|\bl{X})$ is the infinity twistor. Taken as an integral, this can be expressed as a sum of polylogs in four dimensions \cite{Britto:2010xq}. 

Such analytic objects are easier to control by recognizing that the Feynman integrand itself is just a rational differential form in loop momentum space, and the integral over all of $\ell$ space is---in this context---a fundamental pairing. As such, it is equally valid to focus on the differential form

\ieq{
    \includegraphics[valign=c,scale=0.3]{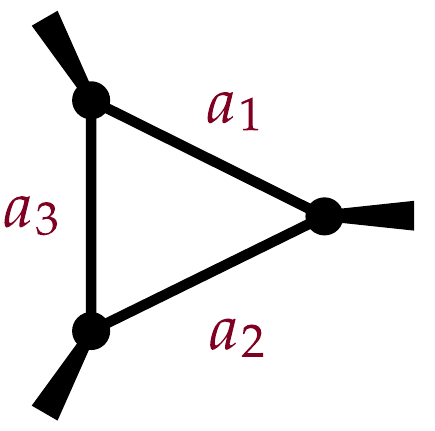} = \frac{\dbar^4\ell}{(\ell|\bur{a_1})(\ell|\bur{a_2})(\ell|\bur{a_3})(\ell|\bl{X})}.
}

The fact that any singularity of the scattering \emph{integrand} $\mathcal{A}$ must be captured by one or more of such fundamental topologies suggests a generic \emph{ansatz} of the form

\ieq{
    \mathcal{A} = \sum_{i}{a}_{i}\mathcal{I}_{i}
}where the $\mathcal{I}_{i}$ are integrands belonging to a master integrand basis $\mathfrak{B}$. Accordingly, the quantities ${a}_{i}$ are loop independent functions of the external kinematics, labelling coefficients in the vector controlled by the size of the basis of master integrands. The question of generalized unitarity reduces to a question of linear algebra: what are the conditions one may impose on this representation of the amplitude integrand that would uniquely determine the ${a}_{i}$? 

In regarding them as rational differential forms in $\ell$ space, Feynman integrands are better thought of as instances of cohomology classes, with equivalent classes related by integration-by-parts\footnote{See \cite{Chetyrkin:1981qh,Kotikov:1990kg,Kotikov:1991pm,Tarasov:1996br,Remiddi:1997ny,Gehrmann:1999as,Laporta:2000dsw,Laporta:2003jz,Lee:2009dh,Grozin:2011mt,Henn:2013pwa,Henn:2014qga,vonManteuffel:2014ixa,Zhang:2016kfo,Kosower:2018obg} for technical details and \cite{Smirnov:2008iw,vonManteuffel:2012np,Lee:2012cn,Maierhofer:2017gsa} for numerical implementations.} identities. Indeed, the basis of the resulting `master integrands' furnish as a result a---perhaps rather complicated---cohomology group, to which there will always be a canonical homological dual corresponding to cycles in $\ell$ space. As such, these can be naturally paired with each other, furnishing loop-independent periods.

Physically, these cycles often correspond to various \emph{cut} conditions, which may set one or more internal propagator on-shell. An example in the case of the triangle is the maximal cut generated by

\ieq{
(\ell|\bur{a_i}) = 0.
}for all $\bur{i}$ . Generically, it is always possible to find---at least in principle (and as we will soon emphasize, in practice as well)---a spanning set of such cuts/cycles, denoted $\mathcal{C}_{i}$ such that the set of equations

\ieq{
    \oint_{\mathcal{C}_{i}}\mathcal{A} = \sum_{j}{a}_{j}\mathbf{M}_{ji}
}where the period matrix---defined according to the following inner product of cycles and forms

\ieq{
    \oint_{\mathcal{C}_{i}}\mathcal{I}_{j} = \mathbf{M}_{ji}
}---is by definition full-rank.

While schematic, these cut equations serve to entirely determine the coefficients ${a}_{i}$, and by implication, the full integrand $\mathcal{A}$. The left hand side of the cut equations are controlled by the quantities

\ieq{
   \mathfrak{a}_i = \oint_{\mathcal{C}_{i}}\mathcal{A}
}which correspond to amplitudes evaluated on the solutions of the cuts defined by the cycles $\mathcal{C}_{i}$. As such, they computationally correspond to products of lower point amplitudes evaluated on the cuts---equivalent to on-shell functions---and serve to encode the behaviour of field theory on the solutions to the cut conditions in loop momentum space.

As a soluble problem, we emphasize that this is always possible in principle. The claim phrased thus is rendered factual due to it being the case that a set of unitarity equations can only be consistently constructed for a \emph{complete} basis of integrands. Choosing a complete basis is often a nontrivial task, but once we have one, \emph{any} suitably generic choice of cuts that serve to pronounce a set of $n$ linearly independent equations will allow us to determine the amplitude correctly.

Ultimately, the linear algebra problem to which generalized unitarity descends owes most of its complexity to what end up being inconvenient choices of one, the cycles $\mathcal{C}_{i}$, and very often, the basis integrands $\mathcal{I}_{i}$ themselves. `Bad' choices of either will often result in very dense period matrices, the diagonalization and inversion of which may turn the problem into a very hard one computationally. Indeed, it is important to underscore the fact that any choice of cuts that furnish a full-rank system will suffice to solve the linear algebra problem as such, despite being computationally hideous. The implication of this is that all full-rank choices are technically equivalent, though some are more morally preferable than others. The dialectic between solving the problem of unitarity equations in principle and finding a \emph{good} solution in practice foreshadows the main topics of our interest in this work.

The issue of specifying a good set of cuts in particular often presents a difficult technical challenge due to several competing reasons; a salient instance of this is when a set of cut conditions denoted by residue operations may not be good enough to completely specify the loop momenta, which results in the question of canonically determining the remaining degrees of freedom. This is usually the case when the maximal cut of a diagram sets fewer propagators to zero than there are loop momenta degrees of freedom, leaving behind undetermined parameters to deal with. Indeed, such situations tend to present in very non-canonical ways, largely due to the freedom in parametrization present when preparing the cut conditions. 

In the present work, we will adumbrate and scrutinize three broad themes emerging when we contend with the general problem of specifying basis integrands and cuts \emph{well}, rather than phrasing them as purely numerical or computational questions. These are, in the order in which we will consider them: performing unitarity and building bases using prescriptive methods, understanding the implications of nonpolylogarithmicity---or \emph{rigidity}---of integrands at higher loops, and the challenge of building prescriptive bases in the presence of rigidity. In doing so, we hope to suggest that a number of technical, conceptual, and aesthetic benefits result by simply rephrasing the question of generalized unitarity in this tone. Let us go over these themes one at a time, speaking to the general issues and challenges we seek to underline.

\paragraph**{\emph{Prescriptive Unitarity and Basis Building. }}The picture of generalized unitarity presented in the preceding discussion highlights the underlying linear-algebraic nature of the issue of determining the coefficients. Indeed, the coefficients are computed by a simple inversion of the period matrix to obtain

\ieq{
    a_{i} = \mathfrak{a}_{j}(\mathbf{M}^{-1})_{ji}.
}The result is if the basis of integrands were not chosen with intent, and an arbitrary set of master integrands were used for the initial \emph{ansatz}, their respective coefficients would always be some linear combinations of on-shell functions, weighted by kinematical constants filling up the period matrix. For dense period matrices, the presence of a determinant factor would result in expressions that obscure simplicities that we often know exist. There is a nice example of this proferred by the case of one-loop scattering amplitudes in $\mathcal{N}=4$ super Yang-Mills. In \cite{Bourjaily:2021hcp}, the authors found a choice of cuts that resulted in the following simple representation of all one-loop amplitudes in this theory

\ieq{
    \mathcal{A}^{\mathcal{N}=4}_{\text{1-loop}} = \sum_{i}\includegraphics[valign = c, scale = 0.3]{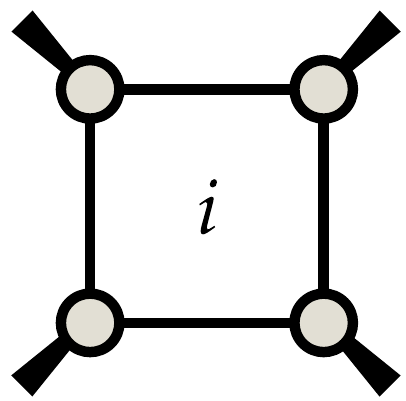}\includegraphics[valign = c, scale = 0.3]{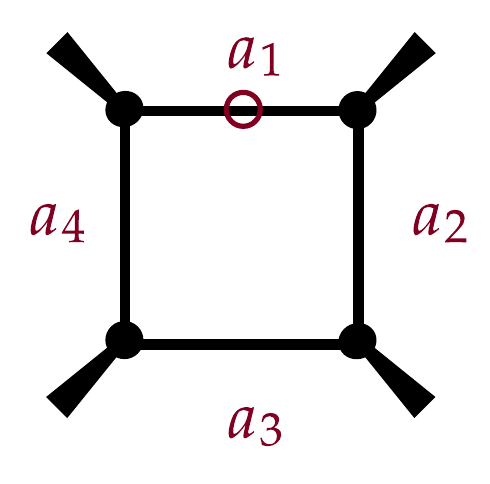}
}where the $i$ spans the two solutions to the quad-cut of the box, on which two box topologies have been normalized. In contrast, using the choice of cuts for the bubble power counting basis we discuss in section \ref{sec:2.2.3}, the same amplitude takes the following schematic representation

\ieq{
\begin{aligned}
     \mathcal{A} =& \sum\underset{{(\var{\alpha},\var{\beta}) = (\var{\alpha^*},\var{\beta^*})}}{\text{eval}}  \includegraphics[valign = c, scale=0.3,trim=0 0 0 0cm]{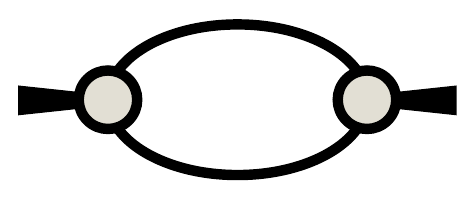}\times\includegraphics[valign = c, scale=0.3,trim=0 0 0 0cm]{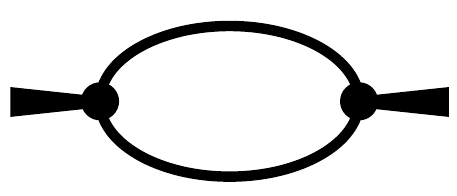}\\& + \sum  \underset{\var{z} = \var{z_{I}}}{\text{eval}}  \includegraphics[valign = c, scale=0.3,trim=0 0 0 0cm]{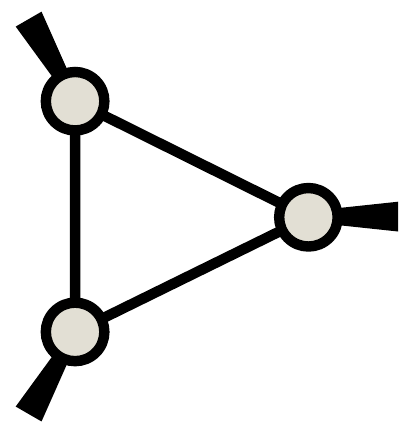}\times\includegraphics[valign = c, scale=0.3,trim=0 0 0 0cm]{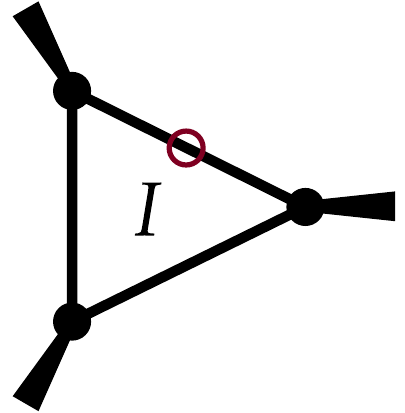}\\
     &+ \sum \includegraphics[valign = c, scale=0.3,trim=0 0 0 0cm]{box-onshell.pdf}\times \includegraphics[valign = c, scale=0.3,trim=0 0 0 0cm]{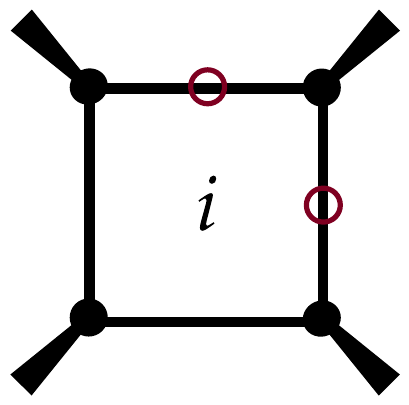}.
\end{aligned}
}Clearly, there is an unnecessary complexity that has been introduced by a poor choice of basis and cuts, obviated in \cite{Bourjaily:2021hcp} by a smarter choice, of basis and cuts.

A very general tone to prepare such questions was developed in  \cite{Bourjaily:2017wjl}, the phrasing of which was the following: any initial \emph{ansatz} that furnished coefficients as linear combinations was not \emph{prescriptive}. In other words, there is no \emph{a priori} canonical map or pairing between the cuts and the integrands. Diagonalizing the column of on-shell functions then becomes equivalent to diagonalizing along homology, where for each integrand, a linear combination of cuts corresponding to the dual in homology is found.

An elegant rewriting of this suggested by the authors was to diagonalize in cohomology, as opposed to homology\footnote{The geometric picture of diagonalizing master integrand bases was studied in a series on recent papers \cite{Mastrolia:2018uzb,Frellesvig:2019kgj,Frellesvig:2019uqt,Mizera:2019vvs,Frellesvig:2020qot,Caron-Huot:2021xqj,Caron-Huot:2021xqj,Chestnov:2022alh,Chestnov:2022xsy,Giroux:2022wav} using twisted intersection matrices, with a formalism that can accommodate dimensional regularization as well.}, instead. Here, once we have specified a set of cuts, we would like to find a basis of integrands $\widetilde{\mathcal{I}}_{i}$ that are canonically dual to the cuts such that the period matrix is the identity

\ieq{
    \oint_{\mathcal{C}_{i}}\widetilde{\mathcal{I}}_{j} = \delta_{ij}
}resulting in the equivalence

\ieq{
    a_{i} = \mathfrak{a}_{i}.
}The implication for integrand bases is that it becomes important to think in terms of \emph{numerator} degrees of freedom spanning the basis. We may illustrate this at one loop using a basis---known to furnish a span for maximally supersymmetric Yang-Mills---containing boxes and triangles as follows:

\ieq{\label{eq:I.11}
    \Bigg\lbrace{{\includegraphics[valign = c, scale = 0.3]{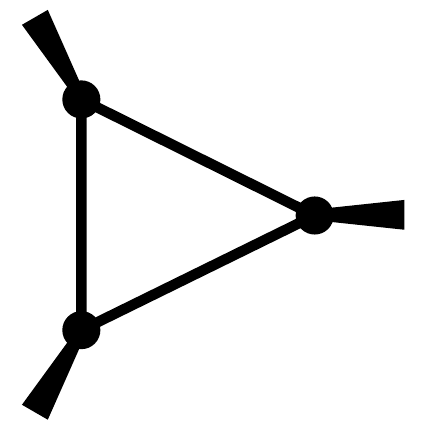}}\;\;,\raisebox{0pt}{\includegraphics[valign = c, scale = 0.3]{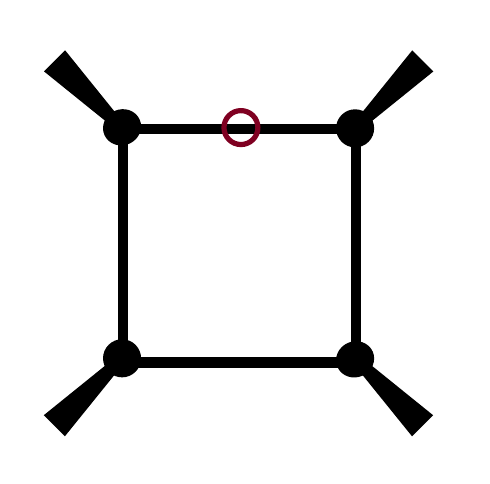}}\Bigg\rbrace}
}where the triangles are spanned by all elements of the form

\ieq{
    \includegraphics[valign=c,scale=0.3]{triangle-labelled.pdf} = \frac{\dbar^4\ell}{(\ell|\bur{a_1})(\ell|\bur{a_2})(\ell|\bur{a_3})(\ell|\bl{X})}
}and boxes by

\ieq{
    \includegraphics[valign=c,scale=0.3]{box_labelled_3gon.pdf} = \frac{(\ell|\bl{N})\dbar^4\ell}{(\ell|\bur{a_1})(\ell|\bur{a_2})(\ell|\bur{a_3})(\ell|\bur{a_4})(\ell|\bl{X})}.
}In four dimensions, the span of all numerators of the form $(\ell|\bl{N})$ is six dimensional (a fact that we will expand upon in section \ref{sec:2}). Four of these however can be spanned---for generic kinematics---by the choices $|\bl{N}) = |\bur{a_i})$ for $i = 1,\dots,4$, corresponding to lower-point triangles. The upshot of this is that any box can be `subtracted out' by any daughter triangle within its spanning space. 

Indeed, once the triangle coefficient is fixed by a choice of cuts, this freedom makes it possible to perform the schematic subtraction

\ieq{
    \raisebox{0pt}{\includegraphics[valign = c, scale = 0.3]{box-3gon.pdf}} - \sum {\includegraphics[valign = c, scale = 0.3]{triangle-3gon}}
}to eliminate all contributions of the spanning contours that fix the triangles on boxes in the basis. This results in a \emph{diagonal} basis of integrands, where each integrand is dressed by a single on-shell function, since any lower-point contour will have zero support on the subtracted, higher point integrands.

A unifying scheme of constructing bases that make prescriptivity manifest at the level of \emph{numerator} subtraction was first delineated in modern form in \cite{Bourjaily:2020qca}, where the notion of building bases was recast as a systematic algorithm. For a given loop order and an integer $p$, this algorithm is an instruction set to develop a basis of integrands that satisfy what the authors call $p$-gon power counting, where in the limit of large loop momenta the integrals obey

\ieq{
    \mathcal{I}(\ell) \sim \frac{1}{(\ell^2)^p}\left(1+\mathcal{O}(\ell^2)\right).
}For example, the basis defined by (\ref{eq:I.11}) is the $3$-gon power counting basis at one loop, where each integrand scales only as badly as triangles. In section \ref{sec:2}, we will review in detail this new method of building bases, exemplifying the basic ideas using one-loop illustrations.

A substantial portion of this part of the present work will deal with the question of simply building a system of full-rank equations that can determine scattering amplitudes at planar two loops, where in the case of $3$-gon power counting one has a basis given by

\ieq{
\begin{aligned}
     \mathfrak{b}_{\text{2-loop}}^{4} = \Bigg\lbrace&\includegraphics[valign = c, scale=0.3,trim=0 0 0 0cm]{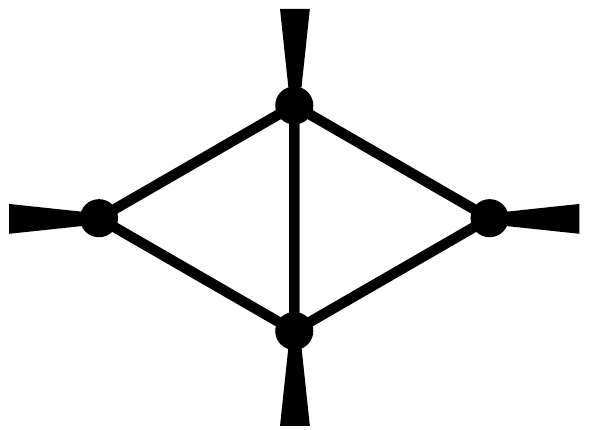}\;,\;\includegraphics[valign = c, scale=0.3,trim=0 0 0 0cm]{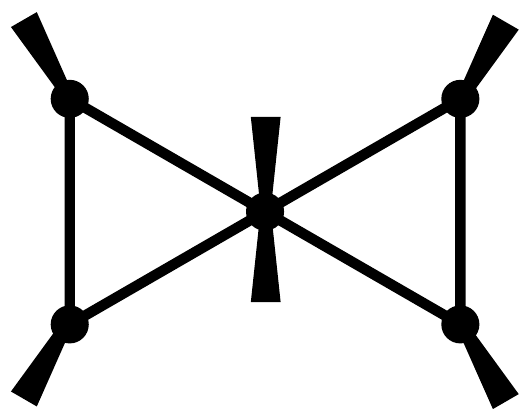}\;,\;\includegraphics[valign = c, scale=0.3,trim=0 0 0 0cm]{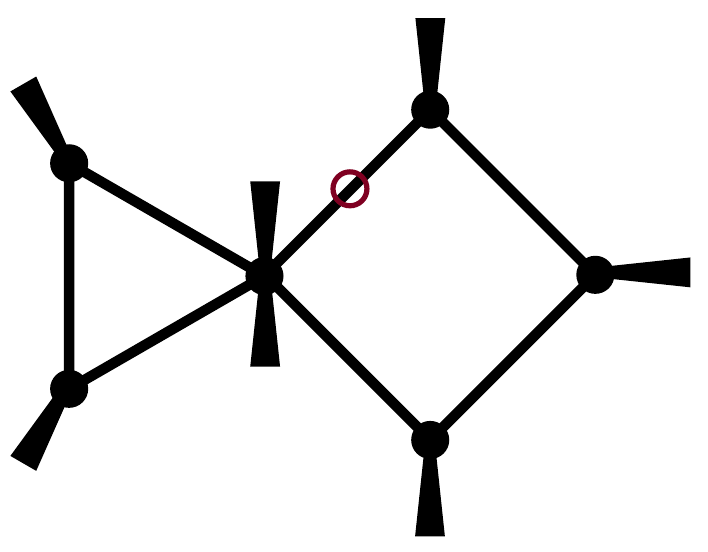}\;,\;\includegraphics[valign = c, scale=0.3,trim=0 0 0 0cm]{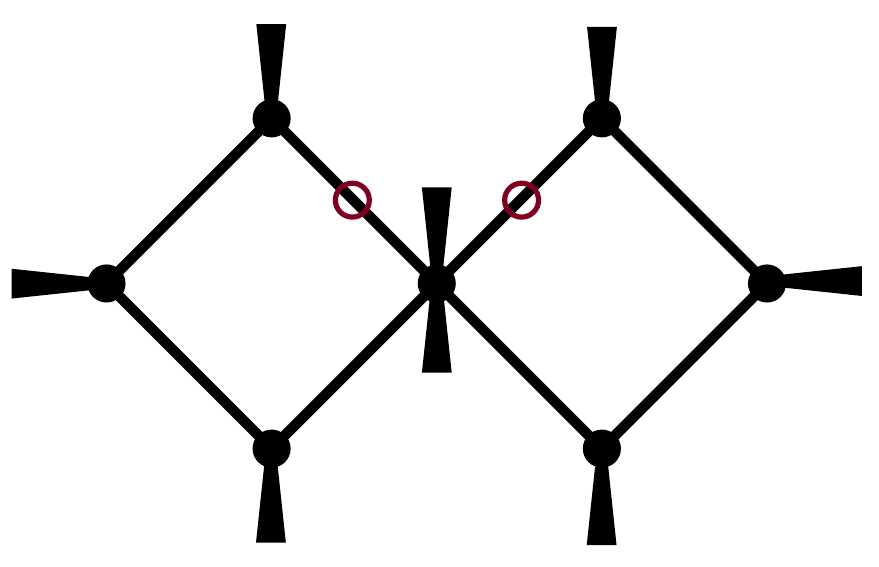}\;,\;\\
    &\includegraphics[valign = c, scale=0.3,trim=0 0 0 0cm]{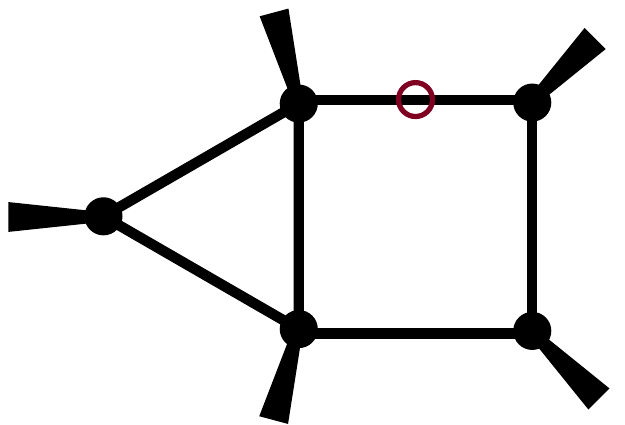}\;,\;\includegraphics[valign = c, scale=0.3,trim=0 0 0 0cm]{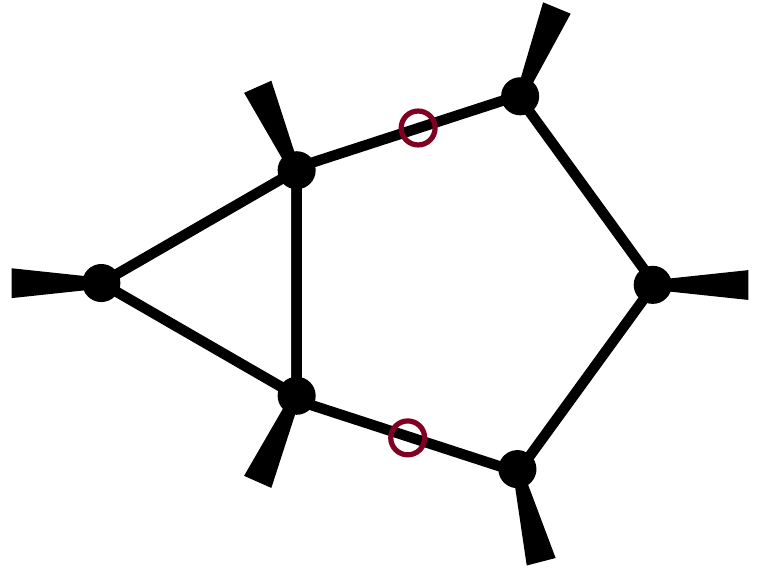}\;,\;\includegraphics[valign = c, scale=0.3,trim=0 0 0 0cm]{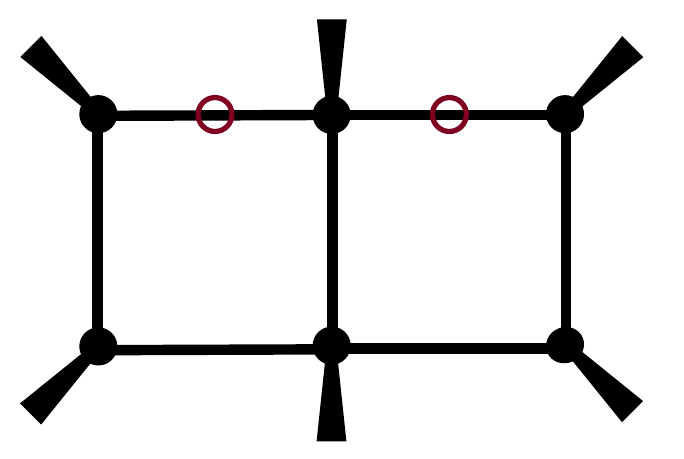}\;,\;\includegraphics[valign = c, scale=0.3,trim=0 0 0 0cm]{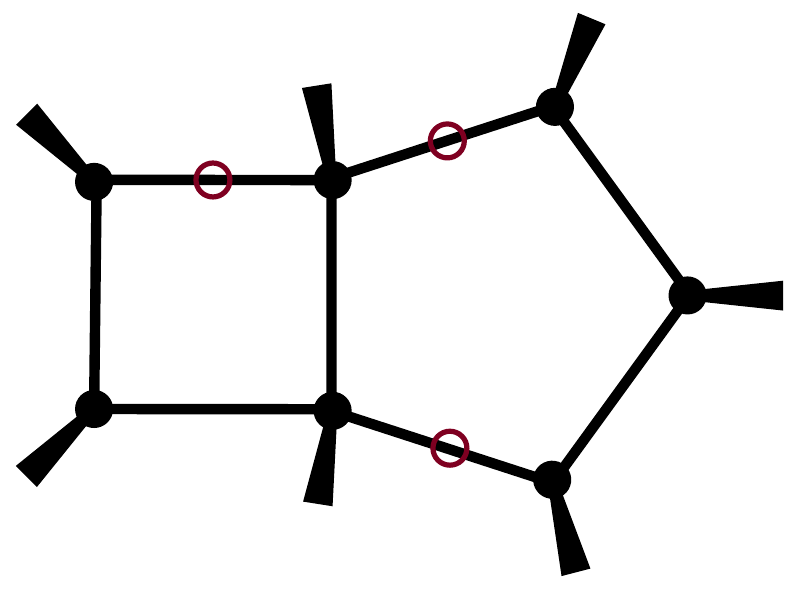} \Bigg\rbrace. 
\end{aligned} 
}Only two of these topologies, namely the kissing boxes and pentaboxes have manifest leading singularities---residue operations that completely localize the loop momenta. Indeed, two of the topologies contain non-polylogarithmic support as well. To illustrate the caveats that naturally accompany these facts, we will carefully construct a very \emph{poor} choice of spanning cuts, which despite their inelegance, will be of full rank, and completely capable of providing a system of equations that completely determine the cut coefficients. The idea will be to emphasize how despite being able to solve the basic problem of unitarity this way, it is often desirable to be more careful is choosing bases and cuts wisely, keeping features of the amplitude we would like to see manifest in mind.

\vspace{10pt}

\paragraph*{\emph{Violating Polylogarithmicity in Maximal Cuts}. }When building bases for quantum field theory at two loops---even in the planar limit---one comes across a number of topologies that do not seem to manifestly support any kind of leading singularity. To see this, we start with the so-called pentabox topology, which in dual coordinates is represented by

\ieq{
    \includegraphics[valign = c, scale=0.3,trim=0 0 0 0cm]{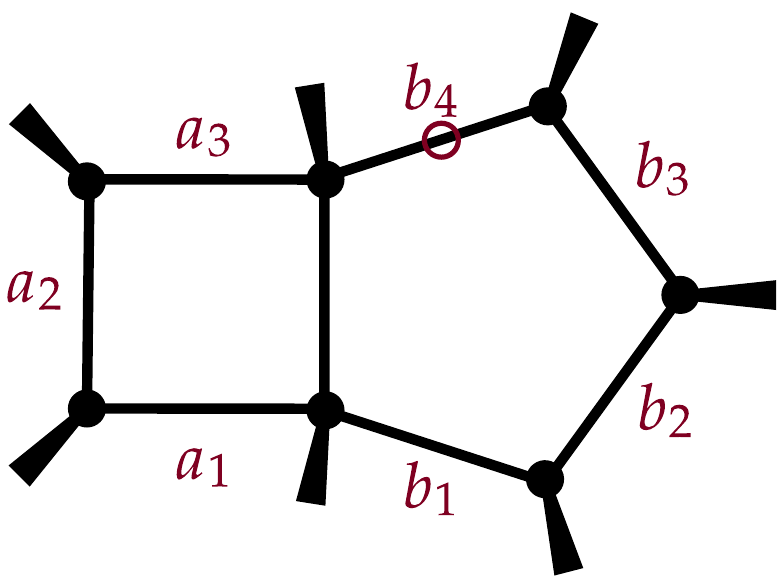} = \frac{(\ell_2|\bl{N})\dbar^4\ell_1\dbar^4\ell_2}{(\ell_1|\bur{a_1})(\ell_2|\bur{a_1})(\ell_3|\bur{a_1})(\ell_1|\ell_2)(\ell_2|\bur{b_1})(\ell_2|\bur{b_2})(\ell_2|\bur{b_3})(\ell_2|\bur{b_4})}. 
}Here, the loop integrand is intrinsically defined---in four dimensions---on a space of dimension $8$, and as such would seem to require at 8 cut conditions to fully localize the loop momenta. Indeed, these are saturated by the following 8 conditions:

\ieq{
    (\ell_1|\bur{a_i}) = (\ell_2|\bur{b_i}) = 0
}and

\ieq{
    (\ell_1|\ell_2) = 0.
}

The quadratic nature of these constraints results in a total of four solutions to the cut equations. Under conditions where such a cut of an integrand has the effect of fully localizing the loop momenta, the convention has been to call such a solution a \emph{leading singularity}. The existence of such singularities generally dramatically simplifies the construction of cut equations.

A notable example, even for $4$-gon power counting at two loops, where we fail to find (what used to be called) a leading singularity is the well-known example of the double box, given by

\ieq{
    \includegraphics[valign = c, scale=0.3,trim=0 0 0 0cm]{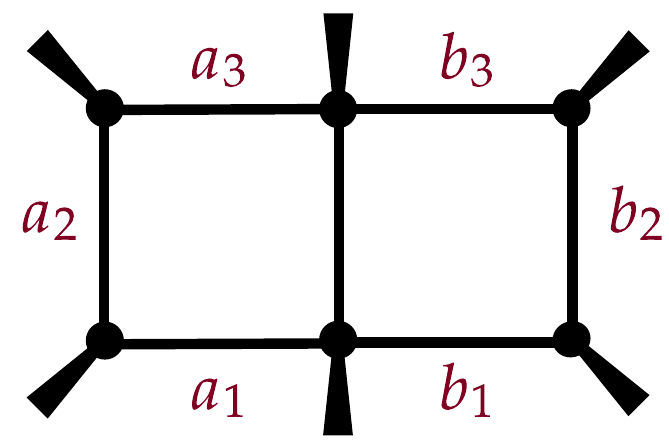} = \frac{\dbar^4\ell_1\dbar^4\ell_2}{(\ell_1|\bur{a_1})(\ell_2|\bur{a_1})(\ell_3|\bur{a_1})(\ell_1|\ell_2)(\ell_2|\bur{b_1})(\ell_2|\bur{b_2})(\ell_2|\bur{b_3})}. 
}The maximal cut---defined as the contour obtained by taking all of the internal propagators on-shell---in this case encircles and enforces the vanishing of a total of seven propagators, spanning a solution space of codimension $1$ in loop momentum space. Indeed, the solution of the double box on the cut conditions results in a differential form that looks like\cite{Bourjaily:2017bsb}

\ieq{
    \underset{\text{seven-cut}}{\oint}\includegraphics[valign = c, scale=0.3,trim=0 0 0 0cm]{doublebox-4-gon-labelled.pdf}  \propto \frac{\dbar\var{\alpha}}{y(\var{\alpha})}
}corresponding to the two solutions of the seven cut. 

Other than the fact that this cut does not succeed in rendering a final contour computing a leading singularity, there is the feature of the denominator, which is algebraic and satisfies

\ieq{
    y(\var{\alpha})^2 = Q(\var{\alpha})
}where $Q(\var{\alpha})$ is a quartic polynomial. Consequently, this curve is that of an elliptic, and violates polylogarithmicity. Indeed, this can also be seen by direct conformal integration of the double box, which results in the following expression

\ieq{
    \int \includegraphics[valign = c, scale=0.3,trim=0 0 0 0cm]{doublebox-4-gon-labelled.pdf} = \int \frac{d\var{\alpha}}{y(\var{\alpha})}H_3(\var{\alpha})
}where $H_3$ is a sum of polylogs of weight 3. The fact that all but one integral can be performed before the breakdown of polylogarithmicity leads one to cite this integral as having a degree of \emph{rigidity}. To be precise, when an integrand is expressed as an iterated integral, its \emph{rigidity} is the minimal integral dimension at which no further polylogarithmic integration can be carried out.

Computationally, elliptic functions are a well-trodden subject in the context of scattering amplitudes---one can see for example \cite{Laporta:2004rb,Adams:2013nia,Bloch:2013tra,Adams:2014vja,Adams:2015gva,Adams:2015ydq,Adams:2016xah,vonManteuffel:2017hms,Bogner:2017vim,Muller:2022gec,Broedel:2019kmn} for various integrated results and \cite{Broadhurst:1993mw,Muller-Stach:2011qkg,Remiddi:2013joa,Adams:2018yfj,Bloch:2014qca,Bloch:2016izu,Adams:2017ejb,Ablinger:2017bjx,Primo:2017ipr,Broedel:2017kkb,Hidding:2017jkk,Lee:2017qql,Lee:2018ojn,Broedel:2018qkq,Weinzierl:2020fyx,Walden:2020odh,Frellesvig:2021hkr} for a number of results on differential equations and mathematical subtleties---but the issue of determining some canonical choice of contour to localize the final degree of freedom left over by the seven-cut largely remained an important unanswered question until recently \cite{Bourjaily:2020hjv,Bourjaily:2021vyj}.

In section \ref{sec:2}, we explain how to bypass this via a \emph{poor} choice of `cut', namely by simply matching to the on-shell function defined by evaluation at a point:

\ieq{
    \underset{\var{\alpha} = \var{\alpha^*}}{\text{eval}} \includegraphics[valign = c, scale=0.3,trim=0 0 0 0cm]{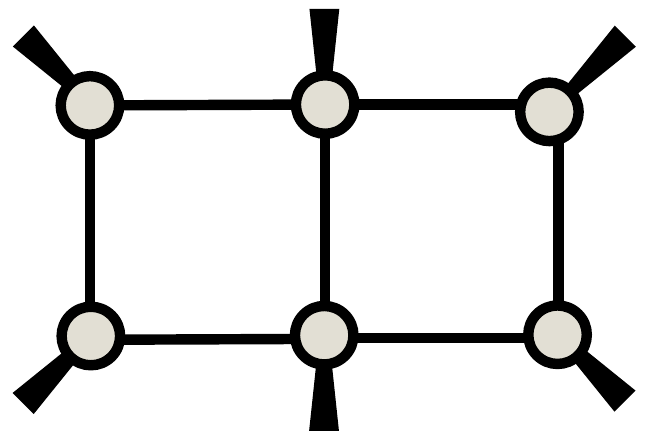}.
}The manifest lack of canonicity in this prescription is then clarified in section \ref{sec:3}, where we expand and elaborate upon the results of \cite{Bourjaily:2020hjv,Bourjaily:2021vyj}, where the question of generalizing a leading singularity in the presence of elliptic structure was studied in depth. 

In short, we demonstrate that it is of value to go back to the idea of a Feynman integrand as being an element (via representative) of a cohomology group, though this time on the support of a cut. We observe that the cohomology spanned by differential forms determining an elliptic curve will be that of the curve, which for generic elliptics is two dimensional. The dual of this cohomology is canonically spanned by the $a$- and $b$-cycles of the elliptic. Given this, it turns out to be valid to `match' the amplitude to the contour presented as the integral over one of these cycles of the on-shell function evaluated on the support of the seven-cut, given by

\ieq{
    \underset{a-\text{cycle}}{\oint}\includegraphics[valign = c, scale=0.3,trim=0 0 0 0cm]{double-box-4-gon-top-level-on-shell.pdf}
}where we made the choice of the $a$-cycle. The remaining elements in the basis are then diagonalized by choosing their numerators such that they vanish on this choice of contour.

The generalization of a leading singularity by making the conceptual shift from

\ieq{
    \text{leading singularity}:= \text{seven-cut}+\text{Residue} \longrightarrow \text{seven-cut}+\text{period integral}
}is what we will refer to as an \emph{elliptic leading singularity}. We will then proceed to expand on two distinct notions of diagonalization on elliptic cuts, which we have labelled as \emph{homological} and \emph{cohomological}. Homological matching is what we have just described, where we focus on diagonalizing by choosing cuts wisely. Contrasting this with cohomological diagonalization, where we match the integrand on the seven cut to the on-shell function on the seven cut as a differential form will be seen to offer certain technical advantages.

\vspace{10pt}

\paragraph*{\emph{Discriminating along Rigidity. }}Building bases and the destruction of polylogarithmicity engage in a rich interplay at two loops, where the issue of constructing bases that manifest rigidity becomes a genuinely nontrivial one. The illustrative example is once again the double box and pentabox. 

The definition of elliptic leading singularities makes no mention of the size of our basis, and as such can be used consistently even at $4$-gon power counting at two loops, where the space of integrands is six dimensional, controlled by two pentaboxes and four double boxes in the most generic kinematic regime. Indeed, the double boxes can be comfortably matched to $a$-cycle period of on-shell functions, while the pentaboxes can be determined by diagonalization on these contours. Unfortunately, the result is that such pentaboxes will be highly impure, and exhibit the general property

\ieq{
    \int \includegraphics[valign = c, scale=0.3,trim=0 0 0 0cm]{pentabox-4-gon-labelled.pdf} = \text{polylogs} + f_{\mathfrak{e}}(\text{elliptic polylogs}).
}This will manifestly \emph{not} be the case for double boxes in this basis, which will satisfy

\ieq{
    \int \includegraphics[valign = c, scale=0.3,trim=0 0 0 0cm]{doublebox-4-gon-labelled.pdf} = (\text{elliptic polylogs})
}and as such, are pure. 

A Feynman integrand is said to have \emph{indefinite rigidity} when it integrates to a combination of quantities with differing rigidity. Clearly, the pentabox in $4$-gon power counting is of indefinite rigidity, and as we will demonstrate, irreducibly so.

A question that presents itself is then the following: is it possible to build a basis at two loops that can accommodate our interest in constructing master integrands that clearly resolve into what we call elements of \emph{definite rigidity}? A basis that manifests this feature will be said to be \emph{stratified in rigidity}. 

The main goal of section \ref{sec:4} will be to present a detailed exposition of \cite{Bourjaily:2022tep}, where it was observed that the generalization to $3$-gon power counting makes it possible to resolve this question. Indeed, the compromise between the size of the basis---which is considerably enlarged upon moving to this choice of power counting---and the need to have a basis of master integrand that manifests purity and rigidity of integrands in a definite sense will be seen to be a notable one, emphasizing how the two phenomena undergo an interplay amongst each other.

It will turn out in the course of demonstrating this that there is little middle ground to be found, and attempts at triangulating between basis enlargement and stratification along rigidity will turn out to be ill-fated. The issue of completeness that plagues such measures will be highlighted and elaborated upon.

\vspace{10pt}

Since the general scope of this work is expository and exploratory in nature, seeking to draw attention to what we find to be an interesting cross-fertilization between these technical problems, some results may appear specialized at times. Indeed, it will be seen to be the case that planar two loops is somewhat special, furnishing a rather wonderful technical laboratory where the problems are complicated enough, but tractability is maintained. The challenges that crop up when any of these assumptions, including and especially planarity and loop order, are relaxed will be discussed in an extended \hyperref[coda]{\emph{Coda}}. The appendices discuss notational and technical aspects of formalism employed in the present work.
\clearpage
\begin{outline}\normalfont
In section \ref{sec:2}, a broad review of generalized and prescriptive approaches to unitarity is provided, alongside an overview of building bases by classifying them along power counting. We exemplify basis building using the case of one loop in section \ref{sec:2.1} and compare and contrast generalized and prescriptive unitarity in section \ref{sec:2.2}. In section \ref{sec:2.3}, we explain how basis building works for planar integrands at two loops, followed by an extended section \ref{sec:2.4}, where we demonstrate the construction of full-rank unitarity cut equations, using a `poorly' chosen set of cuts.

Section \ref{sec:3} deals with the question of \emph{rigidity}---or breakdown of polylogairthmicity---in the context of generalizing leading singularities. In section \ref{sec:3.1}, the origin of elliptic singularities at two loops is discussed at length, which we follow with section \ref{sec:3.2}, where the notion of an elliptic leading singularity is reviewed and expanded upon. Prescriptivity in this context is studied in section \ref{sec:3.3}.

In section \ref{sec:4}, we show how to build bases of master integrands respecting \emph{definite rigidity}. The origin of indefinite rigidity is presented in section \ref{sec:4.1}, which is tailed by an analysis of how stratification is obstructed for $4$-gon power counting at two loops in section \ref{eq:4.2}. In section \ref{sec:4.3}, the extension to $3$-gon power counting is shown to obviate this problem, which we then follow with a discussion of completeness in section \ref{sec:4.4}.

We have attached an extended \hyperref[coda]{\emph{Coda}}, in which a number of future problems are presented from the point of view emphasized in this work, touching again on the three main themes of preparing cuts, rigidity and stratification.

In appendix \ref{app:A}, we review for readers the notation and formalism of dual coordinates used throughout the work, along with momentum twistors and the duality between these. In appendix \ref{app:B}, a general overview of technical details used to study elliptic curves in this work are reviewed. In appendix \ref{app:C}, we provide a walkthrough of a symbolic implementation of one loop integrands on \textsc{Mathematica}.
\end{outline}
\vfill
\begin{acknowledgements}\normalfont
It is a pleasure to thank Mark Alaverdian, Nima Arkani-Hamed, Jacob Bourjaily, Simon Caron-Huot, Hjalte Frellesvig, Holmfridur Hannesdottir, Nigel Higson, Cameron Langer, Andrew McLeod, Sebastian Mizera, Kokkimidis Patatoukos, Michael Plesser, Radu Roiban, Anna Stasto, Jaroslav Trnka, Cristian Vergu, Matt von Hippel, Stefan Weinzierl, Matthias Wilhelm, Chi Zhang and Yaqi Zhang. Partial support for this research is due to ERC Starting Grant (No. 757978) and a grant from the Villum Fonden (No. 15369). This work has been supported in part by the US Department of Energy (No. DE-SC00019066) and a generous fellowship due to Mark and Wendy Willaman.
\end{acknowledgements}
\clearpage
\section{Constructing Bases, Cuts and Unitarity}\label{sec:2}
The interplay of prescriptive unitarity and rigidity becomes apparent for scattering amplitudes at two loops, where we encounter non-polylogarithmic singularities in cuts for the first time. Addressing this question in detail is done best by first understanding how integrand bases and scattering amplitudes are expanded---with a specific emphasis on doing this at two loops---our primary task in this section.

Some of the assumptions we will make during the presentation, unless stated otherwise, are as follows. All Feynman integrands, regardless of loop order, are to be regarded as having massless internal legs. This is not an assumption we will retain however for the external legs, for which we will have to consider generic momentum distributions. In addition, most of the dictionary developed in the present work will apply largely to scattering amplitudes in four dimensions, and considering subtleties due to the effects of dimensional regularization will generally be deferred. We will have some comments to make on this in the context of problems which we believe will be of future interest however. Finally, we will for the most part only consider in the two loop case planar diagrams. To avoid the proliferation of labels, we have decided to, as a matter of convention, refer to the loop momenta on the left loop of such integrands as $\ell_1$ and those on the right side as $\ell_2$.

\subsection{Building Bases and Power Counting}\label{sec:2.1}
We will build bases of Feynman integrands by classifying them using two criteria. The first will be in terms of viewing them as rational functions of loop momenta, correspondingly constrained by the locations of their poles in loop momentum space. The second will be by defining their ultraviolet behaviour, specifically how the integrands scale with the loop momenta when these are taken to infinity. Much of the present section is a review (at least morally) of the point of view first presented in this form in \cite{Bourjaily:2020qca}. The main emphasis in this section though is for the case of one loop; we will generalize much of this to the two loop situation in the later sections.

Building these bases will be in terms of so-called inverse propagators, for which we adopt the compact notation

\begin{equation}
    (\ell|\bl{Q}) = (\ell-\bl{Q})^{2}.
\end{equation}
Here, $\ell$ is some generic loop momentum, while $\bl{Q}$ can refer to any \emph{unintegrated} momentum. Further, the notation itself is dimension agnostic, and makes no reference to specializing to four dimensions. 

We can then define the space of all such inverse propagators for a given loop momenta, or the span of polynomials of the latter form along $\bl{Q}$,

\begin{equation}
    [\ell] = \text{span}_{\bl{Q}\in\mathbb{R}^d}(\ell|\bl{Q}).
\end{equation}
In $d$ dimensions, this span is saturated by $d+2$ choices for $\bl{Q}$:

\begin{equation}
    \text{rank}([\ell]_d) = d+2.
\end{equation}
Specifically, $\bl{Q}$ can be the zero momentum (zero along each of the $d$ directions, in other words), in which case $(\ell|\bl{Q}) = \ell^2$. Further, $\bl{Q}$ can be chosen as one of the $d$ unit vectors along the loop momentum directions. The last degree of freedom is represented by the unit\footnote{We denote the unit by the symbol $1$ in general, but it can be any momentum-dependent monomial, which has units of squared mass. The only constraint is that it is independent of the loop momenta, not that it be unitless.}. Indeed, we have the equivalence

\begin{equation}
    \text{span}_{\bl{Q}\in\mathbb{R}^d}(\ell|\bl{Q}) = \text{span}\lbrace{1,\ell\cdot e_{1},\dots,\ell\cdot e_{d}, \ell^2\rbrace}.
\end{equation}
In four dimensions, which is the subject of our primary concern, this reduces to a six-dimensional space of inverse propagators. Parenthetically, inverse propagators can be chosen to be the building blocks of numerators for our spaces of integrands, so we will often use the two terms interchangeably.

A notational simplification is granted by defining spaces of inverse propagators---or numerators as we will soon emphasize---in this fashion. First, translational invariance is rendered manifest in this notation. Indeed,

\begin{equation}
    [\ell]_d \cong [\ell+\bl{Q}]_d
\end{equation}
for any constant, $d$-dimensional $\bl{Q}$. Second, this means that to define $[\ell]_d$ constructively, we have a relatively vast domain of choice from which we may pick our numerators, so long as they furnish a vector space of dimension six in four dimensions.

The use of this notation allows for compact representations of vector spaces of numerators spanned by higher degree polynomials in the loop momenta. As such, we have the following definition thereof

\begin{equation}
    [\ell]^p_d = \text{span}_{\bl{Q_1},\dots,\bl{Q_p}\in \mathbb{R}^d}(\ell|\bl{Q_1})\dots(\ell|\bl{Q_p})
\end{equation}
with \cite{Bourjaily:2020qca}

\begin{equation}
    \text{rank}([\ell]^p_d) = \binom{d+p}{d}+\binom{d+p-1}{d}.
\end{equation}
A manifest corollary of this notation is a containment of $[\ell]^p_{d}$ in $[\ell]^q_d$ for any $p>q$:

\begin{equation}
    [\ell]^p_{d} \subset [\ell]^q_{d}
\end{equation}
due to the injective embedding of the identity element.

The point of defining numerator spaces of loop momenta in this fashion is to unify the treatment of Feynman integrands that scale in specific ways under the limit of taking the loop momenta to infinity. Indeed, consider a generic one loop Feynman integrand of the form (restricting our attention to the case of one loop)

\begin{equation}
    \mathcal{I} = \frac{\mathcal{N}(\ell)}{(\ell|\bur{a_1})\cdots(\ell|\bur{a_q})}.
\end{equation}
To classify these by their scalings at infinity, we need to only consider the largest power of $\ell^2$ that shows up in the numerator. Indeed, lower powers will be subleading in this limit. Designating this maximum power by $q-p$, we have for the high energy behaviour the following

\begin{equation}
\mathcal{I}(\ell) \sim \frac{1}{(\ell^2)^{q}}\left((\ell^2)^{q-p} + \cdots\right).
\end{equation}
We refer to the number $p$ (which will always be positive in cases we study) as the \emph{power counting} of the Feynman integrand in question. This will be the primary diagnostic we use to construct spaces of Feynman integrands to build or bases.

Lorentz invariance requires here that the numerator of any such integrand be composed only of inverse propagators. As such, any Feynman diagram of the form listed above will belong to the space defined by (being dimensionally agnostic for notational convenience)

\begin{equation}
    \mathfrak{b}^{q}_{p} = \frac{[\ell]^{q-p}}{(\ell|\bur{a_1})\cdots(\ell|\bur{a_{q}})}
\end{equation}
which we will define as the space of Feynman integrands with $p$-gon power counting, upon specification of a particular set of external kinematics. 

Translational invariance of the space of numerators guarantees that we can choose a `pairing' of a specific inverse propagator with one of the propagators in a particular integrand. More concretely, we will choose to represent a propagator decorated with one numerator degree of freedom according to the notation

\begin{equation}
    \raisebox{-2.5pt}{\includegraphics[valign = c, scale = 0.55]{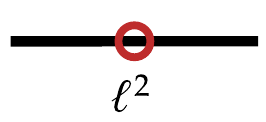}} = \frac{[\ell]}{\ell^2}.
\end{equation}
In this notation, the space $\mathfrak{b}^{q}_{p}$ will be spanned by $q$-gons with $q-p$ of the legs dressed.

We remark as a side note here that this operational definition of power counting is really only consistent when we restrict to planar graphs. Nonplanar graphs bring with them an ambiguity in loop momenta routing, and complicate the definition of power counting. Fortunately, we won't have much occasion to discuss nonplanar diagrams, and have reserved only a short piece on their subtleties towards the end.

A natural class of stratifications of integrands is furnished by performing a their classification by means of power counting. A full discussion of these issues and various exact identities involved were presented in detail in \cite{Bourjaily:2020qca}. Here we'll discuss a summary description.

First, the special case of $q=p$ for $\mathfrak{b}^{p}_{p}$ furnish what we call scalar $p$-gon integrands, and enjoy the best power counting for a fixed $p$. Further, due to the embedding of the identity in $[\ell]$ we have for any $q,q'$ greater than p

\begin{equation}
    \mathfrak{b}^{q'}_{q}\subset \mathfrak{b}^{q}_{p}
\end{equation}
where $q'<q$. This completes a tower of inclusions that describes increasingly loop integrands with increasingly poor ultraviolet behaviour. 

This stratification is controlled by the observation that the element $(\ell|\bur{a_i}) \in [\ell]$ for each $\bur{a_i}$. Accordingly, it is the case that $\mathfrak{b}^{q}_{p}$ contains as a subset the space $\mathfrak{b}^{q}_{p-1}$. More accurately, it will contain $p$ \emph{copies} of $\mathfrak{b}^{q}_{p-1}$, based on which propagator is collapsed. This leads to a division of the space into the contact terms, rendered by $\mathfrak{b}^{q}_{p-1}$, and the \bur{top-level} terms, defined according to

\begin{equation}
    \bur{\widehat{\mathfrak{b}}^{q}_{p}} = \mathfrak{b}^{q}_{p}\backslash\mathfrak{b}^{q}_{p-1}.
\end{equation}
Top-level integrands are spanned by numerator degrees of freedom that are in a precise sense irreducible to daughter topologies. This division into top-level and contact degrees of freedom becomes especially valuable when we contrast generalized and prescriptive methods of unitarity in the next section.

To close this section, let us explicitly work out how basis building works for one loop integrands for $3$- and $2$-gon power counting---often called box and triangle power counting respectively---specified to the case of four dimensions. While this has been done in detail in \cite{Bourjaily:2020qca,Bourjaily:2021ujs}, we include it mainly for the sake of convenience, especially in light of calculations we will do in later sections.

We start with the case of $p=3$, also known as the triangle power counting basis\footnote{The case of $p=d$ is always pathological, for which reason we will avoid it in the case of one loop; we will have little occasion to use it at any rate. Contrasting and comparing this choice against the triangle basis at two loops will form a major part of this work, detailed in section \ref{sec:4}.}. Here, we require that each element in the basis scale only as bad as $(\ell^2)^{-3}$. It is spanned as follows:

\begin{equation}
    \mathfrak{b}^{3} = \mathfrak{b}^{3}_{3}\oplus\mathfrak{b}^{3}_{4} = \text{span}\Bigg\lbrace{\raisebox{0pt}{\includegraphics[valign = c, scale = 0.3]{triangle-3gon}}\;\;,\raisebox{0pt}{\includegraphics[valign = c, scale = 0.3]{box-3gon.pdf}}\Bigg\rbrace}.
\end{equation}
A priori, it would seem odd that we don't need to include pentagons with two inverse propagators, hexagons with three inverse propagators and so on. As it turns out, these topologies are totally reducible to their subtopologies; we'll see how this happens a little later.

Our task now is to count the numerator degrees of freedom here correctly and determine the ranks of the top-level degrees of freedom for each topology. Starting with the triangle topology, it is clear that this is a one dimensional subspace, spanned by some arbitrary scalar that has units of mass square. In other words

\begin{equation}
    \text{rank}(\bur{\widehat{\mathfrak{b}}^3_3}) = \bur{1}
\end{equation}
In the case of the box for $3$-gon power counting, observe that the full numerator space is spanned by all inverse propagators of the form $(\ell|\bl{Q})$, which is of rank $\bl{6}$. However, $4$ of these can be chosen to be one of the internal inverse propagators, a collapse of which would yield a triangle. Indeed, observe that for any three distinct $i,j,k \in \lbrace{1,\dots,4\rbrace}$ we have

\begin{equation}
    \raisebox{2.4pt}{\includegraphics[valign = c, scale = 0.3]{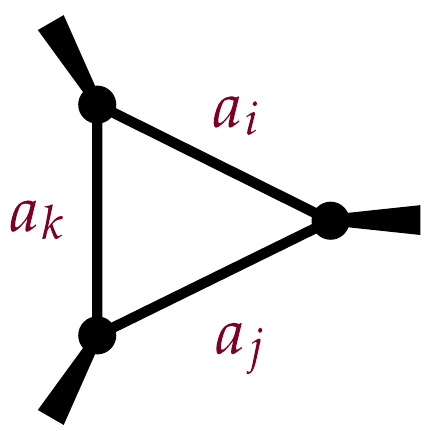}} \in \text{span}\Bigg\lbrace{\includegraphics[valign = c, scale = 0.3]{box_labelled_3gon.pdf}\Bigg\rbrace}
\end{equation}
upon making the choice $\bl{Q} = \bur{a_{f}}$ where $f\neq i,j,k$. Accordingly, the space of numerators of the box in $3$-gon power counting devolves upon four copies of the top-level space for triangles, and top-level terms for the box itself, furnishing the division

\begin{equation}
    \text{rank}({\mathfrak{b}^{3}_{4}}) = \bl{6} = \text{rank}({\bur{\widehat{\mathfrak{b}}^{3}_{4}}}) + 4 = \bur{2} + 4.
\end{equation}
Coming back to the issue of pentagons, we can compute its top-level rank in the same way; it has \bl{20} numerator degrees of freedom, which can be decomposed into $5\times\bur{2} = 10$ top-level boxes, and $10\times \bur{1} = 10$ triangles, which saturate its numerator space, rendering it bereft of any top-level degrees of freedom in $3$-gon power counting in four dimensions. This turns out to be a general feature beyond pentagons as well, leaving the basis spanned entirely by boxes and triangles.

The $2$-gon power counting basis at one loop expands the $3$-gon basis by including scalar bubbles. Here, the basis is spanned by the three topologies as

\begin{equation}
    \mathfrak{b}^{2} = \mathfrak{b}^{2}_{2}\oplus\mathfrak{b}^{2}_{3}\oplus\mathfrak{b}^{2}_{4} = \text{span}\Bigg\lbrace{\includegraphics[valign = c, scale = 0.3]{bubble-2gon.pdf}\;\;,\includegraphics[valign = c, scale = 0.3]{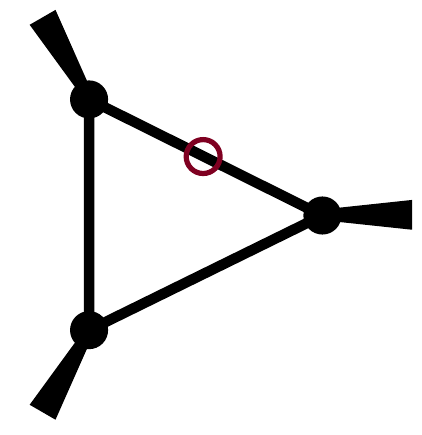}\;\;,\includegraphics[valign = c, scale = 0.3]{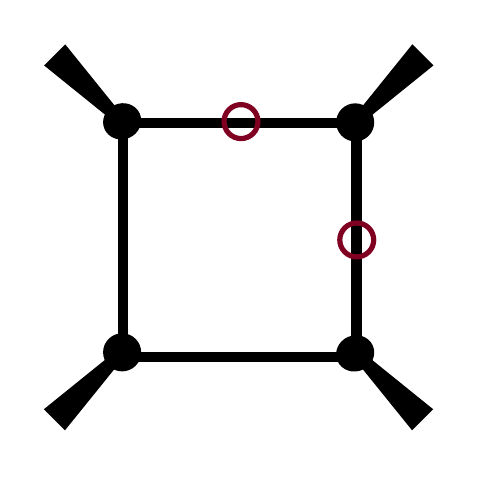}\Bigg\rbrace}.
\end{equation}
The scalar bubbles' numerator space is spanned by a scalar, and as such has dimension (both total as well as top-level) \bur{1}:

\begin{equation}
    \text{rank}(\mathfrak{b}^{2}_{2}) = \bur{1} + 0.
\end{equation}
The triangles have a numerator space spanned by a single inverse propagator, which is $\bl{6}$ dimension. Reducible into $3$ bubbles of top-level dimension \bur{1} each, we have,

\begin{equation}
    \text{rank}(\mathfrak{b}^{2}_{3}) = \underbrace{\bur{3}}_{\text{rank}(\bur{\widehat{\mathfrak{b}}^{2}_{3}})} + \;\;3.
\end{equation}
The boxes in $2$-gon power counting with \bl{20} numerator degrees of freedom are reducible into $4$ triangle subspaces (of dimension \bur{3} each), and $6$ bubble subspaces. Indeed,

\begin{equation}
    \text{rank}(\mathfrak{b}^2_{4}) = \underbrace{\bur{2}}_{\text{rank}(\bur{\widehat{\mathfrak{b}}^{2}_{4}})} + \;\;18.
\end{equation}

The $2$-gon basis can be refined further into $1$-gon, $0$-gon and worse spaces by increasing the sizes of the corresponding numerator spaces. We won't have occasion to discuss these further here however.

The resolution of numerator degrees of freedom into top-level and contact degrees of freedom due to subtopology reductions makes natural a reformulation of traditional generalized unitarity, known as \emph{prescriptive} unitarity, by shifting the burden of diagonalization from cuts to differential forms in loop momentum space. 

\subsection{Generalized versus \emph{Prescriptive} Unitarity}\label{sec:2.2}
The formalism presented in the preceding section lends itself in very elegant fashion to redeveloping generalized unitarity---responsible for much of the state of the art in high precision loop calculations---by moving the emphasis from preparing good sets of unitarity cuts to building effective and `wise' bases of integrands instead. This reformulation is now generally called the \emph{prescriptive} approach to generalized unitarity \cite{Bourjaily:2017wjl}.

The change in perspective offered by prescriptive unitarity can be hard to clarify abstractly in the most general case of relevance to Feynman integrands, so it is of interest to first develop the moral framework involved by pointing to a rudimentary example that would likely be more familiar from elementary complex analysis. After a brief course through this illustration, we'll go over the general picture of prescriptively building and matching bases, to be followed by exemplifying it in the case of one loop. 

\subsubsection{A Toy Model in $\mathbb{P}^1$}
There is an extremely rudimentary yet instructive toy model in one dimension that essentially captures the picture change navigated between generalized and prescriptive approaches to unitarity. For reasons that will be recognized immediately, we'll employ somewhat suggestive notation in dealing with this example.

Suppose we have a complex differential form in $\mathbb{P}^1$, denoted by the `ket' $\ket{f}$, containing only simple poles along $\var{z_{1}}$ and $\var{z_{2}}$\footnote{In this highly simplified version, the residues along the poles $\var{z_{1,2}}$ would have to be equal to each other and opposite in sign due to the global residue theorem if we refuse to consider functions with nontrivial residue at infinity. If however we permit residues at infinity, the analysis in this section will go through, and constraining the residues along the points not at infinity will fix the one at infinity by the GRT.}. This will be expanded according to the \emph{ansatz}

\begin{equation}
    \ket{f} = a_{1}\ket{1} + a_{2}\ket{2}
\end{equation}
where 

\begin{equation}
    \ket{i} = \frac{\dbar\var{z}}{\var{z}-\var{z_i}}.
\end{equation}
Phrased like this, the coefficients $a_{i}$ are determined by residue computations around the two poles. If however we were instead given this problem by stating that the form was expanded using the basis elements

\begin{equation}
\ket{\pm} = \left(\frac{\dbar \var{z}}{\var{z}-\var{z_1}}\pm \frac{\dbar \var{z}}{\var{z}-\var{z_2}}\right)
\end{equation}
we'd have to consider instead a rotated basis of cycles over which to evaluate residues. More precisely, the relevant cohomology group here is spanned by $\ket{\pm}$, with homological duals $\bra{\pm}$ induced according to

\begin{equation}
    \bra{\pm} = \frac{1}{2}\left(\bra{1}\pm\bra{2}\right).
\end{equation}
Now the coefficients $a_{\pm}$ are computed by evaluating the inner products $\braket{\pm|f}$, which involve a total of four residue operations.

By diagonalizing homology, we are performing what amounts to a stripped-down version of generalized unitarity, where we find linear combinations of cuts that isolate the coefficient of each integrand in the basis with which a given scattering amplitude (integrand) is expanded. Geometrically then, prescriptive unitarity is just a change in vocabulary, where we diagonalize cohomology instead of homology. Indeed, if we insisted on retaining our original choice of cycles, we would have found for the inner products the following system of equations

\begin{equation}
    \braket{\pm|f} = \sum_{i = 1,2}\mathbf{M}_{\pm i}a_{i}
\end{equation}
where the period matrix $\mathbf{M}_{ij}$ for $i = \pm$ and $j = 1,2$ is

\begin{equation}
    \mathbf{M}_{ij} = \begin{pmatrix}
\frac{1}{2} & \frac{1}{2} \\
\frac{1}{2} & -\frac{1}{2}
\end{pmatrix}.
\end{equation}
Prescriptive unitarity in this case is then rotating the basis itself by writing down basis elements

\begin{equation}
    \ket{\pm} = \ket{i}(\mathbf{M}^{-1})_{i\pm},
\end{equation}
in effect diagonalizing along cohomology instead. The generalization to $n$ poles instead of $2$ is automatic.

\subsubsection{The General Case}\label{sec:2.2.2}
The general characterization of prescriptive unitarity as a redevelopment of generalized unitarity follows by analogizing Feynman integrands to differential forms, and analogizing cuts to contour prescriptions. Indeed, these aren't really analogies \emph{per se}, since there is a well-defined means of constructing these maps.

That said, the basic picture is as follows. Suppose we have a basis of integrands given by $\mathfrak{B}$, containing Feynman integrands labelled by $\mathcal{I}_{i}$. Expanding a scattering amplitude spanned by such a basis follows by preparing an \emph{ansatz}:

\begin{equation}
    \mathcal{A} = \sum_{\mathcal{I}_{i}\in\mathfrak{B}}a_{i}\mathcal{I}_{i}
\end{equation}
where the task at hand is conveniently specifying the $f_{i}$.

The way this is usually done is simply by finding a spanning set of contours that renders the corresponding set of cut equations full rank. This is always possible; we just need to repeat the calculation done in the last section for the case of Feynman integrands. Each Feynman integrand is naturally regarded as a representative of a cohomology class. Indeed, consider the case of the scalar triangle in four dimensions,

\begin{equation}
    \raisebox{2.5pt}{\includegraphics[valign = c, scale = 0.3]{triangle-labelled-four.pdf}} = \frac{\dbar^4\ell}{(\ell|\bur{a_i})(\ell|\bur{a_j})(\ell|\bur{a_k})(\ell|\bl{X})}
\end{equation}
where in embedding space, the infinity twistor $\bl{X}$ naturally represents the residue at infinity. As a differential form in loop momentum space, this is properly viewed as a representative of a cohomology element, with the derivative operator supplying integration-by-parts (IBP) identities\footnote{I'm grateful to Nigel Higson for a brief discussion on this point.}. In other words, any shift of the triangle integrand of the form renders it invariant under the integral sign

\begin{equation}
\raisebox{2.5pt}{\includegraphics[valign = c, scale = 0.3]{triangle-labelled-four.pdf}}
     \sim \raisebox{2.5pt}{\includegraphics[valign = c, scale = 0.3]{triangle-labelled-four.pdf}} + \left(\partial_{\ell^{\mu}}f^{\mu}(\ell)\right)\dbar^4\ell.
\end{equation}
Given that the basis of Feynman integrands spans \emph{some} cohomology group, we're now required to find a spanning set of cuts which can then by diagonalized in homology.

In practice, this is done by simply selecting a choice of contours $\mathcal{C}_{i}$ that render the period matrix

\begin{equation}
    \oint_{\mathcal{C}_{i}}\mathcal{I}_{j} = \mathbf{M}_{ij}
\end{equation}
full rank. The contours can be specified variously; in the case of the scalar triangle for example we can choose a contour that encircles the kinematic propagators, which supplies a one-parameter form:

\begin{equation}
    \underset{(\ell|\bur{a_{i,j,k}})=0}{\oint}\raisebox{2.5pt}{\includegraphics[valign = c, scale = 0.3]{triangle-labelled-four.pdf}} = \frac{\dbar\var{x}}{J(\var{x})(\ell(\var{x})|\bl{X})},
\end{equation}
where $\ell(\var{x})$ is the loop momentum evaluated on the three cut and $J(\var{x})$ is the Jacobian due to the cut. This can be followed by a number of choices to specify the final contour. One choice would be to simply evaluate at some point $\var{x}=\var{x^*}$. Another would be to compute one of the residues of this form on the poles of $(\ell(\var{x})|\bl{X})$, which furnish two distinct solutions.

Ultimately, so long as we have a sufficiently large number of cuts a subset of which supply a period matrix of full rank, the corresponding cut equations

\begin{equation}
    \oint_{\mathcal{C}_{j}}\mathcal{A} = \sum_{i}\mathbf{M}_{ji}a_{i}
\end{equation}
can be solved by simply inverting the period matrix. The cuts of the amplitudes themselves are computed by stitching together on-shell functions on the solutions of the cut conditions provided by the contour prescriptions.

The inversion of the period matrix amounts to diagonalizing the contours in homology, resulting in the final result being the homological duals of the cohomology spanned by the integrands in the basis. In analogy with the toy model then, \emph{prescriptive} unitarity is then a rotation of the integrands instead as follows,

\begin{equation}
    \widetilde{\mathcal{I}}_{i} = \mathcal{I}_{j}\left(\mathbf{M}^{-1}\right)_{ji}
\end{equation}
which results trivially in 

\begin{equation}
    \oint_{\mathcal{C}_{i}}\widetilde{\mathcal{I}}_{j} = \delta_{ij}.
\end{equation}
In this case, each $a_{i}$ can be computed as a single on-shell function, since the integrands have now been chosen to have been diagonalized in cohomology, given a homology spanned by the choice of contours instead.

A noticeable feature of the example provided by the scalar triangle is the absence of any purely kinematic leading singularities. The maximal cut invariably furnishes a one-form as it cannot constrain all the loop momenta at once, leaving behind another degree of freedom to match. This phenomenon is a generic one, and often irreducibly so, as we will see in the case of elliptic integrands at two loops. To motivate these problems, setting up a basis and the cut conditions at one loop is relatively instructive; let us now move to a systematic presentation of this system.

\subsubsection{Illustration: Prescriptivity at One Loop}\label{sec:2.2.3}
In this section, we'll focus on setting up notation and conventions for defining various inner products between forms and cycles, doing everything in the simplest arena, namely that of one-loop amplitudes. Highlighting challenges and subtleties that show up in doing this will be easier at one loop, and give us enough practice to deal with the more notationally and computationally demanding versions of the same issues at higher loop order.

We start with the case of $3$-gon power counting, which is spanned by scalar triangles and boxes with one inverse propagator:

\begin{equation}
    \mathfrak{b}^3 = \text{span}\Bigg\lbrace{\raisebox{0pt}{\includegraphics[valign = c, scale = 0.3]{triangle-3gon}}\;\;, \raisebox{0pt}{\includegraphics[valign = c, scale = 0.3]{box-3gon}}\Bigg\rbrace}.
\end{equation}
Now we need to find a set of $3$ spanning cuts: \bur{1} to fix the coefficient of each triangle (given a specific leg distribution) and \bur{2} to fix the top-level degrees of freedom of each box. For the case of the triangle, recall that we have the following generic form

\begin{equation}
    \includegraphics[valign = c, scale = 0.3]{triangle-labelled.pdf} = \frac{\dbar^4\ell}{(\ell|\bur{a_1})(\ell|\bur{a_2})(\ell|\bur{a_3})(\ell|\bl{X})} := \ket{\bur{a_1},\bur{a_2},\bur{a_3}}.
\end{equation}
To define the cut, we first take the maximal cut, characterized by the cut conditions

\begin{equation}
    (\ell|\bur{a_i}) = 0
\end{equation}
for $i=1,2,3$. As mentioned in the last section, this will leave us with one degree of freedom left over, with poles controlled by the residue at infinity, and collinear and soft poles contained in the Jacobian factor. An especially convenient choice for the final contour to supply a leading singularity turns out to be taking the residue at infinity, which we can summarize by the symbol

\begin{equation}
     \lbrace{\ell(\var{z})\rightarrow\infty,(\ell|\bur{a_i}) = 0\rbrace}:=\bra{\Omega^{\infty}_{3,\text{triangle}}(\bur{a_{i}})} .
\end{equation}
For the on-shell function to which this is matched, we obtain the following for the triple cut,

\begin{equation}
     \includegraphics[valign = c, scale=0.3,trim=0 0 0 0.2cm]{triangle-onshell.pdf}:=f_{\text{triangle}}(\var{z})
\end{equation}
(where we have simply indicated that it is a function of one parameter $z$). Accordingly, the triangle integrand is supplied with the following coefficient

\begin{equation}
    \frac{1}{\braket{\Omega^{\infty}_{3,\text{triangle}}(\bur{a_{i}})|\bur{a_1},\bur{a_2},\bur{a_3}}}\text{Res}_{(\ell(\var{z})|\bl{X}) = 0}\includegraphics[valign = c, scale=0.3,trim=0 0 0 0cm]{triangle-onshell.pdf}.
\end{equation}

Matching the box in $3$-gon power counting is easy due to the fact that leading singularities of the box saturate the top-level degrees of freedom. Indeed, notice that for a given leg distribution and generic numerator we have for the box

\begin{equation}
    \includegraphics[valign = c, scale=0.3,trim=0 0 0 0cm]{box_labelled_3gon.pdf} = \frac{(\ell|\bl{N_i})\dbar^4\ell}{(\ell|\bur{a_1})(\ell|\bur{a_2})(\ell|\bur{a_3})(\ell|\bur{a_4})(\ell|\bl{X})}:=\ket{\lbrace{\bur{a_i}\rbrace};\bl{N_i}}
\end{equation}
where we have suppressed the leg labels for brevity, we may completely constrain the loop momenta via the two solutions of the quad-cut given by,

\begin{equation}
    \lbrace{(\ell|\bur{a}_{i}) = 0\rbrace}:=\bra{\Omega^{1,2}_{\text{box}}(\bur{a_i})}.
\end{equation}
The first \emph{ansatz} then for the box coefficients would have to be the two functions

\begin{equation}
    f^{1}_{\text{box}} := \frac{1}{\braket{\Omega^{1,2}_{\text{box}}(\bur{a_i})|\lbrace{\bur{a_i}\rbrace};\bl{N_i}}}\includegraphics[valign = c, scale=0.3,trim=0 0 0 0cm]{box-onshell.pdf}.
\end{equation}
This basis isn't diagonal, since each box will have support on its triangle subtopologies. This can be rectified by subtracting out contributions to the cut from each triangle separately, effected by the following shift of the numerator itself

\begin{equation}
    (\ell|\bl{N_i}) \longrightarrow (\ell|\bl{N_i})-\sum_{\lbrace{a_{i_1},\dots,a_{i_3}\rbrace}}\frac{\braket{\Omega^{\infty}_{3,\text{triangle}}(\bur{a_{i_k}})|\lbrace{\bur{a_i}\rbrace},\bl{N_i}}}{\braket{\Omega^{\infty}_{3,\text{triangle}}(\bur{a_{i_k}})|\bur{a_{i_1}},\bur{a_{i_2}},\bur{a_{i_3}}}}(\ell|\bur{a}):=(\ell|\widetilde{\bl{N_i}})
\end{equation}
where the sum is over all subsets of size 3 of $\lbrace{\bur{a_1},\dots,\bur{a_4}\rbrace}$ and $\bur{a}$ is the remaining element given each subset.

Practically then, the choices of $\bl{N_i}$ are entirely arbitrary, so long as they are correctly normalized on the two box cuts, and the contributions from the daughters are correctly subtracted out.

Parenthetically, we can make contact with the notation in previous work by simply pointing out that the convention has generally been to normalize the numerators on the cut and perform subtraction via the numerators themselves. Accordingly, the result for the amplitude is schematically written in the form

\begin{equation}\label{eq:2.48}
    \mathcal{A} = \sum\text{Res}_{(\ell(\var{z})|\bl{X}) = 0}\includegraphics[valign = c, scale=0.3,trim=0 0 0 0cm]{triangle-onshell.pdf}\times\includegraphics[valign = c, scale = 0.3]{triangle-3gon} + \sum\includegraphics[valign = c, scale=0.3,trim=0 0 0 0cm]{box-onshell.pdf}\times\includegraphics[valign = c, scale=0.3,trim=0 0 0 0cm]{box-3gon.pdf}
\end{equation}
where the sum is over all leg distributions and top-level terms and the implication is that the normalization against the cut and subtraction of contact contributions are carried by the numerator degrees of freedom. In this work, we will generally use the older notation and the newer one in terms of kets and bras interchangeably, if only to emphasize their equivalence.

Another approach, though considerably less canonical, is to make reference to the fact that for a choice of power counting, we always work with a basis that is at least complete, if not overcomplete. Due to this fact, what is ultimately required is to simply set up a system of cut equations---regardless of exactly how that is done---which is of full rank. Even if the cuts matched are entirely arbitrary, the completeness of the basis will always guarantee that the amplitude is matched on all cuts once fully consolidated. 

Illustrating this in the $3$-gon power counting basis at one loop again starts with the one-form on the triple cut of the triangle. This time, a completely valid choice would be to simply match the cut on some arbitrary $z = z^{*}$. In other words, we match the amplitude cut to the on-shell function evaluated at a point:

\begin{equation}
    \underset{\var{z} = \var{z^*}}{\text{eval}}\includegraphics[valign = c, scale=0.3,trim=0 0 0 0cm]{triangle-onshell.pdf}
\end{equation}
and normalize the scalar triangle on the cut

\begin{equation}
 \lbrace{\var{z}=\var{z^{*}},(\ell|\bur{a_i}) = 0\rbrace}:=   \bra{\Omega^{\var{z^*}}_{3,\text{triangle}}(\bur{a_i})}
\end{equation}
instead. Now of course, the box has to be subtracted the contributions of the evaluated triple cuts in a manner entirely analogous to the previous example (we omit the long expression). Finally, we have for the amplitude the expression

\begin{equation}
    \mathcal{A} = \sum\underset{\var{z} = \var{z^*}}{\text{eval}}\includegraphics[valign = c, scale=0.3,trim=0 0 0 0cm]{triangle-onshell.pdf}\times\includegraphics[valign = c, scale = 0.3]{triangle-3gon} + \sum\includegraphics[valign = c, scale=0.3,trim=0 0 0 0cm]{box-onshell.pdf}\times\includegraphics[valign = c, scale=0.3,trim=0 0 0 0cm]{box-3gon.pdf}
\end{equation}
and is exactly equivalent to the expansion in equation (\ref{eq:2.48}). 

\paragraph**{Example 2.1. $3$-gon Basis for Four Particles:}
Since the basis is rather small, working out the prescriptive basis for a particular number of particles can be quite instructive. Let us look at the case of four particles, which is nontrivial for $3$-gon power counting. The basis here is given by one box topology, spanned by \bur{2} top level degrees of freedom, and four triangles, spanned by \bur{1} top-level term each:

\begin{equation}
    \Bigg\lbrace{\includegraphics[valign = c, scale = 0.3]{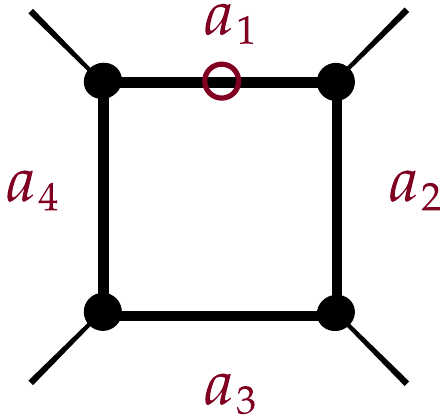}\;\;,\includegraphics[valign = c, scale = 0.3]{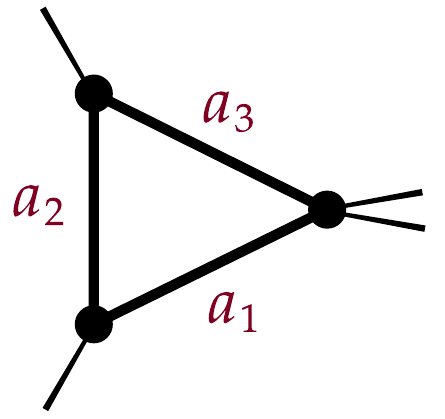}\;\;,\includegraphics[valign = c, scale = 0.3]{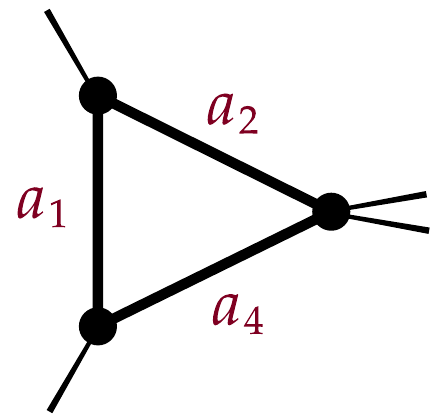}\;\;,\includegraphics[valign = c, scale = 0.3]{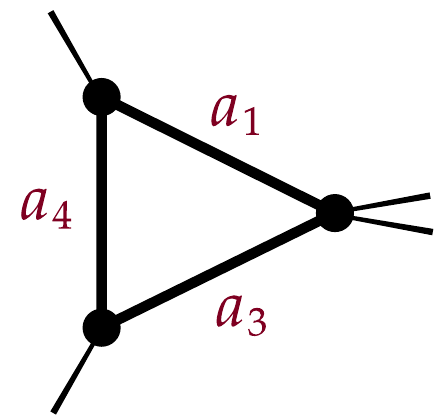}\;\;,\includegraphics[valign = c, scale = 0.3]{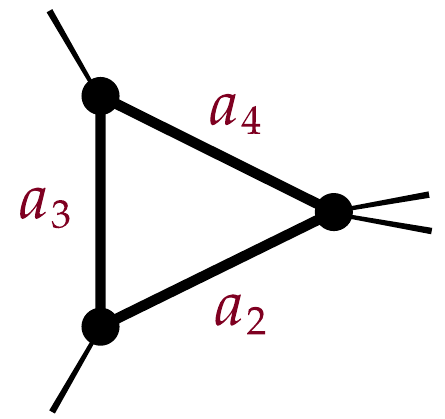}\Bigg\rbrace}
\end{equation}
The contours that we will choose to match are the two quad cuts for the box

\begin{equation}
    \lbrace{\Omega_1,\Omega_2\rbrace}:=\Bigg\lbrace{ \includegraphics[valign = c, scale = 0.3]{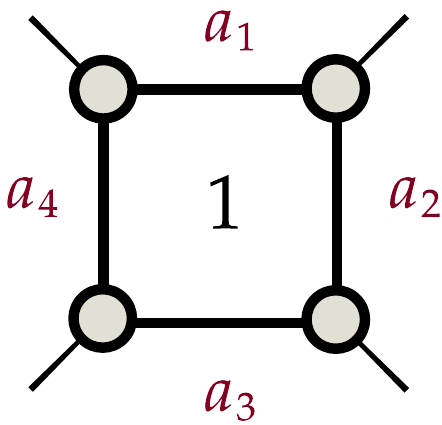}\;\;,\includegraphics[valign = c, scale = 0.3]{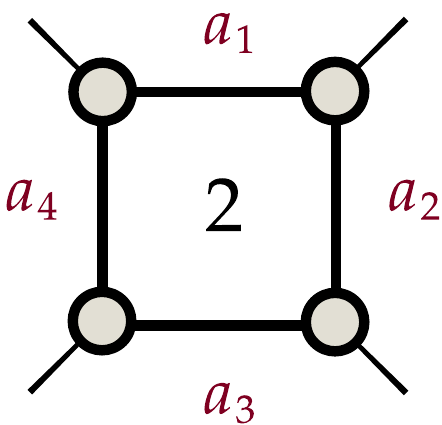}\Bigg\rbrace}
\end{equation}
and the evaluated three particle cuts for the triangles

\begin{equation}
    \lbrace{\Omega_3,\dots,\Omega_6\rbrace}:=\Bigg\lbrace{ \underset{\var{z} = \var{z^*}}{\text{eval}}\includegraphics[valign = c, scale = 0.3]{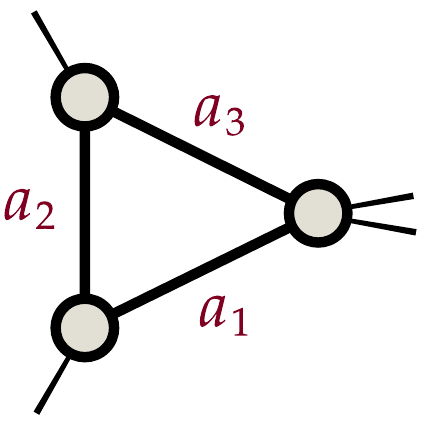}\;,\;\underset{\var{z} = \var{z^*}}{\text{eval}}\includegraphics[valign = c, scale = 0.3]{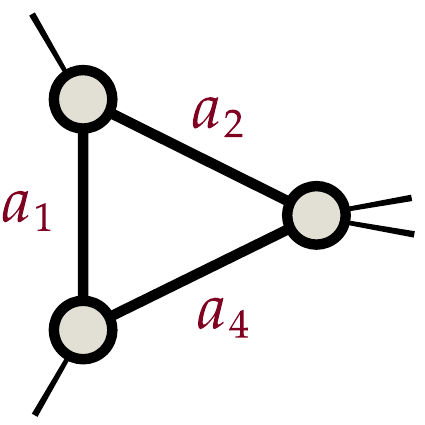}\;,\;\underset{\var{z} = \var{z^*}}{\text{eval}}\includegraphics[valign = c, scale = 0.3]{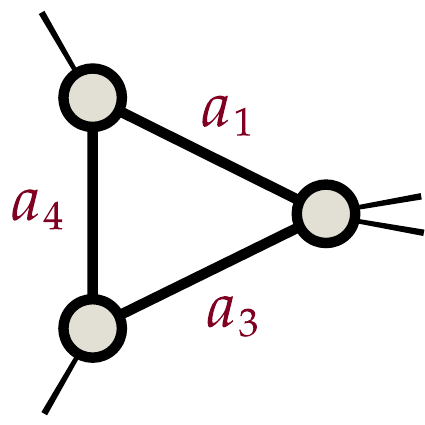}\;,\;\underset{\var{z} = \var{z^*}}{\text{eval}}\includegraphics[valign = c, scale = 0.3]{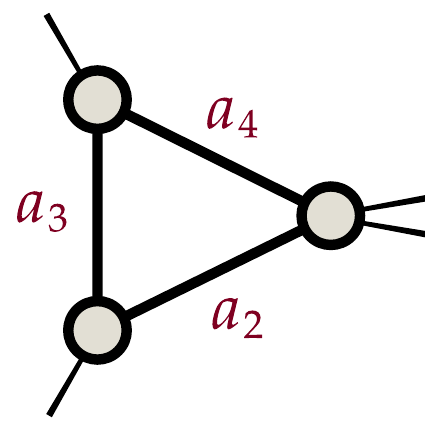}\Bigg\rbrace}
\end{equation}

Given generic numerators for the box and the numerators specified by the contact terms, we find for the period matrix the following upper triangular system

\begin{equation}
    \begin{blockarray}{ccccccc}
 \hspace{1pt}&\Omega_1 & \Omega_2 & \Omega_3 & \Omega_4 & \Omega_5 &\Omega_6\\
\begin{block}{c(cccccc)}
  (\ell|\bl{N_1}) & \bl{f}^{\bur{1}}_{\bl{1}} & \bl{f}^{\bur{2}}_{\bl{1}} & \bl{g}^{\bur{3}}_{\bl{1}} & \bl{g}^{\bur{4}}_{\bl{1}} & \bl{g}^{\bur{5}}_{\bl{1}} &\bl{g}^{\bur{6}}_{\bl{1}}\\
  (\ell|\bl{N_2}) & \bl{f}^{\bur{1}}_{\bl{2}} & \bl{f}^{\bur{2}}_{\bl{2}} & \bl{g}^{\bur{3}}_{\bl{2}} & \bl{g}^{\bur{4}}_{\bl{2}} & \bl{g}^{\bur{5}}_{\bl{2}} &\bl{g}^{\bur{6}}_{\bl{2}}\\
  (\ell|\bur{a_4}) & 0 & 0 & \bl{h}^{\bur{3}}_{\bl{3}} & 0 & 0 &0\\
  (\ell|\bur{a_3}) & 0 & 0 & 0 & \bl{h}^{\bur{4}}_{\bl{4}} & 0 &0\\
  (\ell|\bur{a_2}) & 0 & 0 & 0 & 0 & \bl{h}^{\bur{5}}_{\bl{5}} &0\\
  (\ell|\bur{a_1}) & 0 & 0 & 0 & 0 & 0 &\bl{h}^{\bur{6}}_{\bl{6}}\\
\end{block}
\end{blockarray}
\end{equation}
A short check (even using \textsc{Mathematica}) shows that this is of rank 6. Indeed, the diagonalized numerators will have period matrix equal to the identity (we will not inflict them on the reader).

As an aside, note that the triangle topologies are related to each other by cyclic operations. It was easy here to enumerate them all explicitly, but this rapidly becomes unsustainable for higher multiplicity.
\hfill
\vspace{0.7cm}

The more subtle issue we have glossed over so far is the implication for the numerators once cuts are chosen. It is entirely the case that a `wise' choice of cuts renders the numerators simple in a way more arbitrary ones do not. The best illustration of this is the paradigmatic case of the box, where we can just choose the numerators at the outset to be the solutions to the cut equations themselves---

\begin{equation}
    \bl{N_{i}} = \bl{Q_{i}}
\end{equation}
where the $\bl{Q_{i}}$ are the solutions to the quad cut of the box. Indeed, the numerator $(\ell|\bl{Q_{1}})$ will vanish on the first quad cut, and the other will vanish on the second quad cut. Accordingly, they are automatically diagonalized on the two cuts, saving us the extra work of having to diagonalize the top-level subspace by hand. A common theme of the present work is then that while we will always have some choice of cuts to work with, it is often very interesting to simply ask and answer the question of picking the `morally' right set of cuts instead.

We can go over one more example before moving on the to issue of building bases at two loops, namely the basis of $2$-gon integrands at one loop. This is spanned by scalar bubbles, triangles with one numerator, and boxes with two:

\begin{equation}
     \mathfrak{b}^{2} = \text{span}\Bigg\lbrace{\includegraphics[valign = c, scale = 0.3]{bubble-2gon.pdf}\;\;,\includegraphics[valign = c, scale = 0.3]{triangle-2gon.pdf}\;\;,\includegraphics[valign = c, scale = 0.3]{box-2gon.pdf}\Bigg\rbrace}.
\end{equation}
Starting with the bubbles, note that the general form of the unnormalized bubble in embedding space is as follows

\begin{equation}
    \includegraphics[valign = c, scale=0.3,trim=0 0 0 0cm]{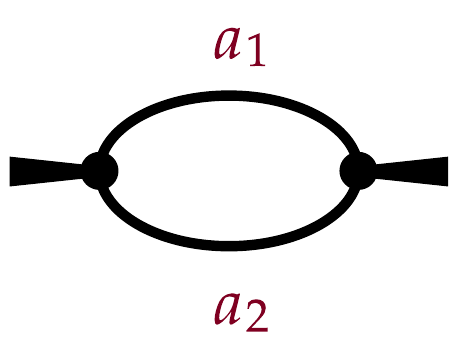} = \frac{\dbar^{4}\ell}{(\ell|\bur{a_1})(\ell|\bur{a_2})(\ell|\bl{X})^2}.
\end{equation}
The natural first step is to take the maximal cut, and set all of the internal propagators to zero;

\begin{equation}
    (\ell|\bur{a_1}) = (\ell|\bur{a_2}) = 0
\end{equation}
yielding a two form as follows

\begin{equation}
    \underset{\substack{(\ell|\bur{a_1})=0\\(\ell|\bur{a_2})=0}}{\oint} \includegraphics[valign = c, scale=0.3,trim=0 0 0 0cm]{bubble_labelled.pdf} = \frac{\dbar\var{\alpha}\dbar\var{\beta}}{J(\var{\alpha},\var{\beta})(\ell(\var{\alpha},\var{\beta})|\bl{X})^2}.
\end{equation}
A natural choice for fixing the two degrees of freedom so obtained would be to analogize the latter example and simply evaluate at $(\var{\alpha},\var{\beta}) = (\var{\alpha^*},\var{\beta^*})$. In this case, the bubble once normalized on the cut is matched to the on-shell function

\begin{equation}
  \underset{{(\var{\alpha},\var{\beta}) = (\var{\alpha^*},\var{\beta^*})}}{\text{eval}}  \includegraphics[valign = c, scale=0.3,trim=0 0 0 0cm]{bubble-onshell.pdf}.
\end{equation}
Modulo the details regarding the number of top-level degrees of freedom, the matching for the triangles and boxes follows the same algorithm. Indeed, the triangles in $3$-gon power counting have \bur{3} top level numerators, and as such they should be chosen by diagonalizing on the three cuts spanned by the maximal cut and evaluations on three distinct $\var{z} = \var{z_{I}}$, corresponding to the on-shell functions

\begin{equation}
  \underset{\var{z} = \var{z_{I}}}{\text{eval}}  \includegraphics[valign = c, scale=0.3,trim=0 0 0 0cm]{triangle-onshell.pdf}.
\end{equation}
The boxes inherit the same properties from the $3$-gon story; they continue to have \bur{2} top level degrees of freedom and ought to be diagonalized on the quad cuts once again. Consolidating these, we have for a scattering amplitude spanned by this basis the schematic form

\begin{equation}
\begin{aligned}
     \mathcal{A} =& \sum\underset{{(\var{\alpha},\var{\beta}) = (\var{\alpha^*},\var{\beta^*})}}{\text{eval}}  \includegraphics[valign = c, scale=0.3,trim=0 0 0 0cm]{bubble-onshell.pdf}\times\includegraphics[valign = c, scale=0.3,trim=0 0 0 0cm]{bubble-2gon.pdf}\\& + \sum  \underset{\var{z} = \var{z_{I}}}{\text{eval}}  \includegraphics[valign = c, scale=0.3,trim=0 0 0 0cm]{triangle-onshell.pdf}\times\includegraphics[valign = c, scale=0.3,trim=0 0 0 0cm]{triangle-top-2-gon.pdf}\\
     &+ \sum \includegraphics[valign = c, scale=0.3,trim=0 0 0 0cm]{box-onshell.pdf}\times \includegraphics[valign = c, scale=0.3,trim=0 0 0 0cm]{box-top-2-gon.pdf}.
\end{aligned}
\end{equation}
It bears mentioning that in the event that for a specific theory (say $\mathcal{N}$=4 super Yang-Mills), both the $3$-gon and $2$-gon bases furnish complete ones, the expansion in either will continue to hold true for the amplitude. The choice of which to use in such a circumstance depends on which property we would like manifest, and as it turns out this phenomenon is presented in an interesting way at two loops, which will be the main subject of section \ref{sec:4}.

Before moving on to generalizing our discussions so far to the pertinent problem of two loops, a brief comment on this problem of fixing unlocalized degrees of freedom in the absence of leading singularities can be made. Going back to the case of the bubble, note that 

\begin{equation}
    (\ell(\var{\alpha},\var{\beta})|\bl{X}) = \var{x_{i}}Q_{ij}\var{x_{j}} + b_{i}\var{x_{i}} + c
\end{equation}
where $\var{x_{i}} \in\lbrace{ \var{\alpha}, \var{\beta}\rbrace}$. A quadric of this form can always be brought into the generic version

\begin{equation}
    (\ell(\var{\alpha},\var{\beta})|\bl{X}) \longrightarrow \var{\alpha}^2 + \var{\beta}^2 + C
\end{equation}
via a linear transformation. Identifying $\var{\alpha}^2 + \var{\beta}^2$ with the form $\var{z}\overline{\var{z}}$ on $\mathbb{CP}^{1}$, the two form obtained from the double cut can always be schematically brought into a top-form on the sphere

\begin{equation}
    \frac{\dbar\var{\alpha}\dbar\var{\beta}}{J(\var{\alpha},\var{\beta})(\ell(\var{\alpha},\var{\beta})|\bl{X})^2} \longrightarrow \frac{\dbar\var{z}\wedge\dbar\overline{\var{z}}}{g(\var{z},\overline{\var{z}})}.
\end{equation}
On the sphere, any noncontractible contour (such as the equator $\var{z} = \overline{\var{z}}$ for example), can be used to pair with this latter form, fixing the last two degrees of freedom in a more---at least at first blush--intrinsic fashion.

Geometrically, what this ultimately amounts to is embedding $\mathbb{CP}^1$ in $\mathbb{CP}^2$ and encircling the equator. Previously, we had instead performed the embedding by making use of the homology class $\text{[pt]}\times\text{[pt]}$ in $\mathbb{CP}^2$. While some embeddings are certainly more endearing than others, it remains an open question as to how exactly we need to choose cycles to fix bubble integrands, especially for power counting worse than $2$-gon, where there is a proliferation of top-level terms for the bubble numerator spaces.

\subsection{Bases of Integrands at Two Loops}\label{sec:2.3}
Building a consistent basis at two loops again demands the compromise between the size of the basis and ease of matching; the $4$-gon and $3$-gon bases, which we will discuss in turn each have their advantages and disadvantages, particularly when we talk about simple quantum field theories like maximally supersymmetric Yang-Mills. 

We will perform the counting for all topologies at two loops for $3$-gon and $4$-gon power counting at two loops, even under threats of excessive repetitiveness, if only to emphasize how divisions into contact and non-contact terms are derived in each case. This will become especially important as fodder for the more detailed issue of stratification that will acquire our interest in section \ref{sec:4}.

For now, let us start with building the two-loop basis (always in four dimensions) for the case of $4$-gon power counting. This basis is spanned by four distinct topologies as follows (although this fact will be refined):

\begin{equation}
    \mathfrak{b}_{\text{2-loop}}^{4} = \Bigg\lbrace{\includegraphics[valign = c, scale=0.3,trim=0 0 0 0cm]{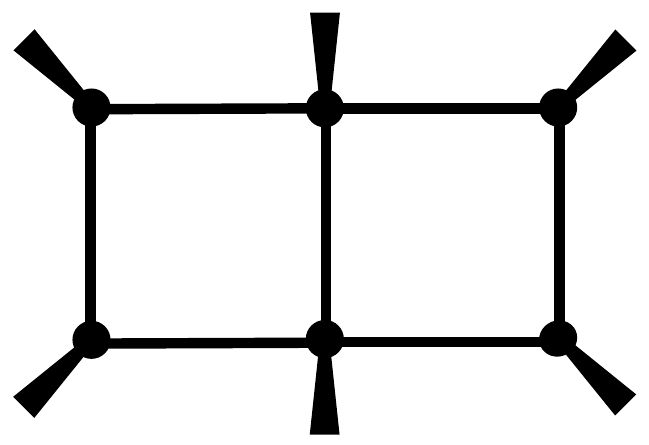}\;,\;\includegraphics[valign = c, scale=0.3,trim=0 0 0 0cm]{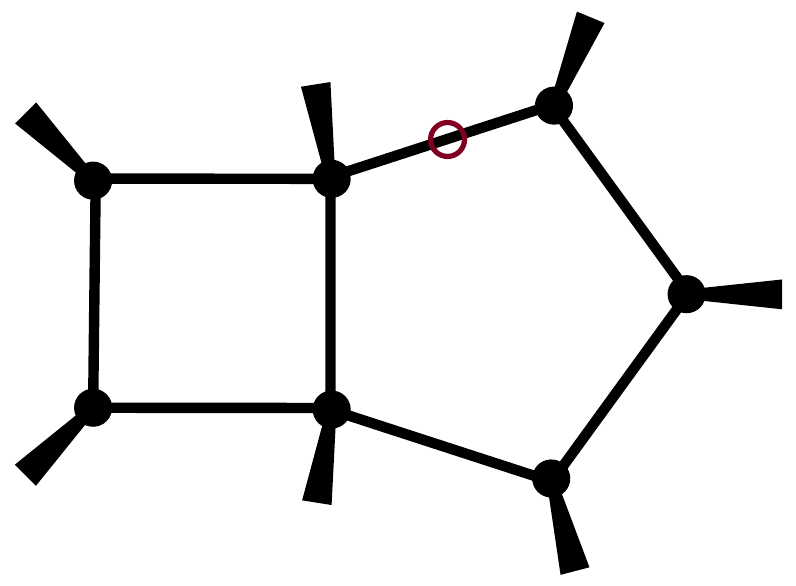}\;,\;\includegraphics[valign = c, scale=0.3,trim=0 0 0 0cm]{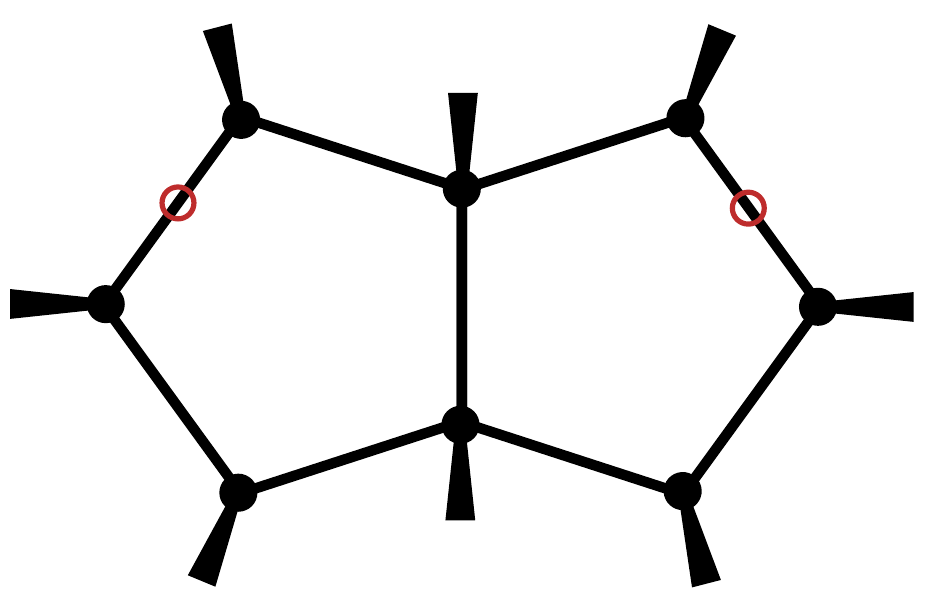}\;,\;\includegraphics[valign = c, scale=0.3,trim=0 0 0 0cm]{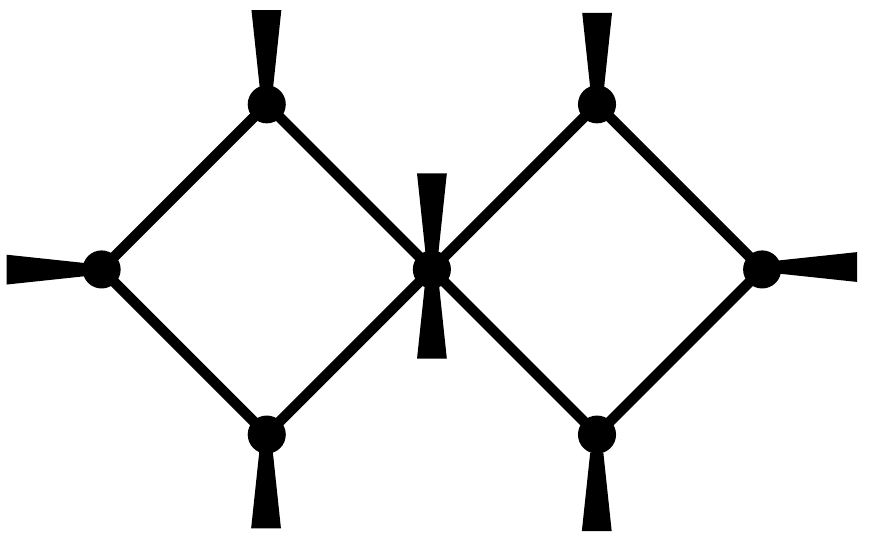}\Bigg\rbrace}. 
\end{equation}
These topologies, from left to right, are referred to as double boxes, pentaboxes, double pentagons and kissing boxes respectively.

With the numerator spaces labelled on the diagrams themselves, we can proceed to evaluate the ranks of their top-level spaces one by one.

Starting with the double box, in $4$-gon power counting, it is just the scalar double box, meaning that its numerator space is just spanned by the scalar $1$. Accordingly, we have that the dimension of this space is just $1$, with \bur{1} top level degree of freedom and no contact terms:

\begin{equation}
\text{rank}\left(\text{span}\Bigg\lbrace{\includegraphics[valign = c, scale=0.3,trim=0 0 0 0cm]{doublebox-4-gon.pdf}\Bigg\rbrace}\right) = \bur{1} + 0.
\end{equation}
The kissing boxes inherit their structure from the one loop case, since they are just product topologies, obtained by simply multiplying two distinct boxes with each other. As such, they have a numerator space spanned by a `product' of scalars, which is one-dimensional with no contact terms:

\begin{equation}
\text{rank}\left(\text{span}\Bigg\lbrace{\includegraphics[valign = c, scale=0.3,trim=0 0 0 0cm]{kissing-box-4-gon.pdf}\Bigg\rbrace}\right) = \bur{1} + 0.
\end{equation}
In the case of the pentabox, we have the numerator space spanned by all numerators of the form $(\ell_2|\bl{Q})$, which in four dimensions is a space of dimension \bl{6}. There are $\binom{4}{1}$ ways in which this can descend into a double box, each of which furnishes \bur{1} top-level degree of freedom. Ultimately, this renders the top-level space of the pentabox in $4$-gon power counting of dimension \bur{2}, supplying

\begin{equation}
\text{rank}\left(\text{span}\Bigg\lbrace{\includegraphics[valign = c, scale=0.3,trim=0 0 0 0cm]{pentabox-4-gon.pdf}\Bigg\rbrace}\right) = \bur{2} + 4.
\end{equation}
The double pentagon is dressed by all numerators of the form

\begin{equation}
    (\ell_1|\bl{N}_{1})(\ell_2|\bl{N}_3)\in\text{span}([\ell_1][\ell_3]\oplus[\ell_1-\ell_2]).
\end{equation}
Contact terms in this case are provided by collapsing the external legs or the internal propagator to supply kissing boxes. Fortunately, the counting of the dimension of this numerator space isn't hard, since $\text{span}[\ell_1-\ell_2] \subset \text{span}([\ell_1][\ell_2])$, which is of dimension $36$. Now in terms of top-level versus contact terms, observe that this topology can degenerate into pentaboxes in $8$ ways, each of which is of top-level dimension $\bur{2}$. Reductions to double boxes can be done in another $4\times 4 = 16$ ways. Finally, there is just one way to recover double boxes. All accounted for, we have the decomposition 

\begin{equation}
\text{rank}\left(\text{span}\Bigg\lbrace{\includegraphics[valign = c, scale=0.3,trim=0 0 0 0cm]{double-pentagon-4-gon.pdf}\Bigg\rbrace}\right) = \bur{3} + 33.
\end{equation}
It turns out---as we will see in the next section---that matters simplify if the kissing box topology is subsumed into the double pentagon instead, and one deals simply with double pentagons. In this case, we use double pentagons with a numerator space spanned by elements of $[\ell_1][\ell_2]$, and we write for the topology

\begin{equation}
\text{rank}\left(\text{span}\Bigg\lbrace{\includegraphics[valign = c, scale=0.3,trim=0 0 0 0cm]{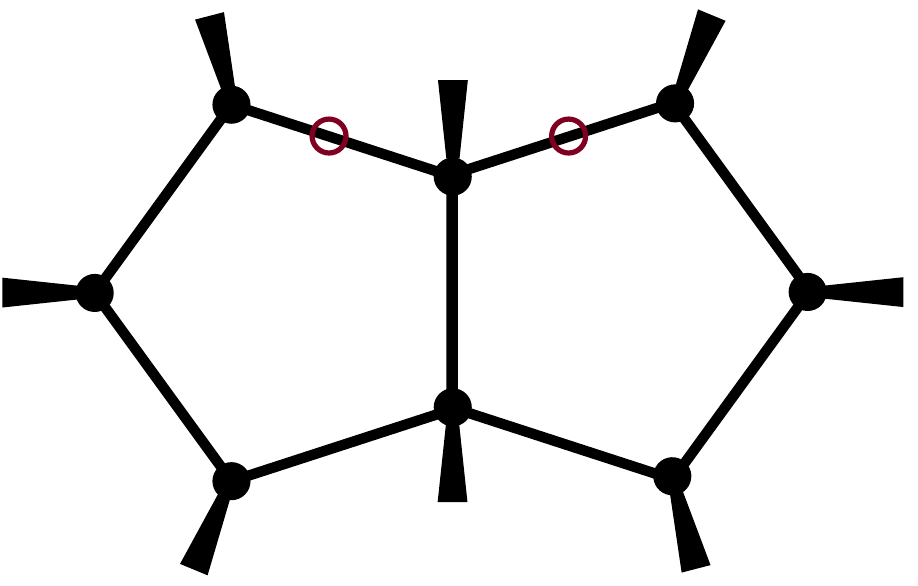}\Bigg\rbrace}\right) = \bur{4} + 32.
\end{equation}
Ultimately, we use the basis spanned by scalar double boxes, pentaboxes and double pentagons in $4$-gon power counting at two loops.

The $3$-gon power counting basis is considerably larger and includes all those elements that scale only as bad as a scalar triangle. This is spanned by all topologies listed below.

\begin{equation}\label{eq:2.74}
\begin{aligned}
     \mathfrak{b}_{\text{2-loop}}^{4} = \Bigg\lbrace&\includegraphics[valign = c, scale=0.3,trim=0 0 0 0cm]{double-triangle-3-gon.pdf}\;,\;\includegraphics[valign = c, scale=0.3,trim=0 0 0 0cm]{kissing-triangle-3-gon.pdf}\;,\;\includegraphics[valign = c, scale=0.3,trim=0 0 0 0cm]{kissing-box-triangle-3-gon.pdf}\;,\;\includegraphics[valign = c, scale=0.3,trim=0 0 0 0cm]{kissing-boxes-3-gon.pdf}\;,\;\\
    &\includegraphics[valign = c, scale=0.3,trim=0 0 0 0cm]{box-triangle-3-gon.pdf}\;,\;\includegraphics[valign = c, scale=0.3,trim=0 0 0 0cm]{pentatriangle-3-gon.pdf}\;,\;\includegraphics[valign = c, scale=0.3,trim=0 0 0 0cm]{doublebox-3-gon.pdf}\;,\;\includegraphics[valign = c, scale=0.3,trim=0 0 0 0cm]{pentabox-3-gon.pdf} \Bigg\rbrace 
\end{aligned} 
\end{equation}
(We have abused notation a little in the cases of the double box and pentabox, which we will explain shortly.) We now have to count the numerator degrees of freedom of each topology and perform their resolutions into contact and top-level degrees of freedom.

The simplest cases are the scalars, namely the double triangle and the kissing triangle. The numerator space of the double triangle is just the span of a single scalar of dimension squared mass (it is the most UV divergent of all integrands in $3$-gon power counting). Correspondingly, we have the simple decomposition

\begin{equation}
\text{rank}\left(\text{span}\Bigg\lbrace{\includegraphics[valign = c, scale=0.3,trim=0 0 0 0cm]{double-triangle-3-gon.pdf}\Bigg\rbrace}\right) = \bur{1} + 0.
\end{equation}
The kissing triangle topology is really just a product topology, of two one-loop triangles multiplied together. Accordingly, it inherits its dimensionality therefrom, and we have the simple correspondence

\begin{equation}
\text{rank}\left(\text{span}\Bigg\lbrace{\includegraphics[valign = c, scale=0.3,trim=0 0 0 0cm]{kissing-triangle-3-gon.pdf}\Bigg\rbrace}\right) = \bur{1} + 0.   
\end{equation}
The remaining topologies on the first line of (\ref{eq:2.74}) are the kissing box-triangle and the kissing box, which once again acquire their dimensions from the daughters at one loop. Indeed, for the kissing box-triangle we have the numerator space spanned by $(\ell_2|\bl{N_1})$, which descends into $4$ kissing triangles and \bur{2} top level degrees of freedom:

\begin{equation}
\text{rank}\left(\text{span}\Bigg\lbrace{\includegraphics[valign = c, scale=0.3,trim=0 0 0 0cm]{kissing-box-triangle-3-gon.pdf}\Bigg\rbrace}\right) = \bur{2} + 4.   
\end{equation}
The kissing boxes are spanned on the numerator space governed by $(\ell|\bl{N_1})(\ell_2|\bl{N_2})$, which is of rank 36. It consists of 8 kissing box triangles and 16 kissing triangles, which combine to supply 32 contact terms, leaving us with

\begin{equation}
\text{rank}\left(\text{span}\Bigg\lbrace{\includegraphics[valign = c, scale=0.3,trim=0 0 0 0cm]{kissing-boxes-3-gon.pdf}\Bigg\rbrace}\right) = \bur{4} + 32.   
\end{equation}

The non-product topologies listed in the second line of (\ref{eq:2.74}) are left, and exhibit more complex decompositions, as they can be reduced to other non-product as well as product topologies.

The box triangle has a numerator space governed by all polynomials of the form $(\ell_2|\bl{N})$, which is of total dimension \bl{6}. Now in terms of contact terms, the middle propagator can't be closed, since that would furnish a bubble, but one of the box legs can be, which can be done in 3 ways. Ultimately, we have the reduction into top-level and contact terms as follows

\begin{equation}
\text{rank}\left(\text{span}\Bigg\lbrace{\includegraphics[valign = c, scale=0.3,trim=0 0 0 0cm]{box-triangle-3-gon.pdf}\Bigg\rbrace}\right) = \bur{3} + 3.  
\end{equation}

The pentatriangle has a numerator space spanned by two inverse propagators in the second loop, of the form $(\ell_2|\bl{N_1})(\ell_2|\bl{N_2})$. Indeed, this is of total dimension \bl{20}, but contains in the span all daughter box-triangles and double triangles. The former can be accessed in a total of $\binom{4}{1} = 4$ ways, with \bur{3} top-level numerators each. The double triangle topology can be reached in a total of $\binom{4}{2} = 6$ ways, with only \bur{1} top-level term each. This yields the break up as

\begin{equation}
\text{rank}\left(\text{span}\Bigg\lbrace{\includegraphics[valign = c, scale=0.3,trim=0 0 0 0cm]{pentatriangle-3-gon.pdf}\Bigg\rbrace}\right) = \bur{2} + 18.  
\end{equation}

It is in the case of the double box and pentabox that we were required to employ some abuse of notation. Specifically, starting with the double box, observe that two distinct routes of degeneration are available. On the one hand, the external legs can be collapsed to recover daughter topologies that are not products of one loop graphs. Alternatively, one can collapse the middle leg can be collapsed alone to obtain a kissing triangle. Indeed, the numerator space is granted by the span of the vector space sum

\begin{equation}
    \text{span}([\ell_1][\ell_2] \oplus [\ell_1-\ell_2]).
\end{equation}
Fortunately, the span of $[\ell_1-\ell_2]$ is contained inside of that of $[\ell_1][\ell_2]$, and as such we may compute the dimension of the latter for the total counting.

Coming to this, we can see that the total dimension is $\bl{36}$, coming out of the product of the two independent loops. For the daughters, we have the following counting: $2\times\binom{3}{1} = 6$ ways of accessing box triangles, $\binom{3}{1}\times \binom{3}{1} = 9$ ways of reaching double triangles and $1$ way of reaching a kissing triangle. The total number of contact numerators are

\begin{equation}
    6\times\bur{3} + 9\times\bur{1} + 1\times\bur{1} = 28
\end{equation}
which results in the division

\begin{equation}
\text{rank}\left(\text{span}\Bigg\lbrace{\includegraphics[valign = c, scale=0.3,trim=0 0 0 0cm]{doublebox-3-gon.pdf}\Bigg\rbrace}\right) = \bur{8} + 28. 
\end{equation}
The pentabox numerator space again arises out of a more complex representation. The choices break down between either choosing to collapse or not to collapse the middle leg. These two choices respectively grant spaces spanned by $[\ell_1][\ell_2]^2$ and $[\ell-\ell_1][\ell_2]$. Indeed, the latter is once again captured by the former, and we have for the total space dimension $\bl{120}$.

The total space fragments into contact terms as follows: there are $\binom{4}{1} = 4$ double box subtopologies, $\binom{3}{1} = 3$ pentatriangles, $\binom{4}{1}\times\binom{3}{1} + \binom{4}{2} = 18$ box triangles, $\binom{3}{1}\times\binom{4}{2} = 18$ double triangles, $1$ kissing box-triangle and $4$ kissing triangles. All accounted for, we have the number of contact degrees of freedom as follows

\begin{equation}
    4\times\bur{8} + 3\times\bur{2} + 18\times\bur{3} + 18\times\bur{1}+1\times\bur{2} + 4\times\bur{1} = 116.
\end{equation}
Accordingly, this means that the total numerator space of the pentabox is naturally cast into \bur{4} top-level terms, and 116 contact degrees of freedom:

\begin{equation}
\text{rank}\left(\text{span}\Bigg\lbrace{\includegraphics[valign = c, scale=0.3,trim=0 0 0 0cm]{pentabox-3-gon.pdf}\Bigg\rbrace}\right) = \bur{4} + 116. 
\end{equation}

\subsection{Choosing Cuts \emph{Poorly}: A Tale of Two Loops}\label{sec:2.4}
Since we have the data for the bases in $4$-gon and $3$-gon power counting at two loops, we can now move on to describing how to construct a set of cut equations in both cases which are of full rank. Further, we can contrast the issues inherent in choosing one kind of power counting versus another, especially the compromise between finding good choices of cuts and dealing with large basis sizes.

In this section, we focus on simply constructing systems of cut equations that are full rank, just to emphasize that \emph{ans\"atze} are solvable not only in principle but also in practice. We will also make it a point to show how our choices of contours to fix the coefficients are not particularly good ones, with an intent to rectify at least some of this in the sections to follow later.

\subsubsection{4-gon Basis at Two Loops}\label{sec:2.4.1}

The $4$-gon power counting basis is especially relevant to the case of maximally supersymmetric Yang-Mills, where it is known to span all planar scattering amplitudes at two and three loops for any multiplicity \cite{Bourjaily:2015jna,Bourjaily:2017wjl}\footnote{Hints of this were first revealed by \cite{Arkani-Hamed:2010pyv}, where local integrand presentations for the MHV and NMHV amplitude at two loops were uncovered, later generalized to all MHV sectors in \cite{Bourjaily:2015jna,Bourjaily:2017wjl}.}. We follow the prescriptions provided in these papers, in large part to motivate the choices we make in the case of $3$-gon power counting in the next section.

Recalling that the basis of $4$-gon power counting is spanned by double boxes, pentaboxes and double pentagons, it's most convenient to find a spanning set of cuts that at least vanish on daughter topologies. Starting then with the double pentagon, it takes the following form in dual momentum space

\begin{equation}
    \includegraphics[valign = c, scale=0.3,trim=0 0 0 0cm]{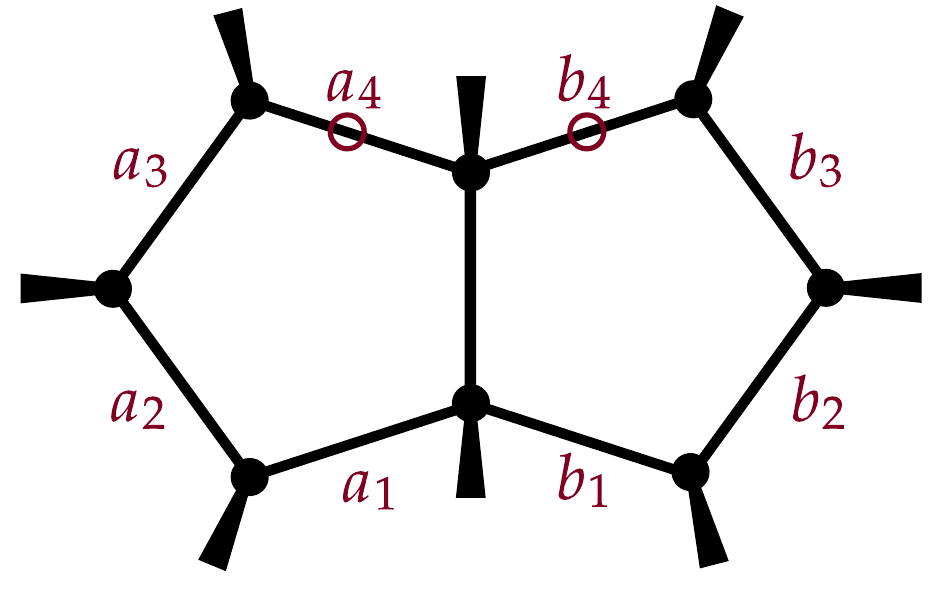} = \frac{(\ell|\bl{N_1})(\ell_2|\bl{N_2})\dbar^{4}\ell_1 \dbar^4\ell_2}{(\ell_1|\bur{a_1})(\ell_1|\bur{a_2})(\ell_1|\bur{a_3})(\ell_1|\bur{a_4})(\ell_1|\ell_2)(\ell_2|\bur{b_1})(\ell_2|\bur{b_2})(\ell_2|\bur{b_1})(\ell_2|\bur{b_4})}
\end{equation}
Now there are two ways of fixing the cuts on this topology. Under the condition that the kissing box is a subtopology, there are a total of $3$ top level degrees of freedom. These can be chosen to be three out of four of the kissing box cuts

\begin{equation}
    \includegraphics[valign = c, scale=0.3,trim=0 0 0 0cm]{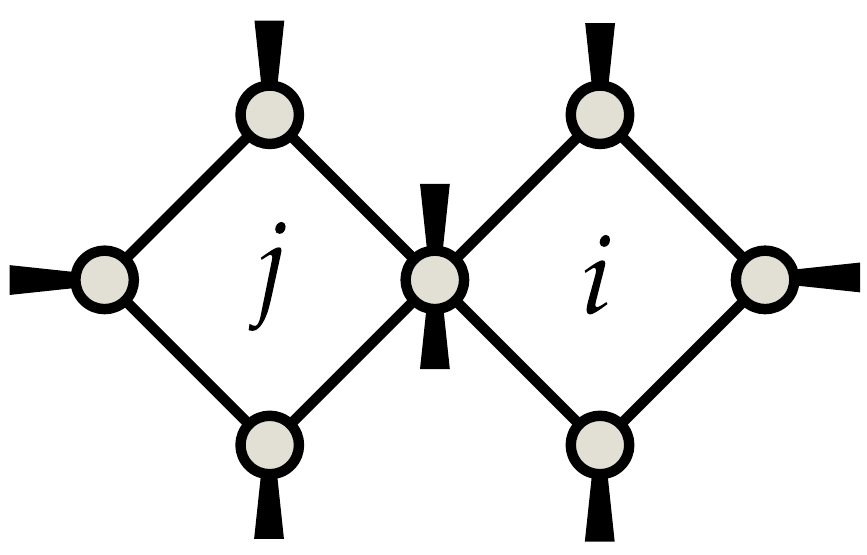}.
\end{equation}
The last one can then be matched to the single degree of freedom provided by the kissing box topology. A simpler route however is to simply subsume the kissing boxes into double pentagons and match the preceding four leading singularities to the top-level degrees of freedom of the double pentagon.

The pentabox in box power counting is a little unpleasant since we only have \bur{2} top level degrees of freedom, but there are a total of 4 leading singularities given a pentabox. Indeed, a generic pentabox

\begin{equation}
    \includegraphics[valign = c, scale=0.3,trim=0 0 0 0cm]{pentabox-4-gon-labelled.pdf} = \frac{(\ell_2|\bl{N_2})\dbar^{4}\ell_1 \dbar^4\ell_2}{(\ell_1|\bur{a_1})(\ell_1|\bur{a_2})(\ell_1|\bur{a_3})(\ell_1|\ell_2)(\ell_2|\bur{b_1})(\ell_2|\bur{b_2})(\ell_2|\bur{b_1})(\ell_2|\bur{b_4})}
\end{equation}
has a total of $8$ internal legs, which completely localize the loop momenta when the maximal cut is computed. Ultimately, this will furnish a total of $4$ solutions to the cut equations, of which we will have to pick \bur{2}---which suffice to fix the top level numerator---such as

\begin{equation}
    \includegraphics[valign = c, scale=0.3,trim=0 0 0 0cm]{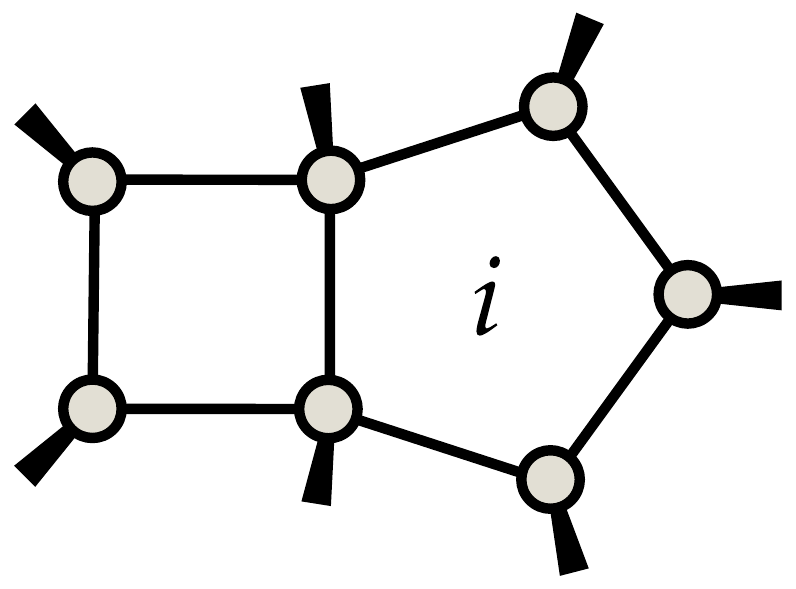}.
\end{equation}
The double box can be fixed in several ways; one natural way is to move to the maximal cut and then fixing at a point. That said, how this has been done in the past is as follows. In \cite{Bourjaily:2015jna}, the local form of the two loop amplitude was verified to be true by comparing against the solution obtained by all-loop recursion \cite{ArkaniHamed:2010kv}. The recursive solution is a rational function given integer-valued momentum twistors, and the maximal cut for the double box for large numbers of particles leads to a proliferation of square roots, making comparison difficult. The solution to this was to instead compute the kissing-triangle topology, which is a two parameter function:

\begin{equation}
     \includegraphics[valign = c, scale=0.3,trim=0 0 0 0cm]{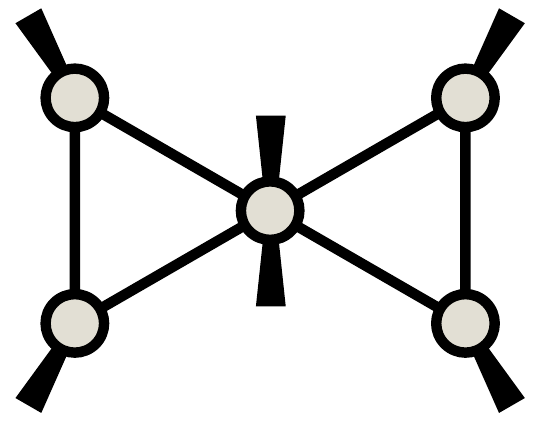} = \frac{d\var{\alpha}d\var{\beta}}{f(\var{\alpha},\var{\beta})}.
\end{equation}
This can then be matched at a fixed point $(\var{\alpha},\var{\beta}) = (\var{\alpha^*},\var{\beta^*})$. 

We consolidate this by mentioning how the numerators are modified by prescriptivity, once they are normalized on the contours chosen above. Starting with the pentaboxes, the normalized numerators will just be $(\ell_2|\bl{Q_{i}})$\footnote{For the sake of convenience and notational simplicity we have assumed that the pentagon is on the second loop, although we do include the case where the pentagon is on the first loop as well. These of course arise as contact terms of the double pentagon anyway.} where $i=1,2$ denote the solutions of the quad cut used to fix the top-level terms. For solution $1$, $i=2$ and vice versa. They are then modified so as to vanish on all the daughter double box topologies as follows

\begin{equation}
    (\ell_2|\bl{Q_i}) \longrightarrow (\ell_2|\bl{Q_{i}}) - \sum_{i=1}^{4}\frac{(\ell_2(\var{\alpha^*},\var{\beta^*})|\bl{Q_i})}{(\ell_1(\var{\alpha^*},\var{\beta^*})|\bur{a_i})}\frac{(\ell_1|\bur{a_i}) }{\mathcal{N}}
\end{equation}
with analogous expressions holding when the collapsed legs are on the second loop instead and $\mathcal{N}$ simply denotes the normalization on the double triangle.

The double box numerators are required to be such that the pentabox cuts as well as the double triangles vanish on them. Indeed, the first shift will have to be according to

\begin{equation}
    \begin{aligned}
        (\ell_1|\bl{N_i})(\ell_2|\bl{N_i})\longrightarrow &(\ell_1|\bl{N_i})(\ell_2|\bl{N_i}) \\&-\sum_{i=1}^{4}\frac{(\ell_1(\bl{Q^{(i)}_{1}})|\bl{N_1})(\ell_2(\bl{Q^{(i)}_{1}})|\bl{N_2})}{(\bl{Q^{(i)}_{2}}|\bur{a_i})}(\ell_1|\bur{a_i})\\&-\sum_{i=1}^{4}\frac{(\ell_1(\bl{\widetilde{Q}^{(i)}_{2}})|\bl{N_1})(\ell_2(\bl{\widetilde{Q}^{(i)}_{2}})|\bl{N_2})}{(\bl{Q^{(i)}_{2}}|\bur{b_i})}(\ell_2|\bur{b_i}) + (\bl{1}\leftrightarrow \bl{2})\\
        &:=(\ell_1|\bl{N'_1})(\ell_2|\bl{N'_2})
    \end{aligned}
\end{equation}
where the exchange indicates the two cut solutions used to fix the pentaboxes, and the sum runs over the various pentabox daughters of the double pentagon. Tildes denote the quad cut solutions for the pentaboxes obtained by collapsing the second loop. The loop momenta are to be evaluated on the solutions to the quad cuts.

Another shift has to be performed to eliminate the contributions of the double boxes:

\begin{equation}
   (\ell_1|\bl{N'_1})(\ell_2|\bl{N'_2}) \longrightarrow (\ell_1|\bl{N'_1})(\ell_2|\bl{N'_2}) - \sum_{i,j=1}^{4}\frac{(\ell_1(\var{\alpha^*},\var{\beta^*})|\bl{N'_1})(\ell_2(\var{\alpha^*},\var{\beta^*})|\bl{N'_2})}{(\ell_1(\var{\alpha^*},\var{\beta^*})|\bur{a_i})(\ell_2(\var{\alpha^*},\var{\beta^*})|\bur{b_j})}(\ell_1|\bur{a_i})(\ell_2|\bur{b_j}). 
\end{equation}
Accordingly, an amplitude spanned by a $4$-gon basis at two loops would admit of the following general expansion in four dimensions

\begin{equation}
\begin{aligned}
    \mathcal{A} = &\sum\underset{(\var{\alpha},\var{\beta})=(\var{\alpha^*},\var{\beta^*})}{\text{eval}}\includegraphics[valign = c, scale=0.3,trim=0 0 0 0cm]{kissing-triangle-3-gon-top-level-on-shell.pdf}\times \includegraphics[valign = c, scale=0.3,trim=0 0 0 0cm]{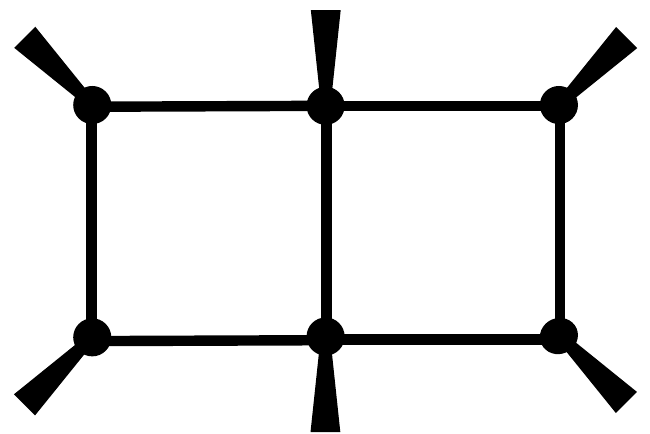} +\\
    &\sum\includegraphics[valign = c, scale=0.3,trim=0 0 0 0cm]{pentabox-4-gon-top-level-on-shell.pdf}\times \includegraphics[valign = c, scale=0.3,trim=0 0 0 0cm]{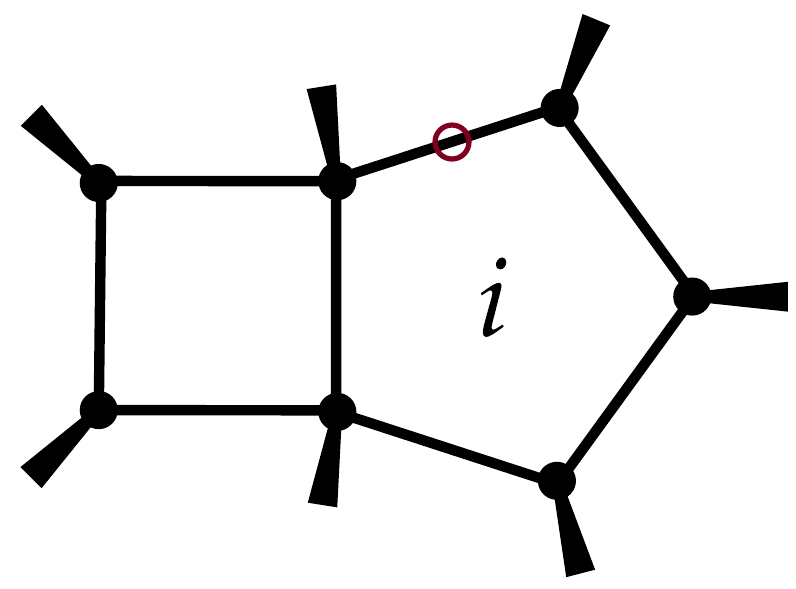}+\\
    &\sum \includegraphics[valign = c, scale=0.3,trim=0 0 0 0cm]{doublepentagon-4-gon-top-level-on-shell.pdf}\times\includegraphics[valign = c, scale=0.3,trim=0 0 0 0cm]{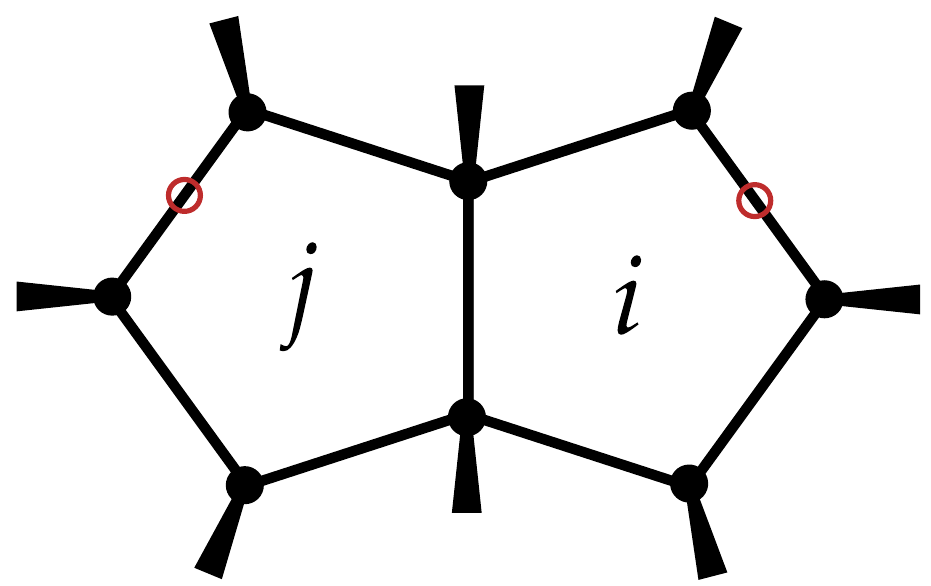}\
\end{aligned}
\end{equation}
where the sums run over the top-level contributions and all possible leg distributions. 

\subsubsection{3-gon Basis at Two Loops}

We'll move on now to the $3$-gon basis but before doing so just mention that the subtraction of contact contributions works the same way there, but is just more cumbersome. As such, we won't repeat the steps we just performed for the $4$-gon basis, but only emphasize the top-level contributions.

The product topologies are just fixed by their descendents from the one loop case; we fix kissing triangles to evaluations:

\begin{equation}
    \underset{(\var{\alpha},\var{\beta})=(\var{\alpha^*},\var{\beta^*})}{\text{eval}}\includegraphics[valign = c, scale=0.3,trim=0 0 0 0cm]{kissing-triangle-3-gon-top-level-on-shell.pdf},
\end{equation}
kissing box-triangles are fixed by

\begin{equation}
    \underset{\var{\alpha}=\var{\alpha^*}}{\text{eval}}\includegraphics[valign = c, scale=0.3,trim=0 0 0 0cm]{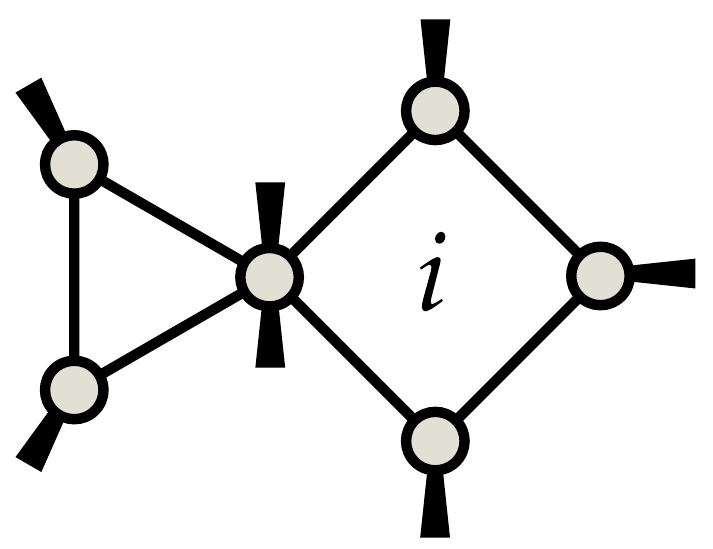}
\end{equation}
and kissing boxes are fixed by

\begin{equation}
    \includegraphics[valign = c, scale=0.3,trim=0 0 0 0cm]{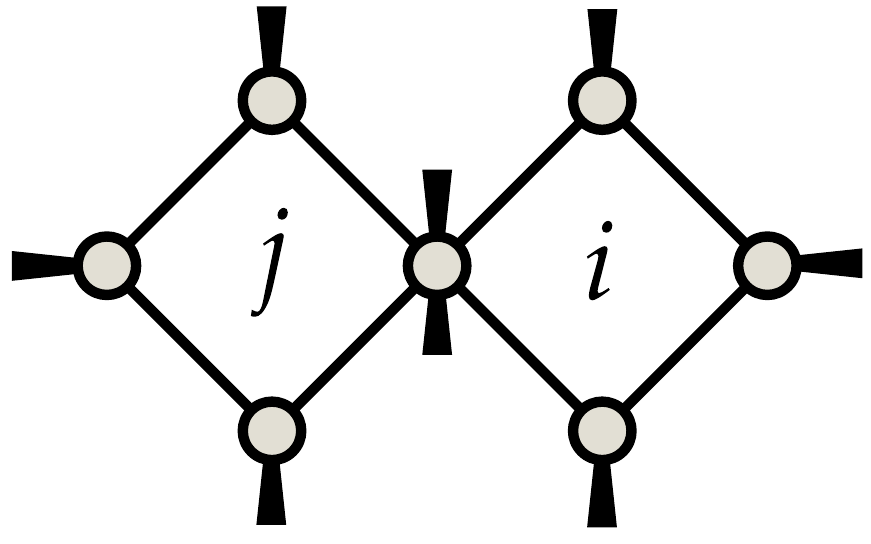}.
\end{equation}
There are five non-product topologies: the double triangle, double box, box triangle, pentatriangle and pentabox. We'll go from the `simplest' to the `hardest' of these in determining the contours used. What we mean by this is just that some of these topologies (the double triangle, pentatriangle and pentabox) have natural choices for leading singularities or maximal contours in a way the other two do not.

Starting with the double triangle, note its form in embedding space

\begin{equation}
    \includegraphics[valign = c, scale=0.3,trim=0 0 0 0cm]{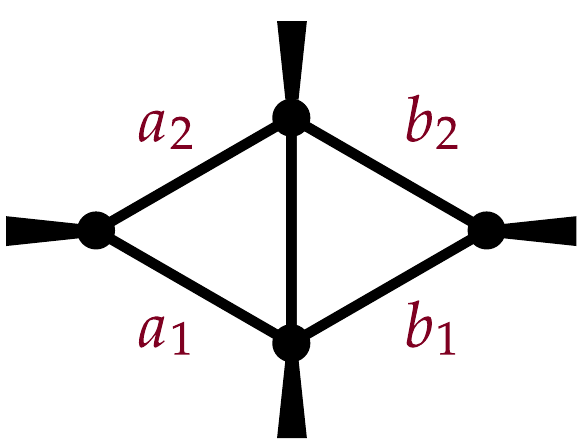} = \frac{\dbar^4\ell_1\dbar^4\ell_2}{(\ell_1|\bur{a_1})(\ell_1|\bur{a_2})(\ell_1|\bl{X})(\ell_1|\ell_2)(\ell_2|\bur{b_1})(\ell_2|\bur{b_2})(\ell_2|\bl{X})}.
\end{equation}
The maximal cut, encircling all of the kinematic legs and the internal propagator is of codimension $3$, and supplies a three-form once computed:

\begin{equation}
    \underset{\substack{(\ell_1|\bur{a_i})=0\\(\ell_2|\bur{b_i})=0\\(\ell_1|\ell_2))}}{\oint} \includegraphics[valign = c, scale=0.3,trim=0 0 0 0cm]{double-triangle-3-gon-labelled.pdf} = \frac{\dbar^3\var{\alpha_i}}{f(\var{\alpha_i})}.
\end{equation}
The relative simplicity of this diagram derives from the fact that it is usually possible to completely localize this function by taking residues. Indeed, the function $f(\var{\alpha_i})$ furnishes three additional poles to encircle, admitting of what we will call a `maximal residue', which conveys residue prescriptions around infinity.

However, it should be emphasized that turns out to not always be exactly true. Leg distributions where $\bur{a_i} = \bur{b_i}$ do not always possess such maximal residues, and may only admit two such further residues, after which one obtains a double pole. There are two ways of handling this: either one can match to a point, or find solutions of the cut equations, possibly in collinear limits, to localize the loop momenta further and access contours of smaller codimension. We will come to such an example shortly.

Absent this subtlety, we will generally state that such topologies are to be matched to these so-called maximal residues. In other words, we match the on-shell function

\begin{equation}
    \text{Res}_{\var{\alpha_i}} \includegraphics[valign = c, scale=0.3,trim=0 0 0 0cm]{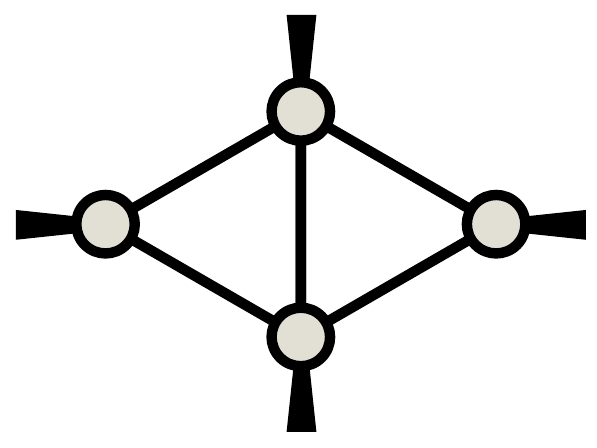}.
\end{equation}

The pentatriangle in triangle power counting has a  numerator space spanned by two inverse propagators; it has the following generic appearance

\begin{equation}
     \includegraphics[valign = c, scale=0.3,trim=0 0 0 0cm]{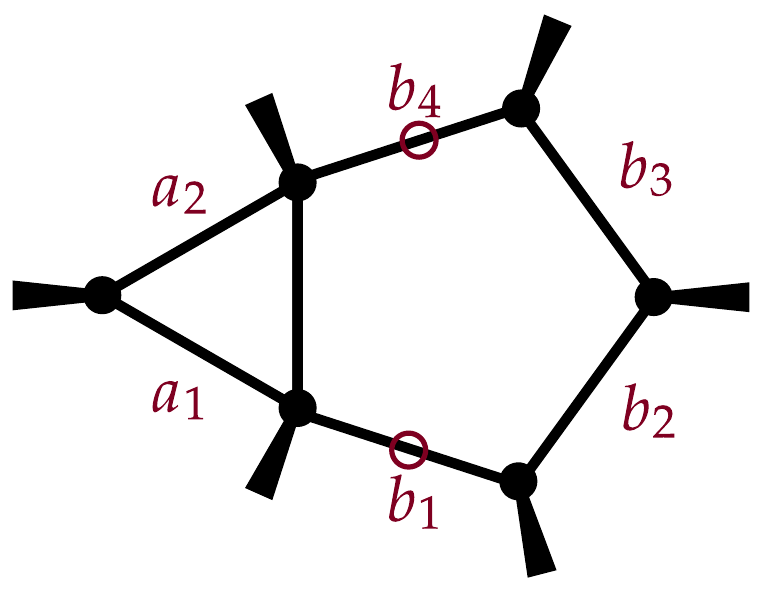} = \frac{(\ell_2|\bl{N_1})(\ell_2|\bl{N_2})\dbar^4\ell_1\dbar^4\ell_2}{(\ell_1|\bur{a_1})(\ell_1|\bur{a_2})(\ell_1|\bl{X})(\ell_1|\ell_2)(\ell_2|\bur{b_1})(\ell_2|\bur{b_2})((\ell_2|\bur{b_3})(\ell_2|\bur{b_4})\ell_2|\bl{X})}.
\end{equation}
The spanning cuts can be done loop by loop; the second loop can be completely localized by cutting each of the $(\ell_2|\bur{b_i})$. Choosing one of the quad cut solutions, this reduces down to the triangle at one loop:

\begin{equation}
    \underset{(\ell_2|\bur{b_i})=0}{\oint} \includegraphics[valign = c, scale=0.3,trim=0 0 0 0cm]{pentatriangle-3-gon-labelled.pdf}\sim\ \includegraphics[valign = c, scale=0.3,trim=0 0 0 0cm]{triangle-labelled.pdf}
\end{equation}
where $\bur{a_3}$ will be one of the quad cut solutions. Cutting $(\ell_1|\ell_2)$ and $(\ell_1|\bur{a_i})$ results in a one form, familiar from the one loop case. We can proceed to evaluate it at a point; in the end we have for the \bur{2} top level terms the following spanning set of cuts

\begin{equation}
     \underset{\var{\alpha} = \var{\alpha^*}}{\text{eval}}\includegraphics[valign = c, scale=0.3,trim=0 0 0 0cm]{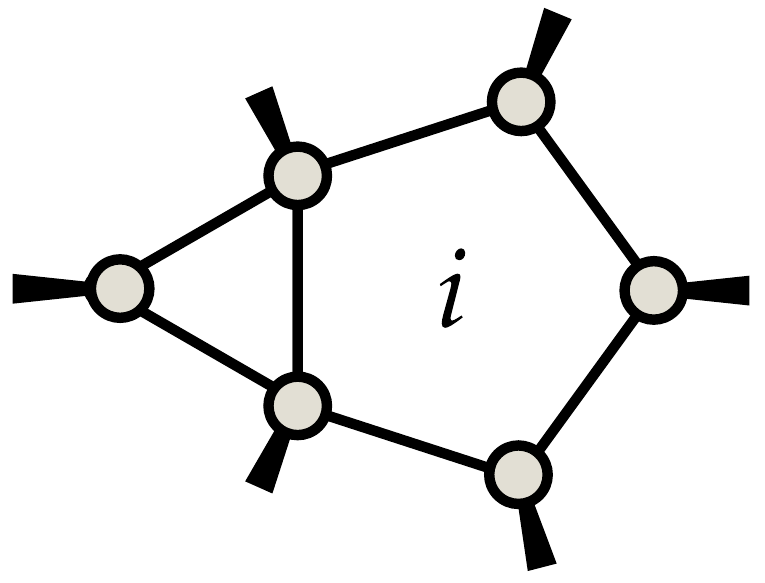}.
\end{equation}

The pentaboxes can be fixed the usual way; since they have \bur{4} top-level degrees of freedom, the leading singularities obtained by taking the maximal cut furnish the four solutions needed to determine them:

\begin{equation}
    \includegraphics[valign = c, scale=0.3,trim=0 0 0 0cm]{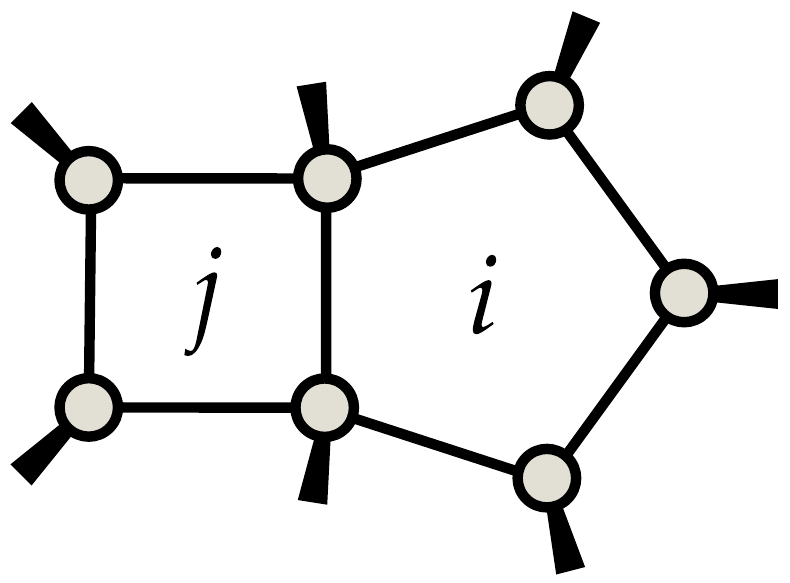}.
\end{equation}

The most complicated diagrams are the box-triangle and the double box. Starting with the box-triangle first, recall its form in dual momentum space

\begin{equation}
    \includegraphics[valign = c, scale=0.3,trim=0 0 0 0cm]{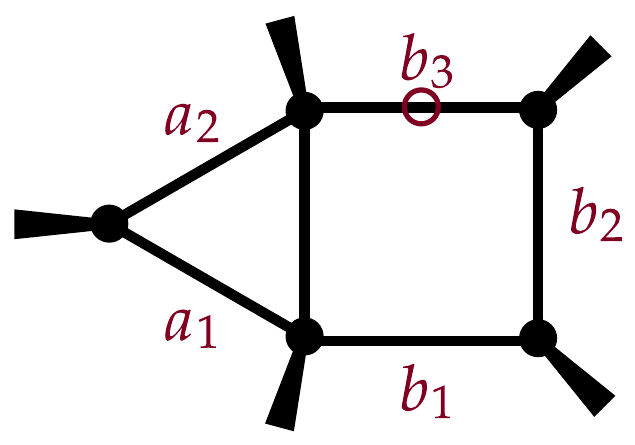} = \frac{(\ell_2|\bl{N_1})\dbar^4\ell_1\dbar^4\ell_2}{(\ell_1|\bur{a_1})(\ell_1|\bur{a_2})(\ell_1|\bl{X})(\ell_1|\ell_2)(\ell_2|\bur{b_1})(\ell_2|\bur{b_2})((\ell_2|\bur{b_3})\ell_2|\bl{X})}.
\end{equation}
There are \bur{3} top level terms that we need to fix, which equals the number of contours we have to find.

The natural step in line with our previous calculations would be to take the na\"ive maximal cut, in which we cut all propagators of the form $(\ell_1|\bur{a_i})$, $(\ell_2|\bur{b_i})$ and $(\ell_1|\ell_2)$. This is a contour that has codimension $2$ in loop momentum space, and will result in a two form therein. 

While we can choose to simply evaluate at two points and be done with it, there is another choice of contour from which more insight can be derived, especially in light of calculations we will be doing in section \ref{sec:4}. This contour involves cutting a single propagator at infinity, namely $(\ell_1|\bl{X})$. This will naturally result in a one-form in momentum space

\begin{equation}
\underset{(\ell_1|\bl{X})=0}{\text{Res}}\underset{\substack{(\ell_1|\bur{a_i})=0\\(\ell_2|\bur{b_i})=0\\(\ell_1|\ell_2)}=0}{\oint} \includegraphics[valign = c, scale=0.3,trim=0 0 0 0cm]{box-triangle-3-gon-labelled.pdf} = \frac{d\var{\alpha}}{f(\var{\alpha})}.
\end{equation}
Generic leg distributions result in $f(\var{\alpha})$ such that the square of this function is a quartic in the variable $\var{\alpha}$, which amounts to a form defined intrinsically on an elliptic curve in $\var{\alpha}$. 

This elliptic \emph{can} degenerate, under certain specific kinematic constraints. First, if a pair of $\bur{a_i}$ and $\bur{a_{i+1}}$ or $\bur{b_i}$ and $\bur{b_{i+1}}$ are light-like separated, the elliptic will degenerate into the product of two quadrics. Two, if $\bur{a_1} = \bur{b_1}$ or $\bur{a_3} = \bur{b_3}$, the same result occurs. The most generic case however, especially for large multiplicity, will always be elliptic.

While a detailed discussion of fixing the final variable in the presence of elliptic degrees of freedom will be the subject of the next section, suffice it to say that the box-triangle is generally a diagram that has \emph{rigidity} 1, namely that we can integrate all but one parameter in the Feynman parametrization before we happen upon non-polylogarithmic structure.

Here we will choose simple evaluations if only to make contact with the next case of the double box. Indeed, the corresponding on-shell diagram amounts to

\begin{equation}
    \underset{\var{\alpha} = \var{\alpha^*_{I}}}{\text{eval}}\underset{(\ell_1|\bl{X})=0}{\text{Res}}\includegraphics[valign = c, scale=0.3,trim=0 0 0 0cm]{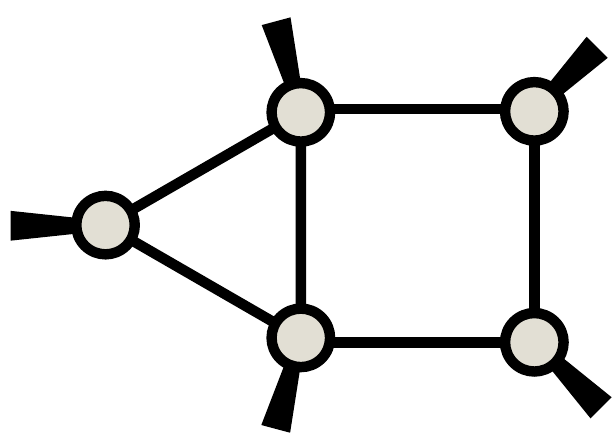}
\end{equation}
where $I$ runs from 1 through 3. 

There is the question of whether or not such a choice will actually be of rank $\bur{3}$ in the space of numerators. Guaranteeing rank $3$ requires taking into account potential Jacobian factors while computing the cuts, and we have done so for the various leg distributions possible in the box-triangle, and verified that choosing $3$ random points on which to evaluate $\var{\alpha}$ indeed furnishes a system of rank \bur{3} in \textsc{Mathematica}.

The double box is matched in analogy with the box-triangle, due to the fact that the subtleties that arise are of exactly the same kind. Indeed, the double box in $3$-gon power counting takes the form

\begin{equation}
    \includegraphics[valign = c, scale=0.3,trim=0 0 0 0cm]{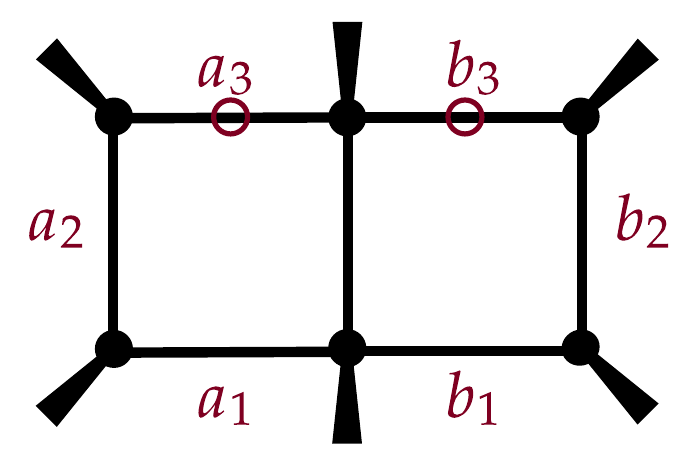} = \frac{(\ell_1|\bl{N_1})(\ell_2|\bl{N_1})\dbar^4\ell_1\dbar^4\ell_2}{(\ell_1|\bur{a_1})(\ell_1|\bur{a_2})(\ell_1|\bur{a_3})(\ell_1|\bl{X})(\ell_1|\ell_2)(\ell_2|\bur{b_1})(\ell_2|\bur{b_2})((\ell_2|\bur{b_3})\ell_2|\bl{X})}.
\end{equation}
This time, one doesn't have occasion to cut the propagators at infinity at the outset; a one form can be recovered by simply cutting all of the kinematic propagators $(\ell_1|\bur{a_i})$, $(\ell_2|\bur{b_i})$ and $(\ell_1|\ell_2)$:

\begin{equation}
    \underset{\substack{(\ell_1|\bur{a_i})=0\\(\ell_2|\bur{b_i})=0\\(\ell_1|\ell_2)}}{\oint}\includegraphics[valign = c, scale=0.3,trim=0 0 0 0cm]{doublebox-3-gon-labelled.pdf} = \frac{d\var{\alpha}}{f(\var{\alpha})}
\end{equation}
where the conditions for $f^2$ to be a quartic are the same as in the case of the box-triangle. 

The numerator space for the double box is of rank \bur{8} on this maximal cut, and as such we need to find eight(!) distinct ways of matching the cut. As it turns out, there is a rich application of this freedom to the problem of what we call \emph{stratification of rigidity}, which we will have occasion to discuss in a later section. For the time being, we pick what is arguably the most `poor' of cuts, namely evaluating at eight distinct points, to match the following on-shell functions

\begin{equation}
    \underset{\var{\alpha} = \var{\alpha^*_J}}{\text{eval}}\includegraphics[valign = c, scale=0.3,trim=0 0 0 0cm]{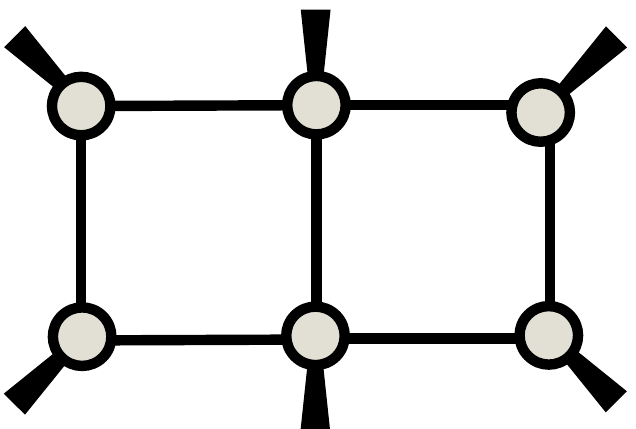}.
\end{equation}
where $J$ runs over \bur{8} distinct choices of pairs $(\bl{N},\bl{N_2})$.

Once again, we note that it is actually required to check that this does furnish a system of rank \bur{8}; there are a total of 48 different leg distributions possible for a double box, when massless, absent and massive legs are taken into account. We have verified that choosing points at random in this fashion does result in a system of full rank.

We should like to emphasize here that this \emph{had} to be the case, for if it were not the basis we have chosen would not be a complete one. Nevertheless, it is a good sanity check on our choice of cuts, to make sure that we are indeed matching distinct contours. For example, if we were to match double boxes to kissing triangles in triangle power counting, we would have actually encountered a system of rank $9$, as we would have obviated the need to view the kissing triangle as a subtopology in this case. An inconsistency would have resulted in trying to match another kissing triangle separately under this condition. 

This completes our choices for spanning sets of cuts in $3$-gon power counting. Let us close this part of the section with a brief discussion of what the numerators would look like. We emphasize that we have picked cuts in perhaps the most obvious, and in some sense most na\"ive way, since choosing cuts in this manner doesn't really tell us much except for the final amplitude itself. It has no effect on resolving the amplitude into say IR finite or divergent parts; it does not classify the amplitude by polylogarithmicity; nor do technical clarifications we might want to see in the exact form of an amplitude find themselves manifest. Accordingly, it is the case that the numerators obtained by diagonalizing on these cuts would be far from elegant, and rather unilluminating at that. Improving choices of this kind is the reason to ask how we should choose cuts `wisely', as even if they can be chosen in a simple fashion as we have, the result may not be too revealing.

We'll complete the section by going over an explicit example of building the basis and preparing the cuts for the case of four particles in $3$-gon power counting.

\paragraph*{Example 2.2. 4-Particles in Triangle Power Counting: }
The simplest nontrivial case of four particles is itself involved in $3$-gon power counting, as we have a number of topologies to care about. These are: one double box, three box triangles, four double triangles, one kissing triangle, and their inequivalent cyclic counterparts. We enumerate these cyclic seeds below:

\begin{equation}
\lbrace{\mathcal{I}^{J}_{1},\mathcal{I}_{2}\rbrace} = \Bigg\lbrace{\includegraphics[valign = c, scale=0.3,trim=0 0 0 0cm]{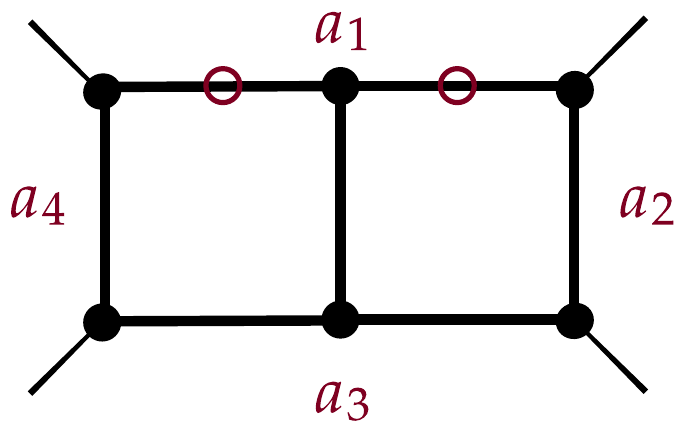},\includegraphics[valign = c, scale=0.3,trim=0 0 0 0cm]{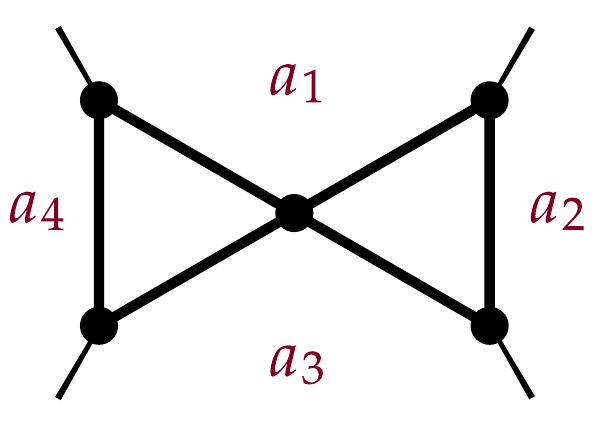}\Bigg\rbrace},
\end{equation}

\begin{equation} 
\begin{aligned}
&\lbrace{\mathcal{I}^{I}_{3},\dots,\mathcal{I}^{I}_{5}\rbrace} =
\Bigg\lbrace{\includegraphics[valign = c, scale=0.3,trim=0 0 0 0cm]{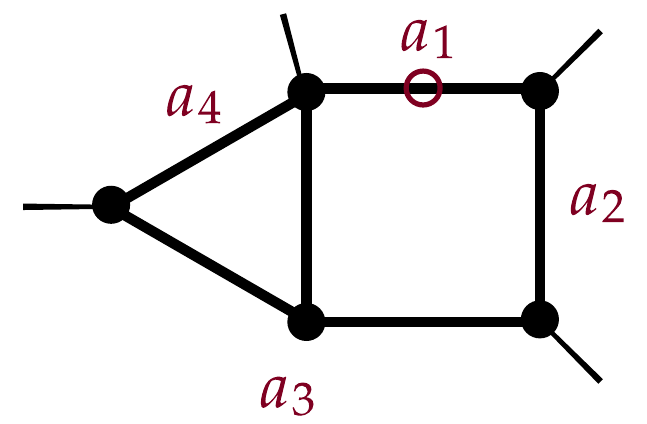}\;,\;\includegraphics[valign = c, scale=0.3,trim=0 0 0 0cm]{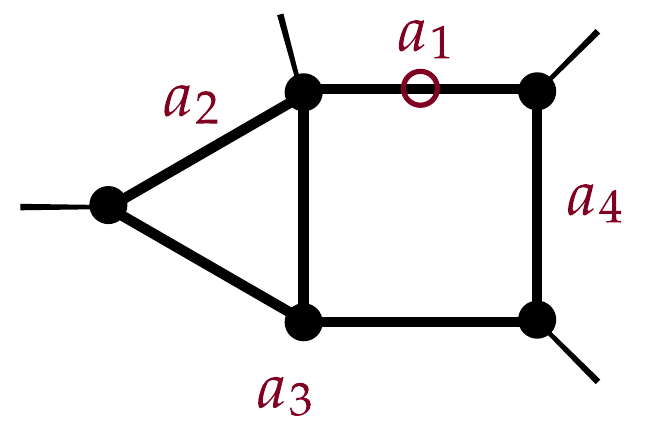}\;,\;\includegraphics[valign = c, scale=0.3,trim=0 0 0 0cm]{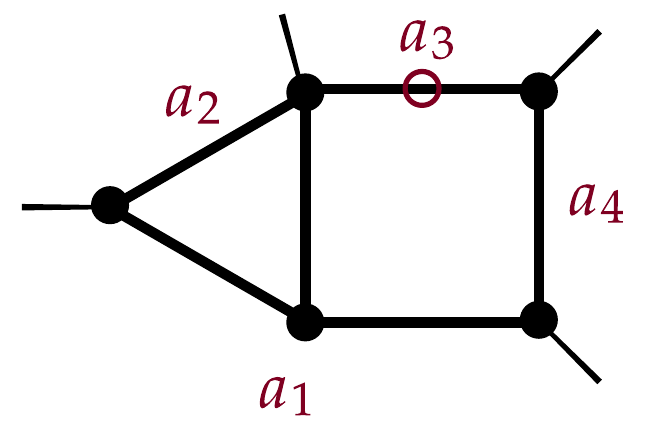}\Bigg\rbrace}.
\end{aligned}
\end{equation}

\begin{equation}  \lbrace{\mathcal{I}_{3},\dots,\mathcal{I}_{6}\rbrace} = \Bigg\lbrace{\includegraphics[valign = c, scale=0.3,trim=0 0 0 0cm]{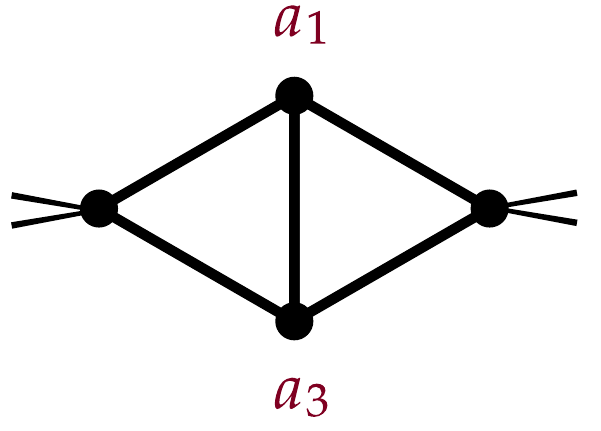}\;,\;\includegraphics[valign = c, scale=0.3,trim=0 0 0 0cm]{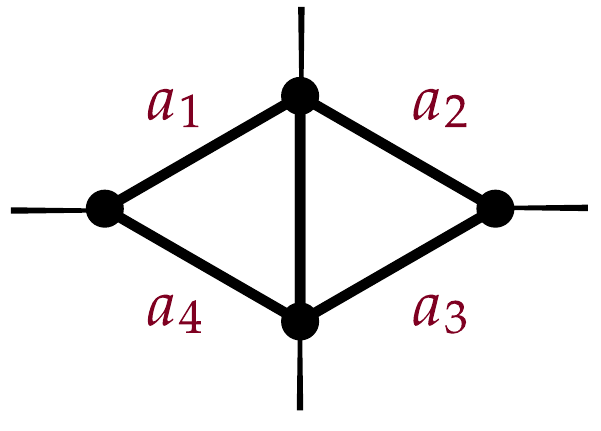}\;,\;\includegraphics[valign = c, scale=0.3,trim=0 0 0 0cm]{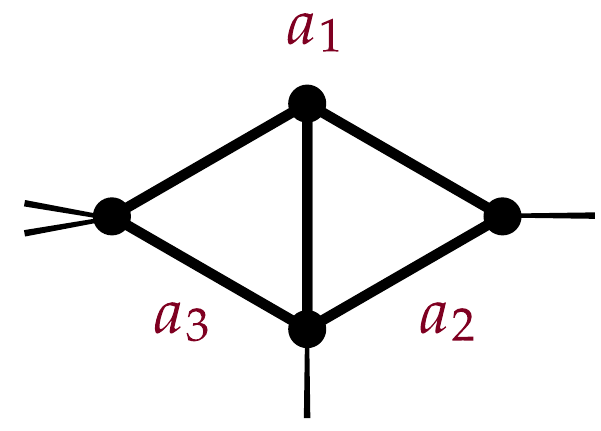}\;,\;\includegraphics[valign = c, scale=0.3,trim=0 0 0 0cm]{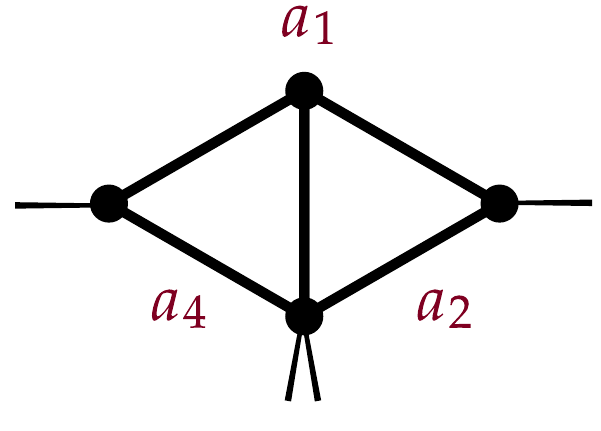}\Bigg\rbrace},
\end{equation}
where $J$ runs from $1$ through $8$ spanning the top-level terms of the double box and $I$ runs from 1 through 3 spanning the top-level terms of the box triangles. We can now enumerate the cuts (at the risk of looking repetitive, but this is for sake of completeness): for the double box we can simply pick \bur{8} points on which to evaluate the maximal cut to obtain the spanning set of contours

\begin{equation}
    \lbrace{\Omega^J_{1}\rbrace} = \Bigg\lbrace\underset{\var{\alpha}=\var{\alpha^*_1}}{\text{eval}}\includegraphics[valign = c, scale=0.3,trim=0 0 0 0cm]{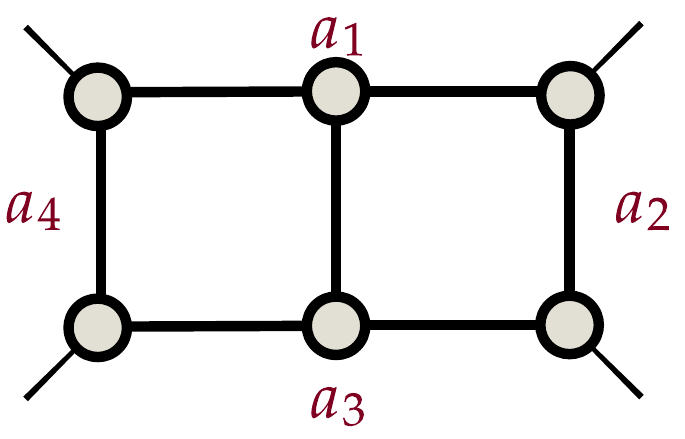},\dots, \underset{\var{\alpha}=\var{\alpha^*_8}}{\text{eval}}\includegraphics[valign = c, scale=0.3,trim=0 0 0 0cm]{double-box-3-gon-four-particles-on-shell.pdf}\Bigg\rbrace
\end{equation}
we have checked numerically on \textsc{Mathematica} that this is indeed of rank \bur{8}.

There is a subtlety regarding the box triangles, which we have presently chosen to avoid, namely those containing massless triangle subtopologies. This issue will be discussed at length in the concluding chapter of the paper, but for reasons we will discuss therein, we may exclude them from the basis. As such, we choose to match the remaining topologies on their maximal cuts along with residues drawn at infinity at points. For example,

\begin{equation}
    \lbrace{\Omega^I_{3}\rbrace} = \Bigg\lbrace\underset{\var{\alpha}=\var{\alpha^*_1}}{\text{eval}}\underset{(\ell|\bl{X})=0}{\text{Res}}\includegraphics[valign = c, scale=0.3,trim=0 0 0 0cm]{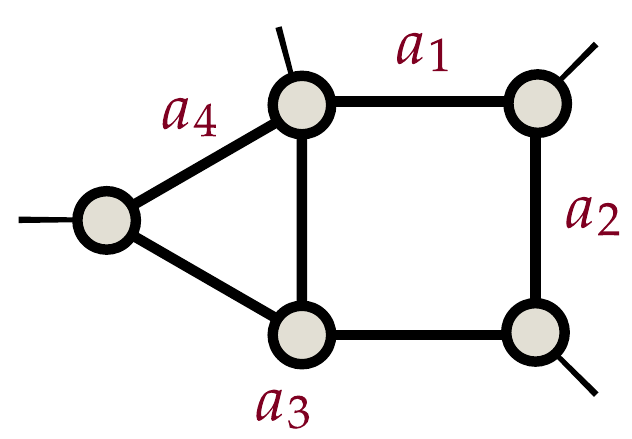},\dots,\underset{\var{\alpha}=\var{\alpha^*_3}}{\text{eval}}\underset{(\ell|\bl{X})=0}{\text{Res}}\includegraphics[valign = c, scale=0.3,trim=0 0 0 0cm]{box-triangle-3-gon-four-particles-1-on-shell.pdf}\Bigg\rbrace.
\end{equation}
We have made sure that systems defined in this fashion are indeed of rank \bur{3}.

The kissing triangle is easy, we just match onto the contour evaluating at two points:

\begin{equation}
    \Omega_2 = \underset{(\var{\alpha},\var{\beta})=(\var{\alpha^*_1},\var{\beta^*_1})}{\text{eval}}\includegraphics[valign = c, scale=0.3,trim=0 0 0 0cm]{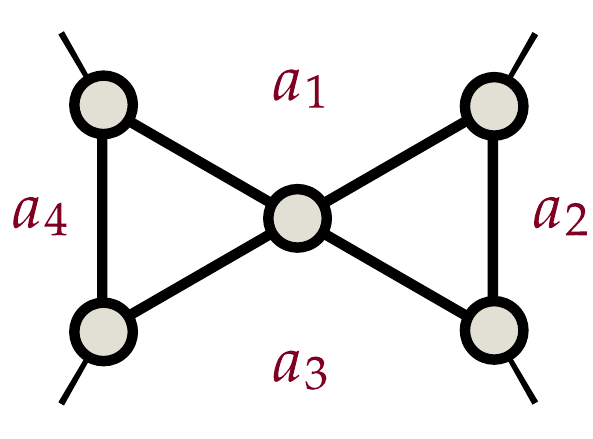}.
\end{equation}
We could have also chosen to match to the final residue at infinity. The double triangles present a small challenge to importing the results of the preceding discussion. Specifically, we can point out the case of the following codimension 3 cut

\begin{equation}
    \underset{\substack{(\ell_1|\bur{a_i})=0\\(\ell_2|\bur{b_i})=0\\(\ell_1|\ell_2)=0}}{\oint} \includegraphics[valign = c, scale=0.3,trim=0 0 0 0cm]{double-triangle-3-gon-4-particles-1.pdf} = \frac{\dbar^3\var{\alpha_j}}{f(\var{\alpha_j})}.
\end{equation}
Nominally, this is of the form we discussed previously, and it should be possible to take a maximal residue. This turns out to be difficult with all of the solutions to the cut equations, but there is a solution to the five cut conditions that actually localizes six of the loop momenta. It follows from the soft-collinear limit obtained by taking one of the internal vertices collinear

\begin{equation}
    \includegraphics[valign = c, scale=0.3,trim=0 0 0 0cm]{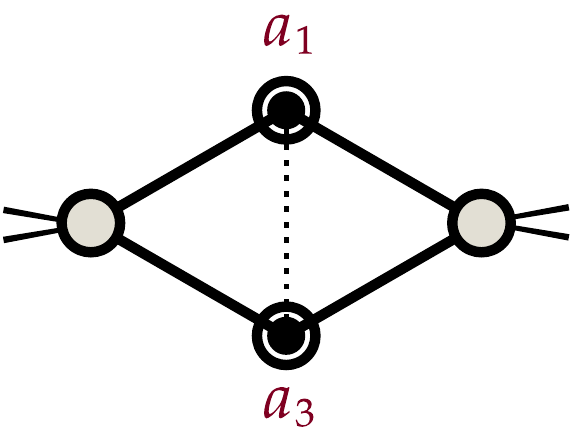}. 
\end{equation}
As it turns out, this cut will reveal a double pole corresponding to the two-loop momenta accessing infinity. There appears to be no simple way to avoid this, as any choice of cuts will eventually manifest this fact; as such, it turns out to be easiest to simply match onto a point\footnote{Topologies of this kind are inconvenient, at least in the context of $\mathcal{N}=4$, where they imply a violation of maximal transcendentality; I'm grateful to Jaroslav Trnka for emphasizing to me. The reader may also see for example \cite{Bourjaily:2021hcp} for a discussion of this when dealing with six-particle amplitudes.}:

\begin{equation}
    \underset{\var{\alpha_1}=\var{\alpha^*}}{\text{eval}}\;\includegraphics[valign = c, scale=0.3,trim=0 0 0 0cm]{doubletriangle-3-gon-4-particles-soft-collinear.pdf}.
\end{equation}
The other topologies also manifest collinear limits that make it easy to compute maximal residues, but they also have simple poles at infinity to which we can match the integrands. We choose to match to the following cuts

\begin{equation}
\lbrace{\mathcal{I}_{4},\mathcal{I}_5,\mathcal{I}_6\rbrace} = \Bigg\lbrace{\underset{\var{\alpha_i}}{\text{Res}}\includegraphics[valign = c, scale=0.3,trim=0 0 0 0cm]{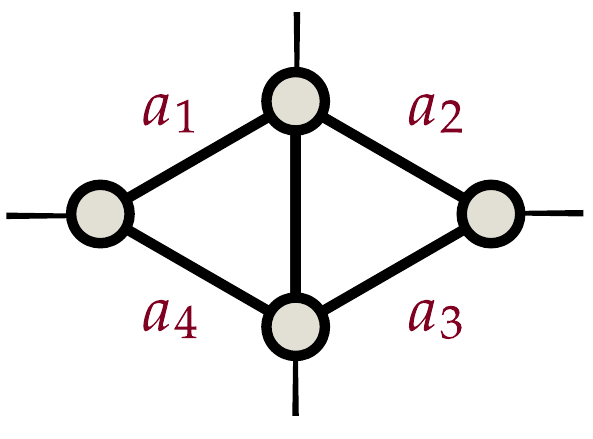}\;,\;\underset{\var{\alpha_i}}{\text{Res}}\includegraphics[valign = c, scale=0.3,trim=0 0 0 0cm]{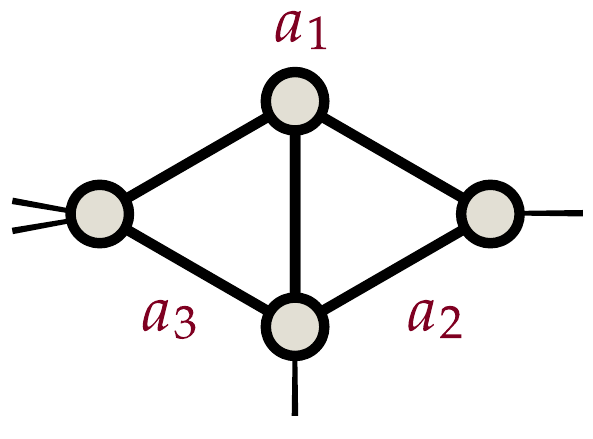}\;,\;\underset{\var{\alpha_i}}{\text{Res}}\includegraphics[valign = c, scale=0.3,trim=0 0 0 0cm]{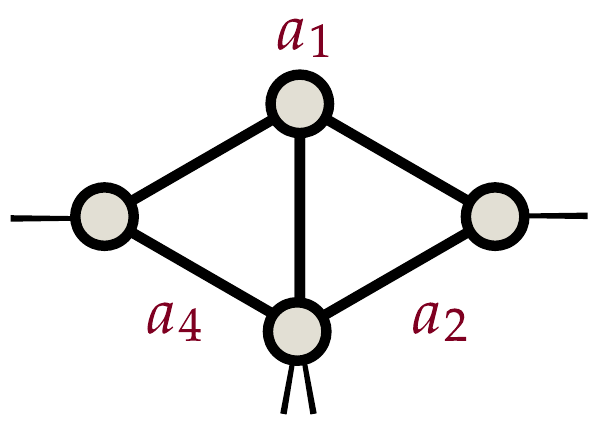}\Bigg\rbrace}
\end{equation}
keeping in mind that the maximal residues encircle poles at infinity. This choice offers a simplification for $\mathcal{N}=4$ sYM, in which amplitudes vanish at infinity. 

This completes our enumeration of cycles for this set of integrands. 
\hfill
\vspace{0.7cm}

Our emphasis in this part of the paper has been to clarify the general procedure of building a basis given conditions of ultraviolet behaviour. We looked at $4$-gon bases, which classify all poles that do not appear at infinity, and the $3$-gon basis, which takes into account all leading singularities and cuts, including the ones that occur at infinity. We matched the cuts by simply demanding the system have full rank but placed no conditions of simplicity to restrain these choices as such. Our focus in the following sections will be to remedy this indiscretion and shed some light on two main problems---how to intrinsically define the notion of a leading singularity when the maximal cut has nonzero codimension, and building master integrand basis that can account for the integrands exhibiting algebraic structures that go beyond polylogarithmicity.

\clearpage
\section{Rigidity and Leading Singularities}\label{sec:3}
The word rigidity---first defined in \cite{Bourjaily:2018yfy}---refers to the general phenomenon observed at two loops and beyond, where polylogarithms no longer suffice to integrate a given master integrand in a basis. Operationally, for iterated integrals, it is the smallest number of integrals past which polylogarithmic integration can no longer be performed. Our primary task in the present section will be to investigate instances of rigidity at planar two loops and discuss approaches to prescriptive unitarity that can accommodate rigidity in the integrand basis.

To start, we may recall a simple avatar of this phenomenon, observed at two loops, for the (now generally named) elliptic double box with scalar numerators in four dimensions:

\begin{equation}
    \includegraphics[valign = c, scale=0.3,trim=0 0 0 0cm]{doublebox-4-gon.pdf}
\end{equation}
which was variously conjectured \cite{Paulos:2012nu,Caron-Huot:2012awx,Nandan:2013ip} to be expressible in the algebraic form

\begin{equation}
    \int\includegraphics[valign = c, scale=0.3,trim=0 0 0 0cm]{doublebox-4-gon.pdf} = \int \dbar\var{\alpha}\frac{1}{\sqrt{Q(\var{\alpha})}}\text{PL}_{3}.
\end{equation}
Here, $Q(\var{\alpha})$ was expected to be a quartic in $\var{\alpha}$, which represents an elliptic curve, and $\text{PL}_3$ is a collection of polylogs of weight 3.

This expectation was confirmed by direct evaluation in \cite{Bourjaily:2017bsb}, where it was found that this is indeed the case; $\text{PL}_3$ was evaluated in terms of Goncharov polylogs \cite{Goncharov:1998kja} and verified to be of weight 3. To the extent that this integral can be performed by iterated evaluations over simple poles before reaching a non-polylogarithmic component, we have one parameter left over. Accordingly, it is said that this integrand has \emph{rigidity} 1.

As stated earlier, this integrand can be matched by making no reference to elliptics---or indeed to square roots arrived at in taking the 7-cut---by simply matching it, at least for $4$-gon power counting, to kissing triangles. More interesting however is the question of whether or not a suitable redefinition of the notion of a leading singularity exists in the case of non-codimension zero cuts, or for the elliptic case in particular, where the generic on-shell function

\begin{equation}
    \includegraphics[valign = c, scale=0.3,trim=0 0 0 0cm]{double-box-3-gon-top-level-on-shell.pdf}  
\end{equation}
certainly contains elliptic support for general leg distributions.

Our main task in the present section will be to elaborate upon this question, and discuss the analysis of this problem presented in \cite{Bourjaily:2020hjv,Bourjaily:2021vyj}. The main idea here is to have recourse to the homological versus cohomological picture of building bases, and view the matching of cuts as really a question of allocating cycle contributions based on the underlying geometry of the integrand. 

We will start with a general discussion of how and where elliptic curves arise at two loops, in the planar case in particular. We will then move on to presenting the notion of a \emph{period integral}, familiar from mathematical literature, and the pertinence thereof to generalizing the concept of a leading singularity. After moving on to actually applying this to the box power counting basis and determining a prescriptive set of cuts, we will close the section with applications to maximally supersymmetric Yang-Mills.

\subsection{Rigid and Polylogarithmic Structures at Two Loops}\label{sec:3.1}
In this section, we will attempt to summarize in some detail the conditions under which a double box will be rendered non-polylogarithmic and acquire nonzero rigidity. In doing so, we will be rather comprehensive, and emphasize how this is an unavoidable fact for large enough multiplicity of the external states.

It turns out to be most convenient to do in momentum twistor space, the most important features of which are reviewed for readers that so desire in appendix \ref{app:A}. In general, since we label dual momentum coordinates according to $\bur{a_1},\dots$, $\bur{b_1},\dots$ and so on, we will associate with the momentum labels, which refer to lines in twistor space, the twistor labels

\begin{equation}
    (\bur{a}_1) = (\bur{Aa}),
\end{equation}

\begin{equation}
    (\bur{a_2}) = (\bur{Bb})
\end{equation}
and so on.

Given this, we can start with the double box for generic external momenta, which in momentum twistor space we represent by

\begin{equation}
    \includegraphics[valign = c, scale=0.3,trim=0 0 0 0cm]{doublebox-4-gon-labelled.pdf} = \frac{\mathfrak{N} \dbar^4\ell_1\dbar^4\ell_2}{\lab{(\ell_1)\bur{Aa}}\lab{(\ell_1)\bur{Bb}}\lab{(\ell_1)\bur{Cc}}\lab{(\ell_1)(\ell_2)}\lab{(\ell_2)\bur{Dd}}\lab{(\ell_2)\bur{Ee}}\lab{(\ell_2)\bur{Ff}}}
\end{equation}
where $\mathfrak{N}$ denotes a kinematic factor that is responsible for normalization in twistor space. We would like to manifest the elliptic nature of the seven-cut. The most thorough way of doing this would be to compute the Jacobian of the equations controlling this cut, but that turns out to be rather cumbersome in a general chart. A quicker way is to use a more convenient chart spanned by the external momenta.

We do this as follows. To compute the seven-cut, we re-express them first in terms of the six-cut described by twistors according to the following four-bracket conditions

\begin{equation}
\lab{(\ell_1)\bur{Aa}}=\lab{(\ell_1)\bur{Bb}}=\lab{(\ell_1)\bur{Cc}}=0
\end{equation}
and
\begin{equation}
    \lab{(\ell_2)\bur{Dd}}= \lab{(\ell_2)\bur{eE}}= \lab{(\ell_2)\bur{fF}}=0.
\end{equation}

The geometry of twistor space trivializes these relations in purely rational terms; following the formulae expressed in appendix \ref{app:A}, noting that the latter conditions demand that the line $(\ell_1)$ intersect the lines $(\bur{Aa})$, $(\bur{Bb})$ and $(\bur{Cc})$ and $(\ell_2)$ the lines $(\bur{Dd})$, $(\bur{Ee})$ and $(\bur{Ff})$, we may write down the following six-cut solution

\begin{equation}
    (\ell_1) = (\bur{\widehat{A}\widehat{C}})\text{ and }(\ell_2) = (\bur{\widehat{D}\widehat{F}})
\end{equation}
where the hatted points are defined according to

\begin{equation}
    \bur{\widehat{A}}= \bur{a} + \var{\alpha_1}\bur{A},
\end{equation}
and

\begin{equation}
    \bur{\widehat{C}} = (\bur{cC})\cap(\bur{bBa})+\var{\alpha_1}(\bur{cC})\cap(\bur{bBA}).
\end{equation}
for the first loop. For the second loop, we have the analogous definitions

\begin{equation}
    \bur{\widehat{D}} = \bur{d} + \var{\alpha_2}\bur{D},
\end{equation}
and

\begin{equation}
   \bur{\widehat{F}} = (\bur{fF})\cap(\bur{eEd})+\var{\alpha_2}(\bur{fF})\cap(\bur{eED}).
\end{equation}
The seven cut is now administered by the condition that sets the propagator $(\ell_1|\ell_2)$ to vanish, which in terms of momentum twistors boils down to the algebraic equation as follows

\begin{equation}
\lab{\bur{\widehat{A}\widehat{C}\widehat{D}\widehat{F}}}=0,
\end{equation}
where the hatted variables are defined as above.

Naturally, this is a quadric in $\var{\alpha_1}$ and $\var{\alpha_2}$, and \emph{a priori} we have no guidance as to which of these variables we have to choose. We remark first that this is the origin of the elliptic. Notice that say we solve for $\var{\alpha_1}$, then the solutions $\var{\alpha_1^{\pm}}$ are controlled by the two solutions of the corresponding quadratic equation. The discriminant will be a quartic in $\var{\alpha_2}$, which is the source of ellipticity. A similar story holds of course for solutions in terms of $\var{\alpha_2}$.

The question now is whether or not these two elliptics are the same, since any alleged equivalence will by no means be manifest in the two respective charts. Still, this happens to be the case and should be verified. To see this, we compute the following two odd contours with respect to $\var{\alpha_1}$ and $\var{\alpha_2}$,

\begin{equation}
    \left(\underset{\var{\alpha_1}=\var{\alpha_1^+}}{\oint}-\underset{\var{\alpha_1}=\var{\alpha_1^-}}{\oint}\right)\underset{\substack{(\ell_1|\bur{a_i})=0\\(\ell_2|\bur{b_i})=0}}{\oint}\includegraphics[valign = c, scale=0.3,trim=0 0 0 0cm]{doublebox-4-gon-labelled.pdf}
\end{equation}
and

\begin{equation}
    \left(\underset{\var{\alpha_2}=\var{\alpha_2^+}}{\oint}-\underset{\var{\alpha_2}=\var{\alpha_2^-}}{\oint}\right)\underset{\substack{(\ell_1|\bur{a_i})=0\\(\ell_2|\bur{b_i})=0}}{\oint}\includegraphics[valign = c, scale=0.3,trim=0 0 0 0cm]{doublebox-4-gon-labelled.pdf}.
\end{equation}

For the first of these, we obtain the one form (relabelling $\var{\alpha_2} = \var{\alpha}$)

\begin{equation}
    -i\frac{c\dbar\var{\alpha}}{y_1(\var{\alpha})}
\end{equation}
where

\begin{equation}
    \begin{aligned}
        y_1(\var{\alpha})  = &\left(\lab{\bur{A}((\bur{cC})\cap(\bur{bBa}))\bur{\widehat{D}}\bur{\widehat{F}}}+\lab{\bur{a}((\bur{cC})\cap(\bur{bBA}))\bur{\widehat{D}}\bur{\widehat{F}}}\right)^2 - \\
        &4\lab{\bur{A}((\bur{cC})\cap(\bur{bBA}))\bur{\widehat{D}}\bur{\widehat{F}}}\lab{\bur{a}((\bur{cC})\cap(\bur{bBa}))\bur{\widehat{D}}\bur{\widehat{F}}}.
    \end{aligned}
\end{equation}
Due to the quadratic nature in $\var{\alpha}$ of $\bur{\widehat{C}}\bur{\widehat{D}}$, this is clearly a quartic.

Repeating this exercise for the residue around the second variable instead, we find that we once again have the form

\begin{equation}
    -i\frac{c\dbar\var{\alpha}}{y_2(\var{\alpha})}
\end{equation}
where this time we find

\begin{equation}
    \begin{aligned}
          y_1(\var{\alpha})  = &\left(\lab{\bur{d}((\bur{fF})\cap(\bur{eED}))\bur{\widehat{A}}\bur{\widehat{C}}}+\lab{\bur{D}((\bur{fF})\cap(\bur{eEd}))\bur{\widehat{A}}\bur{\widehat{C}}}\right)^2 - \\&4\lab{\bur{D}((\bur{fF})\cap(\bur{eED}))\bur{\widehat{A}}\bur{\widehat{C}}}\lab{\bur{d}((\bur{fF})\cap(\bur{eEd}))\bur{\widehat{A}}\bur{\widehat{C}}}.
    \end{aligned}
\end{equation}

It isn't manifest, but these two elliptics are equivalent to each other by a birational transformation. This had to be true; the elliptic curve owes itself to the Jacobian of the seven cut, which is not dependent on the choice of chart. Nevertheless, it is instructive to verify that this is indeed the case. This proceeds by converting each of the elliptics to the so called Weierstrass form

\begin{equation}
    Q(\var{\alpha}) \longrightarrow  \var{\alpha}^3 - g_2 \var{\alpha}-g_3
\end{equation}
where the $g_2$ and $g_3$ end up being functions of the coefficients of the quartic (see appendix \ref{app:B}). The $j$-invariant is then defined according to

\begin{equation}
    j = \frac{g_2^3}{g_2^3-27g_3^2}.
\end{equation}
If the two $j$-invariant computing are identical, the two elliptic curves are as well. This can be tested for arbitrarily chosen momentum twistors using the {\tt{two\_loop\_amplitudes.m}} \cite{Bourjaily:2015jna} package for example\footnote{I am grateful to Cameron Langer for teaching me how to do this.}, and one can verify it analytically by simply computing the Jacobian instead.

There are a couple of degenerate cases, in which the elliptic breaks down into a square of quadratics or admits of another residue (or both). These cases occur when it becomes possible to apply another contour to move down to codimension zero in loop momentum space.

As such, this takes place when there are massless vertices, permitting access to a collinear limit, where both angle and square brackets of incident momenta go to zero. Indeed, this takes place under one of two conditions. The first of these is when one of the corners becomes massless, characterized by $\bur{a_i}$ and $\bur{a_{i+1}}$ or $\bur{b_i}$ and $\bur{b_{i+1}}$ become lightlike separated. A paradigmatic case is furnished by

\begin{equation}
    \includegraphics[valign = c, scale=0.3,trim=0 0 0 0cm]{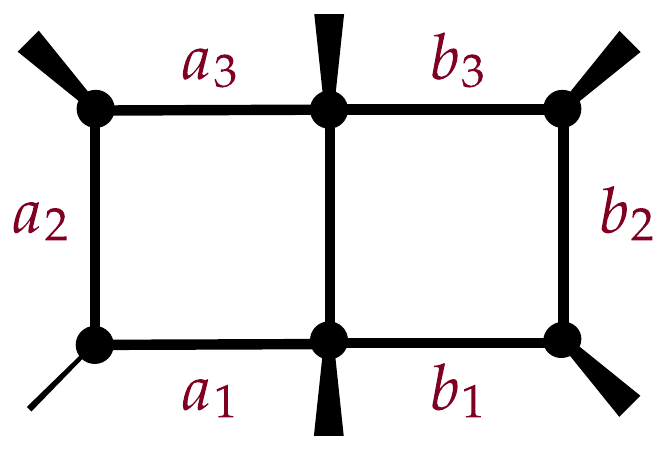} = \frac{\mathfrak{N}\dbar^4\ell_1\dbar^4\ell_2}{\lab{(\ell_1)\bur{Aa}}\lab{(\ell_1)\bur{ab}}\lab{(\ell_1)\bur{Cc}}\lab{(\ell_1)(\ell_2)}\lab{(\ell_2)\bur{Dd}}\lab{(\ell_2)\bur{Ee}}\lab{(\ell_2)\bur{Ff}}}
\end{equation}
which has been obtained by setting $\bur{B} = \bur{a}$ in momentum twistor space. It is most convenient now to use the form of the elliptic obtained by integrating out $\var{\alpha_1}$, and applying the limit thereto. Indeed, what we obtain as a consequence of this is the following,

\begin{equation}
    \lab{\bur{A}((\bur{cC})\cap(\bur{bBa}))\bur{\widehat{D}}\bur{\widehat{F}}} \underset{\bur{B}\rightarrow\bur{a}}{\longrightarrow}0,
\end{equation}
\begin{equation}
   \lab{\bur{a}((\bur{cC})\cap(\bur{bBa}))\bur{\widehat{D}}\bur{\widehat{F}}}\underset{\bur{B}\rightarrow\bur{a}}{\longrightarrow}0.
\end{equation}
This turns the quartic into the square of a quadratic identically

\begin{equation}
    y_{1}(\var{\alpha}) \underset{\bur{B}\rightarrow\bur{a}}{\longrightarrow} \lab{\bur{a}((\bur{cC})\cap(\bur{baA}))\bur{\widehat{D}}\bur{\widehat{F}}}^2.
\end{equation}

Factored out, we can take the even and odd residues with respect to the last variable due to the roots of the quadratic. 

This final set of residues is due to encircling the collinear limit of the three-point function spanned by the massless leg and its adjacent propagators; we represent this limit by a transparent circle in the corresponding on-shell function

\begin{equation}
    \includegraphics[valign = c, scale=0.3,trim=0 0 0 0cm]{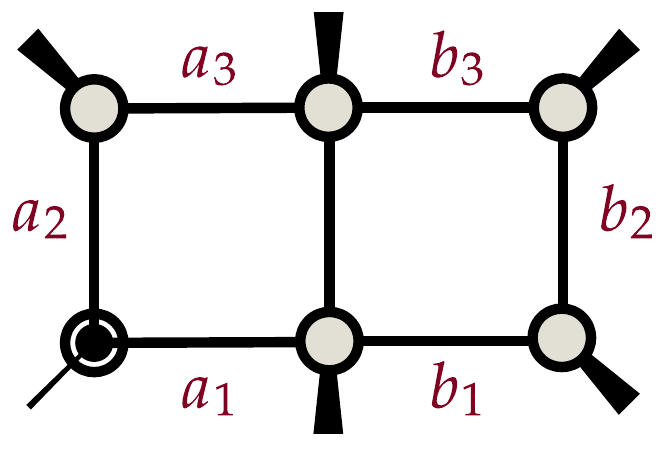}.
\end{equation}

A massless internal corner is recovered when one of two conditions is satisfied, namely the equivalence of either $\bur{a_1}$ and $\bur{b_1}$ or $\bur{a_3}$ and $\bur{b_3}$. The first case is representative and corresponds to writing $(\bur{Dd}) = (\bur{Aa})$. 

It is easiest to access this limit by simply setting $\bur{D} = \bur{A}$ and $\bur{d} = \bur{a}$; although the equivalence is ultimately between the corresponding lines spanned by the two pairs. This can be expressed in integrand form as given below

\begin{equation}
    \includegraphics[valign = c, scale=0.3,trim=0 0 0 0cm]{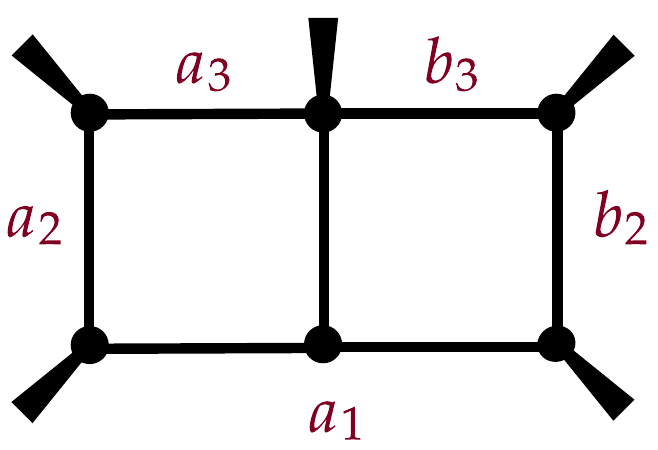} = \frac{\mathfrak{N}\dbar^4\ell_1\dbar^4\ell_2}{\lab{(\ell_1)\bur{Aa}}\lab{(\ell_1)\bur{Bb}}\lab{(\ell_1)\bur{Cc}}\lab{(\ell_1)(\ell_2)}\lab{(\ell_2)\bur{Aa}}\lab{(\ell_2)\bur{Ee}}\lab{(\ell_2)\bur{Ff}}}.
\end{equation}
This time, recourse may be had to the original condition for the seven cut, since it will be seen to deliver the correct residue prescription more naturally. Indeed, note that the seven cut prescription in this case turns out to enforce the following condition

\begin{equation}
\lab{\bur{\widehat{A}}\bur{\widehat{C}}\bur{\widehat{D}}\bur{\widehat{F}}} = \var{\alpha_2}\lab{\bur{A}\bur{\widehat{C}}\bur{a}\bur{\widehat{F}}}+\var{\alpha_1}\lab{\bur{a}\bur{\widehat{C}}\bur{A}\bur{\widehat{F}}} 
\end{equation}
which results in the factored form

\begin{equation}
 \underset{\text{seven cut}}{\oint}\includegraphics[valign = c, scale=0.3,trim=0 0 0 0cm]{doublebox-4-gon-massless-corner.pdf} =   \frac{\dbar^2\var{\alpha_i}}{(\var{\alpha_1}-\var{\alpha_2})\lab{\bur{A}\bur{\widehat{C}}\bur{a}\bur{\widehat{F}}}}
\end{equation}
which has the effect of rendering the encirclement of another residue trivial.

Once again, this choice of leading singularity can be represented by an on-shell diagram computing the maximal cut; we indicate the collinear limit by placing a transparent circle on the internal vertex according to

\begin{equation}
    \includegraphics[valign = c, scale=0.3,trim=0 0 0 0cm]{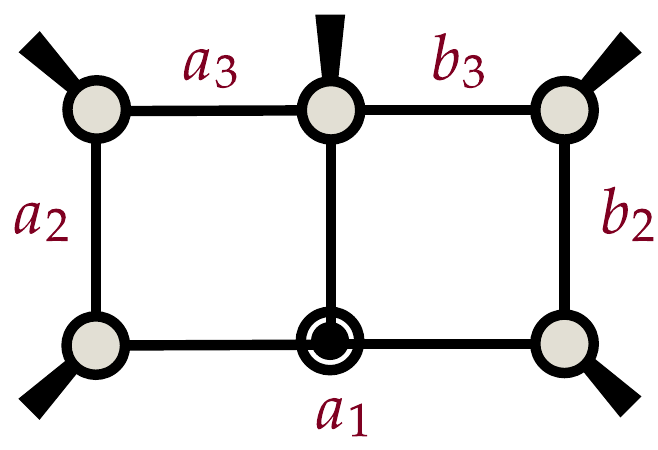}
\end{equation}
As a stray comment, we record here that the collinear limit can give way to a soft limit instead when two adjacent legs are massless. Indeed, the internal leg being on-shell and massless enforces the collinearity of the adjacent vertex as well. As such, we represent the matching of such limits in terms of the following on-shell diagram

\begin{equation}
    \includegraphics[valign = c, scale=0.3,trim=0 0 0 0cm]{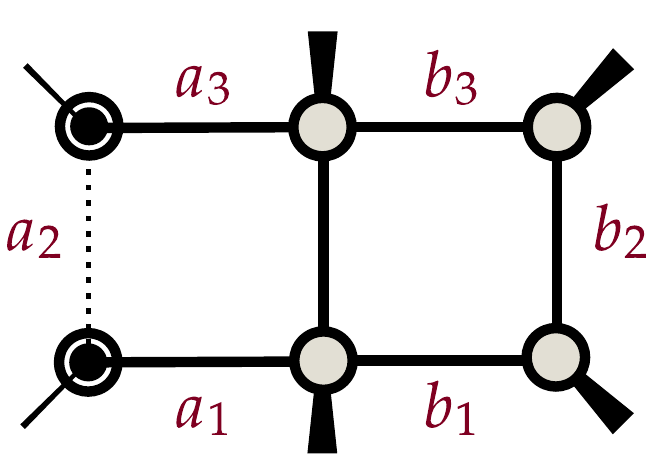}
\end{equation}
where it is to be understood that a dotted leg indicates the corresponding soft limit. 

Another point to note is that although the elliptic double box is the only case of interest (in the sense of containing elliptic singularities) in $4$-gon power counting, the situation is more complicated when we generalize to $3$-gon power counting, where we have box-triangles. To see that they contain elliptic curves as well, note their form in momentum twistor space

\begin{equation}
    \includegraphics[valign = c, scale=0.3,trim=0 0 0 0cm]{box-triangle-3-gon-labelled.pdf} = \frac{\mathfrak{N}\lab{(\ell_2)(\bl{N})}\dbar^4\ell_1\dbar^4\ell_2}{\lab{(\ell_1)\bur{Aa}}\lab{(\ell_1)\bur{Bb}}\lab{(\ell_1)\bl{Xx}}\lab{(\ell_1)(\ell_2)}\lab{(\ell_2)\bur{Dd}}\lab{(\ell_2)\bur{Ee}}\lab{(\ell_2)\bur{Ff}}\lab{(\ell_2)\bl{Xx}}}.
\end{equation}
The elliptic curve may be reached in this case by cutting instead of $\lab{(\ell_1)\bur{cC}}$ the propagator $\lab{(\ell_1)\bl{xX}}$ (where we have used the line $\bl{xX}$, with two arbitrary points $\bl{x}$ and $\bl{X}$---abusing notation in the process---to represent it). Note also that the box-triangle topology leads to encounters with elliptic singularities before we have occasion to see them in the double box; already at 8 particles, there is a leg distribution such that no collinear limit is permitted.

The primary takeaway in all of this is absent certain specific leg distributions, we are going to have to deal with the rise of elliptic singularities, especially as we move to higher and higher multiplicities. Accordingly, it becomes important to canonically determine how to prepare contours that can capture in some intrinsic fashion such singularities, without compromising a fair contention with them, as evaluation onto points would have. In the next section, we will motivate and expand upon the necessary generalization of leading singularities to their elliptic counterparts.

\subsection{Period Integrals and Elliptic Leading Singularities}\label{sec:3.2}
In this section, we will present a few technical results on elliptic functions and period integrals in order to prepare the necessary framework and notation for the main topic of discussion, namely the idea of an \emph{elliptic leading singularity}.

We start with the elliptic curve on $\mathbb{CP}^{1}$, on which we use a chart tracked by the inhomogeneous coordinate $\var{\alpha}$. We use the quartic form of the curve, in the case of which it is governed by the equation

\begin{equation}
   E: y^{2} = Q(\var{\alpha})
\end{equation}
where 

\begin{equation}
    Q(\var{\alpha}) = (\var{\alpha}-\bur{r_1})(\var{\alpha}-\bur{r_2})(\var{\alpha}-\bur{r_3})(\var{\alpha}-\bur{r_4}).
\end{equation}
We will present the technical results of elliptic functions we require for general choices of the roots, but for positive kinematics, they roots are always in complex conjugate pairs; we canonically take these pairs to be $\bur{r_{1,2}}$ and $\bur{r_{3,4}}$. There is also a canonical ordering of the roots we have chosen; $\mathfrak{R}(\bur{r_1})>\mathfrak{R}(\bur{r_3})$ and $\mathfrak{I}(\bur{r_{1,3}})>0$, which allow for a definition of a canonical cross-ratio which is always real:

\begin{equation}
    \varphi = \frac{(\bur{r_2}-\bur{r_1})(\bur{r_3}-\bur{r_4})}{(\bur{r_2}-\bur{r_3})(\bur{r_1}-\bur{r_4})}
\end{equation}

Due to the fact that we have for the elliptic curve $E$ a middle dimensional homology group of dimension $2$:

\begin{equation}
    H^1(E) = 2
\end{equation}
any differential form defined on the elliptic curve can be naturally paired with one of two cycles, generally called the $a$ and $b$ cycle. In our case, we use the convention that the $a$-cycle encircles the roots $\bur{r_1}$ and $\bur{r_2}$ and the $b$-cycle encircles the roots $\bur{r_1}$ and $\bur{r_3}$; specifically, they traverse the branch cut present---on account of the square root---between these pairs. This convention leads to the following intersection number $\lab{|}:H^{1}(E)\times H^{1}(E)\rightarrow \mathbb{C}$ between the cycles, denoted by $\mathfrak{A}$ and $\mathfrak{B}$ respectively

\begin{equation}
    \lab{\mathfrak{A}|\mathfrak{B}} = 1
\end{equation}
with any other pair resulting in zero.

This rank $2$ period matrix can be dualized in cohomology by pairing the cycles with forms instead, which will enable us to make contact with the differential form picture of local integrands in amplitude bases. 

A general, non-diagonal basis of forms can be obtained by simply taking the following two cohomology representatives\footnote{As in the case of Feynman integrands, these cohomology representatives are defined according to an equivalence relation that shifts them by a total derivative. Such shifts evaluate to zero when intersected with the (closed) $a$- and $b$-cycles.} on the curve $E$

\begin{equation}
    \omega(\var{\alpha}) = \frac{\dbar\var{\alpha}}{y(\var{\alpha})}
\end{equation}
and

\begin{equation}
    \omega_{r}(\var{\alpha}) = \frac{y(\bur{r})\dbar\var{\alpha}}{(\var{\alpha}-\bur{r})y(\var{\alpha})}.
\end{equation}
The period matrix generated by this choice of cycles and forms is generally dense, and is filled with entries built out of so-called \emph{elliptic functions}. For the $a$-cycle, to take the first case, we find the following two functions

\begin{equation}\label{eq:3.40}
    \underset{a\text{-cycle}}{\oint}\omega(\var{\alpha}) = \frac{2i}{\pi\sqrt{\bur{r_{32}}\bur{r_{41}}}}K[\varphi]:=\braket{\mathfrak{A}|\omega}
\end{equation}
and

\begin{equation}\label{ee:3.41}
    \underset{a\text{-cycle}}{\oint}\omega_{r}(\var{\alpha})  = \frac{2i}{\pi\sqrt{\bur{r_{32}}\bur{r_{41}}}}\frac{y(\bur{r})}{(\bur{r_4}-\bur{r})}\left(K[\varphi]+\frac{\bur{r_{42}}}{(\bur{r_2}-\bur{r})}\Pi\Bigg[\frac{(\bur{r_4}-\bur{r})\bur{r_{21}}}{(\bur{r_2}-\bur{r})\bur{r_{41}}};\varphi\Bigg]\right):=\braket{\mathfrak{A}|\omega_{r}}.
\end{equation}
Here $K[]$ is a complete elliptic function of the first kind and $\Pi[;]$ is a complete elliptic function of the second kind. The $b$-cycle integrals follow more or less analogously, with the replacements

\begin{equation}
    \varphi \longrightarrow 1-\varphi
\end{equation}
and

\begin{equation}
    \bur{r_2}\longleftrightarrow\bur{r_3}.
\end{equation}
The integrals that we have just described evaluate what are called \emph{periods} of the defined differential forms, and are accordingly referred to as period integrals generally.

Generic choices of the root $\bur{r}$ will always render the period matrix full rank, and as such any differential form supported on the elliptic curve can be expanded in a diagonalized basis given by the rotation of the vector $(\ket{y},\ket{r})$ by its inverse. In the context of Feynman integrals, this means that any cut which is known to have elliptic support must evaluate to a nonzero number along one of the cycles and be expressible in terms of the basis vectors we have just derived. Conversely, any integrand that fails to obey this will necessarily be polylogarithmic, in a rigorous sense.

We underline these points to emphasize that the dual homology of a given Feynman integrand can be gleaned by studying their period integrals. Further, they provide a natural generalization of a leading singularity by analogizing the calculation of a residue to the computation of a period integral instead.

The analogy is derived from mapping a leading singularity to a period by relating similar cases in the double box. Going back to the double box seven cut containing a collinear limit

\begin{equation}
    \underset{\text{seven cut}}{\oint}\includegraphics[valign = c, scale=0.3,trim=0 0 0 0cm]{doublebox-4-gon-massless-corner.pdf} =   -i\frac{\dbar\var{\alpha}}{\lab{\bur{a}\bur{\widehat{C}}(\var{\alpha})\bur{A}\bur{\widehat{F}}(\var{\alpha})}}
\end{equation}
the cohomology group is spanned by a $d\log$ form supported on one of the poles in the denominator in $\var{\alpha}$ (the residue on the other is determined by the GRT in this example). The leading singularity is then a homological pairing between this form and the cycle encircling the pole.

Consistent with this notion of a leading singularity, we are motivated to construct the following general notion of an \emph{elliptic leading singularity} by the two period integrals:

\begin{equation}
    \mathfrak{e}_{a} = \underset{a\text{-cycle}}{\oint}\includegraphics[valign = c, scale=0.3,trim=0 0 0 0cm]{double-box-4-gon-top-level-on-shell.pdf} 
\end{equation}
on the $a$-cycle and 

\begin{equation}
    \mathfrak{e}_{b} = \underset{b\text{-cycle}}{\oint}\includegraphics[valign = c, scale=0.3,trim=0 0 0 0cm]{double-box-4-gon-top-level-on-shell.pdf} 
\end{equation}
on the $b$-cycle.

Since the elliptic leading singularities are defined using general on-shell functions, they will have different concrete expressions depending on the theory used, but a lot of common features can be derived. First, any on-shell function corresponding to this seven cut will have the following analytic expansion

\begin{equation}
    \includegraphics[valign = c, scale=0.3,trim=0 0 0 0cm]{double-box-4-gon-top-level-on-shell.pdf} = \frac{\dbar\var{\alpha}}{y(\var{\alpha})}\left(\mathfrak{b}_{0} + \sum_{\var{\alpha^{(i)}}} \frac{y(\var{\alpha^{(i)}})}{y(\var{\alpha})(\var{\alpha}-\var{\alpha^{(i)}})} \includegraphics[valign = c, scale=0.3,trim=0 0 0 0cm]{pentabox-4-gon-top-level-on-shell.pdf}\right)
\end{equation}
where in this case we have used $i$ schematically, to represent all possible further factorizations of the double box that result in the production of pentabox leading singularities, which are pure numbers. We remark that each term in this second sum has been appropriately normalized, so as to issue the correct pentabox leading singularity upon evaluation. An illustrative example of this fact is provided by the ten-point on-shell function due to the double box (that shows up for example in the N$^3$MHV sector in maximally sYM \cite{Bourjaily:2020hjv})

\begin{equation}
    \includegraphics[valign = c, scale=0.3,trim=0 0 0 0cm]{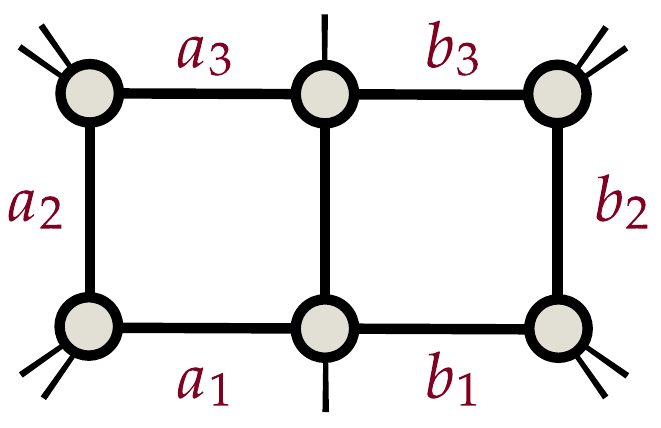}.
\end{equation}
In this ten point case specifically, we can find contained in this on-shell diagram a total of $24$ factorization channels, owing to four leading singularities for each of the factorizations possible in the diagram. An example can be found by taking the following case: the vertex between $\bur{b_3}$ and $\bur{b_2}$ once factorized results a total of four solutions; schematically, the on-shell functions take on the following appearance:

\begin{equation}
    \Bigg\lbrace{\includegraphics[valign = c, scale=0.3,trim=0 0 0 0cm]{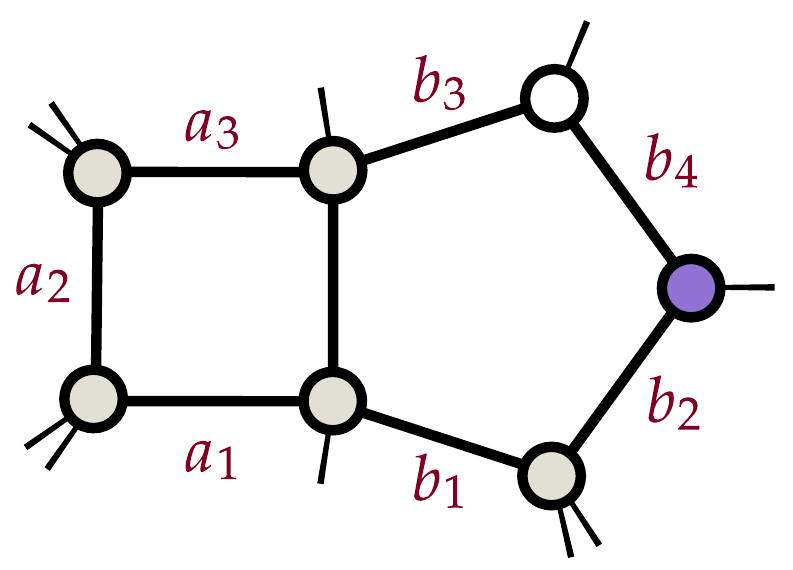}\;,\;\includegraphics[valign = c, scale=0.3,trim=0 0 0 0cm]{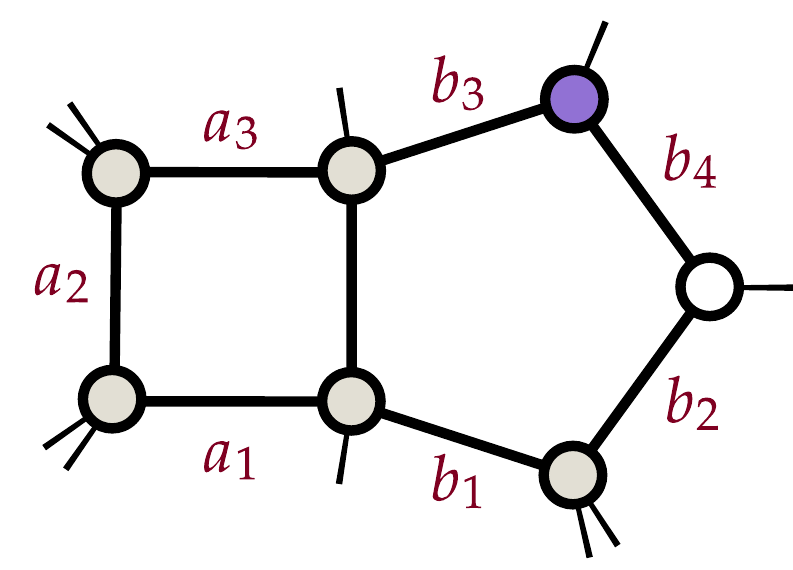}\Bigg\rbrace}
\end{equation}
where the colors indicate the parities of the solutions on the pentagon side of the integrand, each of which will then give rise to two solutions on the derived quad cut of the box.  Generically of course, the number of factorization channels will depend on the factorizations of all the subordinate amplitudes building the on-shell function.

We emphasize that this fact is theory-agnostic; regardless of the specific model at play, the double box subleading singularity will contain all of these possible degenerations, and as such the $a$-cycle period integral will be privy to this information. 

Indeed, the same kind of expansion holds in the case of box-triangles. Here, since the seven-cut was reached by cutting infinity, the terms in the expansion in terms of residues will include in reality all double boxes and pentatriangles that can be reached from the box-triangle, with the additional constraint being that the propagator at infinity is cut as well, rendering all of these `double boxes' and `pentatriangles' leading singularities.

Another common feature is the nature of the factor $\mathfrak{b}_{0}$. Accounting for poles at infinity in addition to those arising due to kinematic factorizations, expanding the elliptic cut in this manner means that although the quantity $\mathfrak{b}_{0}$ (in the double box and the box triangle by analogy) can always be written---trivially, yet importantly---as the following difference

\begin{equation}
    \mathfrak{b}_{0} = y(\var{\alpha})\includegraphics[valign = c, scale=0.3,trim=0 0 0 0cm]{double-box-4-gon-top-level-on-shell.pdf} - \sum_{i}y(\var{\alpha})\omega_{i}(\var{\alpha})\includegraphics[valign = c, scale=0.3,trim=0 0 0 0cm]{pentabox-4-gon-top-level-on-shell.pdf}).
\end{equation}
As a complex function of $\var{\alpha}$, since every pole of the double box is accounted for by one of the $\omega_{i}$, the quantity $\mathfrak{b}_{0}$---as a direct consequence of Liouvilles's theorem---must be independent in $\var{\alpha}$. Pedantically, this means that $\mathfrak{b}_{0}$ is \emph{always} equal to 

\begin{equation}
    \mathfrak{b}_{0} = \underset{\var{\alpha}=\var{\alpha^*}}{\text{eval}}y(\var{\alpha})\includegraphics[valign = c, scale=0.3,trim=0 0 0 0cm]{double-box-4-gon-top-level-on-shell.pdf} - \sum_{i}y(\var{\alpha})\omega_{i}(\var{\alpha^*})\includegraphics[valign = c, scale=0.3,trim=0 0 0 0cm]{pentabox-4-gon-top-level-on-shell.pdf})
\end{equation}
regardless of the choice of $\var{\alpha^*}$. We express such a choice by writing $\mathfrak{b}_{0}(\var{\alpha}\rightarrow\var{\alpha^*})$ in the expression for the elliptic leading singularities. An example of convenient choice turns out to be $\var{\alpha} = \bur{r_4}$, which has the effect of removing all terms of the form $K[\varphi]\mathfrak{pb}_{i}$ for all pentaboxes $\mathfrak{pb}_i$.

By using period integrals to transfer the technology of leading singularities to integrands which don't support loop momenta, the next step would be to generalize our the prescriptive representation of two loop amplitudes, first in $4$-gon power counting, making use of this new definition of leading singularities in the presence of elliptic curves.

\subsection{Diagonalizing on Elliptic Cuts}\label{sec:3.3}
The decay of the double box into purely elliptic (definite rigidity) and polylogarithmic pieces alongside the new definition of elliptic leading singularities suggest that amplitudes can be matched in one of two ways. Either one can match double boxes to the elliptic LS and prescriptively constrain the higher topologies, or we can match term by term. This amounts to either matching the period integrals, or the differential forms directly, which we have referred to as homological and cohomological diagonalization respectively.

Both methods are instructive to perform concretely, and we'll go over them in turn. We will continue to stick to planar two loops and restrict our attention for now to the $4$-gon basis.

\subsubsection{Prescriptivity by Homology}
Since elliptic leading singularities are pure numbers, entirely similar in this regard to ordinary leading singularities, an integrand that has support on the corresponding elliptic curve can be normalized along this singularity. A somewhat rudimentary, but illustrative example is provided even at the level of ordinary $\varphi^4$ theory, where the elliptic double box is just a term in the Feynman expansion (which coincides with the unitarity basis in this case).

The elliptic leading singularity (say along the $a$-cycle) here---in the case of the paradigmatic ten-particle graph of the preceding section---amounts to the following period integral

\begin{equation}
 \underset{a\text{-cycle}}{\oint}\includegraphics[valign = c, scale=0.3,trim=0 0 0 0cm]{ten-particle-doublebox-onshell-example-1.pdf}\Bigg|_{\varphi^{4}} = -i\underset{a\text{-cycle}}{\oint}\frac{\dbar \var{\alpha}}{y(\var{\alpha})} :=\mathfrak{e}^{\varphi^4}_{a}.
\end{equation}
The monomial nature of the on-shell function is due to the fact that the four point vertex doesn't factorize in the theory we've chosen. Indeed, the corresponding scalar graph will simply have to be normalized by a pure number

\begin{equation}
    \includegraphics[valign = c, scale=0.3,trim=0 0 0 0cm]{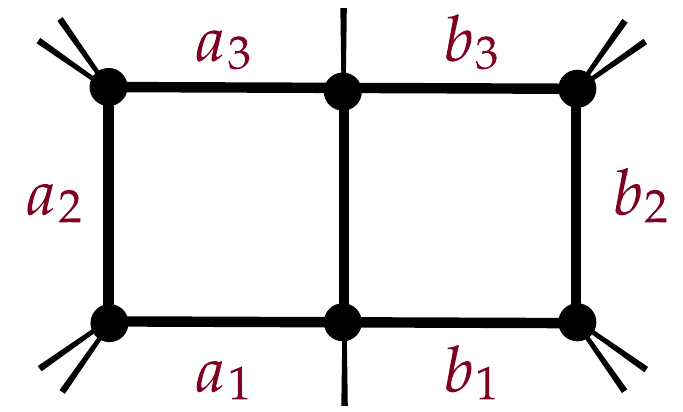}\longrightarrow \frac{1}{\mathfrak{e}^{\varphi^4}_{a}}\includegraphics[valign = c, scale=0.3,trim=0 0 0 0cm]{ten-particle-doublebox-scalar-example-1.pdf.pdf}
\end{equation}
and its coefficient will have to be chosen to be $\mathfrak{e}^{\varphi^4}_{a}$.

This example seems to trivialize the problem, but it captures the essence of what we want to do. Each double box is to be first normalized by its period integral on the $a$-cycle, which is simply the scalar quantity we have just computed in the preceding discussion. Then, it is to be dressed with the period integral of the corresponding \emph{on-shell function}, so as to ensure matching to field theory.

To do this, we start again with the $4$-gon basis at two loops, spanned by scalar boxes, pentaboxes with one numerator and double pentagons with two

\begin{equation}
    \Bigg\lbrace{\includegraphics[valign = c, scale=0.3,trim=0 0 0 0cm]{doublebox-4-gon.pdf}\;,\;\includegraphics[valign = c, scale=0.3,trim=0 0 0 0cm]{pentabox-4-gon.pdf}\;,\;\includegraphics[valign = c, scale=0.3,trim=0 0 0 0cm]{doublepentagon-full-4-gon.pdf}\Bigg\rbrace}.
\end{equation}
Only the double boxes present a challenge with elliptic singularities. Accordingly, nothings changes vis a vis the analysis in section \ref{sec:2.4.1} with regard to the double pentagons and the pentaboxes. All the matching is done in the same way in these two cases. 

Starting with the double pentagons, they are matched as usual to the following four kissing box on-shell functions, corresponding to four solutions of the cut equations:

\begin{equation}
    \includegraphics[valign = c, scale=0.3,trim=0 0 0 0cm]{doublepentagon-4-gon-top-level-on-shell.pdf}.
\end{equation}

Once again, despite there not being enough degrees of freedom to match all of the pentabox leading singularities to pentabox integrands, we can choose two arbitrarily as follows

\begin{equation}
     \includegraphics[valign = c, scale=0.3,trim=0 0 0 0cm]{pentabox-4-gon-top-level-on-shell.pdf}.
\end{equation}

Given these cuts, it is important for the next step that they are done prescriptively. The double pentagon numerators must be chosen so that they vanish on all pairs of pentabox cuts their daughters are matched onto. Assuming we have done this, we introduce the following notation

\begin{equation}
    \begin{aligned}
        \mathcal{A}_{\text{pb+db}} = &\sum  \includegraphics[valign = c, scale=0.3,trim=0 0 0 0cm]{pentabox-4-gon-top-level-on-shell.pdf}\times\includegraphics[valign = c, scale=0.3,trim=0 0 0 0cm]{pentabox-4-gon-top-level.pdf}  + \\
        &\sum \includegraphics[valign = c, scale=0.3,trim=0 0 0 0cm]{doublepentagon-4-gon-top-level-on-shell.pdf} \times \includegraphics[valign = c, scale=0.3,trim=0 0 0 0cm]{doublepentagon-4-gon-top-level.pdf}
    \end{aligned}
\end{equation}
where we emphasize that the double boxes \emph{have not} been fixed yet. The notation introduced is just to simplify writing for the next step, which will fix the double boxes.

To compute the double box coefficient and prescriptively match the entire amplitude, we take any generic double box in the basis and normalize it against the $a$-cycle of its seven-cut by writing

\begin{equation}
    \includegraphics[valign = c, scale=0.3,trim=0 0 0 0cm]{doublebox-4-gon.pdf} \longrightarrow \frac{1}{\lab{\mathfrak{A}|\omega}} \includegraphics[valign = c, scale=0.3,trim=0 0 0 0cm]{doublebox-4-gon.pdf}
\end{equation}
where we have just schematically used $\omega$ to denote the form on the seven-cut. The amplitude is matched to field theory by dressing this box with the period integral

\begin{equation}
    \underset{a-\text{cycle}}{\oint}\includegraphics[valign = c, scale=0.3,trim=0 0 0 0cm]{double-box-4-gon-top-level-on-shell.pdf} = \mathfrak{b}_{0}\lab{\mathfrak{A}|\omega} + \sum_{i}\lab{\mathfrak{A}|\omega_r}\mathfrak{pb}_{i}
\end{equation}
where $\mathfrak{pb}_i$ has been used to denote all pentaboxes that can be recovered from the particular seven cut under consideration. The next step is to render the matching to each $a$-cycle prescriptive. 

To do this, we note that once a specific seven cut is chosen, it isolates the contribution of that particular double box. As such, any subtractions may be performed at this stage in terms of the corresponding $\omega$-forms, as they are in one-to-one correspondence with double box elements in the basis. We emphasize that this would not have been the case if say we wanted to subtract out pentabox integrands while making reference to seven cuts, as given any seven cut, a large number of pentaboxes would be supported thereon.

That said, let us see what happens when a specific seven cut is taken of the amplitude so far. Indeed, we obtain the following schematic expression

\begin{equation}
\begin{aligned}
     &\frac{1}{\lab{\mathfrak{A}|\omega}}\left(\mathfrak{b}_{0}\braket{\mathfrak{A}|\omega} + \sum_{i}\braket{\mathfrak{A}|\omega_{r_i}}\mathfrak{pb}_{i}\right)\ket{\omega} + \underset{\text{seven-cut}}{\oint}\mathcal{A}_{\text{pb+db}} =  \\
     &\frac{1}{\lab{\mathfrak{A}|\omega}}\left(\mathfrak{b}_{0}\braket{\mathfrak{A}|\omega} + \sum_{i}\braket{\mathfrak{A}|\omega_{r_i}}\mathfrak{pb}_{i}\right)\ket{\omega} + \sum_{i}\ket{\omega_{r_i}}\mathfrak{pb}_{i}.
\end{aligned}
\end{equation}
In this form, the subtraction simply devolves upon shifting the quantity $\mathcal{A}_{\text{pb+db}}$ by a number of contact terms corresponding to double boxes, suitably normalized to cancel their pentabox factorizations:

\begin{equation}
    \mathcal{A}_{\text{pb+db}} \longrightarrow \mathcal{A}_{\text{pb+db}} - \sum \frac{1}{\lab{\mathfrak{A}|\omega}} \includegraphics[valign = c, scale=0.3,trim=0 0 0 0cm]{doublebox-4-gon.pdf}\sum_{i}\lab{\mathfrak{A}|\omega_{r_i}}\mathfrak{pb}_{i}.
\end{equation}
The notation may appear cumbersome, but we have simply performed a summation over all double box topologies, each of which is dressed with on-shell functions of each pentabox factorization, and normalization factors designed to eliminate the contributions of pentaboxes and double pentagons to the $a$-cycle of each. Indeed, the immediate implication of the latter shift is to render the seven cut of a particular double box of the form

\begin{equation}
\begin{aligned}
     &\frac{1}{\lab{\mathfrak{A}|\omega}}\left(\mathfrak{b}_{0}\braket{\mathfrak{A}|\omega} + \sum_{i}\braket{\mathfrak{A}|\omega_{r_i}}\mathfrak{pb}_{i}\right)\ket{\omega} + \underset{\text{seven-cut}}{\oint}\mathcal{A}_{\text{pb+db}} =  \\
     &\frac{1}{\lab{\mathfrak{A}|\omega}}\left(\mathfrak{b}_{0}\braket{\mathfrak{A}|\omega} + \sum_{i}\braket{\mathfrak{A}|\omega_{r_i}}\mathfrak{pb}_{i}\right)\ket{\omega} + \sum_{i}\left(\ket{\omega_{r_i}} - \frac{\lab{\mathfrak{A}|\omega_{r_i}}}{\lab{\mathfrak{A}|\omega}}\ket{\omega}\right)\mathfrak{pb}_{i}.
\end{aligned}
\end{equation}
Evaluating the $a$-cycle period by dotting with $\bra{\mathfrak{A}}$, one can immediately check that all pentabox and double pentagon contributions are exterminated.

Prescriptivity and completeness conspire to guarantee that other period integrals are matched as well. An understanding of this morally can be reached by making reference to the previous case of the pentabox leading singularities. Although we could only match two out of four leading singularities to numerators of the pentaboxes themselves, correctly matching the double pentagons ensured that the sum would always result in the right leading singularity when taken as a whole. This is just a reflection of the fact that the basis is complete (illustrated for example in the package {\texttt{two\_loop\_amplitudes.m}} for $\mathcal{N}$=4 sYM).

Indeed, we can check that the computation of the $b$-cycle period integral, obtained by dotting with $\bra{\mathfrak{B}}$ results in the following

\begin{equation}
    \frac{1}{\lab{\mathfrak{A}|\omega}}\left(\mathfrak{b}_{0}\braket{\mathfrak{A}|\omega} + \sum_{i}\braket{\mathfrak{A}|\omega_{r_i}}\mathfrak{pb}_{i}\right)\braket{\mathfrak{B}|\omega} + \sum_{i}\left(\braket{\mathfrak{B}|\omega_{r_i}} - \frac{\lab{\mathfrak{A}|\omega_{r_i}}}{\lab{\mathfrak{A}|\omega}}\braket{\mathfrak{B}|\omega}\right)\mathfrak{pb}_{i}
\end{equation}
which simplifies to

\begin{equation}
    \mathfrak{b}_{0}\braket{\mathfrak{B}|\omega} + \sum_{i}\braket{\mathfrak{B}|\omega_{r_i}}\mathfrak{pb}_i = \mathfrak{e}_b
\end{equation}
as required.

This completes our discussion of diagonalization by homology, which amounts to a simple generalization of prescriptive unitarity as such to our new notion of elliptic leading singularities. Before moving on to the case of diagonalization along cohomology, we will go over our simple ten particle example again, just for sake of illustration.

\paragraph*{Example 3.1. } Consider the following six-dimensional subspace of the basis for ten particles at $4$-gon power counting:

\begin{equation}
    \lbrace{\mathcal{I}_1,\mathcal{I}_2\rbrace} = \Bigg\lbrace\includegraphics[valign = c, scale=0.3,trim=0 0 0 0cm]{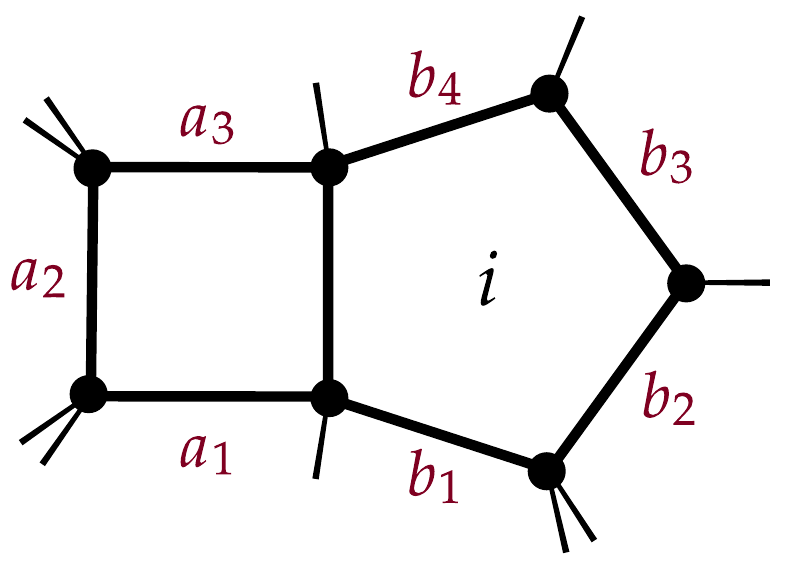}\Bigg\rbrace
\end{equation}
and

\begin{equation}
    \lbrace{\mathcal{I}_3,\dots,\mathcal{I}_6\rbrace} = \Bigg\lbrace\includegraphics[valign = c, scale=0.3,trim=0 0 0 0cm]{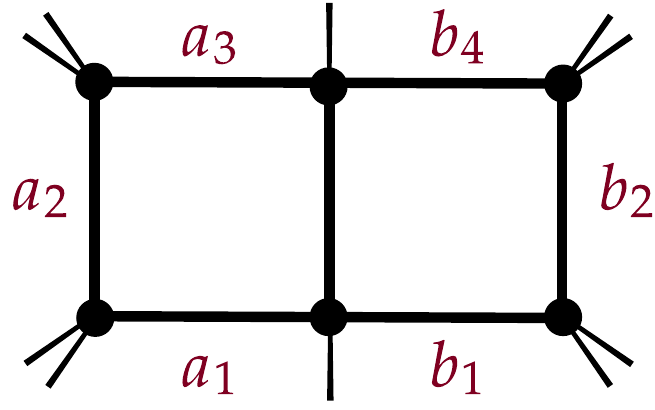}\;,\;\includegraphics[valign = c, scale=0.3,trim=0 0 0 0cm]{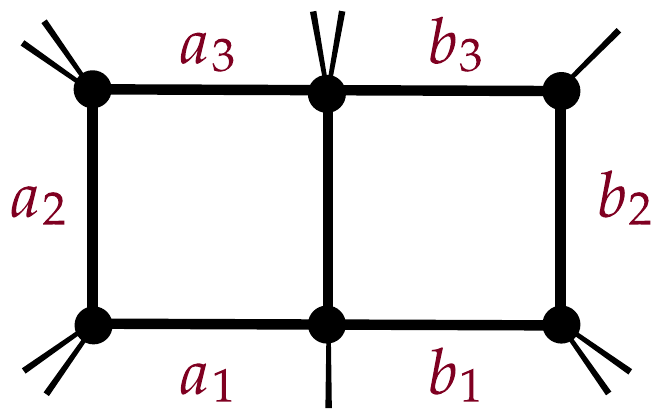}\;,\;\includegraphics[valign = c, scale=0.3,trim=0 0 0 0cm]{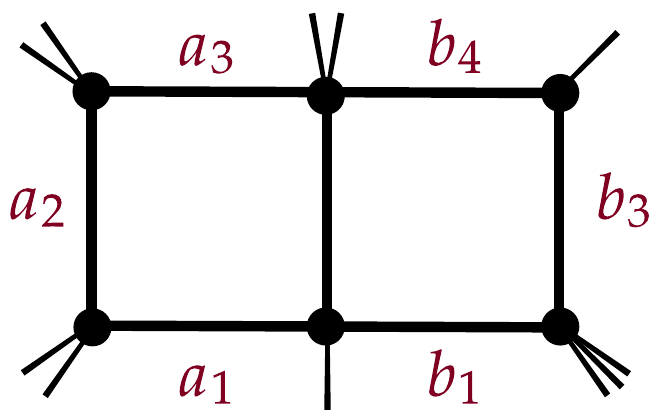}\;,\;\includegraphics[valign = c, scale=0.3,trim=0 0 0 0cm]{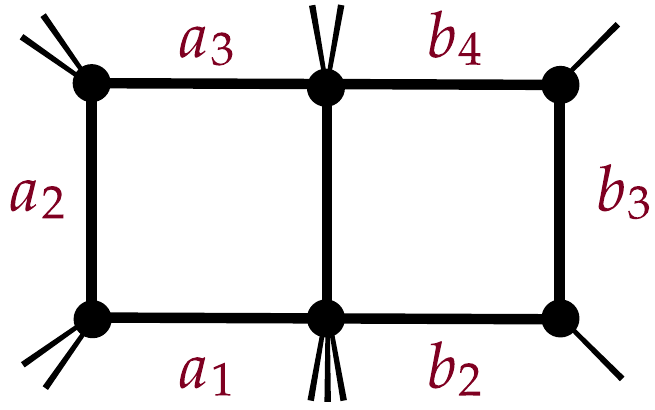}\Bigg\rbrace.
\end{equation}

To match the pentaboxes, we choose the following two solutions to the octacut equations

\begin{equation}
    \lbrace{\Omega_1,\Omega_2\rbrace} = \Bigg\lbrace\includegraphics[valign = c, scale=0.3,trim=0 0 0 0cm]{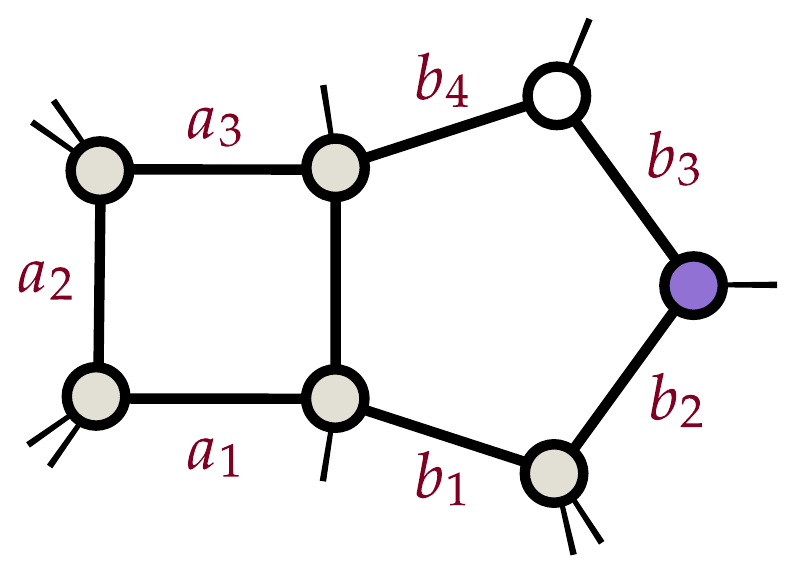}\;,\;\includegraphics[valign = c, scale=0.3,trim=0 0 0 0cm]{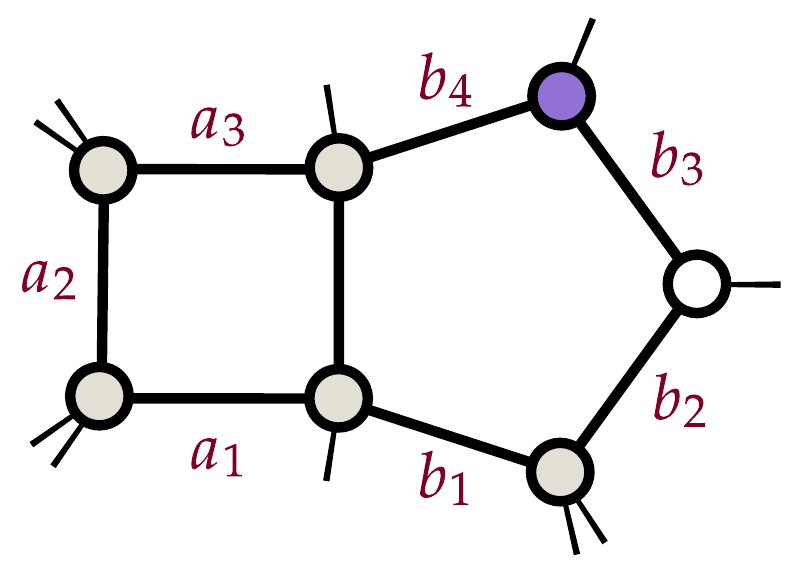}\Bigg\rbrace
\end{equation}
so that the collinear solutions are matched by making reference to completeness via kissing boxes. Now for the double boxes, note that of the integrands $\mathcal{I}_3$ through $\mathcal{I}_4$, only the first contains elliptics. Indeed, the others admit of collinear or soft limits, and we pick accordingly 

\begin{equation}
    \lbrace{\Omega_3,\dots,\Omega_6\rbrace} = \Bigg\lbrace\underset{a-\text{cycle}}{\includegraphics[valign = c, scale=0.3,trim=0 0 0 0cm]{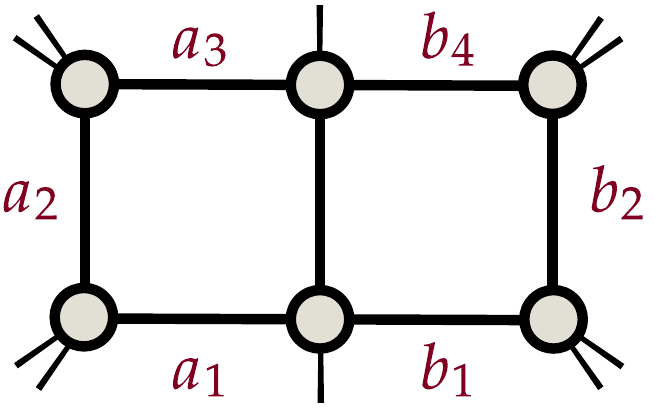}}\;,\;\includegraphics[valign = c, scale=0.3,trim=0 0 0 0cm]{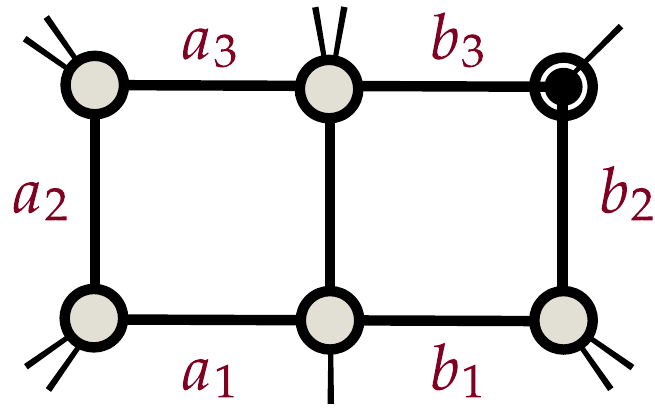}\;,\;\includegraphics[valign = c, scale=0.3,trim=0 0 0 0cm]{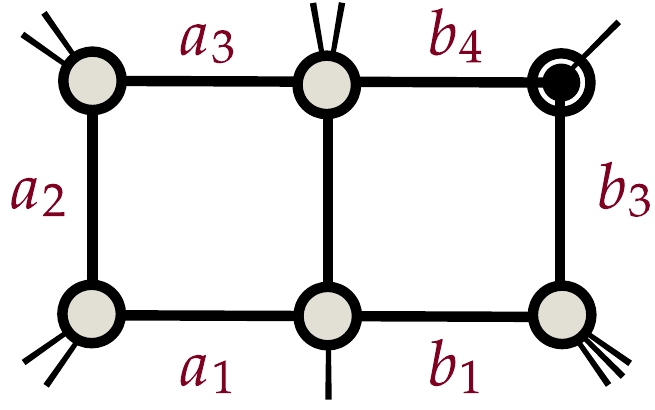}\;,\;\includegraphics[valign = c, scale=0.3,trim=0 0 0 0cm]{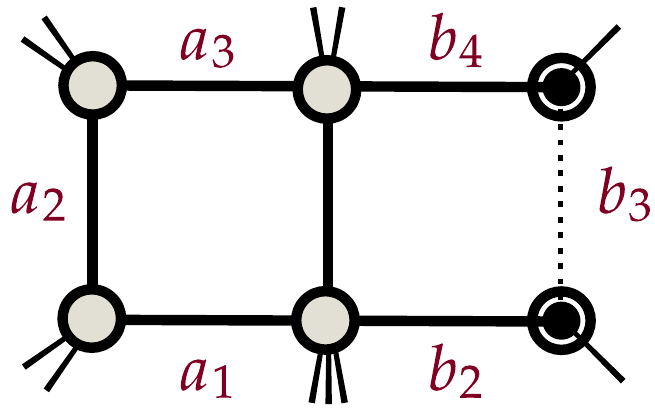}\Bigg\rbrace
\end{equation}
where we have just represented the $a$-cycle period by a subscript.

The period matrix here (with the pentabox numerators chosen to diagonalize the cuts at the outset) turns out to be

\begin{equation}
    \begin{blockarray}{ccccccc}
 \hspace{1pt}&\Omega_1 & \Omega_2 & \Omega_3 & \Omega_4 & \Omega_5 &\Omega_6\\
\begin{block}{c(cccccc)}
  \mathcal{I}_1 & \bl{f}^{\bur{1}}_{\bl{1}} & 0 & \bl{g}^{\bur{3}}_{\bl{1}} & \bl{g}^{\bur{4}}_{\bl{1}} & \bl{g}^{\bur{5}}_{\bl{1}} &\bl{g}^{\bur{6}}_{\bl{1}}\\
  \mathcal{I}_2 & 0 & \bl{f}^{\bur{2}}_{\bl{2}} & \bl{g}^{\bur{3}}_{\bl{2}} & \bl{g}^{\bur{4}}_{\bl{2}} & \bl{g}^{\bur{5}}_{\bl{2}} &\bl{g}^{\bur{6}}_{\bl{2}}\\
  \mathcal{I}_3 & 0 & 0 & \bl{h}^{\bur{3}}_{\bl{3}} & 0 & 0 &0\\
  \mathcal{I}_4 & 0 & 0 & 0 & \bl{h}^{\bur{4}}_{\bl{4}} & 0 &0\\
  \mathcal{I}_5 & 0 & 0 & 0 & 0 & \bl{h}^{\bur{5}}_{\bl{5}} &0\\
  \mathcal{I}_6 & 0 & 0 & 0 & 0 & 0 &\bl{h}^{\bur{6}}_{\bl{6}}\\
\end{block}
\end{blockarray}
\end{equation}
very easily seen to be of full rank, and hence diagonalizable.

\hfill
\vspace{0.7cm}

The choice of the $a$-cycle is consistent with the contour prescriptions encircling the collinear limits when external masses are taken to zero. Indeed, the $a$-cycle contour degenerates into a residue prescription that correctly encloses the poles formed from the quartic, which ultimately amount to the collinear limits. We could have simply written down `$a$-cycle' for each of the diagrams in the last example, just remembering what that means when the poles of the quartic collide.

Parenthetically, we remark that in this particular case, the choice to match one of the pentaboxes to another of the elliptic cycles could have been made, thus diagonalizing the entire space into rigid and non-rigid parts. While we will go over this solution in greater detail in the next section, this is a general phenomenon of interest, but only appears as a marginal case here. Diagonalizing along rigidity will form the major point of discussion of section \ref{sec:4}.

\subsubsection{Prescriptivity by Cohomology}
Diagonalizing by prescribing the correct matching onto cuts ensures---given completeness---that a spanning set of cuts will always supply an amplitude that is matched on all cuts. While this is guaranteed to work as long as the basis is a complete (or overcomplete) one, there is a way of making this matching precise by matching the forms directly. We'll briefly go over how this is to be done.

In homological diagonalization, we matched the \emph{period integral} of the on-shell function corresponding to a given seven-cut. Cohomological diagonalization proceeds instead by matching the on-shell function as a differential form instead. Suppose we have already matched all of the pentaboxes and the double pentagons, so that all codimension zero residues are accounted for. This will ensure that we have the equivalence

\begin{equation}
    \underset{\text{seven-cut}}{\oint}\mathcal{A}_{\text{pb+db}} = \sum_{i}\ket{\omega_{r_{i}}}\mathfrak{pb}_{i},
\end{equation}
which actually matches the entire pole-dependent part of the double box on-shell function:

\begin{equation}
    \includegraphics[valign = c, scale=0.3,trim=0 0 0 0cm]{double-box-4-gon-top-level-on-shell.pdf}  = \mathfrak{b}_{0}\ket{\omega} + \sum_{i}\ket{\omega_{r_{i}}}\mathfrak{pb}_{i}.
\end{equation}
This means that in order to match the on-shell function \emph{at every point} in $\var{\alpha}$ space, we just have to ensure that the double boxes appear with coefficients that guarantee matching with the pole-independent piece upon evaluation on the seven cut. Indeed, for a scalar double box (suppressing kinematic and Jacobian normalization factors), we can ensure that this is the case by simply writing

\begin{equation}
    \includegraphics[valign = c, scale=0.3,trim=0 0 0 0cm]{double-box-4-gon-top-level.pdf} \longrightarrow \mathfrak{b}_{0}\includegraphics[valign = c, scale=0.3,trim=0 0 0 0cm]{double-box-4-gon-top-level.pdf}
\end{equation}
which is a consistent statement due to the previous observation; the numerator $\mathfrak{b}_{0}$ is independent of $\var{\alpha}$ due to Liouville's theorem.

The evident simplification offered by this means of diagonalization is apparent in that very little effort is required to make it prescriptive. Since the matching essentially descends from matching higher point topologies, only the coefficient of the double box itself is left undetermined, a demand that is met by the analytic structure of the on-shell function.

That said, cohomological diagonalization obscures other features of the coefficients one may want to highlight. First, the coefficients determined in this manner render the double box impure in the sense of \cite{Broedel:2018qkq}, in that their derivatives post-integration cannot be expressed in terms of pure functions. Further, in the case of maximally supersymmetric Yang-Mills, the double box coefficient will not be Yangian invariant in cohomological diagonalization. In the former case, since the elliptic periods simply encircle compact contours inside the Grassmannian, Yangian invariance of the $a$- and $b$-cycle period integrals are guaranteed identically.

Before moving on, we should like to point out that all of this technology would be expected to import in similar analytic fashion even for non-elliptic singularities. There, the main bottleneck would not be in generalizing a leading singularity \emph{per se}, but in actually being able to compute the period integrals themselves.
\clearpage
\section{Stratifying Bases by Rigidity}\label{sec:4}
In the last section, we studied cuts of codimension greater that 0 in the presence of non-polylogarithmic singularities, with an emphasis on so-called elliptic singularities which are known to arise at two loops. By exploiting the duality between unitarity cuts and cohomology in loop momentum space, we generalized the notion of a leading singularity to integrands where codimension 0 contours did not exist in a strict sense.

One of the upshots of using rigidity as a means of classifying Feynman integrands is that it offers another means to discriminate subspaces of bases from one another and resolve them in this fashion. Take for example the issue of IR finiteness. At one loop, it is possible to write down a set of contours (see for example \cite{Bourjaily:2021ujs} for a detailed discussion of these contours) such that \emph{any} element in the master integrand basis is either IR finite or IR divergent

Since there is no overlap between these sectors---a fact guaranteed by prescriptivity---any one loop amplitude that has at worst $2$-gon power counting can always be expressed in the schematic form

\begin{equation}
    \mathcal{A}^{\text{1-loop}} = \sum f_{\text{fin}}\mathcal{I}_{\text{fin}} + \sum f_{\text{div}}\mathcal{I}_{\text{div}}.
\end{equation}

The completeness of this basis and the separation wrought between finite and divergent parts makes it possible to write down local integrand representations for any quantity at one loop in a way that manifests if and how they express IR divergence. The question we would now like to ask is whether or not this kind of stratification and resolution is possible instead on the basis of \emph{rigidity}. In other words, can we write down a master integrand basis that clearly separates into two disjoint subspaces: one that contains only elements of pure rigidity 0, and one that has elements of pure rigidity 1 (staying at planar two loops)? 

The primary goal of the present section will be to emphasize the interplay between bases building and contour prescriptions that results from trying to address this question, and to highlight the compromise that has to be accepted between picking a small or otherwise simple basis versus enlarging it if one wishes to manifest resolutions along rigidity. 

\subsection{Local Integrands and Multiple Elliptic Curves}\label{sec:4.1}
Local integrands forms of amplitudes at two loops make manifest the fact that rigidity is often expressed in terms of more than one elliptic curve.

In previous examples, we focused on specific cuts, each of which furnished a single elliptic curve. Working backward however, it is easy to see that an integrand that supports an elliptic curve will generally have more than one. The contrast between an integrand supporting one versus several elliptic curves can be seen by comparing the scalar double box

\begin{equation}\label{eq:4.2}
    \includegraphics[valign = c, scale=0.3,trim=0 0 0 0cm]{doublebox-4-gon-labelled.pdf} = \frac{\dbar^4\ell_1 \dbar^4\ell_2}{(\ell|\bur{a_1})(\ell|\bur{a_2})(\ell|\bur{a_3})(\ell_1|\ell_2)(\ell_2|\bur{b_1})(\ell_2|\bur{b_2})(\ell_2|\bur{b_3})}
\end{equation}
to the `scalar' pentabox

\begin{equation}
    \includegraphics[valign = c, scale=0.3,trim=0 0 0 0cm]{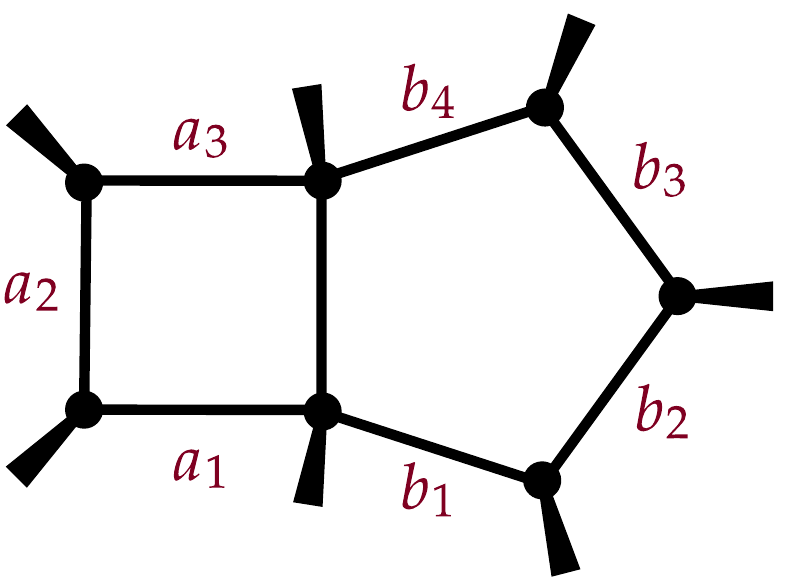} = \frac{\dbar^4\ell_1 \dbar^4\ell_2}{(\ell|\bur{a_1})(\ell|\bur{a_2})(\ell|\bur{a_3})(\ell_1|\ell_2)(\ell_2|\bur{b_1})(\ell_2|\bur{b_2})(\ell_2|\bur{b_3})(\ell_2|\bur{b_4})}.
\end{equation}
Notice that in the case of the scalar double box, the absence of any poles at infinity guarantees that it will not have more than a single elliptic curve. Indeed, there is a single odd residue of codimension seven, which is reached by encircling the contours 

\begin{equation}
    (\ell_1|\bur{a_i}) = (\ell_2|\bur{b_i}) = 0
\end{equation}
and

\begin{equation}
    (\ell_1|\ell_2) = 0.
\end{equation}

The scalar pentabox on the other hand supports---in its most generic form---a maximum of four different elliptic curves, corresponding to the four contours

\begin{equation}
    (\ell_1|\bur{a_i}) = (\ell_2|\bur{b_{\sigma(i)}}) = 0
\end{equation}
and

\begin{equation}
    (\ell_1|\ell_2) = 0
\end{equation}
where $\sigma()$ has been used to denote the $\binom{4}{3}$ choices of $3$ out of the four $\bur{b_i}$.

The key point here is not so much that there are several elliptic curves, which presents a computational challenge, but that any resolution by rigidity will have to make concessions to this feature, and render any master integrand basis containing such an integrand stratified along \emph{each curve} separately.

A technical implication of this fact is that upon integration over the loop momenta, the general form of any such integrand will not be pure. Specifically, notice that upon kinematic normalization, we can integrate expression (\ref{eq:4.2}) to obtain

\begin{equation}
    \int \includegraphics[valign = c, scale=0.3,trim=0 0 0 0cm]{doublebox-4-gon-labelled.pdf} = \text{elliptic polylogs}
\end{equation}
which is pure in the sense of \cite{Broedel:2018qkq}. In contrast, this \emph{will not} be true in the case of the scalar pentabox, which has to integrate to the more complex form

\begin{equation}\label{eq:4.9}
    \int\includegraphics[valign = c, scale=0.3,trim=0 0 0 0cm]{pentabox-scalar-labelled.pdf} = \text{polylogs} + \sum \bl{f}_{\mathfrak{e}}(x_{\bur{ij}})\times(\text{elliptic polylogs})
\end{equation}
where $\bl{f}_{\mathfrak{e}}$ is a scalar function of the external kinematics. The `elliptic polylogs' addend may itself decompose into sectors dependent on the various underlying elliptic curves.

The destruction of purity implied by the multiple elliptic curve substructure of the pentabox is a diagnostic of the fact that it is not stratified by rigidity. In other words, there isn't a sense in which the scalar pentabox admits of an expansion in master integrands which distinguish themselves from each other by their polylogarithmicity or ellipticity. The question now is whether or not such a construction is possible in principle, and if so, whether or not it is possible in practice.

The asymptotics of $4$-gon power counting eventually ensure that stratification along rigidity becomes prohibitive, as the number of leg distributions supporting more elliptic curves than can be stratified grows. We'll move now to laying out how this takes place in detail.

\subsection{Indefinite Rigidity and Obstructions to Stratification}\label{sec:4.2}
We have argued that a generic integrated pentabox is not polylog for arbitrary leg distributions but is `contaminated' with elliptic pieces. Further, these are accompanied by kinematic prefactors that render them impure, in that they cannot be expressed as integrals of pure descendants. When this is the case, the integrated expression is said to have indefinite rigidity, where it cannot be said, in any precise sense, to have a fixed rigidity, but can only be expressed as a sum of quantities that individually have fixed---or \emph{definite}---rigidity.

The issue of stratification is redefined as the problem of constructing a basis of master integrands where each element has definite rigidity. As alluded to already, we cannot achieve this in $4$-gon power counting. Our task in the present section will be to demonstrate the points of obstruction and the combinatorial principles at play that render this question impossible to resolve under conditions of box power counting.

Our strategy will be to enumerate all of the pentabox leg distributions that are potential parents of elliptic subtopologies, and indicate the extent to which attempts at stratification are confounded by each. For topologies containing no legs attached to the internal propagator we have

\begin{equation}
    \begin{aligned}
        \Bigg\lbrace &\underset{\mathcal{I}_1}{ \includegraphics[valign = c, scale=0.3,trim=0 0 0 0cm]{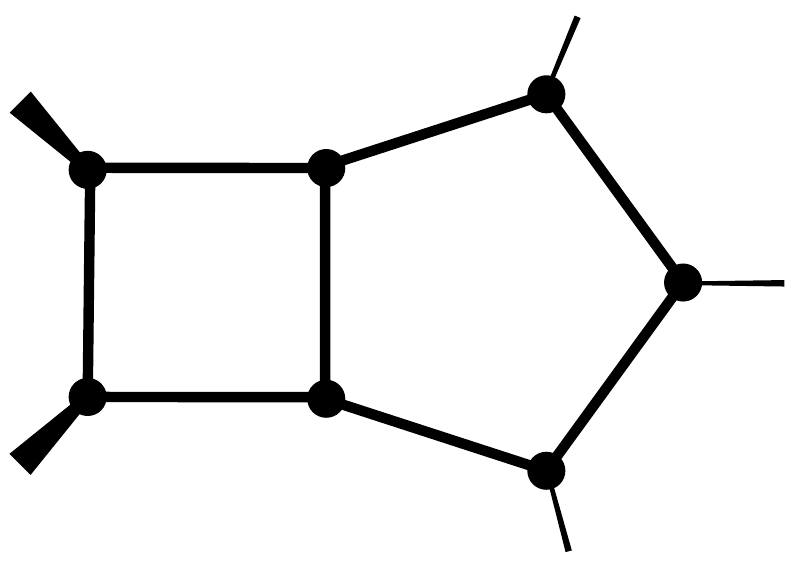}}\;,\;\underset{\mathcal{I}_2}{ \includegraphics[valign = c, scale=0.3,trim=0 0 0 0cm]{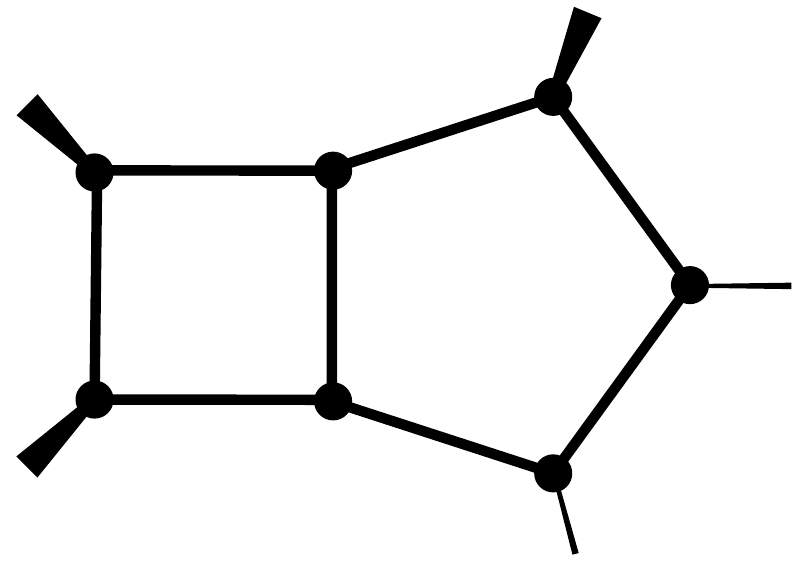}}\;,\;\underset{\mathcal{I}_3}{ \includegraphics[valign = c, scale=0.3,trim=0 0 0 0cm]{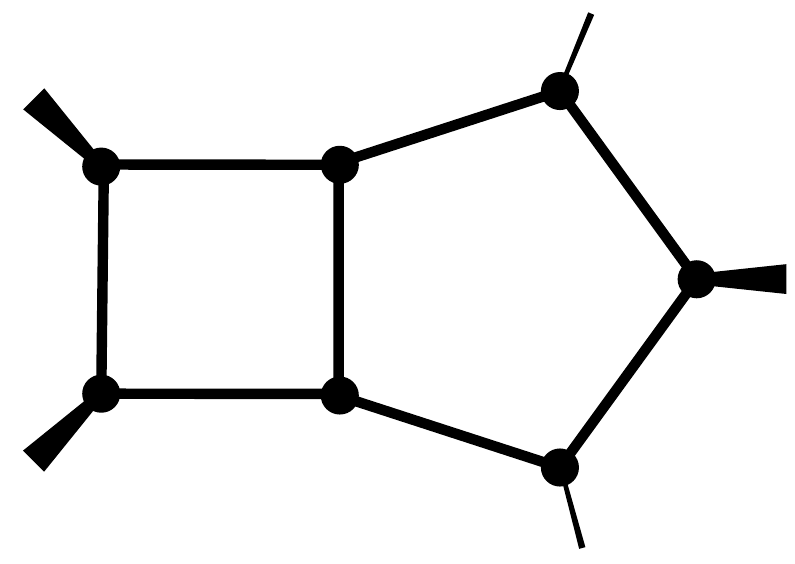}}\;,\;\\
        &\underset{\mathcal{I}_4}{ \includegraphics[valign = c, scale=0.3,trim=0 0 0 0cm]{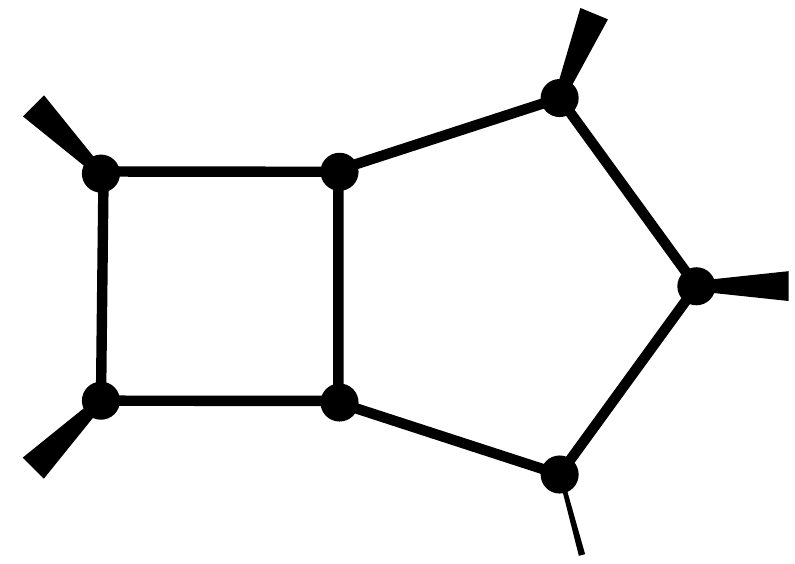}}\;,\;\underset{\mathcal{I}_5}{ \includegraphics[valign = c, scale=0.3,trim=0 0 0 0cm]{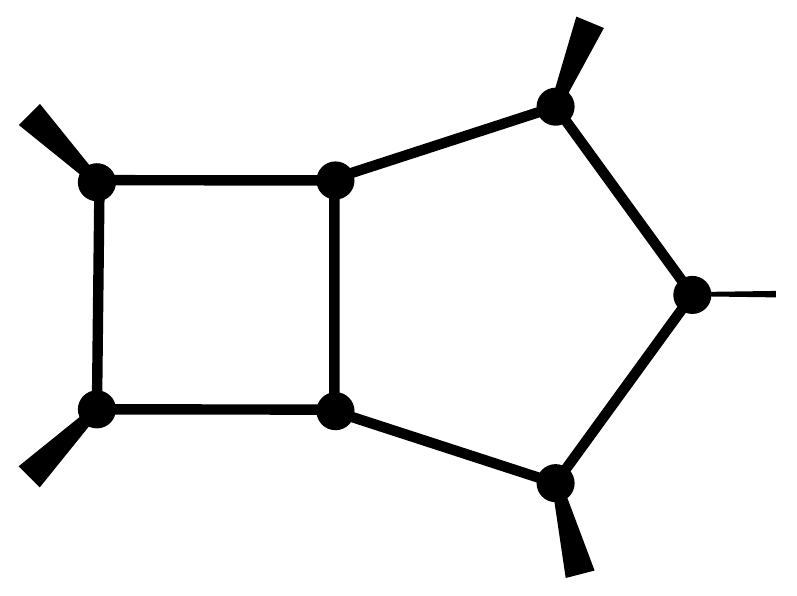}}\;,\;\underset{\mathcal{I}_6}{ \includegraphics[valign = c, scale=0.3,trim=0 0 0 0cm]{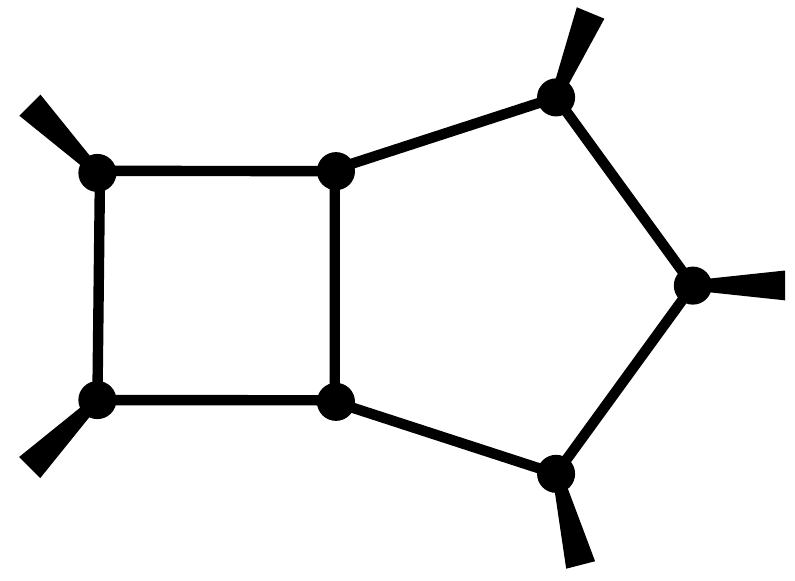}}\Bigg\rbrace
    \end{aligned}
\end{equation}
for the ones with one leg attached,
\begin{equation}
    \begin{aligned}
        \Bigg\lbrace &\underset{\mathcal{I}_7}{ \includegraphics[valign = c, scale=0.3,trim=0 0 0 0cm]{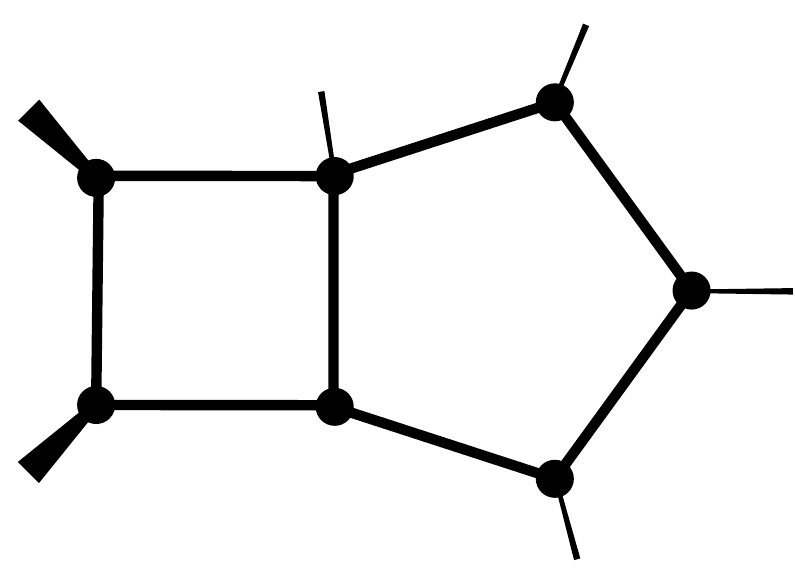}}\;,\;\underset{\mathcal{I}_8}{ \includegraphics[valign = c, scale=0.3,trim=0 0 0 0cm]{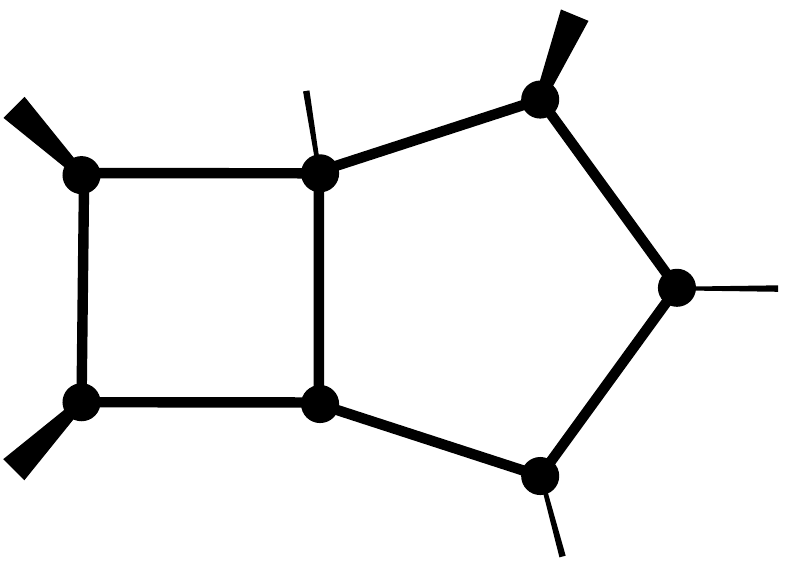}}\;,\;\underset{\mathcal{I}_9}{ \includegraphics[valign = c, scale=0.3,trim=0 0 0 0cm]{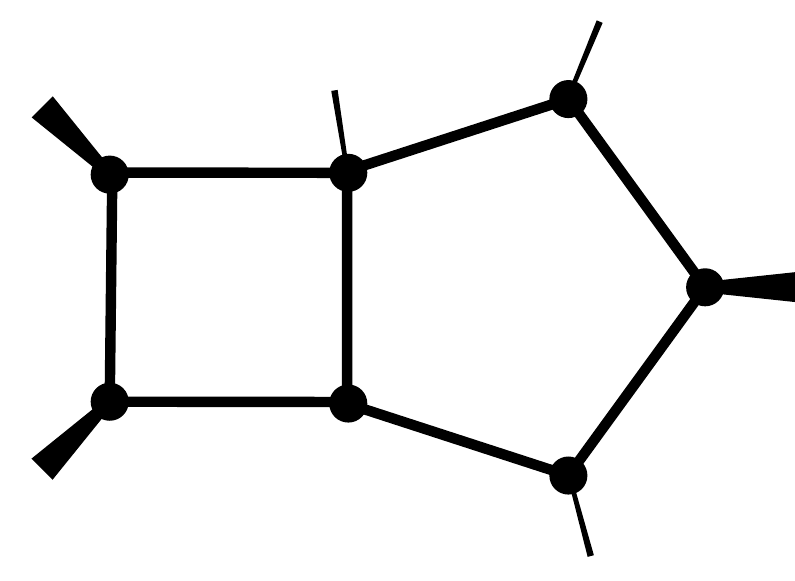}}\;,\;\underset{\mathcal{I}_{10}}{ \includegraphics[valign = c, scale=0.3,trim=0 0 0 0cm]{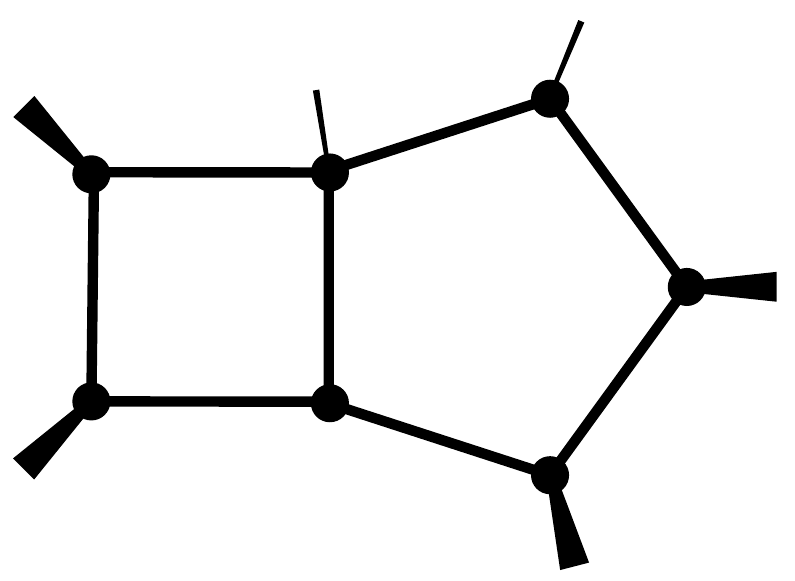}}\;,\;\\
        &\underset{\mathcal{I}_{11}}{ \includegraphics[valign = c, scale=0.3,trim=0 0 0 0cm]{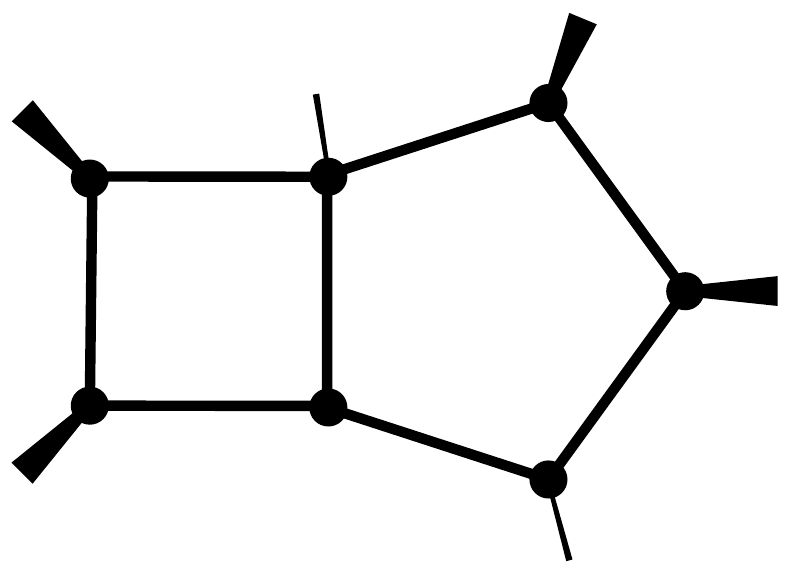}}\;,\;\underset{\mathcal{I}_{12}}{ \includegraphics[valign = c, scale=0.3,trim=0 0 0 0cm]{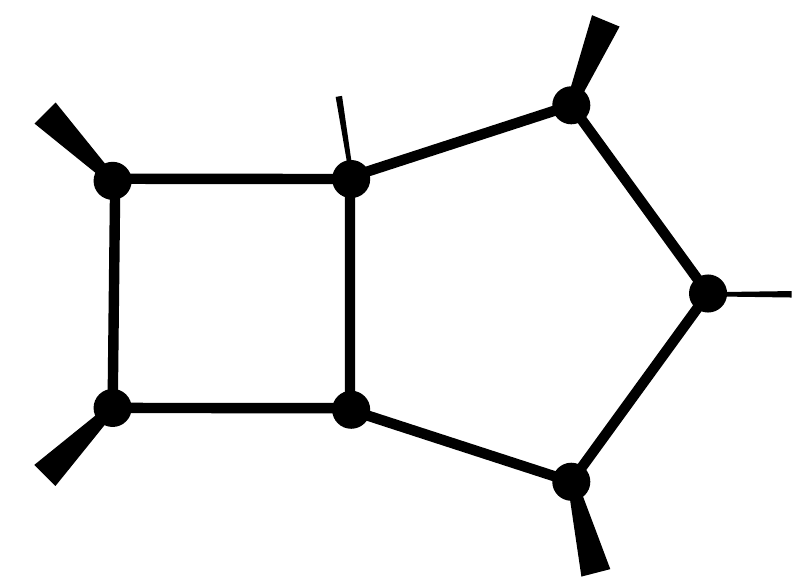}}\;,\;\underset{\mathcal{I}_{13}}{ \includegraphics[valign = c, scale=0.3,trim=0 0 0 0cm]{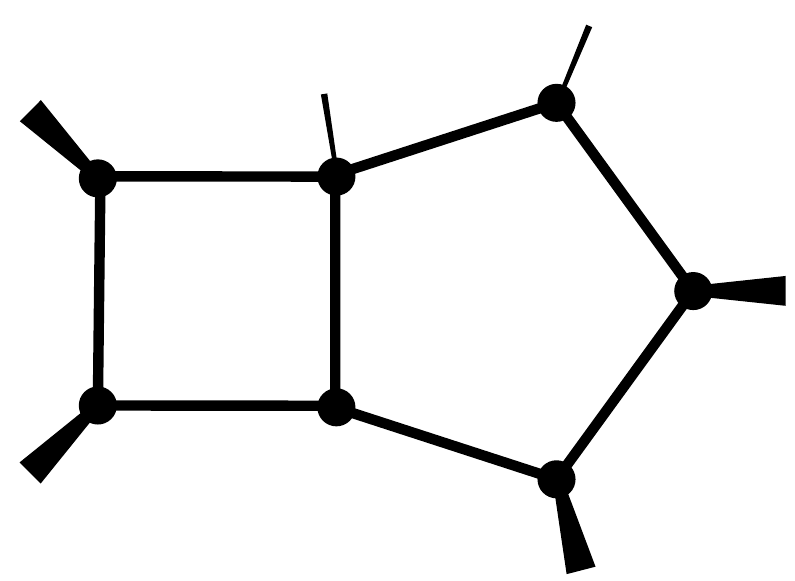}}\;,\;\underset{\mathcal{I}_{14}}{ \includegraphics[valign = c, scale=0.3,trim=0 0 0 0cm]{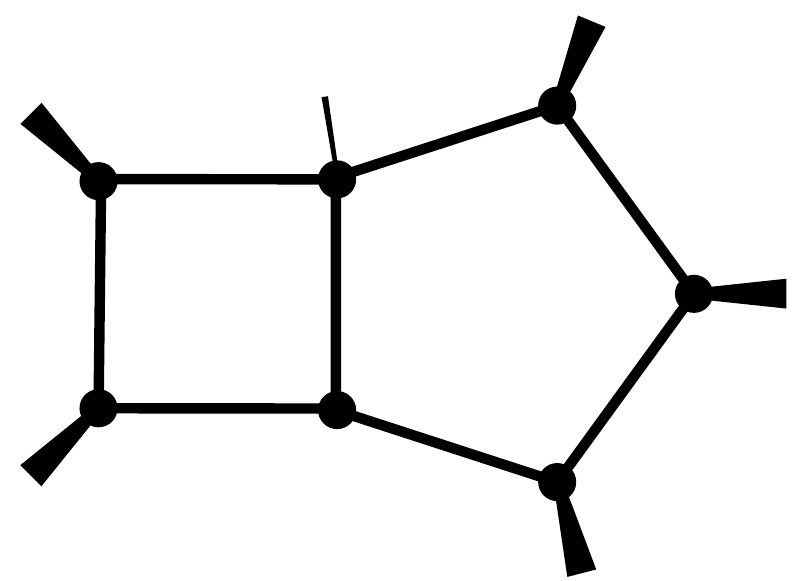}}\Bigg\rbrace
    \end{aligned}
\end{equation}
and the ones with two legs attached we find

\begin{equation}
    \begin{aligned}
        \Bigg\lbrace &\underset{\mathcal{I}_{15}}{ \includegraphics[valign = c, scale=0.3,trim=0 0 0 0cm]{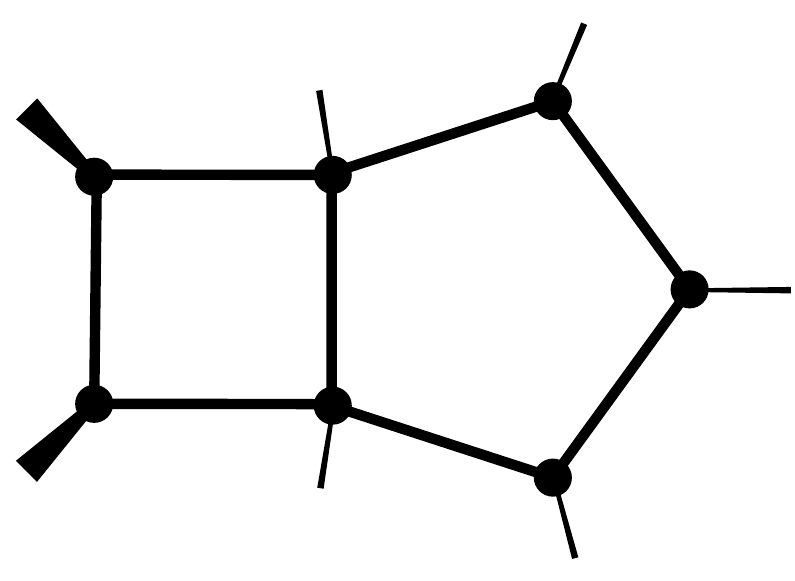}}\;,\;\underset{\mathcal{I}_{16}}{ \includegraphics[valign = c, scale=0.3,trim=0 0 0 0cm]{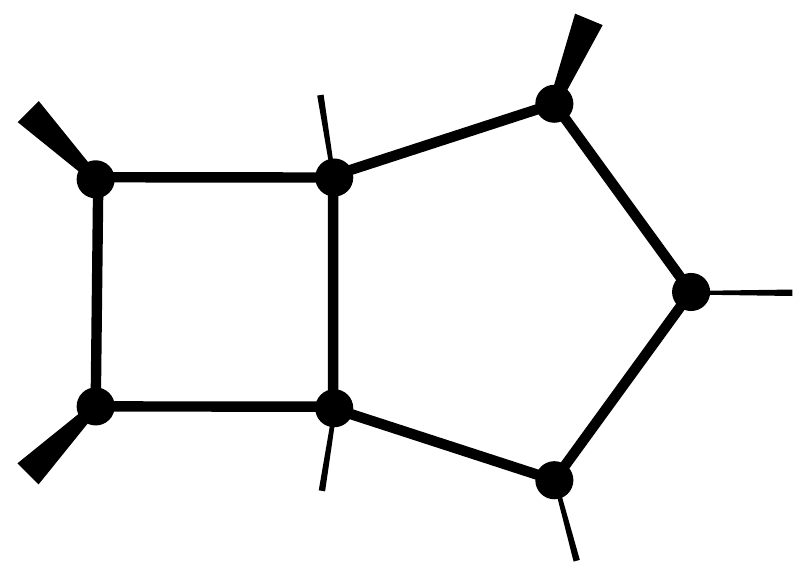}}\;,\;\underset{\mathcal{I}_{17}}{ \includegraphics[valign = c, scale=0.3,trim=0 0 0 0cm]{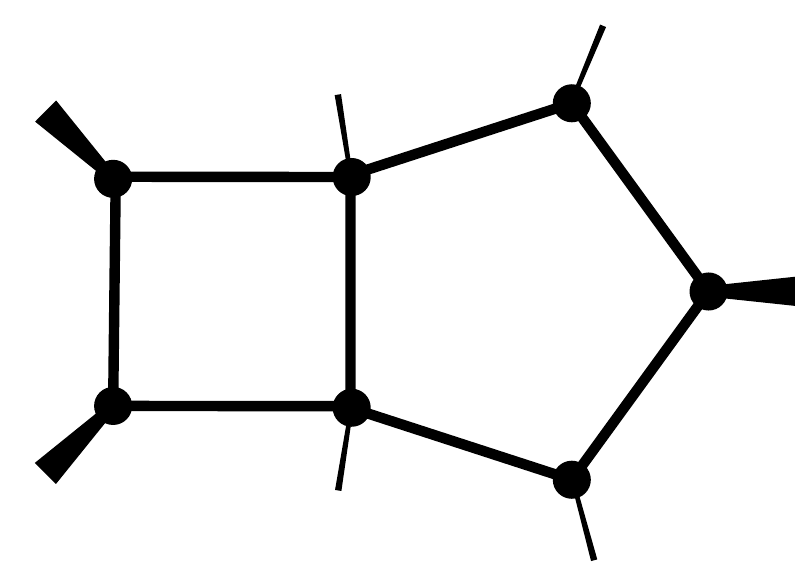}}\;,\;\\
        &\underset{\mathcal{I}_{18}}{ \includegraphics[valign = c, scale=0.3,trim=0 0 0 0cm]{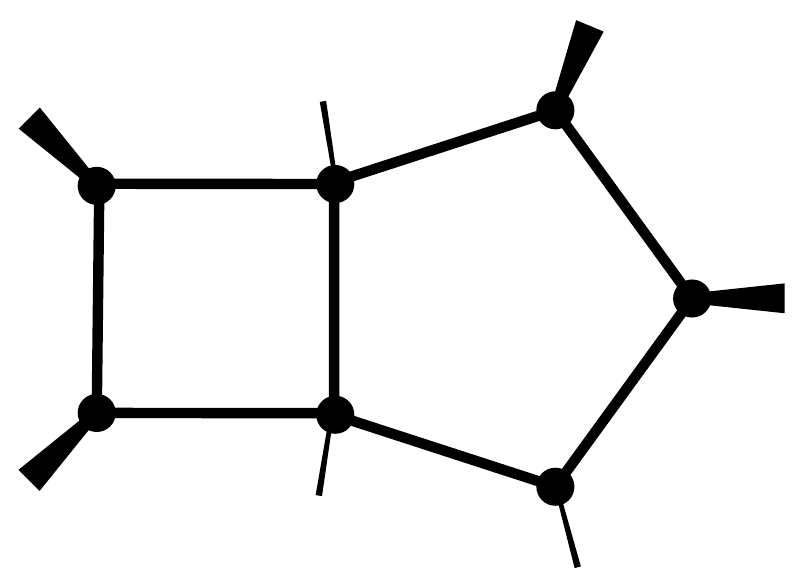}}\;,\;\underset{\mathcal{I}_{19}}{ \includegraphics[valign = c, scale=0.3,trim=0 0 0 0cm]{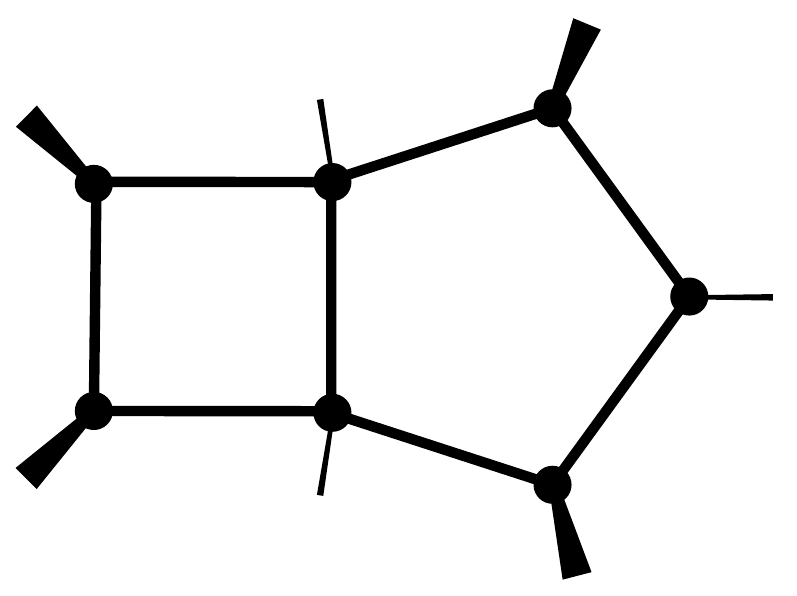}}\;,\;\underset{\mathcal{I}_{20}}{ \includegraphics[valign = c, scale=0.3,trim=0 0 0 0cm]{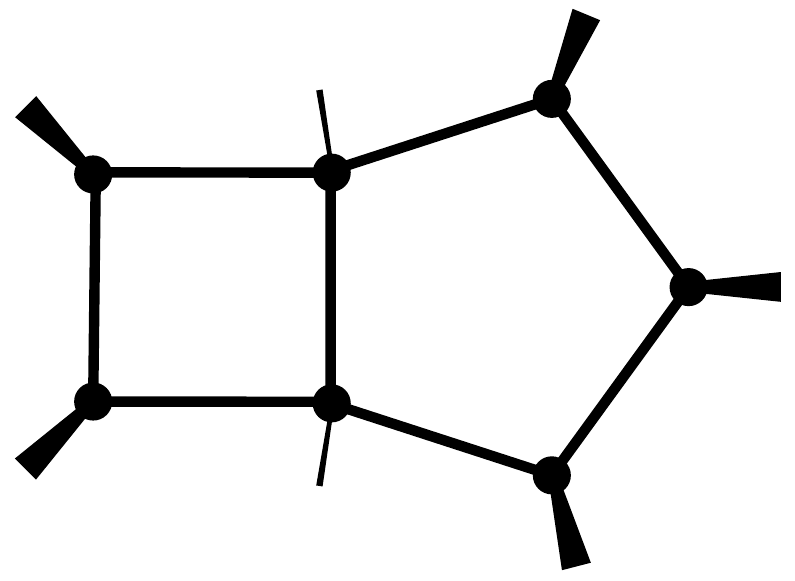}}\Bigg\rbrace.
    \end{aligned}
\end{equation}

Integrands $\mathcal{I}_1$ through $\mathcal{I}_6$ will have a massless internal vertex on $(\ell_1|\ell_2)$ regardless of leg collapses. Accordingly, the quartic controlling any putative elliptic will always factorize, rendering them polylogarithmic. Indeed, any such graph will only contain polylogarithmic daughters.

In the second class, integrands $\mathcal{I}_{11}$ and $\mathcal{I}_{14}$ each support one elliptic curve, obtained by degenerating the lowest leg of the pentagon. The rest support zero elliptic curves. 

Finally, in the third class, we find that $\mathcal{I}_{16}$ supports 1 elliptic curve, $\mathcal{I}_{18}$ and $\mathcal{I}_{19}$ support 2 each and $\mathcal{I}_{20}$ supports $4$. The other two have polylogarithmic daughters.

We will now explore each case independently, to illustrate how the inclusion of multiple elliptic curves systematically obstructs stratification, elaborating finally on how distributions furnishing the topology $\mathcal{I}_{20}$ make even diagonalization along the elliptic part impossible. This fact, along with a brief combinatorial argument, will strengthen and underline why high multiplicity totally forecloses any hope of stratification in $4$-gon power counting. 

\paragraph*{Example 4.1. No Elliptic Curve.} The case of topology $\mathcal{I}_{5}$, which first shows up at $9$ particles, is an example of a pentabox which supports no elliptic curve. This open subset of integrands in box power counting contains master integrals spanned by (relabelling)

\begin{equation}
    \lbrace{\mathcal{I}_1,\mathcal{I}_2\rbrace} = \Bigg\lbrace\includegraphics[valign = c, scale=0.3,trim=0 0 0 0cm]{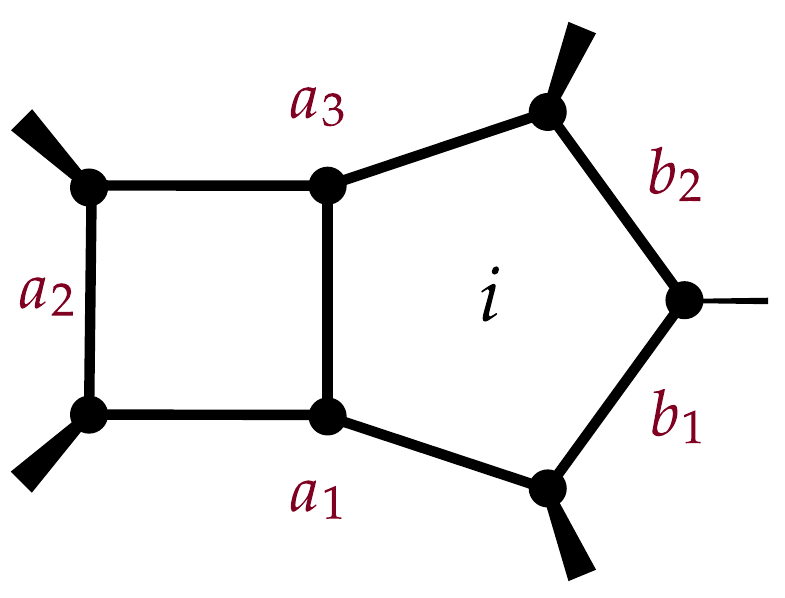} \Bigg\rbrace
\end{equation}
and

\begin{equation}
\lbrace{\mathcal{I}_3,\dots,\mathcal{I}_6\rbrace} = \Bigg\lbrace\includegraphics[valign = c, scale=0.3,trim=0 0 0 0cm]{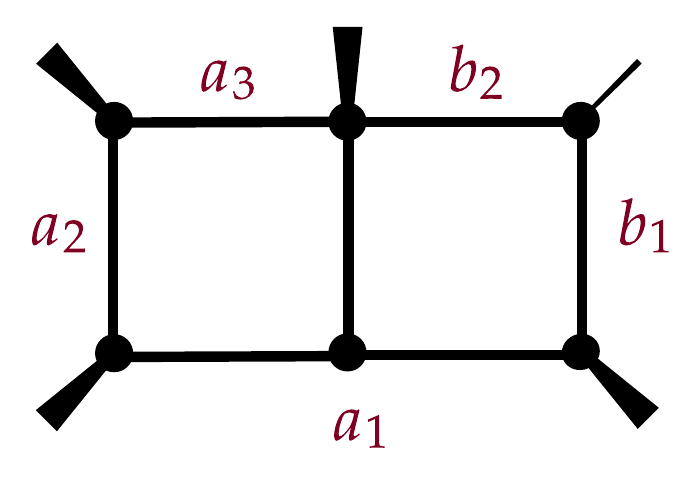}\;,\;\includegraphics[valign = c, scale=0.3,trim=0 0 0 0cm]{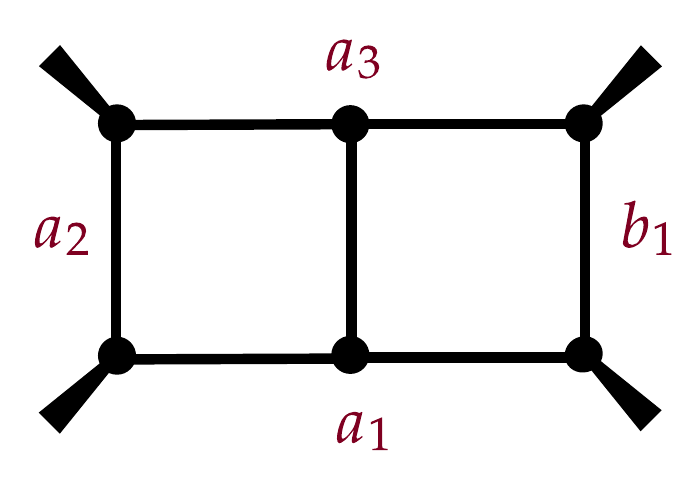}\;,\;\includegraphics[valign = c, scale=0.3,trim=0 0 0 0cm]{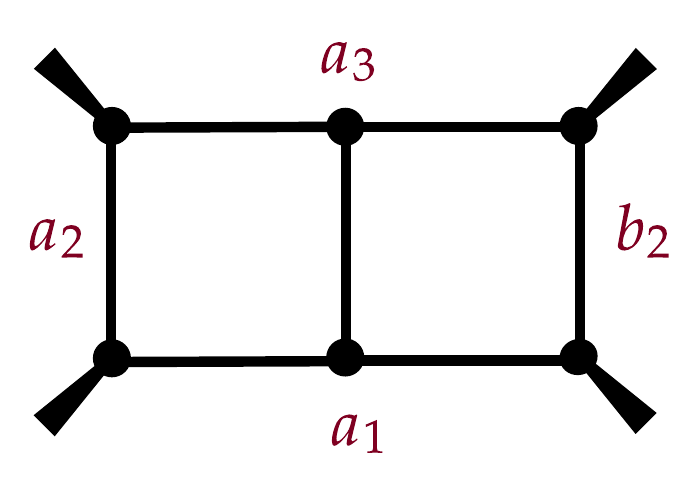}\;,\;\includegraphics[valign = c, scale=0.3,trim=0 0 0 0cm]{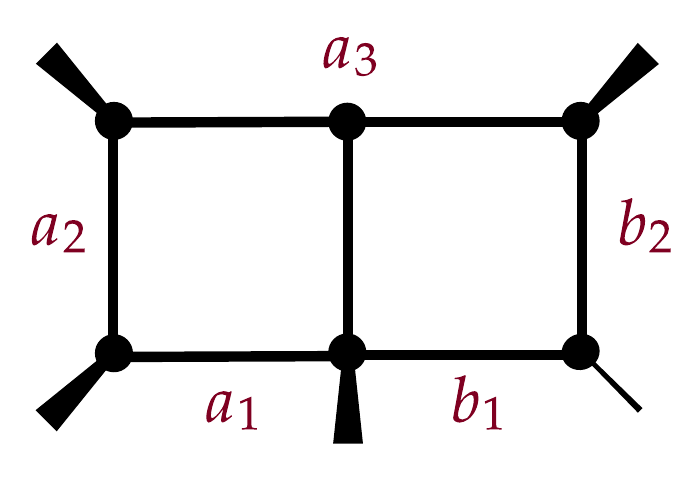}\Bigg\rbrace.
\end{equation}
Indeed, none of the daughter double boxes contain any elliptics, and always support another residue on the maximal cut to render it a leading singularity due to collinear regions. This can be made concrete by the following choices of spanning contours:

\begin{equation}
    \lbrace{\Omega_1,\Omega_2\rbrace} = \Bigg\lbrace\includegraphics[valign = c, scale=0.3,trim=0 0 0 0cm]{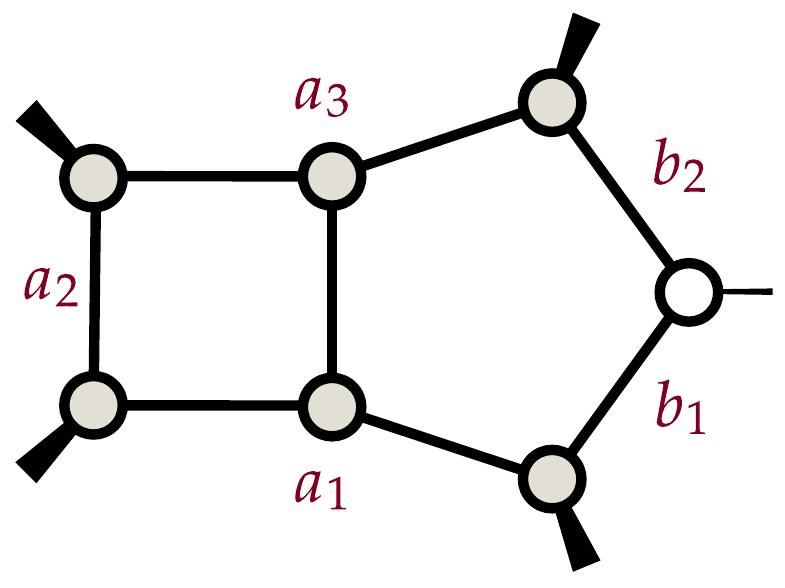}\;,\;\includegraphics[valign = c, scale=0.3,trim=0 0 0 0cm]{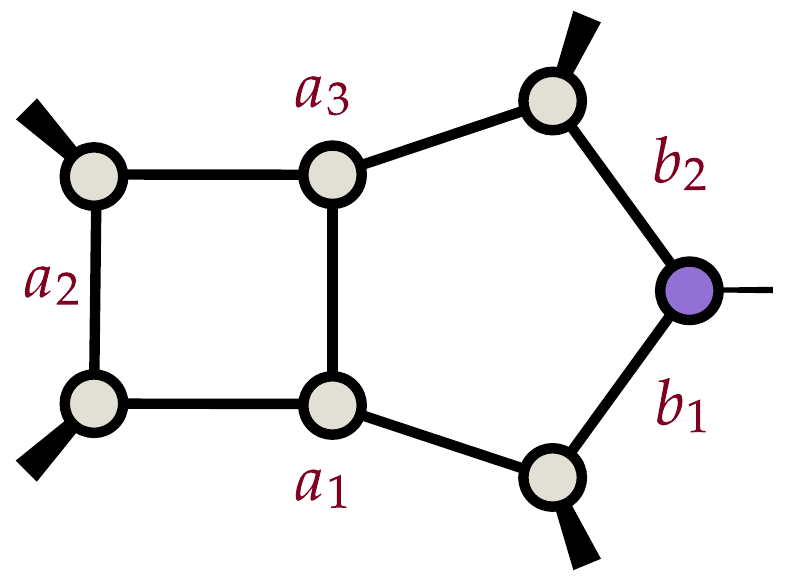}\Bigg\rbrace
\end{equation}
and

\begin{equation}
    \lbrace{\Omega_3,\dots,\Omega_6\rbrace} = \Bigg\lbrace  \includegraphics[valign = c, scale=0.3,trim=0 0 0 0cm]{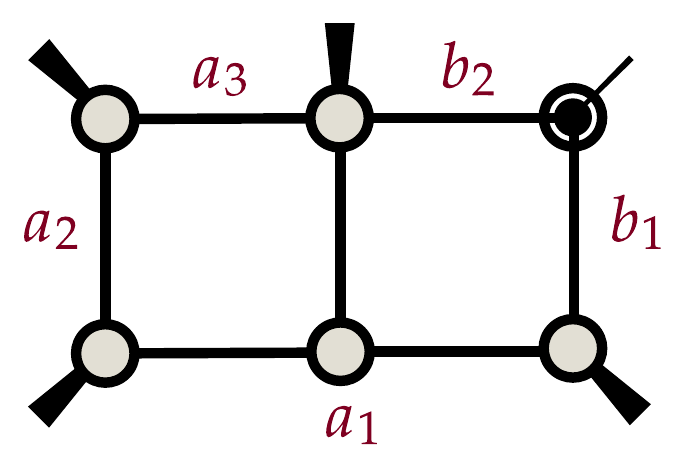}\;,\;\includegraphics[valign = c, scale=0.3,trim=0 0 0 0cm]{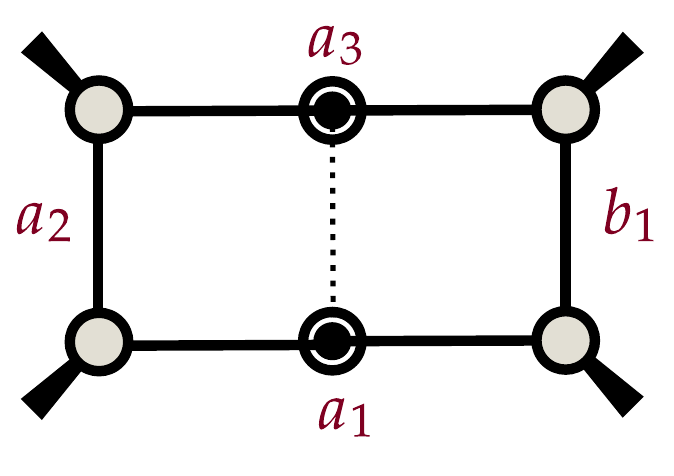}\;,\;\includegraphics[valign = c, scale=0.3,trim=0 0 0 0cm]{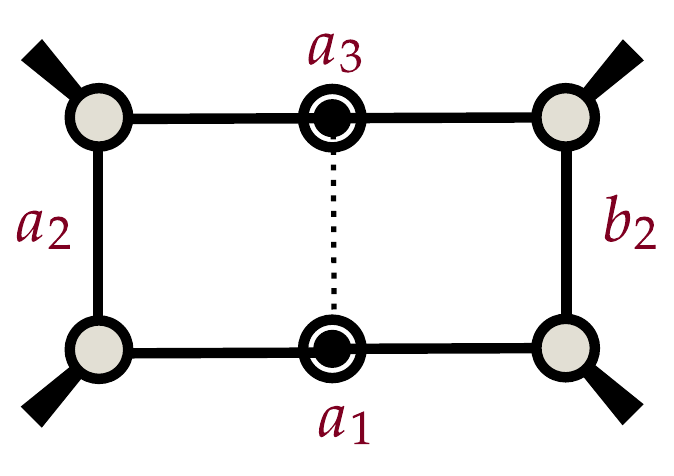}\;,\;\includegraphics[valign = c, scale=0.3,trim=0 0 0 0cm]{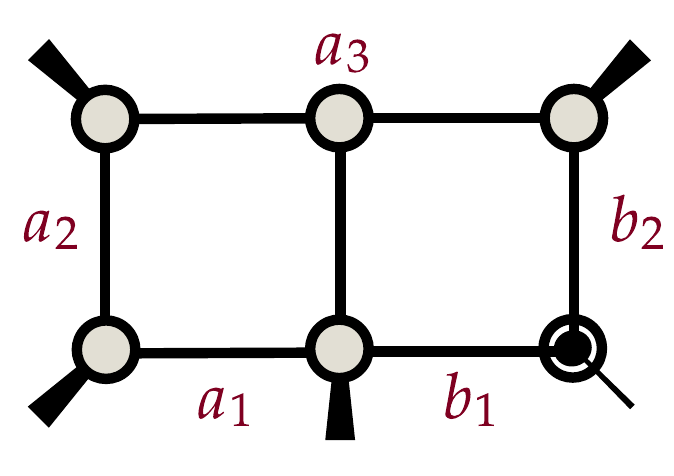}\Bigg\rbrace.
\end{equation}
We have already seen the period matrix for this choice; choosing the pentagon numerators to diagonalize their subspace renders it upper triangular:

\begin{equation}
    \begin{blockarray}{ccccccc}
 \hspace{1pt}&\Omega_1 & \Omega_2 & \Omega_3 & \Omega_4 & \Omega_5 &\Omega_6\\
\begin{block}{c(cccccc)}
  \mathcal{I}_1 & \bl{f}^{\bur{1}}_{\bl{1}} & 0 & \bl{g}^{\bur{3}}_{\bl{1}} & \bl{g}^{\bur{4}}_{\bl{1}} & \bl{g}^{\bur{5}}_{\bl{1}} &\bl{g}^{\bur{6}}_{\bl{1}}\\
  \mathcal{I}_2 & 0 & \bl{f}^{\bur{2}}_{\bl{2}} & \bl{g}^{\bur{3}}_{\bl{2}} & \bl{g}^{\bur{4}}_{\bl{2}} & \bl{g}^{\bur{5}}_{\bl{2}} &\bl{g}^{\bur{6}}_{\bl{2}}\\
  \mathcal{I}_3 & 0 & 0 & \bl{h}^{\bur{3}}_{\bl{3}} & 0 & 0 &0\\
  \mathcal{I}_4 & 0 & 0 & 0 & \bl{h}^{\bur{4}}_{\bl{4}} & 0 &0\\
  \mathcal{I}_5 & 0 & 0 & 0 & 0 & \bl{h}^{\bur{5}}_{\bl{5}} &0\\
  \mathcal{I}_6 & 0 & 0 & 0 & 0 & 0 &\bl{h}^{\bur{6}}_{\bl{6}}\\
\end{block}
\end{blockarray}
\end{equation}
and by immediate implication full rank. This subspace is then naturally stratified; all of the elements are simply of rigidity 0.

\hfill
\vspace{0.7cm}

\paragraph*{Example 4.2. One Elliptic Curve. } The topology $\mathcal{I}_{14}$ has one daughter topology that is elliptic. We find this topology relevant starting at 10 particles, and it is one of two examples where elliptic stratification is possible for box power counting. The master integrand basis is given by

\begin{equation}
    \lbrace{\mathcal{I}_1,\mathcal{I}_2\rbrace} = \Bigg\lbrace{\includegraphics[valign = c, scale=0.3,trim=0 0 0 0cm]{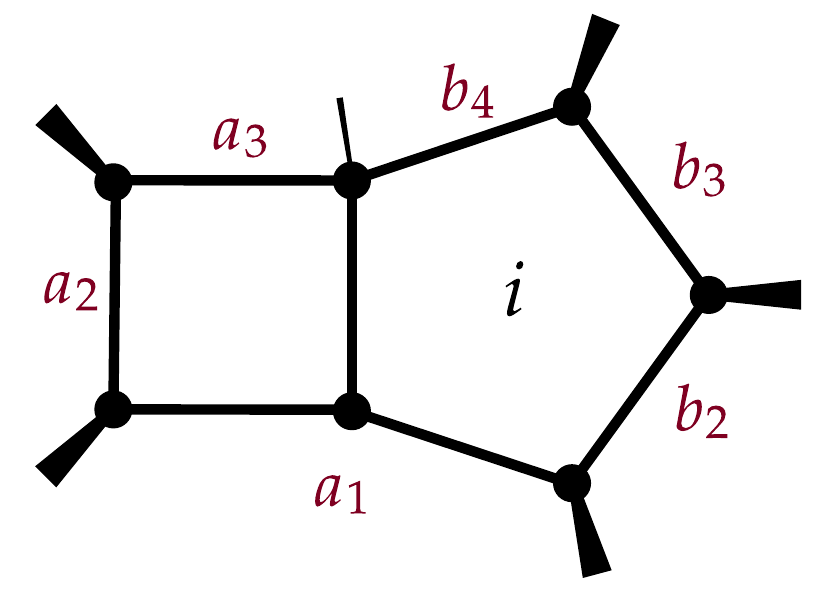}\Bigg\rbrace}
\end{equation}
and

\begin{equation}
\lbrace{\mathcal{I}_{3},\dots,\mathcal{I}_{6}\rbrace} = \Bigg\lbrace\includegraphics[valign = c, scale=0.3,trim=0 0 0 0cm]{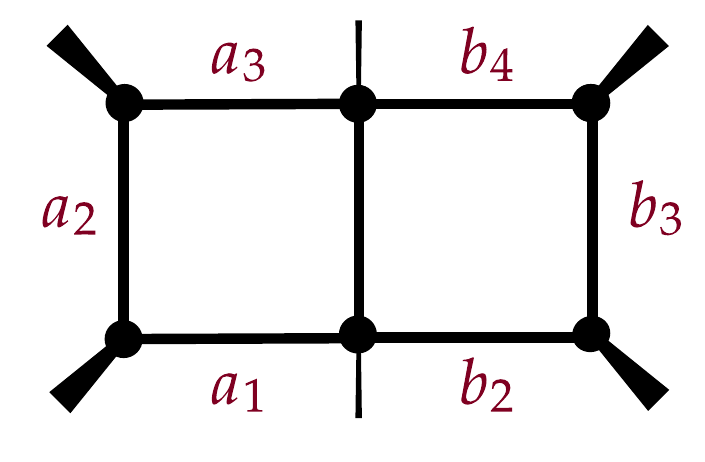}\;,\;\includegraphics[valign = c, scale=0.3,trim=0 0 0 0cm]{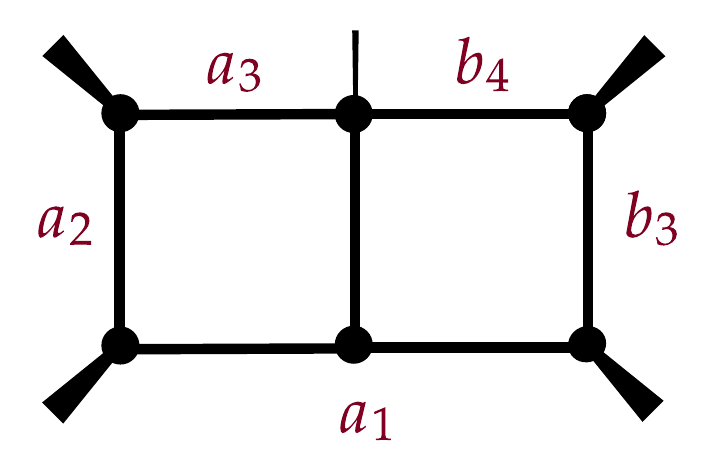}\;,\;\includegraphics[valign = c, scale=0.3,trim=0 0 0 0cm]{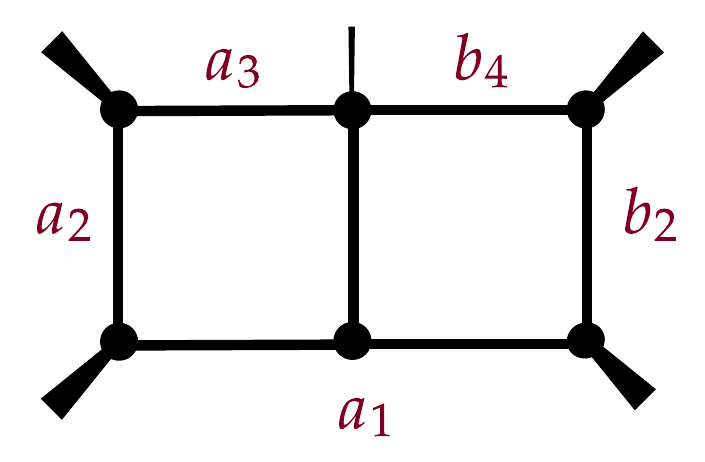}\;,\;\includegraphics[valign = c, scale=0.3,trim=0 0 0 0cm]{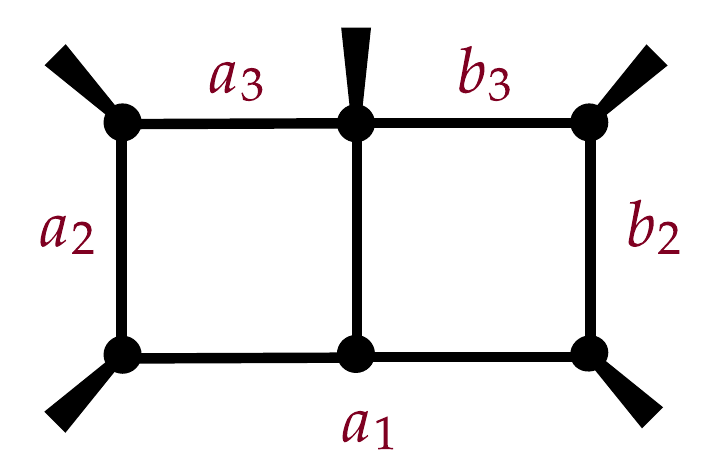}\Bigg\rbrace 
\end{equation}
Of the daughters, only $\mathcal{I}_3$ supports an elliptic curve. The rest are all pure polylog. The strategy to stratify this is as follows. The double box is matched as usual to the $a$-cycle period. For the pentabox however, one of the numerators is chosen so as to be diagonal to the $b$-cycle period. This way, the six-dimensional subspace is decomposed into two masters which span the elliptic cut, and four which are of rigidity zero. We have

\begin{equation}
    \lbrace{\Omega_1,\Omega_2\rbrace} = \Bigg\lbrace \includegraphics[valign = c, scale=0.3,trim=0 0 0 0cm]{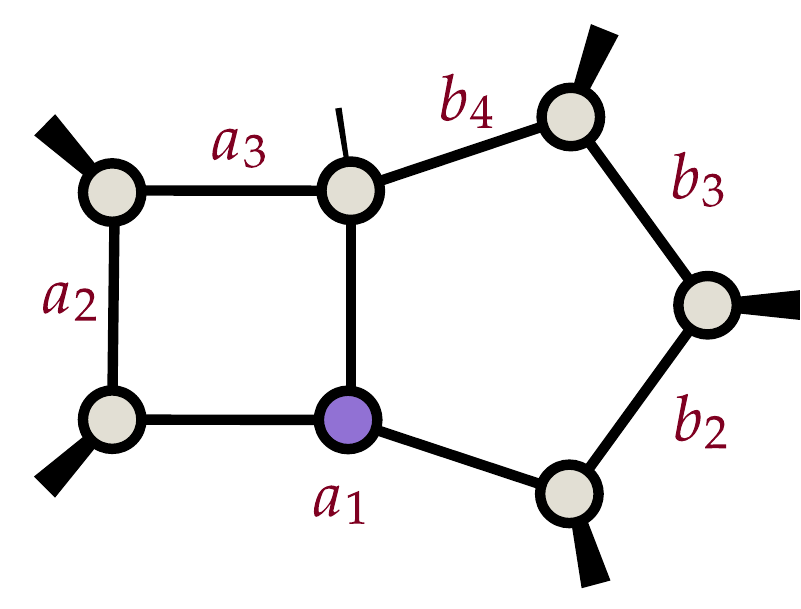}\;,\;\underset{b\text{-cycle}}{\includegraphics[valign = c, scale=0.3,trim=0 0 0 0cm]{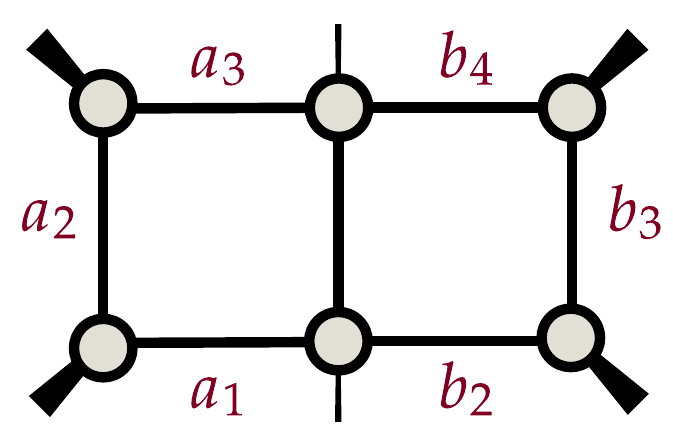}} \Bigg\rbrace
\end{equation}
and

\begin{equation}
   \lbrace{\Omega_3,\dots,\Omega_6\rbrace} = \Bigg\lbrace\underset{a\text{-cycle}}{\includegraphics[valign = c, scale=0.3,trim=0 0 0 0cm]{one_elliptic_pentabox_integrand_on-shell_daughter_1.pdf}}\;,\;\includegraphics[valign = c, scale=0.3,trim=0 0 0 0cm]{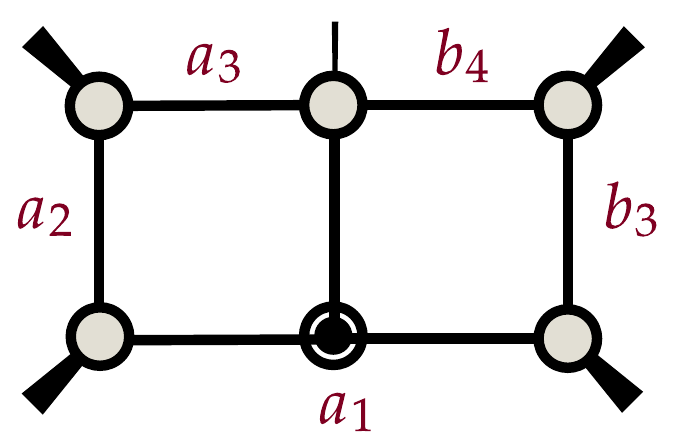}\;,\;\includegraphics[valign = c, scale=0.3,trim=0 0 0 0cm]{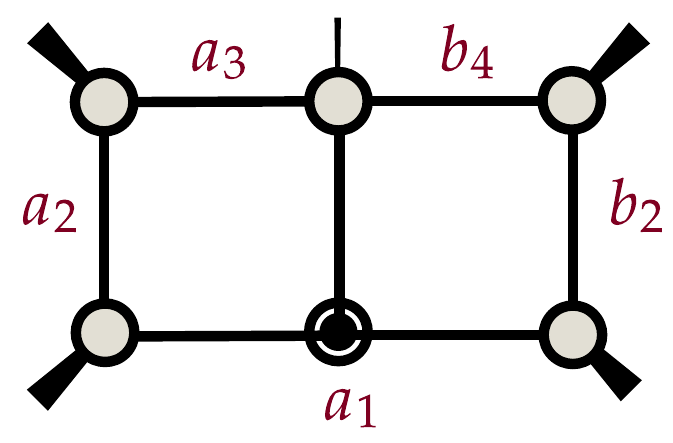}\;,\;\includegraphics[valign = c, scale=0.3,trim=0 0 0 0cm]{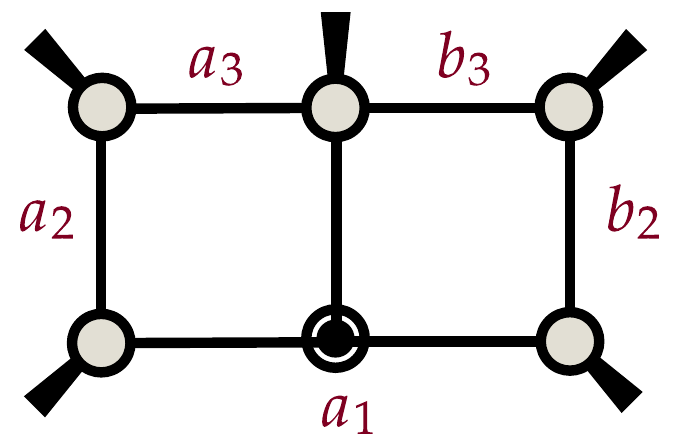}\Bigg\rbrace.
\end{equation}
This time, the period matrix takes the form

\begin{equation}
    \begin{blockarray}{ccccccc}
 \hspace{1pt}&\Omega_1 & \Omega_2 & \Omega_3 & \Omega_4 & \Omega_5 &\Omega_6\\
\begin{block}{c(cccccc)}
  \mathcal{I}_1 & \bl{f}^{\bur{1}}_{\bl{1}} & \bl{f}^{\bur{2}}_{\bl{1}} & \bl{g}^{\bur{3}}_{\bl{1}} & \bl{g}^{\bur{4}}_{\bl{1}} & \bl{g}^{\bur{5}}_{\bl{1}} &\bl{g}^{\bur{6}}_{\bl{1}}\\
  \mathcal{I}_2 & \bl{f}^{\bur{1}}_{\bl{2}} & \bl{f}^{\bur{2}}_{\bl{2}} & \bl{g}^{\bur{3}}_{\bl{2}} & \bl{g}^{\bur{4}}_{\bl{2}} & \bl{g}^{\bur{5}}_{\bl{2}} &\bl{g}^{\bur{6}}_{\bl{2}}\\
  \mathcal{I}_3 & 0 & \bl{h}^{\bur{2}}_{\bl{3}} & \bl{h}^{\bur{3}}_{\bl{3}} & 0 & 0 &0\\
  \mathcal{I}_4 & 0 & 0 & 0 & \bl{h}^{\bur{4}}_{\bl{4}} & 0 &0\\
  \mathcal{I}_5 & 0 & 0 & 0 & 0 & \bl{h}^{\bur{5}}_{\bl{5}} &0\\
  \mathcal{I}_6 & 0 & 0 & 0 & 0 & 0 &\bl{h}^{\bur{6}}_{\bl{6}}\\
\end{block}
\end{blockarray}
\end{equation}
which is full rank.

\hfill
\vspace{0.7cm}

In this example, we may observe that $\widetilde{\mathcal{I}}_2$ actually has mixed rigidity. Indeed, although it has been diagonalized so as to only have elliptic support along the $b$-cycle of the daughter elliptic, there are $3$ polylog contours corresponding to the three remaining leading singularities of the pentabox along which $\widetilde{\mathcal{I}}_2$ will have support as well. Already we find at this level that at least one of the elements of the basis must be of indefinite rigidity. This fact is more evident in the next example.

\paragraph*{Example 4.3. Two Elliptic Curves. }
In topology $\mathcal{I}_{18}$, encountered first at 11-particles, we find an example of a parent pentabox having two elliptic curves, obtained by collapsing one of the legs containing the massless particles. 

The basis of master integrals is given by the sets

\begin{equation}
    \lbrace{\mathcal{I}_1,\mathcal{I}_{2}\rbrace} = \Bigg\lbrace\includegraphics[valign = c, scale=0.3,trim=0 0 0 0cm]{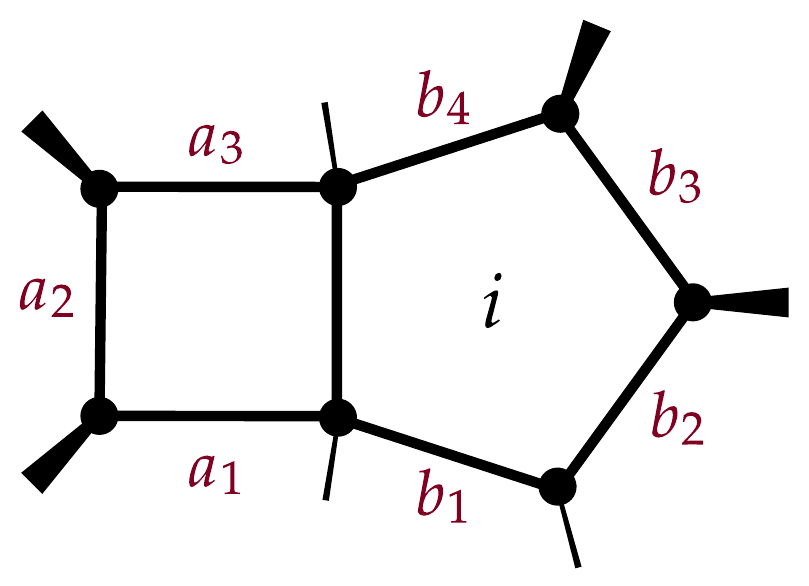}\Bigg\rbrace
\end{equation}
and

\begin{equation}
\lbrace{\mathcal{I}_{3},\dots,\mathcal{I}_{6}\rbrace} = \Bigg\lbrace\includegraphics[valign = c, scale=0.3,trim=0 0 0 0cm]{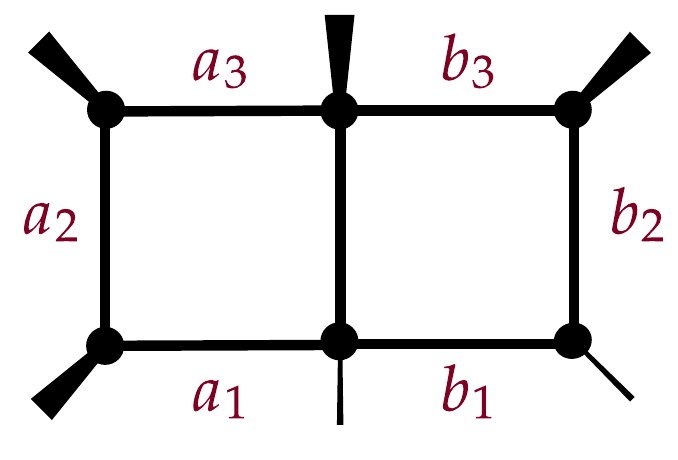}\;,\;\includegraphics[valign = c, scale=0.3,trim=0 0 0 0cm]{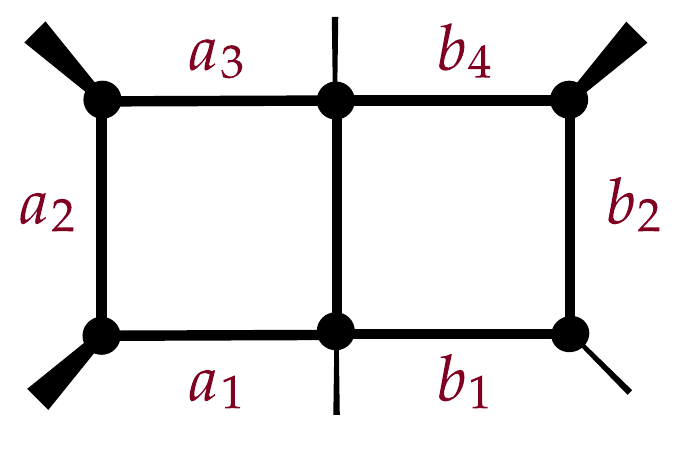}\includegraphics[valign = c, scale=0.3,trim=0 0 0 0cm]{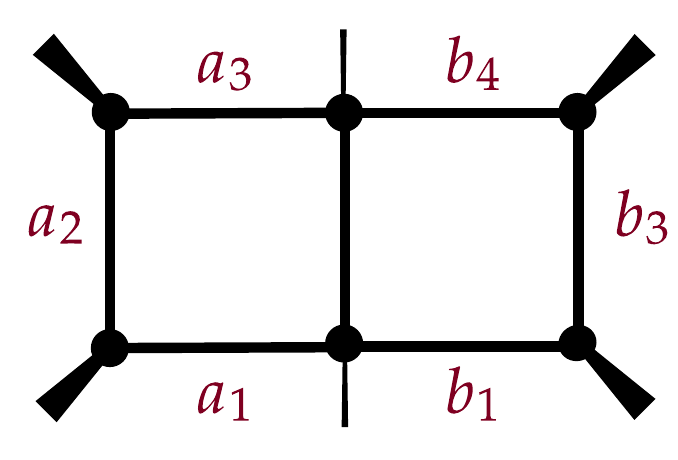}\;,\;\includegraphics[valign = c, scale=0.3,trim=0 0 0 0cm]{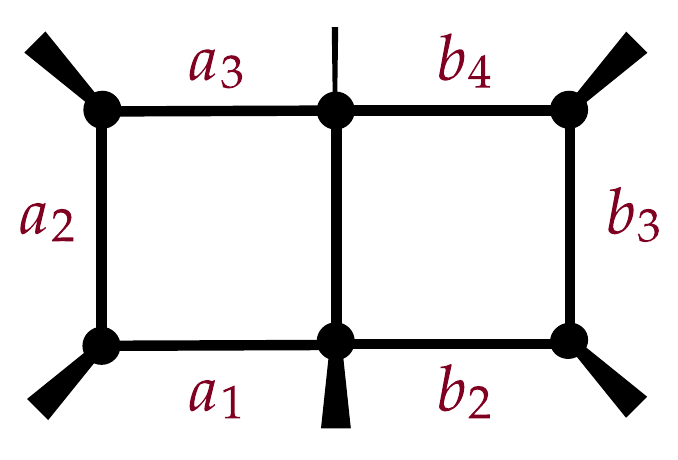} \;,\;\Bigg\rbrace.
\end{equation}
We can see that $\mathcal{I}_{5}$ and $\mathcal{I}_{6}$ have rigidity 1. 

The stratification strategy generalizes that of the previous example: we simply match the two pentaboxes to the $b$-cycles of the daughter elliptics, and each elliptic double box to their respective $a$-cycle. We have

\begin{equation}
    \lbrace{\Omega_1,\Omega_2\rbrace} = \Bigg\lbrace  \underset{b-\text{cycle}}{\includegraphics[valign = c, scale=0.3,trim=0 0 0 0cm]{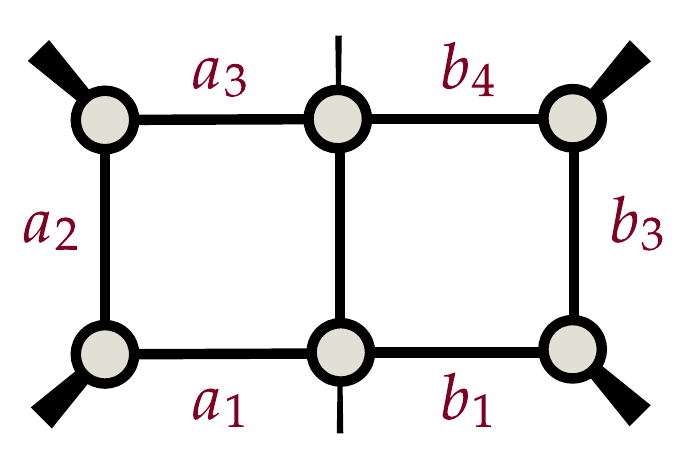}}\;,\;\underset{b-\text{cycle}}{\includegraphics[valign = c, scale=0.3,trim=0 0 0 0cm]{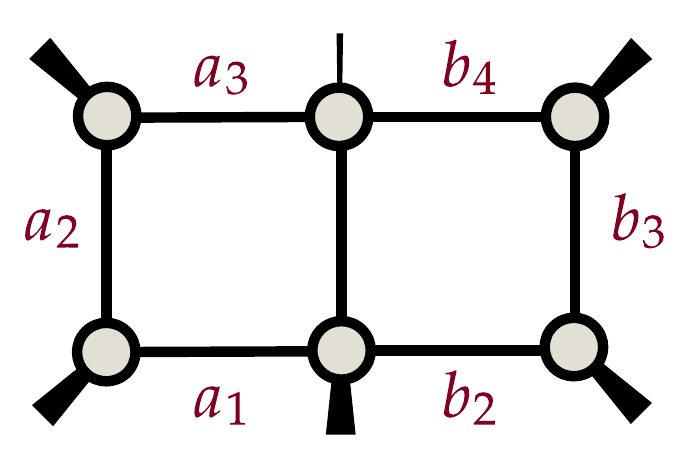}}\Bigg\rbrace
\end{equation}
and

\begin{equation}
    \lbrace{\Omega_3,\dots,\Omega_6\rbrace} = \Bigg\lbrace \includegraphics[valign = c, scale=0.3,trim=0 0 0 0cm]{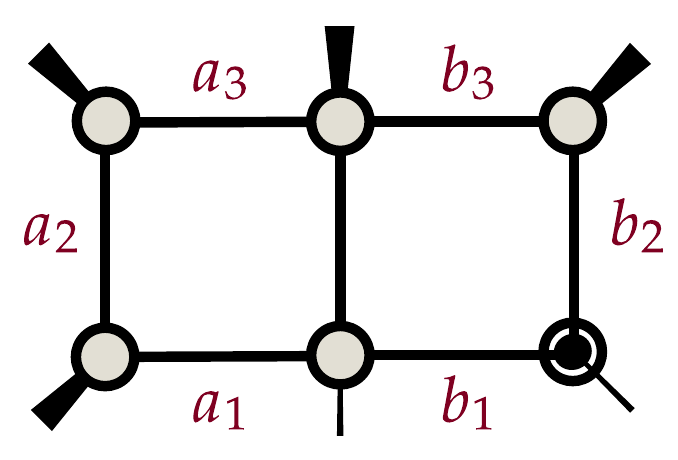}\;,\;\includegraphics[valign = c, scale=0.3,trim=0 0 0 0cm]{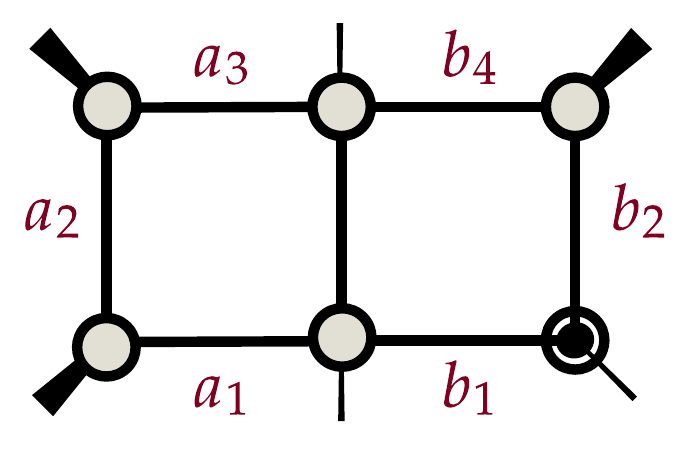}\;,\;\underset{a-\text{cycle}}{\includegraphics[valign = c, scale=0.3,trim=0 0 0 0cm]{archetype-2-4-doublebox-3-onshell.pdf}}\;,\;\underset{a-\text{cycle}}{\includegraphics[valign = c, scale=0.3,trim=0 0 0 0cm]{archetype-2-4-doublebox-4-onshell.pdf}} \Bigg\rbrace. 
\end{equation}
The period matrix that results is given by

\begin{equation}
    \begin{blockarray}{ccccccc}
 \hspace{1pt}&\Omega_1 & \Omega_2 & \Omega_3 & \Omega_4 & \Omega_5 &\Omega_6\\
\begin{block}{c(cccccc)}
  \mathcal{I}_1 & \bl{f}^{\bur{1}}_{\bl{1}} & \bl{f}^{\bur{2}}_{\bl{1}} & \bl{g}^{\bur{3}}_{\bl{1}} & \bl{g}^{\bur{4}}_{\bl{1}} & \bl{g}^{\bur{5}}_{\bl{1}} &\bl{g}^{\bur{6}}_{\bl{1}}\\
  \mathcal{I}_2 & \bl{f}^{\bur{1}}_{\bl{2}} & \bl{f}^{\bur{2}}_{\bl{2}} & \bl{g}^{\bur{3}}_{\bl{2}} & \bl{g}^{\bur{4}}_{\bl{2}} & \bl{g}^{\bur{5}}_{\bl{2}} &\bl{g}^{\bur{6}}_{\bl{2}}\\
  \mathcal{I}_3 & 0 & 0 & \bl{h}^{\bur{3}}_{\bl{3}} & 0 & 0 &0\\
  \mathcal{I}_4 & 0 & 0 & 0 & \bl{h}^{\bur{4}}_{\bl{4}} & 0 &0\\
  \mathcal{I}_5 & \bl{h}^{\bur{1}}_{\bl{5}} & 0 & 0 & 0 & \bl{h}^{\bur{5}}_{\bl{5}} &0\\
  \mathcal{I}_6 & 0 & \bl{h}^{\bur{2}}_{\bl{6}} & 0 & 0 & 0 &\bl{h}^{\bur{6}}_{\bl{6}}\\
\end{block}
\end{blockarray}.
\end{equation}
This is full-rank, and as such can be diagonalized. This results in a stratified basis of four master integrands of (at most) rigidity 1 and two of rigidity 0. 
\hfill
\vspace{0.7cm}

Again we find that some of the diagonalized integrands are necessarily of indefinite rigidity. Specifically, we require that $\widetilde{\mathcal{I}}_{1,2}$ will have indefinite rigidity, as not only will they contain $b$-cycle support of their respective dual elliptic contours, but must also support \emph{all four} leading singularities of the pentabox. Clearly, even if we were to diagonalize the basis by separating out elliptic contours from non-elliptic ones, imposing definite rigidity of the basis will not be possible.

Although there are no pentaboxes that have three elliptic curves, it is the case that any such event would also obstruct stratification. The reason is that there will not be a second element in the initial basis to support the $b$-cycle of the third elliptic, due to there being only two pentabox numerators. The $b$-cycle period would vanish on any other topology (excluding its double box), rendering the period matrix less-than-full rank. For four elliptic curves, the problem is more fundamental; the basis itself being only \bl{6} dimensional simply cannot furnish the minimally eight degrees of freedom needed to diagonalize all of its elliptics.

We conclude this part of the section by a combinatorial exercise. For large multiplicity, the topology $\mathcal{I}_{20}$ dominates over all of the others exponentially swamping out the others. The consequence is that stratification become \emph{arbitrarily impossible} for large multiplicity in $4$-gon power counting. Our main task in the remaining part of this section will be to demonstrate how this problem is obviated---in a sense, trivially---by the extension to $3$-gon power counting.

\subsection{Diagonalization along Definite Rigidity}\label{sec:4.3}
In this section, we aim to show that the choice of permitting $3$-gon power counting will have the effect of stratifying our basis along rigidity for a certain choice of spanning contours. Since there are a few working parts that require the use of unfamiliar notation, we will go back to a simple toy model in one dimension, and motivate the general case, to be discussed in section \ref{sec:4.3.2}.

\subsubsection{Another Toy Model in $\mathbb{P}^1$}
Stratification along ridigity finds a more familiar avatar in one dimension, where it is easier to concretely demonstrate the cohomological nature of the diagonalization procedure. Suppose we have a general elliptic differential form, given an elliptic curve, and a non-elliptic cohomology spanned by a single pole $\var{z} = \var{z_{1}}$. In other words, the cohomology group here is three-dimensional, dual to two elliptic cycles and one residue operation around $\var{z_{1}}$. 

In the notation used in the previous discussions, we have

\begin{equation}
   \lbrace{C_1,C_2,C_3\rbrace} = \lbrace{a-\text{cycle}, b-\text{cycle}, \underset{\var{z} = \var{z_1}}{\text{Res}}\rbrace}.
\end{equation}

The dual problem is to find a cohomology basis dual to this set of contours (or cuts) in which the differential form can be expanded. We choose for this basis a first \emph{ansatz} given by 

\begin{equation}
    \lbrace{f_{1},f_2,f_3\rbrace} = \bigg\lbrace \frac{\dbar\var{z}}{y(\var{z})}, \frac{\dbar\var{z}}{y(\var{z})(\var{z}-\var{z_1})}, \frac{\dbar\var{z}}{\var{z} - \var{z_{1}}} \bigg\rbrace
\end{equation}
where $y^2$ is the quartic governing the elliptic curve. The computation of the period matrix then follows the conventions set in section \ref{sec:3.2}, so that we have

\begin{equation}
    \begin{blockarray}{cccc}
 \hspace{1pt}&C_1 & C_2 & C_3\\
\begin{block}{c(ccc)}
  f_1 & \lab{\mathfrak{A}|\omega} & \lab{\mathfrak{B}|\omega} & 0 \\
  f_2 & \lab{\mathfrak{A}|\omega_{\var{z_1}}} & \lab{\mathfrak{B}|\omega_{\var{z_1}}} & y(\var{z_1})^{-1} \\
  f_3 & 0 & 0 & 1 \\
\end{block}
\end{blockarray}:=\mathbf{M}_{ij}.
\end{equation}
The matrix is block diagonal, and for generic elliptic curves can be verified (both numerically and analytically) to be of full rank. The diagonalized basis functions are then given by

\begin{equation}
    \widetilde{f}_{i} = f_{j}(\mathbf{M}^{-1})_{ji},
\end{equation}
which results in the following final result to the `cut' conditions 

\begin{equation}
    \widetilde{f}_{1} = \frac{\lab{\mathfrak{B}|\omega_{\var{z_1}}}f_{2} - \lab{\mathfrak{B}|\omega}f_{1} + \lab{\mathfrak{B}|\omega}y(\var{z_1})^{-1}f_3}{\lab{\mathfrak{A}|\omega}\lab{\mathfrak{B}|\omega_{\var{z_1}}}-\lab{\mathfrak{B}|\omega}\lab{\mathfrak{A}|\omega_{\var{z_1}}}},
\end{equation}

\begin{equation}
    \widetilde{f}_{2} = \frac{-\lab{\mathfrak{A}|\omega_{\var{z_1}}}f_{1} + \lab{\mathfrak{A}|\omega}f_{2}-\lab{\mathfrak{A}|\omega}y(\var{z_1})^{-1}f_3}{\lab{\mathfrak{A}|\omega}\lab{\mathfrak{B}|\omega_{\var{z_1}}}-\lab{\mathfrak{B}|\omega}\lab{\mathfrak{A}|\omega_{\var{z_1}}}},
\end{equation}
and

\begin{equation}
    \widetilde{f}_3 = f_{3}.
\end{equation}

The diagonalization ensures that the first two basis forms are pure elliptic, and have no polylog pollution. Due to the \emph{ansatz} picked, the polylog piece is automatically stratified and has no elliptic support. 

Note that this would not have been possible had we only been allowed two basis forms. Diagonalizing on the elliptics would have introduced polylog support, and vice versa., Ultimately, at least as many basis elements rigid contours are required; more if stratification is to be done consistently. This is the essential idea behind the entire procedure of stratification we will develop in the following; a large enough basis will guarantee---so long as we have a full rank period matrix---stratification along rigidity.

\subsubsection{Planar Two Loops}\label{sec:4.3.2}
We argue in this section that the extension to $3$-gon power counting renders the stratification problem almost trivial, with the cavil that the cases of the double box and box triangle have to be studied a little more closely than the others. We demarcate this analysis into two parts, based on the nature of the maximal cut of a given topology. For those topologies whose maximal cut contains elliptics, it becomes necessary to stratify the top-level degrees of freedom. 

Our strategy will follow an expanded version of the arguments in \cite{Bourjaily:2022tep}, where we work our way up from the simplest to most complicated topologies, showing that each can be expanded in a master integrand (sub)basis that is manifestly stratified along rigidity. When we will show this for the pentabox, it will amount to the statement that there does exist a basis of master integrands at two loops which are diagonalized along rigidity. 

\paragraph*{The Double Triangle.}
The double triangle in $3$-gon power counting,

\begin{equation}
     \includegraphics[valign = c,scale = 0.3]{double-triangle-3-gon-labelled.pdf} = \frac{\dbar^4\ell_1\dbar^4\ell_1}{(\ell_1|\bur{a_1})(\ell_1|\bur{a_2})(\ell_1|\bl{X}) (\ell_1|\ell_2)(\ell_2|\bur{b_1})(\ell_2|\bur{b_2})(\ell_2|\bl{X})}
\end{equation}
is automatically stratified. Seeing this involves two steps; the first is that since the numerator space is spanned by a scalar and there are no daughter topologies, there is just \bur{1} top-level degree of freedom to match. If the double triangle is itself of zero rigidity, we are done, since the top-level numerator can be matched by choosing some maximal cut surface.

Integrated expressions (using for example {\texttt{HyperInt}} \cite{Panzer:2014caa}) for the double triangle are available \cite{chiPersonal}. An easier way of seeing this however is as a degeneration of the general scalar double box

\begin{equation}
    \includegraphics[valign = c,scale = 0.3]{doublebox-4-gon-labelled.pdf}
\end{equation}
where $\bur{a_1} = \bur{b_1} = \bl{X}$, where $\bl{X}$ is the infinity twistor. The result is the factorization of the corresponding elliptic curve into 

\begin{equation}
\lab{\bur{\widehat{A}}\bur{\widehat{C}}\bur{\widehat{D}}\bur{\widehat{F}}} = \var{\alpha_2}\lab{\bl{x}\bur{\widehat{C}}\bl{X}\bur{\widehat{F}}}+\var{\alpha_1}\lab{\bl{X}\bur{\widehat{C}}\bl{x}\bur{\widehat{F}}}.
\end{equation}
On the contour $\var{\alpha_1} = \var{\alpha_2}$, the latter quadratic takes the form

\begin{equation}
    \begin{aligned}
0 = & \lab{\bl{x}((\bur{cC}\cap\bur{bB}\bl{x}))\bl{X}((\bur{fF})\cap(\bur{eE}\bl{x}))}\\
&\var{\alpha}\lab{\bl{x}((\bur{cC}\cap\bur{bB}\bl{x}))\bl{X}((\bur{fF})\cap(\bur{eE}\bl{X}))}\\
&\var{\alpha}\lab{\bl{x}((\bur{cC}\cap\bur{bB}\bl{X}))\bl{X}((\bur{fF})\cap(\bur{eE}\bl{x}))}\\
&\var{\alpha}^2\lab{\bl{x}((\bur{cC}\cap\bur{bB}\bl{X}))\bl{X}((\bur{fF})\cap(\bur{eE}\bl{X}))}
    \end{aligned}
\end{equation}
The top-level term could be matched onto one of two cuts---the residues of which add up to zero due to the GRT---the even or odd combinations of the two poles of the latter quadratic. The factorization of the elliptic guarantees that there are no other sources of rigidity, and as such, the double triangle is naturally stratified.

\paragraph*{The Box-Triangle. }
The box triangle is a rather nontrivial case, and we will go through it in some detail. The general box triangle in $3$-gon power counting has a numerator space spanned by one inverse propagator:

\begin{equation}
    \includegraphics[valign = c,scale = 0.3]{box-triangle-3-gon-labelled.pdf} = \frac{(\ell_2|\bl{N})\dbar^4\ell_1\dbar^4\ell_1}{(\ell_1|\bur{a_1})(\ell_1|\bur{a_2})(\ell_1|\bl{X}) (\ell_1|\ell_2)(\ell_2|\bur{b_1})(\ell_2|\bur{b_2})(\ell_2|\bur{b_3})(\ell_2|\bl{X})}.
\end{equation}
$3$ contact terms are spanned by the daughter double triangles obtained by writing $|\bl{N}) = |\bur{b_i})$, leaving \bur{3} top-level terms. The daughters have already been seen to be stratified, and each is expanded in integrands with zero rigidity. Our task now is to show that there exists a basis of \bur{3} master integrand numerators that diagonalize the top-level degrees of freedom along rigidity.

\emph{A priori}, there isn't a combinatorial obstruction; the box triangle contains one elliptic on the seven cut obtained by taking the maximal cut surface followed by $(\ell_2|\bl{X})=0$. A polylogarithmic contour in concert with the two cycles of this elliptic could in principle saturate the top-level terms, which can then be diagonalized. We will show that this does work in practice.

We parametrize the six cut the usual way, with the $\ell_1$ loop spanned by $\var{\alpha_1}$ and the $\ell_2$ loop spanned by $\var{\alpha_2}$. Eliminating $\var{\alpha_1}$ by encircling the odd residue of 

\begin{equation}
    (\ell_1(\var{\alpha_1})|\ell_2(\var{\alpha_2})) = 0
\end{equation}
we find for a general double box the one form (relabelling $\var{\alpha_2}=\var{\alpha}$)

\begin{equation}
    \underset{\text{seven-cut}}{\oint}\includegraphics[valign = c,scale = 0.3]{box-triangle-3-gon-labelled.pdf} = -i\frac{(\ell_2(\var{\alpha})|\bl{N})\dbar\var{\alpha}}{(\ell_2(\var{\alpha})|\bl{X})y(\var{\alpha})}
\end{equation}
where

\begin{equation}
   \begin{aligned}
        y_1(\var{\alpha})  = &\left(\lab{\bl{X}((\bur{cC})\cap(\bur{bB}\bl{X}))\bur{\widehat{C}}\bur{\widehat{D}}}+\lab{\bl{x}((\bur{cC})\cap(\bur{bB}\bl{X}))\bur{\widehat{C}}\bur{\widehat{D}}}\right)^2 - \\
        &4\lab{\bl{X}((\bur{cC})\cap(\bur{bB}\bl{X}))\bur{\widehat{C}}\bur{\widehat{D}}}\lab{\bl{x}((\bur{cC})\cap(\bur{bB}\bl{x}))\bur{\widehat{C}}\bur{\widehat{D}}}.
    \end{aligned}
\end{equation}
The form on the seven cut also contains two poles due to the propagator at infinity, with solutions $\var{\alpha^{\pm}}$ at

\begin{equation}
\begin{aligned}
    0 =& \lab{\bur{d}((\bur{fF})\cap(\bur{eEd}))\bl{xX}}+\\
    &\var{\alpha}\left(\lab{\bur{d}((\bur{fF})\cap(\bur{eED}))\bl{xX}}+\lab{\bur{D}((\bur{fF})\cap(\bur{eEd}))\bl{xX}}\right)+\\
    &\var{\alpha}^2\lab{\bur{D}((\bur{fF})\cap(\bur{eED}))\bl{xX}}.
\end{aligned}
\end{equation}

A convenient \emph{ansatz} for the initial choice of numerators turns out be 

\begin{equation}
\lbrace{|\bl{N_1}),|\bl{N^0_2}),|\bl{N^0_2})\rbrace} = \lbrace{|\bl{X}), |\bl{Q_{\text{odd}}}), |\bl{Q_{\text{even}}})\rbrace}
\end{equation}

where $|\bl{Q_{\text{odd,even}}})$ correspond to numerators normalized on the octacuts formed by the even and odd residues of the previous quartic respectively\footnote{The vagaries of the notation can be mystifying; the numerator $|\bl{Q_{\text{odd}}})$ \emph{vanishes} on the odd contour.}.

If we represent the roots of $(\ell_2(\var{\alpha})|\bl{Q_{\text{even,odd}}})$ by $\var{q^{\pm}_{\text{even,odd}}}$, we obtain for the partial fractioned representations of the one forms

\begin{equation}
\frac{(\ell_2(\var{\alpha})|\bl{Q_{\text{even,odd}}})\dbar\var{\alpha}}{(\ell_2(\var{\alpha})|\bl{X})y(\var{\alpha})} = \frac{1}{y(\var{\alpha})} + \frac{\Delta(\var{\alpha^+};\var{q^{\pm}_{\text{even,odd}}})}{y(\var{\alpha})(\var{\alpha}-\var{\alpha^+})}-\frac{\Delta(\var{\alpha^-};\var{q^{\pm}_{\text{even,odd}}})}{y(\var{\alpha})(\var{\alpha}-\var{\alpha^-})}
\end{equation}
where

\begin{equation}
    \Delta(\var{\alpha};\var{q^{\pm}_{\text{even,odd}}}) = \frac{\var{\alpha}^2 - \var{\alpha}(\var{q^{+}_{\text{even,odd}}} + \var{q^{-}_{\text{even,odd}}})-\var{q^{+}_{\text{even,odd}}}\var{q^{-}_{\text{even,odd}}}}{\var{\alpha^+}-\var{\alpha^-}}.
\end{equation}
The shifts

\begin{equation}
    |\bl{Q_{\text{even,odd}}})\longrightarrow |\bl{Q_{\text{even,odd}}}) + |\bl{X}) = |\bl{N_{1,2}})
\end{equation}
eliminate the pure elliptic pieces from the corresponding forms.

The final set of spanning contours are defined (once the seven cut has been computed) by the $a$-cycle, $b$-cycle and even residue around the poles of $(\ell_2(\var{\alpha})|\bl{X})$. This results in the following period matrix

\begin{equation}
    \begin{blockarray}{cccc}
 \hspace{1pt}&\Omega_1 & \Omega_2 & \Omega_3\\
\begin{block}{c(ccc)}
  |\bl{N_1})& \bl{J}^{\bur{a}}_{\bur{1}} & \bl{J}^{\bur{b}}_{\bur{1}} & 0 \\
  |\bl{N_2}) & \bl{J}^{\bur{a}}_{\bur{2}} & \bl{J}^{\bur{b}}_{\bur{2}}& 0 \\
  |\bl{N_3}) &\bl{J}^{\bur{a}}_{\bur{3}} & \bl{J}^{\bur{b}}_{\bur{3}} & 1 \\
\end{block}
\end{blockarray}:=\mathbf{M}_{ij}.
\end{equation}
where

\begin{equation}
    \bl{J}^{\bur{a,b}}_{\bur{1}} = \underset{a,b-\text{cycle}}{\oint}\frac{\dbar\var{\alpha}}{y(\var{\alpha})}
\end{equation}

\begin{equation}
    \bl{J}^{\bur{a,b}}_{\bur{2}} = \underset{a,b-\text{cycle}}{\oint}\dbar\var{\alpha}\left(\frac{\Delta(\var{\alpha^+};\var{q^{\pm}_{\text{even}}})}{y(\var{\alpha})(\var{\alpha}-\var{\alpha^+})}-\frac{\Delta(\var{\alpha^-};\var{q^{\pm}_{\text{even}}})}{y(\var{\alpha})(\var{\alpha}-\var{\alpha^-})}\right),
\end{equation}
and
\begin{equation}
    \bl{J}^{\bur{a,b}}_{\bur{3}} = \underset{a,b-\text{cycle}}{\oint}\dbar\var{\alpha}\left(\frac{\Delta(\var{\alpha^+};\var{q^{\pm}_{\text{odd}}})}{y(\var{\alpha})(\var{\alpha}-\var{\alpha^+})}-\frac{\Delta(\var{\alpha^-};\var{q^{\pm}_{\text{odd}}})}{y(\var{\alpha})(\var{\alpha}-\var{\alpha^-})}\right)
\end{equation}
The period matrix is full-rank. Accordingly, the numerators can be diagonalized using the inverse.

The consequence of this is a basis of master integrands for \emph{any} box-triangle in $3$-gon power counting that is naturally stratified along rigidity. Two of the master integrands span all of the elliptic contributions, while the last one is pure polylog. This will be true for any box-triangle due to the nature of the topology---since there is a elliptic curve contained in it, diagonalization along that elliptic suffices to diagonalize the entire top-level set of numerators.

\paragraph*{The Double Box.} One of the consequences of $3$-gon power counting is to leave the elliptic double box dependent on a total of \emph{seven} elliptic curves. To see this, we start with the form of the double box in dual conformal coordinates:

\begin{equation}
    \includegraphics[valign = c,scale = 0.3]{doublebox-3-gon-labelled.pdf} = \frac{(\ell_1|\bl{N})(\ell_2|\bl{N'})\dbar^4\ell_1\dbar\ell^4}{(\ell_1|\bur{a_1})(\ell_1|\bur{a_2})(\ell_1|\bur{a_3})(\ell_1|\bl{X})(\ell_1|\ell_2)(\ell_2|\bur{b_1})(\ell_2|\bur{b_2})(\ell_2|\bur{b_3})(\ell_2|\bl{X})}.
\end{equation}
The seven elliptic curves come from the following combinatorial exercise. The numerator space of the double box is \bl{36} dimensional. Ignoring double triangle contributions for the time being, note that we can have a total of

\begin{equation}
    \binom{3}{1} + \binom{3}{1} = 6
\end{equation}
daughter box-triangles, amounting to a collapse of one external side of one of the two boxes. Each of these box-triangles contains a single elliptic curve, and as such its maximal cut is supported by its parent double box. Additionally, the double box itself supports an intrinsic elliptic curve along the contours spanned by

\begin{equation}
    (\ell_1|\bur{a_{i}}) = (\ell_2|\bur{b_{i}}) = 0
\end{equation}
and

\begin{equation}
    (\ell_1|\ell_2) = 0.
\end{equation}
This results in a total of seven elliptics along which we need to stratify the master integrand basis. 

The $\bl{36}$ dimensional space of inverse propagators spanning the numerator space of the double box in triangle power counting decomposes as follows. We have 9 contact terms giving rise to double triangles, naturally spanned by a master integrand basis that is identically stratified by rigidity. We have a further 18 contact terms due to box-triangle daughters, which are spanned by a set of 6 \bur{3} dimensional subbases which we have already shown to be stratified. 1 contact term is just the kissing triangle. What we have left is an \bur{8} dimensional space of top-level terms, which we will now show will suffice to stratify the basis along the remaining elliptic curve.

Showing that there exists an eight-dimensional space of contours that achieves this follows the same methods as that of the box-triangle, but is much more involved technically. Accordingly, we'll be somewhat telegraphic, and simply describe how it is done. Starting with a generic numerator $(\ell_1|\bl{N_1})(\ell_2|\bl{N_2})$, we can compute the even and odd contours corresponding to the seven-cut of the double box (no propagators at infinity cut). The results are as follows

\begin{equation}
    \underset{\text{even 7-cut}}{\oint}\includegraphics[valign = c,scale = 0.3]{doublebox-3-gon-labelled.pdf} = -i\frac{P_{4}(\var{\alpha})\dbar\var{\alpha}}{P_{8}(\var{\alpha})}
\end{equation}
and

\begin{equation}
    \underset{\text{odd 7-cut}}{\oint}\includegraphics[valign = c,scale = 0.3]{doublebox-3-gon-labelled.pdf} = -i\frac{P_{4}(\var{\alpha})\dbar\var{\alpha}}{y(\var{\alpha})P_{8}(\var{\alpha})}
\end{equation}
where $P_4$ and $P_8$ are quartic and octic polynomials in the undetermined variable $\var{\alpha}$, while we have as usual denoted the elliptic by $y(\var{\alpha})$.

We see that in addition to the two choices for elliptic contours, we have a total of \emph{eight} polylogarithmic contours we may evaluate to arrive at a leading singularity, furnishing a total of 16 contour choices with which we can saturate the \bur{8} top-level degrees of freedom. 

To ensure stratification, we pick two of these as the $a$- and $b$-cycles following the odd-seven cut. For the remaining six, the fact that they are all polylog gives us some freedom. We just mention that we have ensured that there is a full-rank choice we can make out of the remaining 14 contours. Ultimately, the \bur{8} top-level terms can be resolved into a stratified basis of master integrands, two of which are of rigidity 1 and six of which are of rigidity 0.

The result is that the \bl{36} dimensional space of double boxes is completely stratified, and is always expressible in terms of master integrands which are of definite rigidity.

\paragraph*{The Pentatriangle.} The pentatriangle does not have double box subtopologies but does contain a number of box triangles. Indeed, it contains a total of $4$ box-triangle daughters, formed by contracting one of the pentagon edges, seen from the form in dual momentum space

\begin{equation}
    \includegraphics[valign = c,scale = 0.3]{pentatriangle-3-gon-labelled.pdf} = \frac{(\ell_2|\bl{N})(\ell_2|\bl{N'})\dbar^4\ell_1\dbar\ell^4}{(\ell_1|\bur{a_1})(\ell_1|\bur{a_2})(\ell_1|\bl{X})(\ell_1|\ell_2)(\ell_2|\bur{b_1})(\ell_2|\bur{b_2})(\ell_2|\bur{b_3})(\ell_2|\bur{b_4})(\ell_2|\bl{X})}.
\end{equation}
Any assignment setting $|\bl{N}) = |\bur{b_i})$ yields a box-triangle.

The result of this is that the generic integrated form of the pentatriangle will be of mixed rigidity, controlled by elliptic polylogs formed out of \emph{four}, generally speaking, distinct elliptic curves. The job at hand then is to show that any pentatriangle can be spanned by a master integrand basis where each element is of mixed rigidity. 

The problem is rendered almost trivial by the fact that all of the elliptic curves are governed by contact terms. The space of contact degrees of freedom spanned by the box-triangle subtopologies is 

\begin{equation}
    \bur{3}\times \binom{4}{1} = 12
\end{equation}
dimensional. In accordance with the stratification proved for the box-triangle, this subspace can always be written in terms of a basis of master integrands that resolves into 8 elements of definite rigidity 1 and 4 of definite rigidity 0, saturating all of the elliptic degrees of freedom contained in the general pentatriangle.

We also have a $6$ dimensional subspace spanned by double-triangle topologies, which is of definite rigidity. Left behind are a total of \bur{2} contact degrees of freedom, which can be diagonalized along the following contours of definite rigidity

\begin{equation}
    \underset{(\ell_1|\bl{X})=0}{\text{Res}}\includegraphics[valign = c,scale = 0.3]{pentatriangle-3-gon-top-level-on-shell.pdf}
\end{equation}
where $i = 1,2$.

The consequence of this counting exercise is that for any generic numerator, the pentatriangle always admits of a diagonalized basis expansion of dimension \bl{20}, running through a span of 8 integrands of definite rigidity 1 and 12 integrands of rigidity 0.

\paragraph*{The Pentabox. } Stratifying pentaboxes ultimately generalizes the process used for pentatriangle, since the salient features remain largely unchanged. A general pentabox in triangle power counting,

\begin{equation}
    \includegraphics[valign = c,scale = 0.3]{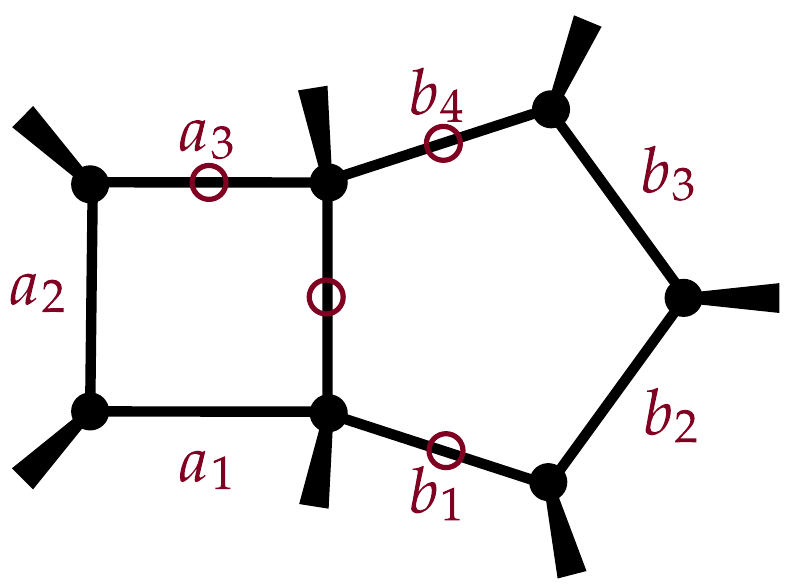} = \frac{(\ell_1|\bl{N_1})(\ell_2|\bl{N_2})(\ell_2|\bl{N_3})\dbar^4\ell_1\dbar\ell^4}{(\ell_1|\bur{a_1})(\ell_1|\bur{a_2})(\ell_1|\bur{a_3})(\ell_1|\bl{X})(\ell_1|\ell_2)(\ell_2|\bur{b_1})(\ell_2|\bur{b_2})(\ell_2|\bur{b_3})(\ell_2|\bur{b_4})(\ell_2|\bl{X})}
\end{equation}
has a numerator space that is \bl{120} dimensional, spanned by 116 contact terms and \bur{4} contact degrees of freedom. The pentabox itself admits of 22 different elliptic contours. The counting of these come from the contact terms; there are a total of $4$ double box subtopologies and a total of 

\begin{equation}
    \binom{4}{1}\binom{3}{1} + \binom{4}{2} = 18
\end{equation}
box-triangle subtopologies. Each of these brings with it one top-level elliptic, adding up to a total of 22.

As in the case of the pentatriangle, each of these elliptic curves is contained entirely within the numerator space spanned by contact degrees of freedom. Specifically, we have for each of the double boxes \bur{8} top-level terms, amounting to 

\begin{equation}
    \bur{8}\times 4 = 32
\end{equation}
master integrands, which can be developed in a basis of 8 integrands of definite rigidity 1 and 24 of definite rigidity 0, accounting for four out of 22 elliptics.

The box-triangle subtopologies come equipped with \bur{3} top-level master integrands each, furnishing a total of 

\begin{equation}
    \bur{3}\times 18 = 54
\end{equation}
master integrands. These are resolved into a basis of 18 integrands with rigidity 1 and 36 polylogarithmic integrands. 

The remaining $30$ contact degrees of freedom are spanned by $6$ pentatriangle master integrands, $2$ kissing box-triangle master integrands, $4$ kissing triangle master integrands and $18$ double triangles, all of which are purely polylogarithmic by the arguments of the preceding discussions. Accordingly, we are left with the \bur{4} top-level pentaboxes, which we expand in terms of \bur{4} master integrands of definite rigidity 0 by matching to the four maximal cuts

\begin{equation}
    \includegraphics[valign = c,scale = 0.3]{pentabox-3-gon-top-level-on-shell.pdf}.
\end{equation}

A key point to emphasize here is that prescriptivity guarantees that all of these arguments go through, since the master integrands are defined by construction to be diagonal with respect to contours which discriminate---\emph{by construction}---elements of differing rigidity. By preparing a decomposition into subspaces that are stratified ensures the entire basis is stratified as well, since diagonalization guarantees that the spanning cuts of one subspace have zero support on any other.

\subsection{Numerators, Stratification and Incomplete Bases}\label{sec:4.4}
A preeminent theme of our approach to resolving the problem of stratifying bases was the enlargement of the bases by dressing the scalar integrands with inverse propagators. Doing this gave us a much larger set of master integrands to work with, and by making good choices of contour prescriptions, we were able to classify the bases in a manner that respected definite rigidity.

In that context, there is the problem of whether or not this is the optimal basis size to work with. In other words, would there have been numerator bases that were smaller, yet capable of accommodating a division between its elements guided by rigidity? Answering this question amounts to having recourse to the interplay between completeness and basis size. We'll keep this discussion somewhat schematic, and try to emphasize the basis building---rather than technical or analytic---aspects involved.

In the case of the box-triangle, the only alternative to a numerator space spanned by a single inverse propagator would have been one spanned by a scalar, exemplified by integrands of the form

\begin{equation}
    \includegraphics[valign = c,scale = 0.3]{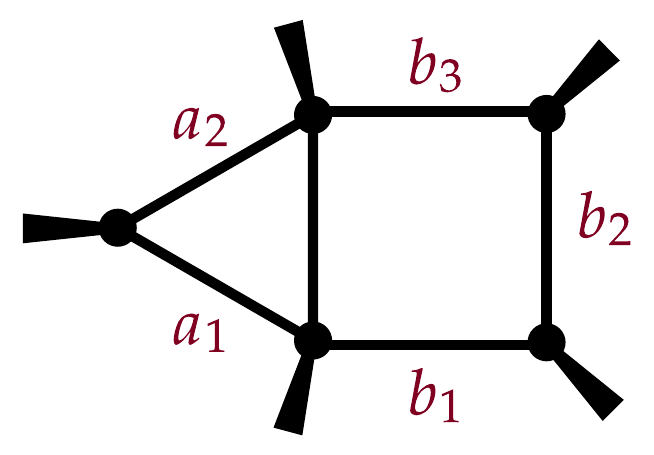} = \frac{\dbar^4\ell_1\dbar^4\ell_1}{(\ell_1|\bur{a_1})(\ell_1|\bur{a_2})(\ell_1|\bl{X}) (\ell_1|\ell_2)(\ell_2|\bur{b_1})(\ell_2|\bur{b_2})(\ell_2|\bur{b_3})}.
\end{equation}

Since the numerator space is spanned by a scalar, it is \bl{1} dimensional, with no contact degrees of freedom. The result is a space of top-level numerators which is of dimension \bur{1}, which is at least short one further dimension required for stratification. The upshot of this fact is that we \emph{require} at least one numerator factor, which automatically introduces double triangles into our basis. 

If we look now at double boxes, the $3$-gon power counting requirement endowed them with two inverse propagators. The case of a scalar numerator is just that of $4$-gon power counting, which we know can't be stratified. If instead we have one inverse propagator, we find general elements of the form

\begin{equation}
    \includegraphics[valign = c,scale = 0.3]{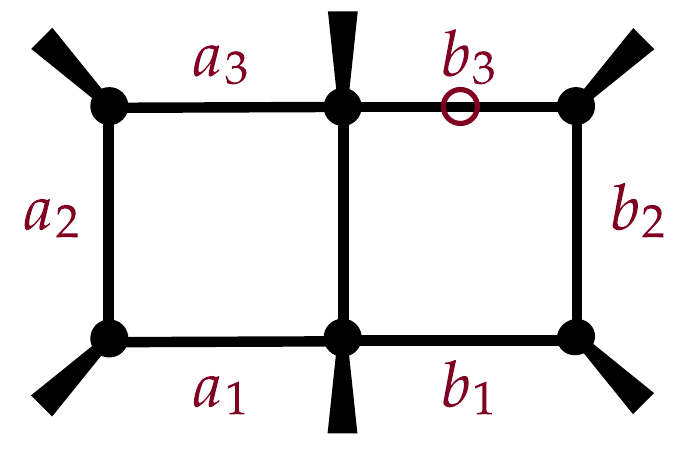} = \frac{(\ell_2|\bl{N})\dbar^4\ell_1\dbar^4\ell_1}{(\ell_1|\bur{a_1})(\ell_1|\bur{a_2})(\ell_1|\ell_2)(\ell_2|\bur{b_1})(\ell_2|\bur{b_2})(\ell_2|\bur{b_3})(\ell_2|\bl{X})}.
\end{equation}
A counting exercise reveals the presence of \emph{4} elliptic curves, due to the freedom of choosing three out of four propagators to cut on the second loop. The eight elliptic contours spanning the rigidity-1 part of the homology cannot be spanned by the master integrand basis controlled by this topology, which is \bl{6} dimensional. Once again, this space of master integrands cannot support a stratified set of contour prescriptions.

An entirely similar circumstance presents itself with pentatriangles with fewer than two inverse propagators; a generic element is given by the following in dual momentum space

\begin{equation}
    \includegraphics[valign = c,scale = 0.3]{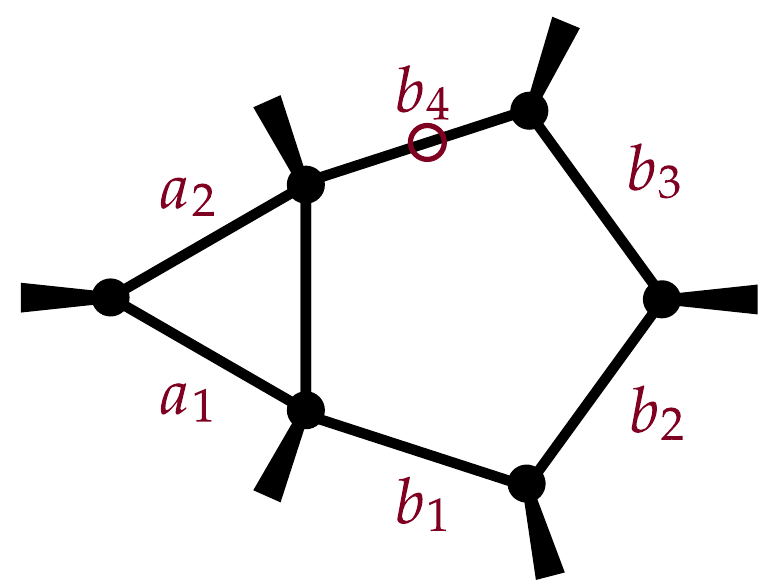} = \frac{(\ell_2|\bl{N})\dbar^4\ell_1\dbar^4\ell_1}{(\ell_1|\bur{a_1})(\ell_1|\bur{a_2})(\ell_1|\ell_2)(\ell_2|\bur{b_1})(\ell_2|\bur{b_2})(\ell_2|\bur{b_3})(\ell_2|\bur{b_4})}.
\end{equation}
There are a total of \emph{4} elliptic curves in the general integrated expression, courtesy of the four propagators on the pentagon side (which may be gleaned by analogizing it to the double box by the prescription $|\bl{X}) = |\bur{b_4})$). The space of \bl{6} numerators spanned by the single inverse propagator is again far too small; we would need at least 8 to saturate the elliptic degrees of freedom alone.

We won't belabor this case by considering pentaboxes with one numerator, which exhibit the same pathology we have just observed with the cases of the double box and pentatriangle. The case of pentaboxes with two numerators is amusing---if nothing else after some thought---so it merits some discussion\footnote{I'm grateful to Jacob Bourjaily for several clarifications on this matter.}. Consider the case of the pentabox with a numerator space spanned by elements in $[\ell_1][\ell_2]$

\begin{equation}
    \includegraphics[valign = c,scale = 0.3]{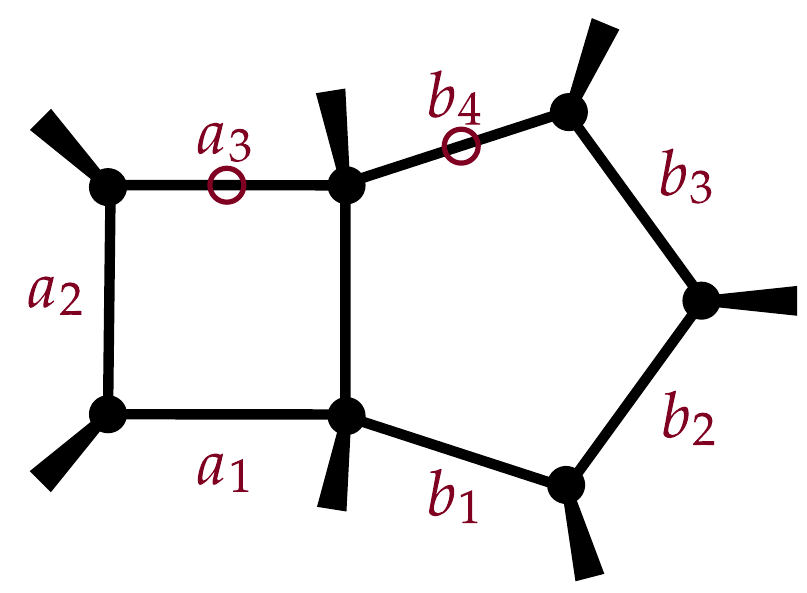} = \frac{(\ell_1|\bl{N_1})(\ell_2|\bl{N_2})\dbar^4\ell_1\dbar^4\ell_1}{(\ell_1|\bur{a_1})(\ell_1|\bur{a_2})(\ell_1|\bur{a_3})(\ell_1|\bl{X})(\ell_1|\ell_2)(\ell_2|\bur{b_1})(\ell_2|\bur{b_2})(\ell_2|\bur{b_3})(\ell_2|\bur{b_4})}.
\end{equation}
This diagram has a total of 

\begin{equation}
    \binom{4}{3}\times \binom{4}{3} = 16
\end{equation}
contours that potentially support elliptic curves, which are all distinct for generic kinematics. The result is a space of 32 contours spanning the part of the homology dual to the span of masters that exact a rigidity of 1. 

\emph{A priori} it would seem that we're in luck since the space of numerators is \bl{36} dimensional and the arguments applied in the marginal cases of 1 or 2 elliptic curves in $4$-gon power counting would work. In other words, 32 out of the \bl{36} numerators would be matched to elliptic contours, while the remaining 4 would be matched by polylog residues, regardless of the daughters' inability to do so on their own.

This appears to be a phenomenon that re-exerts itself for pentaboxes drawing numerators from the span of $[\ell_2]^2$:

\begin{equation}
    \includegraphics[valign = c,scale = 0.3]{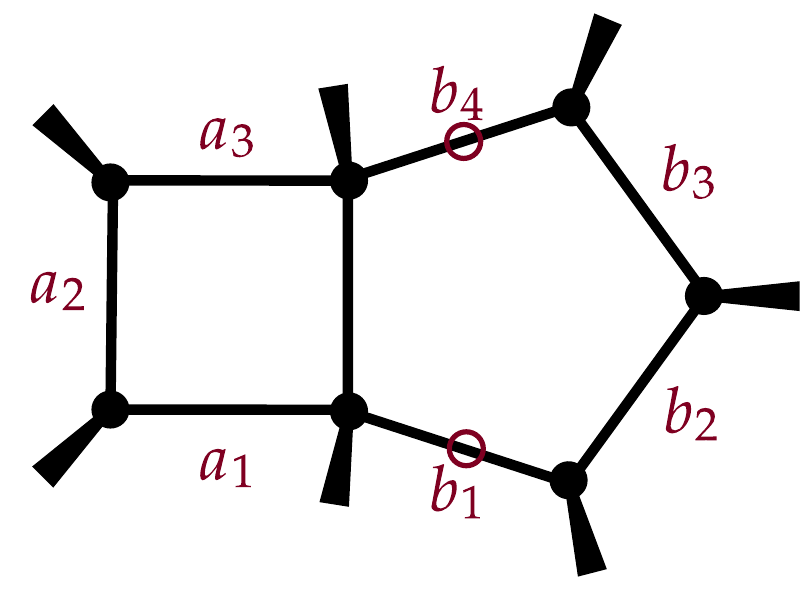} = \frac{(\ell_2|\bl{N_1})(\ell_2|\bl{N_2})\dbar^4\ell_1\dbar^4\ell_1}{(\ell_1|\bur{a_1})(\ell_1|\bur{a_2})(\ell_1|\bur{a_3})(\ell_1|\ell_2)(\ell_2|\bur{b_1})(\ell_2|\bur{b_2})(\ell_2|\bur{b_3})(\ell_2|\bur{b_4})(\ell_2|\bl{X})}.
\end{equation}
Such integrands support a total of 

\begin{equation}
    \binom{5}{3} = 10
\end{equation}
contours that render the cuts elliptic. Again, the \bl{20} dimensional space of master integrands so obtained seem to be ideal for performing the task of stratification, as they potentially saturate all of the elliptic contours.

A more consistent approach would be of course to consider the total span of numerators in $[\ell_1][\ell_2]\oplus[\ell_2]^2$, which is \bl{50} dimensional. In this case, the number of cuts supporting elliptic singularities is \emph{22}, matching the most general pentabox in $3$-gon power counting. The increase in the number of elliptic curves doesn't appear immediately prohibitive; they furnish a total of 44 contours in the homology that exhibit rigidity $1$, and as such may appear to be potentially diagonalizable, making use of 44 out of the \bl{50} master integrands rendered by the choice of two, rather than three inverse propagators.

This is where we are required to make mention of a complete versus incomplete basis\footnote{The language here is misleading, but conventional. By \emph{basis} we just refer to the set of master integrands, which may or may not be \emph{complete}. This is distinct from the use of the word in normal linear algebra. I'm grateful to Cameron Langer for pointing out this confusion; we don't at the moment have a better choice of words.}. Although the na\"ive counting exercise seems to suggest that this numerator space is sufficient to diagonalize all pentaboxes, the problem, in this case, is not so much the size of the basis as it is the fact that such a basis is very likely \emph{incomplete}. 

Indeed, one way of seeing this---in a somewhat extreme avatar---is to compare it to simply truncating our basis in $4$-gon power counting to just contain integrands that support at most 2 elliptic curves, recalling that it was a single archetypal topology of the form

\begin{equation}
    \includegraphics[valign = c, scale=0.3,trim=0 0 0 0cm]{pentabox-archetype-1-8.pdf}
\end{equation}
that contained \emph{4} elliptic curves in its homology.

This is manifestly an unreasonable insistence, since we would have to eliminate all integrands with generic leg distributions, which are the vast majority of integrands---in the sense of asymptotics---as the number of particles becomes large.

Such a basis would also be obviously incomplete, as it ignores---asymptotically---\emph{all} elements in the total span in favour of those that can be stratified. Restricting our attention to pentaboxes with two numerators is the analogous prescription in $3$-gon power counting. Ultimately, stratifying the basis must be done in a manner that is consistent with having one that is both \emph{complete} as well as large enough that stratification is possible. The former condition often has the effect of significantly enlarging the basis once looser power-counting requirements are demanded, but that appears to be the minimum price to enable resolutions according to definite rigidity.
\clearpage
\phantomsection
\label{coda}
\addcontentsline{toc}{section}{\emph{Coda}}
\section*{\emph{Coda}}
\vspace{-15pt}{\color{lapis}\rule{\textwidth}{.6pt}}

\vspace{10pt}
In this work, we have explored and expanded upon a view of insisting that there is a unifying script that makes it possible to ask and answer very interesting questions about scattering amplitudes to which exact results already exist in some---albeit perhaps not the most elegant---form. As mentioned in the introduction, we have exemplified this view by focusing on three aspects of amplitudes at two loops: building bases of integrands and cycles, diagonalizing along \emph{cohomology}, and rigidity (or non-polylogarithmicity). Furthermore, the restriction to planar two loops provided an exceptionally convenient laboratory to test this view, given the interplay between complexity and tractability rendered at this level.

Since we have been concerned with these broad themes, a number of open problems present themselves once any conditions assumed so far are relaxed. As such, in this extended conclusion to the present work, we will highlight and strengthen what we believe are the most interesting of such directions, emphasizing which may be tackled using methods currently available and which likely require mathematical---as opposed to computational---developments before they are likely to give way.

\vspace{10pt}

\paragraph*{\emph{Homological Challenges \& Subleading Maximal Cuts}. }The notion of an elliptic leading singularity solves more than anything else the rudimentary problem of localizing a final degree of freedom left unfixed by the maximal cut---in the paradigmatic case---of the elliptic double box

\ceq{
    \includegraphics[valign=c,scale=0.3]{doublebox-4-gon-labelled.pdf}
}in $4$-gon power counting, which \emph{does not} admit of \emph{any} leading singularities in the absence of poles at infinity for generic kinematics. Notably, this problem does not show up until one descends to bubble power counting at one loop. Indeed, the triangle in $3$-gon power counting

\ceq{
    \includegraphics[valign=c,scale=0.3]{triangle-labelled.pdf} = \frac{\dbar^4\ell}{(\ell|\bur{a_1})(\ell|\bur{a_2})(\ell|\bur{a_3})(\ell|\bl{X})}
}does support two leading singularities once it is permitted to access infinity, even for generic kinematics that do not encode codimension zero cycles in collinear subspaces.

In the case of the bubble diagram in $2$-gon power counting however, a similar problem is encountered, due to the presence of a double pole at infinity

\ceq{
    \includegraphics[valign=c,scale=0.3]{bubble_labelled.pdf} = \frac{\dbar^4\ell}{(\ell|\bur{a_1})(\ell|\bur{a_2})(\ell|\bl{X})^2}.
}On the codimensional two cut spanned by the contours $(\ell|\bur{a_i}) = 0$, one finds a double pole encircling the twistor at infinity. It is an unpleasant fact that there is no residue supported by the attendant contour, rendering the integrand bereft of any canonical choice of the cycle around which it may be matched. This problem was ignored in our discussion of preparing cycles at one loop by contending ourselves with matching at a point, a procedure rendered technically correct only on account of completeness.

One solution suggested recently\footnote{I am grateful to Nima Arkani-Hamed and Simon Caron-Huot for discussions on this matter.} has been to weaken the insistence on \emph{holomorphic} contours and search for cycles in real subspaces of complex momentum space. The so-called spherical contour is obtained by the canonical cycle contained in the embedding of 

\ceq{
    \mathbb{CP}^1\times\mathbb{CP}^1/\text{diag}
}in the two-dimensional complex $\mathbb{P}^2$ spanned by the two-parameter subspace enclosed by the cut conditions. This solution is a very elegant way of dealing with this problem in the case of bubble power counting, since the bubble coefficient is matched by one contour due to being of rank \bur{1}. Unfortunately, this is not the case for worse power counting, which is definitely encountered for theories of serious interest, including pure Yang-Mills and gravity at one loop.

Another avatar of the double pole phenomenon of bubbles is observed for topologies at two loops, even at $3$-gon power counting where one has a massless triangle. To see an archetype of this, simply consider the massless triangle at one loop, where for each $i$ (cyclicity assumed) we have

\ceq{
    (\bur{a_{i+1}}|\bur{a_i}) = 0.
}The problem with this topology is that any condition which sets two legs on-shell renders the third on-shell as well, furnishing a so-called \emph{transverse residue}. Indeed, this is the reasoning behind dropping topologies of the form

\ceq{
    \includegraphics[valign=c,scale=0.3]{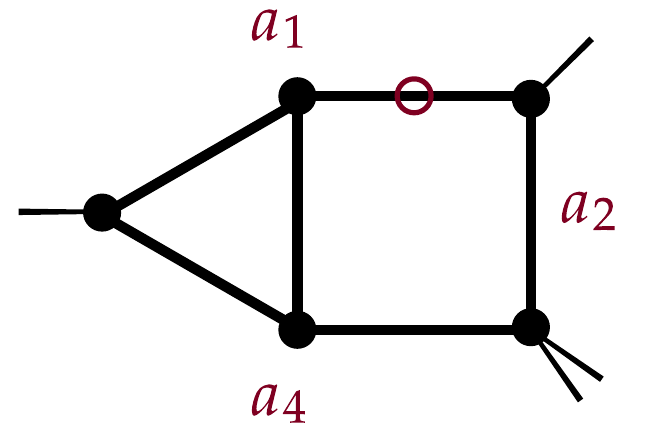}
}
and

\ceq{
 \includegraphics[valign=c,scale=0.3]{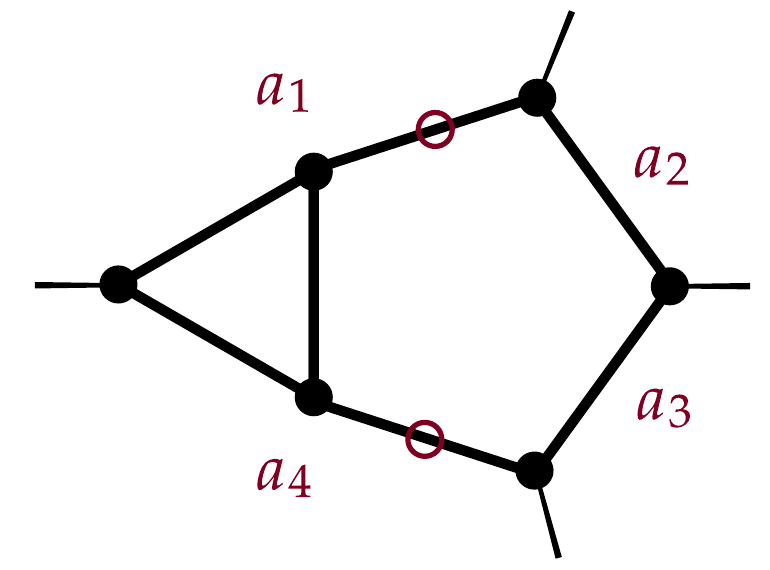}
}when preparing a $3$-gon power counting basis for four-particle scattering at one loop\footnote{We point out that a similar choice was made by the authors of \cite{Bourjaily:2021hcp,Bourjaily:2021iyq} for the case of six-particles, where all topologies with transverse residues were excluded from the basis.}. There simply isn't any maximal cut that yields a non-zero residue onto which one can match the amplitude. 

One possible solution for this rarefied case is simply that any cut that furnishes a residue proportional to a massless triangle must be excluded due to it amounting to a simple renormalization of the three-point function on this contour. This is somewhat unappealing, and a more intrinsic and canonical geometric scheme for matching cycles in the absence of any obvious choices for leading singularities such as is the case here would be preferable.

\vspace{10pt}

\paragraph*{\emph{Diagonalizing along Characteristic Divergences}. }Prescriptive methods of unitarity trivialize the computation of coefficients of the integrands once they are diagonalized, since each coefficient corresponds to a single on-shell function computed on the cut defined by the dual cycle. The upshot of this when dealing with the question of rigidity is that whenever a set of integrands dual to a set of cycles organized by rigidity is found, the master integrands are such that each has what we have called \emph{definite rigidity}, where polylogarithmic and non-polylogarithmic functions have no occasion to mix. 

A derivative of this fact is the realization that it ought to be possible to prepare a spanning set of contours that discriminate not on the basis of rigidity, but on the nature of the divergence exhibited by various classes of integrands. For example---at planar two loops---purely collinear divergences are captured by diagrams with massless legs attached to massive corners, such as

\ceq{
    \includegraphics[valign = c, scale=0.3,trim=0 0 0 0cm]{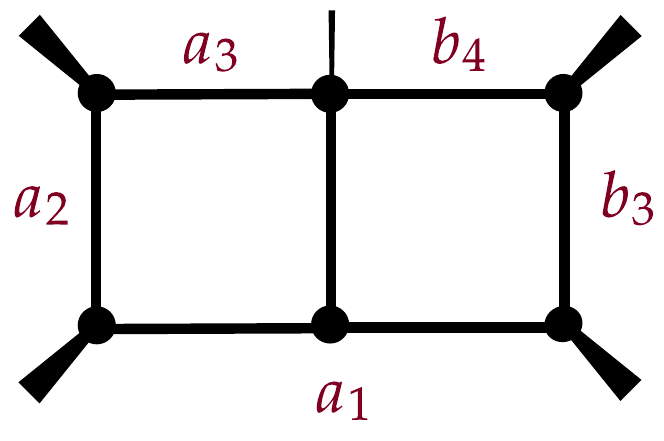}
}whose divergence can be isolated by the contour choice

\ceq{
    \includegraphics[valign = c, scale=0.3,trim=0 0 0 0cm]{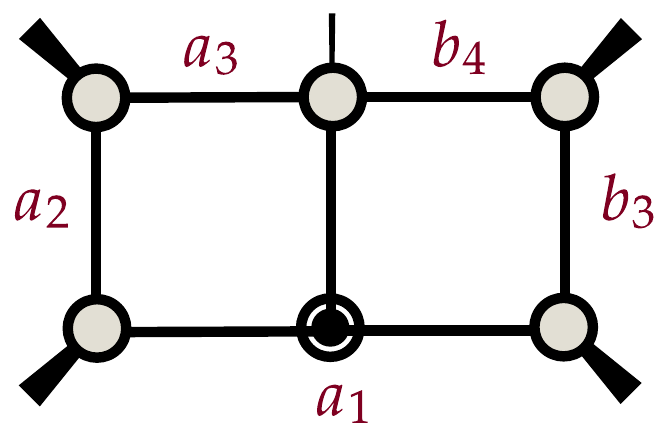}.
}Soft-collinear divergences on the other hand are exhibited by diagrams with adjacent three-point vertices, such as

\ceq{
    \includegraphics[valign = c, scale=0.3,trim=0 0 0 0cm]{no_elliptic_pentabox_pentabox_integrand_daughter_2.pdf}
}which can be extracted in a diagonalizable prescription by the contour choice granted by

\ceq{
    \includegraphics[valign = c, scale=0.3,trim=0 0 0 0cm]{no_elliptic_pentabox_pentabox_on-shell_daughter_2.pdf}.
}A proximate---and perhaps short-term---set of problems that this suggests is to systematically construct both a basis and a spanning set of cuts at two loops which can exhibit the division into IR finite and IR divergent integrands in a format that is manifest. 

Dealing with divergences deep in the ultraviolet are perhaps more involved, mainly when one wants to reveal them in dimensional regularization (DimReg). The problem here is that the scale $\mu_{i}$ introduced in each loop during DimReg must be isolated via an introduction of a new dimension for each loop variable, requiring a basis defined intrinsically in $4 + L$ dimensions, rather that in just $4$.

A downstream effect of this is the appearance of worsening power counting of a theory. To see this, one can use the exemplifying case of one loop pure Yang-Mills, where we know that amplitudes at one loop may be expanded in a basis given by\footnote{I'm grateful to Jacob Bourjaily for making me aware of this fact.} 

\ceq{
   \text{span}\left(\includegraphics[valign = c, scale=0.3,trim=0 0 0 0cm]{box-2gon.pdf}\right) + \mu^2\times\text{span}\left(\includegraphics[valign = c, scale=0.3,trim=0 0 0 0cm]{box-2gon.pdf}\right). 
}The problem with this is that the dimensionful parameter $\mu^2$---defined intrinsically in five, not four, dimensions---is contained in the span

\ceq{
    \mu^2\in [\ell]^2_{d=5}.
}The result is that if we wanted to construct a basis to express pure Yang-Mills amplitudes in DimReg in a dimensionally agnostic way, we would have to expand the amplitude in the span

\ceq{
    \text{span}\left(\includegraphics[valign = c, scale=0.3,trim=0 0 0 0cm]{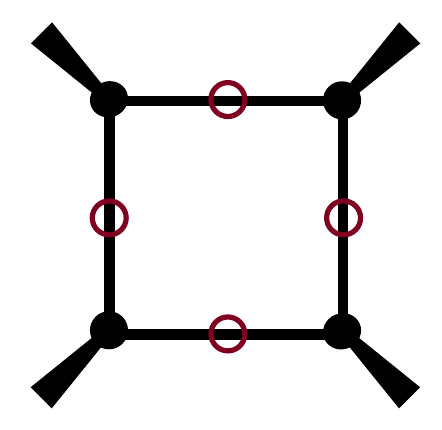}\right)
}
in $d+1$ dimensions. 

The interplay between IR stratification and the basis enlargement due to DimReg is interesting already at one loop---and likely very nontrivial at two---in light of the recent work \cite{Bourjaily:2021ujs}. Here, IR stratification was rendered manifest by an elegant choice of contours that made the resolution into IR finite and divergent pieces clear at the level of cuts. The integrated expressions, done using dimensional regularization, were provided, even though the integrand numerators---defined intrinsically in four dimensions---were not constructed using the enlarged basis. It would be very interesting to generalize that paper to accommodate this.

\vspace{10pt}

\paragraph*{\emph{Manifesting Rigidity and Higher Periods}. }Going back to our basis of integrands at planar two loops, we can recap the essence of what we have demonstrated. Specifically, we have shown that for each integrand topology and their master integrands, the basis always decomposes into two disjoint subsets, where one is built out of integrands of rigidity 1 and the second of rigidity zero. In terms of the box-triangle, to be concrete, we find that for the topology

\ceq{
    \includegraphics[valign = c, scale=0.3,trim=0 0 0 0cm]{box-triangle-3-gon-labelled.pdf} = \frac{(\ell_2|\bl{N_i})\dbar^4\ell_1\dbar^4\ell_2}{(\ell_1|\bur{a_1})(\ell_1|\bur{a_2})(\ell_1|\bl{X})(\ell_1|\ell_2)(\ell_2|\bur{b_1})(\ell_2|\bur{b_2})(\ell_2|\bur{b_3})(\ell_2|\bur{b_4})(\ell_2|\bl{X})}
}we can develop the basis in terms of two numerators $|\bl{N}) \in \lbrace{|\bl{Q_1}),|\bl{Q_2})\rbrace}$ which render the integrand purely elliptic, and a third that renders it purely polylog.

The main issue with this manner of framing the result is that the explicit forms of the numerator have no way of manifesting the rigidity of the final result. Indeed, for the case of the polylogarithmic integrand, the integrand itself is expanded into a sum of terms that have, at least in our representation, elliptic coefficients.

Unfortunately, it doesn't seem to be the case that direct evaluation---such as using {\tt{HyperInt}} for example---will be of much use in this regard. Ultimately, the representation of our numerators is dependent on some initial choice of masters that were diagonalized along certain cycle prescriptions. The elliptic character of two of these choices naturally rendered the final result dependent on the elliptic periods. Absent a clever guess of the resulting numerator that obviates the need to perform this diagonalization, it is unlikely that direct integration will be illuminating. 

A possible way out of this may be to compute the symbol algebra instead. A rigorous list of the symbol alphabet on which the integrated expression depends would furnish essentially a full proof of rigidity zero since none of the letters in the alphabet would contain elliptic terms. 

Beyond the case of rigidity 1, at the time of writing, period integrals aren't well understood, particularly for the proximate Feynman integrands of interest. Specifically, relaxing the conditions of planarity and two-loop order results in the rise of three diagrams in particular:

\ceq{
    \Bigg\lbrace{\includegraphics[valign = c, scale=0.3,trim=0 0 0 0cm]{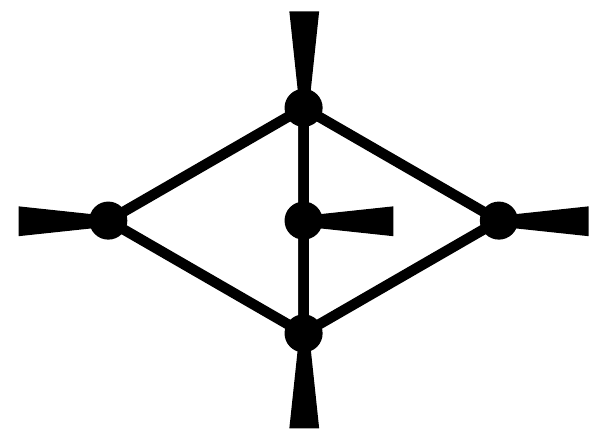}\;,\;\includegraphics[valign = c, scale=0.3,trim=0 0 0 0cm]{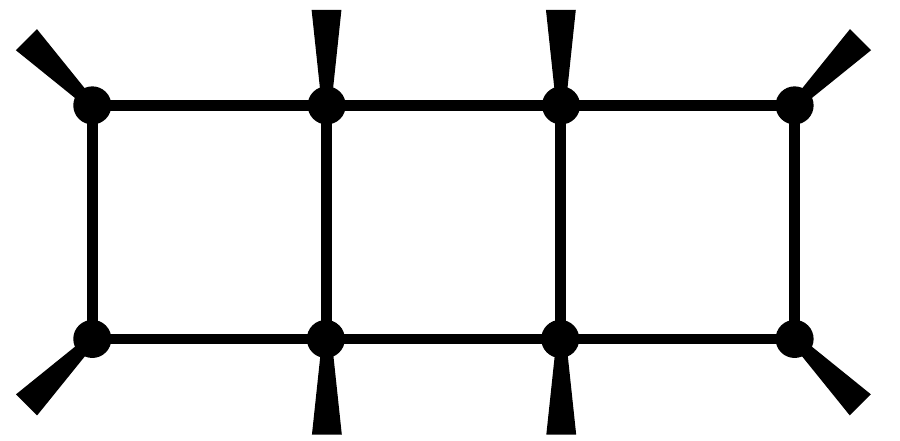}\;,\;\includegraphics[valign = c, scale=0.3,trim=0 0 0 0cm]{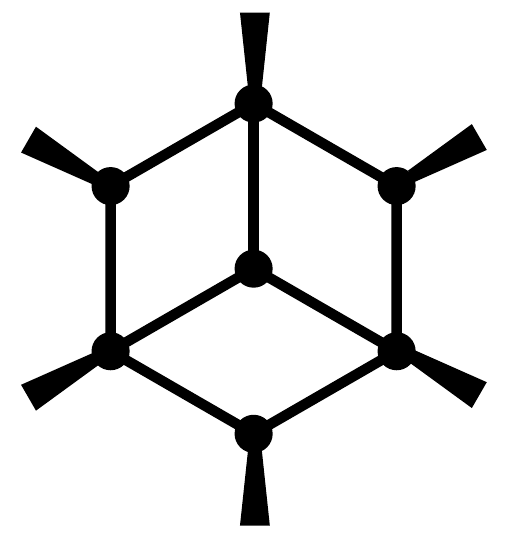}\Bigg\rbrace}
}known---from left to right---as the tardigrade\footnote{This name---despite heroic efforts---has been assiduously successful in evading me.}, the triple box and the three-loop wheel. Each of these contains in its maximal cut (involving a transverse residue for the three-loop wheel) a $K3$ surface, which is generally of 22 dimensions in homology. Counting arguments of basis sizes down to bubble power counting suggests that this may not be the rank of the homology for the K3 surfaces that we require, as we expect them to be rather singular. At the moment of writing, the period integrals of these surfaces remain unknown. It will be of immense interest to evaluate them, from both a mathematical as well as physical perspective.
\clearpage
\appendix
\titleformat{name=\section}[display]
{\normalfont}
{\footnotesize\textsc{Appendix \thesection}}
{0pt}
{\Large\bfseries}
[\vspace{-10pt}\color{lapis}\rule{\textwidth}{0.6pt}]
\section{Dual Momentum Space, Momentum Twistors and Dual Conformal Integration}\label{app:A}
Throughout the various sections of this work, we have---at times interchageably so---made use of two distinct, but closely related formalisms. These are dual momentum coordinates and momentum twistors, which are tools to trivialize momentum conservation and masslessness respectively. An additional pair of upshots is that their use has the effect of dramatically simplifying otherwise technically challenging algebraic calculations, namely those involving Feynman integration and calculating the solutions of cut conditions. In this appendix, we will review the salient features of these formalisms, with an emphasis on how they simplify and trivialize computations, rather than detailed mathematical proofs.

We start with dual coordinates, which are a means of manifesting momentum conservation given any $n$-tuple of points in Minkowski space. They are defined, for a particle $p_{i}$ by the implicit relation

\begin{equation}
    x_{i+1} - x_{i} = p_{i}
\end{equation}
where the labeling (assuming, say, $n$ particles) is to be regarded as cyclic.

The consequence of this condition is that one only needs to specify $n$ point such that each consecutive pair is lightlike separated. This automatically yields a configuration of four-vectors that identically ensure momentum conservation. The additional simplification offered by this prescription is that sums of momentum are expressed in terms of simple differences; for example, we have

\begin{equation}
    x_{i} - x_{j} = p_{i-1}+p_{i-2}+\dots+p_{j}
\end{equation}
in accordance with which

\begin{equation}
    x^2_{ij}:=(x_i-x_j)^2 = s_{i-1\dots j}
\end{equation}
where it is to be understood that the right-hand side is the generalized Mandelstam variable.

Loop momenta can be incorporated by simply assigning to each loop---making use of the presence of invariance under translations in the process---an independent four-vector $x_{\ell}$, which then allows us to represent propagators using the Lorentz-invariant quantities

\begin{equation}
    (\ell|\bur{i}):=(x_{\ell}-x_{i})^2 
\end{equation}
which is the notation that we have used in the main body of this work. 

Treating Feynman integrands as being rational functions in $x$-space rather than in ordinary momentum space reveals an underlying symmetry not visible conventionally, namely dual conformal invariance---or just conformal invariance in $x$-space---for integrands that scale like boxes for large $x_{\ell}$ in four dimensions. Indeed, taking as an illustration the box integrand

\begin{equation}
    \mathcal{I}_{\text{box}} = \frac{\dbar^{4}\ell}{(\ell|\bur{a_1})(\ell|\bur{a_2})(\ell|\bur{a_3})(\ell|\bur{a_4})}
\end{equation}
where it is to be assumed that the loop integral is in $x$-space, we can check that under

\begin{equation}
    x_{\ell}\longrightarrow \frac{x_{\ell}}{x_{\ell}^2}
\end{equation}
we have

\begin{equation}
    (\ell|\bur{a_{i}})\longrightarrow \frac{1}{x_{\ell}^2}(\ell|\bur{a_i})
\end{equation}
and

\begin{equation}
    \dbar^4\ell \longrightarrow \frac{1}{(x^2_\ell)^4}\dbar^4\ell
\end{equation}
rendering the integrand invariant. This is also the case for general conformal transformations in $x$-space \cite{Drummond:2006rz,Bern:2008ap,Drummond:2008aq}. 

Dual conformal invariance suggests the possible utility of an embedding space formalism where this invariance may be trivialized. Indeed, this can be done by reassigning to each dual coordinate $x_i$ a vector in $D+2$ (in the general case) dimensions according to

\begin{equation}
    X^{M}_i = \begin{pmatrix}
        x^\mu_i\\
        x^2_i\\
        1
    \end{pmatrix}
\end{equation}
with a metric defined by the following

\begin{equation}
    g^{MN} = \begin{pmatrix}
        -2\eta^{\mu\nu}&0&0\\
        0&0&1\\
        0&1&0
    \end{pmatrix}.
\end{equation}
The result of doing this is that inner products between dual coordinates in $D$ dimensions, are naturally linearized in embedding space

\begin{equation}
    (\bur{i}|\bur{j}) = (x_{i}-x_{j})^2 = h^{MN}X^{M}_iX^N_j.
\end{equation}

In this formalism, loop momenta are embedded using the same principle

\begin{equation}
    X^{M}_\ell = \begin{pmatrix}
        x^\mu_\ell\\
        x^2_{\ell}\\
        1
    \end{pmatrix}
\end{equation}
with the cavil that the measure of loop integration needs to be appropriately renormalized to take note of the lightlike character of momentum in embedding space

\begin{equation}
    \dbar^D\ell \longrightarrow \dbar^{D+2}X_{\ell}\delta((\ell|\ell)).
\end{equation}
An important point here is that in embedding space, a loop integrand must have scaling weight $d$ in $d$ dimensions in each loop separately ($d=4$ in our discussion hereafter). Integrands that once translated to embedding space do not satisfy this will have to be rendered in accordance with scaling by including a compensatory number of factors of $(\ell|\bl{X})$, where $|\bl{X})$ corresponds to the point at infinity.

The linearization of the inner product has a rather pleasant effect on the question of loop integration via Feynman parametrization for general integrands. Indeed, going back to the case of the double box, observe that we may Feynman parametrize the denominator according to

\begin{equation}
    \sum_{i}\var{\alpha_i}h^{MN}X_{\ell}^M X_{i}^{M} 
\end{equation}
which can be demystified by defining

\begin{equation}
    |\bl{Q}) = \var{\alpha_1}|\bur{1})+\dots + \var{\alpha_4}|\bur{4})
\end{equation}
giving us

\begin{equation}
    \mathcal{I}_{\text{box}} = 2\Gamma(4)\int[d^3\var{\alpha}]\frac{\dbar^4\ell}{(\ell|\bl{Q})^4}.
\end{equation}
where we have indicated the projective nature of the Feynman parametrization. When brought into this particular generic form, the loop momenta can be identically integrated by making application of the following relation---which we will accept as an identity for the sake of brevity---in $n$ dimensions

\begin{equation}
    2\Gamma(n)\int[d^{n-1}\var{\alpha_i}]\int d^{n+2}X_{\ell}\delta((\ell_\ell))\frac{1}{(\ell|\bl{Q})^{n}} = \frac{\Gamma(n/2)}{(2\pi)^{n/2}}\int[d^{n-1}\var{\alpha_i}]\frac{1}{(\bl{Q}|\bl{Q})^{n/2}}.
\end{equation}
Since we will only invoke this identity in four dimensions, we can list it, along with its derivatives for the record. Specifically, higher powers in the numerator can be reached by acting on this formula using the operator

\begin{equation}
    \left(\frac{d}{d\bl{Q}}\bigg|\bl{N}\right).
\end{equation}
We list the first few such identities, specifically the ones we require for our purposes as follows

\begin{equation}\label{eq:A19}
    2\Gamma(4)\int \frac{\dbar^4\ell}{(\ell|\bl{Q})^4} = \frac{1}{(2\pi)^2}\frac{1}{(\bl{Q}|\bl{Q})^2},
\end{equation}

\begin{equation}\label{eq:A20}
    2\Gamma(5)\int \frac{(\ell|\bl{N_1})\dbar^4\ell}{(\ell|\bl{Q})^5} = \frac{3}{(2\pi)^2}\frac{(\bl{Q}|\bl{N_1})}{(\bl{Q}|\bl{Q})^3},
\end{equation}

\begin{equation}\label{eq:A21}
    2\Gamma(6)\int \frac{(\ell|\bl{N_1})(\ell|\bl{N_2})\dbar^4\ell}{(\ell|\bl{Q})^5} = \frac{3}{(2\pi)^2}\left(\frac{3(\bl{Q}|\bl{N_1})(\bl{Q}|\bl{N_2})}{(\bl{Q}|\bl{Q})^4} - \frac{(\bl{N_1}|\bl{N_2})}{(\bl{Q}|\bl{Q})^3}\right)
\end{equation}

\begin{equation}\label{eq:A22}
\begin{aligned}
     &2\Gamma(7)\int \frac{(\ell|\bl{N_1})(\ell|\bl{N_2})(\ell|\bl{N_3})\dbar^4\ell}{(\ell|\bl{Q})^6} = \\
     &\frac{3}{(2\pi)^2}\left(4\frac{(\bl{Q}|\bl{N_1})(\bl{Q}|\bl{N_2})(\bl{Q}|\bl{N_3})}{(\bl{Q}|\bl{Q})^4} - \frac{(\bl{N_1}|\bl{N_2})(\bl{Q}|\bl{N_3})}{(\bl{Q}|\bl{Q})^3}-\frac{(\bl{N_2}|\bl{N_3})(\bl{Q}|\bl{N_1})}{(\bl{Q}|\bl{Q})^3}-\frac{(\bl{N_3}|\bl{N_1})(\bl{Q}|\bl{N_2})}{(\bl{Q}|\bl{Q})^3}\right).
\end{aligned}
\end{equation}
These relations make it possible to algorithmically integrate---in a manner that can be automated \cite{bourjailyCode}---any integral that enjoys dual conformal invariance. A number of such examples were provided in \cite{Bourjaily:2019jrk}, but for the purposes of illustration, we'll go over the most general such integral that we would require for the case of planar two loops that we considered, namely that of the pentabox in $3$-gon power counting.

\paragraph*{Example A.1. Manifestly Dual Conformal Integration of the $3$-gon Pentabox. }We start with the full dual conformally invariant form of the generic pentabox in triangle power counting

\begin{equation}
    \includegraphics[valign=c,scale=0.3]{pentabox-3-gon-labelled.pdf} = \frac{(\ell_1|\bl{N_1})(\ell_2|\bl{N_2})(\ell_2|\bl{N_3})\dbar^4\ell_1\dbar^4\ell_2}{(\ell_1|\bur{a_1})(\ell_1|\bur{a_2})(\ell_1|\bur{a_3})(\ell_1|\bl{X})(\ell_1|\ell_2)(\ell_2|\bur{b_1})(\ell_2|\bur{b_2})(\ell_2|\bur{b_3})(\ell_2|\bur{b_4})(\ell_2|\bl{X})}.
\end{equation}
The integration over the loop momenta is done in steps, first integrating out $\ell_1$ followed by an integration over $\ell_2$. The first step is reached by defining the Feynman parametrized quantity

\begin{equation}
    |\bl{R_1}) = \var{\alpha_1}|\bur{a_1})+\var{\alpha_2}|\bur{a_2})+\var{\alpha_3}|\bur{a_3})+\var{\alpha_4}|\bl{X}) + \bur{\gamma_1}|\ell_2) := |\bl{Q_1}) + \bur{\gamma_1}|\ell_2)
\end{equation}
yielding the parametrized integrand

\begin{equation}
    \begin{aligned}
        &\frac{(\ell_1|\bl{N_1})(\ell_2|\bl{N_2})(\ell_2|\bl{N_3})\dbar^4\ell_1\dbar^4\ell_2}{(\ell_1|\bur{a_1})(\ell_1|\bur{a_2})(\ell_1|\bur{a_3})(\ell_1|\bl{X})(\ell_1|\ell_2)(\ell_2|\bur{b_1})(\ell_2|\bur{b_2})(\ell_2|\bur{b_3})(\ell_2|\bur{b_4})(\ell_2|\bl{X})}\\
        &=\int [d^3\var{\alpha_i}]d\bur{\gamma_1}\frac{(\ell_1|\bl{N_1})(\ell_2|\bl{N_2})(\ell_2|\bl{N_3})\dbar^4\ell_1\dbar^4\ell_2}{(\ell_1|\bl{R_1})^5(\ell_2|\bur{b_1})(\ell_2|\bur{b_2})(\ell_2|\bur{b_3})(\ell_2|\bur{b_4})(\ell_2|\bl{X})}.
    \end{aligned}
\end{equation}
Performing the integral over $\ell_1$ using (\ref{eq:A20}) results in the following expression

\begin{equation}
    \begin{aligned}
        &\int [d^3\var{\alpha_i}]d\bur{\gamma_1}\frac{(\ell_1|\bl{N_1})(\ell_2|\bl{N_2})(\ell_2|\bl{N_3})\dbar^4\ell_1\dbar^4\ell_2}{(\ell_1|\bl{R_1})^5(\ell_2|\bur{b_1})(\ell_2|\bur{b_2})(\ell_2|\bur{b_3})(\ell_2|\bur{b_4})(\ell_2|\bl{X})}\\
        &=\frac{3}{(2\pi)^2}\int[d^3\var{\alpha_i}]d\bur{\gamma_1}\frac{(\bl{R_1}|\bl{N_1})}{(\bl{R_1}|\bl{R_1})^3}\frac{(\ell_2|\bl{N_2})(\ell_2|\bl{N_3})\dbar^4\ell_2}{(\ell_2|\bur{b_1})(\ell_2|\bur{b_2})(\ell_2|\bur{b_3})(\ell_2|\bur{b_4})(\ell_2|\bl{X})}.
    \end{aligned}
\end{equation}

In this form, the integrand no longer looks manifestly dual conformal invariant in the variable $\ell_2$. This can be remedied by noticing that we have the following equivalences

\begin{equation}
    (\bl{R_1}|\bl{N_1}) = (\bl{Q_1}|\bl{N_1}) + \bur{\gamma_1}(\ell_2|\bl{Q_1}),
\end{equation}

\begin{equation}
    (\bl{R_1}|\bl{R_1}) = (\bl{Q_1}|\bl{Q_1}) + 2\bur{\gamma_1}(\ell_2|\bl{Q_1})
\end{equation}
which allows us to make application of the fact

\begin{equation}
    \int d\bur{\gamma_1}\left(\frac{(\bl{Q_1}|\bl{N_1}) + \bur{\gamma_1}(\ell_2|\bl{Q_1})}{\left((\bl{Q_1}|\bl{Q_1}) + 2\bur{\gamma_1}(\ell_2|\bl{Q_1})\right)^3}\right) = \frac{2 \left(\bl{Q_1}|\bl{N_1}\right)+\left(\bl{Q_1}|\bl{Q_1}\right)}{8 \left(\bl{Q_1}|\bl{Q_1}\right)^{2}
   \left(\ell_2|\bl{Q_1}\right)}.
\end{equation}
The result of this is to render the following integrated expression as a function of $\ell_2$

\begin{equation}
   :=\frac{3}{8(2\pi)^2}\int[d^3\var{\alpha_i}]\left(\frac{2 \left(\bl{Q_1}|\bl{N_1}\right)+\left(\bl{Q_1}|\bl{Q_1}\right)}{\left(\bl{Q_1}|\bl{Q_1}\right)^{2}}\right)\frac{(\ell_2|\bl{N_2})(\ell_2|\bl{N_3})\dbar^4\ell_2}{\left(\ell_2|\bl{Q_1}\right)(\ell_2|\bur{b_1})(\ell_2|\bur{b_2})(\ell_2|\bur{b_3})(\ell_2|\bur{b_4})(\ell_2|\bl{X})}. 
\end{equation}
The $\ell_2$ integral is done analogously, by first performing Feynman parametrization according to

\begin{equation}
    \frac{(\ell_2|\bl{N_2})(\ell_2|\bl{N_3})\dbar^4\ell_2}{\left(\ell_2|\bl{Q_1}\right)(\ell_2|\bur{b_1})(\ell_2|\bur{b_2})(\ell_2|\bur{b_3})(\ell_2|\bur{b_4})(\ell_2|\bl{X})} = 2\Gamma(6)\int [d^{5}\var{\beta_i}]d\bur{\gamma_2}\frac{(\ell_2|\bl{N_2})(\ell_2|\bl{N_3})\dbar^4\ell_2}{(\ell_2|\bl{R_2})^6}
\end{equation}
where

\begin{equation}
    |\bl{R_2}) = \var{\beta_1}|\bur{b_1})+\var{\beta_2}|\bur{b_2})+\var{\beta_3}|\bur{b_3})+\var{\beta_4}|\bur{b_4}) + \var{\beta_5}|\bl{Q_1})+\bur{\gamma_2}|\bl{X}):=|\bl{Q_2})+\bur{\gamma_2}|\bl{X}).
\end{equation}
We have not chosen to deprojective the integrand by setting one of the $\var{\beta_i}$ to as the authors in \cite{Bourjaily:2019jrk}. Performing the loop integration results in the expression

\begin{equation}
    \begin{aligned}
        \frac{3}{(2\pi)^2}\int [d^{5}\var{\beta_i}]d\bur{\gamma_2}\left(\frac{3(\bl{R_2}|\bl{N_2})(\bl{R_2}|\bl{N_3})}{(\bl{R_2}|\bl{R_2})^4} - \frac{(\bl{N_1}|\bl{N_2})}{(\bl{R_2}|\bl{R_2})^3}\right)
    \end{aligned}
\end{equation}
where

\begin{equation}
    (\bl{R_2}|\bl{N_i}) = (\bl{Q_2}|\bl{N_i}) + \bur{\gamma_2}(\bl{X}|\bl{N_i})
\end{equation}
and

\begin{equation}
    (\bl{R_2}|\bl{R_2}) = (\bl{Q_2}|\bl{Q_2}) + 2\bur{\gamma_2}(\bl{X}|\bl{Q_2}).
\end{equation}
Once again, the $\bur{\gamma_2}$ integral can be carried out analytically, and results in the expression that follows.

\begin{equation}
    \begin{aligned}
       & \frac{9}{8(2\pi)^4}\int[d^3\var{\alpha_i}][d^{5}\var{\beta_i}]d\bur{\gamma_2}\frac{(2 \left(\bl{Q_1}|\bl{N_1}\right)+\left(\bl{Q_1}|\bl{Q_1}\right))}{\left(\bl{Q_1}|\bl{Q_1}\right)^{2}}\left(\frac{3(\bl{R_2}|\bl{N_2})(\bl{R_2}|\bl{N_3})}{(\bl{R_2}|\bl{R_2})^4} - \frac{(\bl{N_1}|\bl{N_2})}{(\bl{R_2}|\bl{R_2})^3}\right)\\
       &\frac{9}{16(2\pi)^4}\int[d^3\var{\alpha_i}][d^{5}\var{\beta_i}]\frac{(2g_1+1)}{f_1f_2f_3}\left(\frac{n_2}{f_2}+\frac{n_{22}}{f_2^2}+\frac{n_{23}}{f_2f_3}+\frac{n_{33}}{f_3^2}\right)
    \end{aligned}
\end{equation}
where

\begin{equation}
\begin{aligned}
    f_1 &= (\bl{Q_1}|\bl{Q_1})\\
    f_2 &= (\bl{X}|\bl{Q_2})\\
    f_3 &= (\bl{Q_2}|\bl{Q_2})\\
    g_1&=\frac{(\bl{Q_1}|\bl{N_1})}{f_1},
\end{aligned}
\end{equation}
and

\begin{equation}
    \begin{aligned}
        n_2 & = -2(\bl{N_2}|\bl{N_2})\\
        n_{22}&= (\bl{X}|\bl{N_2})(\bl{X}|\bl{N_3})\\
        n_{23}& = (\bl{X}|\bl{N_2})(\bl{Q_1}|\bl{N_3}) + (\bl{X}|\bl{N_3})(\bl{Q_1}|\bl{N_2})\\
        n_{33} &= 4(\bl{Q_1}|\bl{N_2})(\bl{Q_1}|\bl{N_3}).
    \end{aligned}
\end{equation}
Dual conformal invariance can be manifested by recording the fact that we have for $f_{2}$

\begin{equation}
    f_{2} = \sum_{i=1}^{4}\var{\alpha_i}(\bl{X}|\bur{a_i})+\sum_{i=1}^{4}\var{\beta_5}\var{\beta_i}(\bl{X}|\bur{b_i}).
\end{equation}
Indeed, this can be homogenized by performing the replacements

\begin{equation}
    \var{\alpha_i}\longrightarrow\frac{1}{(\bl{X}|\bur{a_i})}\var{\alpha_i}
\end{equation}
and

\begin{equation}
    \var{\beta_i}\longrightarrow\frac{1}{(\bl{X}|\bur{b_i})}\var{\beta_i}
\end{equation}
which results in

\begin{equation}
    f_{2} = \var{\alpha_1}+\var{\alpha_2}+\var{\alpha_3}+\var{\alpha_4} + \var{\beta_5}\var{\beta_1}+\var{\beta_5}\var{\beta_2}+\var{\beta_5}\var{\beta_3}+\var{\beta_5}\var{\beta_4}.
\end{equation}
An exercise in exhaustion reveals the entire integrand to be manifestly DCI in this fashion as well, but we won't repeat the steps here. The full formulas for the interested reader are available upon request.

\hfill
\vspace{0.7cm}

The simplifications provided by dual coordinates in the matters of momentum conservation and loop integration are echoed by similar simplifications offered by momentum twistors, which trivialize both the problems of masslessness and enforcement of cut conditions. Initially introduced by Hodges in \cite{Hodges:2009hk}, the basic idea is to identify each $x_{i}$ with a line in twistor space $\mathbb{P}^3$ according to the incidence relation \cite{Witten:2003nn}

\begin{equation}
    \mu^{\dot{a}} =  x^{\dot{a}a}_{i}\lambda^{i}_{a}
\end{equation}
where the $\lambda$ is the usual spinor variable associated to momentum $p_i$. That this is a line in twistor space can be gleaned by the fact that an arbitrary spinor encodes an embedding of $\mathbb{P}^1$ when expressed in homogenized coordinates, immediately determining its corresponding $\mu$ counterpart via the incidence relation.

The benefit of drawing this equivalence is due to the general freedom in prescribing a line is twistor space, which by definition determines a corresponding dual coordinate. Indeed, working in the opposite direction, since any line is uniquely characterized by a choice of points $(Z_1,Z_2)$ in $\mathbb{P}^3$, a set of dual coordinates is automatically furnished by preparing $2n$ points in twistor space.

The technical dictionary between the two can be worked out by identifying

\begin{equation}
    Z_{i} = \begin{pmatrix}
        \mu_{\dot{a}}\\
        \lambda_{a}
    \end{pmatrix}
\end{equation}
where the components are understood to satisfy the incidence relation. It can be verified that \cite{Bourjaily:2010wh}

\begin{equation}
    x^2_{ij} = \frac{\lab{Z_{i-1}Z_{i}Z_{j}Z_{j+1}}}{\lab{Z_{i-1}Z_{i}(\bl{X})}\lab{Z_{j}Z_{j+1}(\bl{X})}}
\end{equation}
where $(\bl{X})$ is the line corresponding to the infinity twistor and the angle brackets denote the four-dimensional determinant.

Lightlike separation then corresponds to the vanishing of the four bracket in the latter expression. Due to the fact that $(i\; i-1)$ and $(j\;j+1)$ are just lines, and the vanishing of the determinant is due, and only due, to the linear dependence of one of the rows on the other, this corresponds to the lines intersecting.

Naturally then, picking any two points in twistor space constructs a line that automatically corresponds to a lightlike vector in ordinary Minkowski space. Furthermore, cut conditions are readily trivialized into geometric ones. We illustrate this by demonstrating how the use of twistors dramatically simplifies computing the solutions to the four-cut of the ordinary scalar double box in four dimensions.

\paragraph*{Example A.2. Quad Cut of the Scalar Box. } Consider the general form of the scalar box in four dimensions

\begin{equation}
    \mathcal{I}_{\text{box}} = \frac{\dbar^4\ell}{(\ell|\bur{a_1})(\ell|\bur{a_2})(\ell|\bur{a_3})(\ell|\bur{a_4})}
\end{equation}
where the $\bur{a_i}$ are coordinates in dual momentum space. We rewrite this in momentum twistor space by making the following associations

\begin{equation}
    \begin{aligned}
        (\bur{a_1}) &= (\bur{Aa})\\
        (\bur{a_2}) &= (\bur{Bb})\\
        (\bur{a_3}) &= (\bur{Cc})\\
        (\bur{a_4}) &= (\bur{Dd})
    \end{aligned}
\end{equation}
and $(\ell) = ({\widehat{A}\widehat{B}})$. Dropping purely kinematical factors, the Feynman integrand devolves upon an expression involving four determinant factors

\begin{equation}
    \frac{1}{GL(2)}\frac{\dbar^2\bur{\widehat{A}}\dbar^2\bur{\widehat{B}}}{\lab{\bur{\widehat{A}}\bur{\widehat{B}}\bur{Aa}}\lab{\bur{\widehat{A}}\bur{\widehat{B}}\bur{Bb}}\lab{\bur{\widehat{A}}\bur{\widehat{B}}\bur{Cc}}\lab{\bur{\widehat{A}}\bur{\widehat{B}}\bur{Dd}}}
\end{equation}
where the general linear group mods out the redundancy in describing a line using two points.

The cut conditions then devolve upon four geometric statements, namely that we have

\begin{equation}
    \lab{(\ell)\bur{Aa}} =  \lab{(\ell)\bur{Bb}} =  \lab{(\ell)\bur{Cc}} =  \lab{(\ell)\bur{Dd}} = 0
\end{equation}
implying that the line $(\ell)$ intersects the four lines $(\bur{Aa}), \dots , (\bur{Dd})$. The first of these can be easily enforced by simply placing $\bur{\widehat{A}}$ on the line $(\bur{Aa})$ by writing

\begin{equation}
    \bur{\widehat{A}} = \bur{a} + \var{\alpha}\bur{A}.
\end{equation}
The second and third can be guaranteed by making reference to the geometric problem. While two points determine a line, three determine a plane, and as such, it makes sense to compute the quantity intersecting a plane and line as follows

\begin{equation}
    (\bur{bB})\cap(\bur{cC\widehat{A}})
\end{equation}
which is expanded in two ways

\begin{equation}
   := \bur{b}\lab{\bur{BcC\widehat{A}}} + \bur{B}\lab{\bur{cC\widehat{A}B}}
\end{equation}
and

\begin{equation}
    -\bur{c}\lab{\bur{C\widehat{A}bB}}-\bur{C}\lab{\bur{c\widehat{A}bB}}-\bur{\widehat{A}}\lab{\bur{cCbB}},
\end{equation}
which are equivalent due to a Schouten identity in four dimensions. This renders the solution of the second and third cut conditions manifest due to the expansions, leaving us with the fourth to deal with.

This turns out to be a quadratic constraint. Indeed, observe that we have

\begin{equation}
    \begin{aligned}
        \lab{\bur{\widehat{A}\widehat{B}Dd}} = &\lab{\bur{a((\bur{bB})\cap(\bur{cCa}))Dd}} + \\&\var{\alpha}\left(\lab{\bur{a((\bur{bB})\cap(\bur{cCA}))Dd}} + \lab{\bur{A((\bur{bB})\cap(\bur{cCa}))Dd}})\right) +\\
        &\var{\alpha}^2\lab{\bur{A((\bur{bB})\cap(\bur{cCa}))Dd}}
    \end{aligned}
\end{equation}

The result is two solutions for $\var{\alpha}$, which correspond to the fact that the cut conditions are quadratic constraints, and furnish two different solutions.

\hfill
\vspace{0.7cm}

We remark that it is possible to implement such cut conditions numerically as well. Indeed, any set of kinematics can be generated by simply using random numbers to set up the $Z_{i}$s. For the loop momenta, the $GL(2)$ invariance can be used to pick the convenient chart of

\begin{equation}
(\bur{\widehat{A}\widehat{B}}) = \begin{pmatrix}
\var{\alpha_1}& \var{\alpha_3}\\
 0& 1\\
 \var{\alpha_2}& \var{\alpha_4}\\
 1&0
\end{pmatrix}
\end{equation}
This reveals all of the determinants characterizing the propagators as functions of four independent variables, in terms of which the cut solutions can be determined. The change to the basis used in the preceding example amounts to a global (constant) determinant.
\clearpage
\section{Aspects of Elliptic Curves}\label{app:B}
A number of results in the present paper rely heavily on various technical features of elliptic curves. The purpose of this appendix is to collect these facts in one location, largely for readers unfamiliar with the notation and simple mathematical details involved. Here, as in the preceding appendix, we will generally eschew proofs of the more involved mathematical claims, but seek to concretely present the most relevant and important formulae required to understand the results in the body of the paper.

An elliptic curve is the specialization to $g=1$ of a generic family of complex curves governed by equations of the following form

\begin{equation}
    y(\var{z})^2 = (\var{z}-\var{z_1})(\var{z}-\var{z_2})\dots(\var{z}-\var{z_{2g+1}})(\var{z}-\var{z_{2g+2}})
\end{equation}
known as curves of \emph{hyperelliptic} type. A generic hyperelliptic curve spans a surface which is of genus $g$, and is as such isomorphic to a genus-$g$ Riemann surface. 

The first homology group of a Riemann surface of genus $g$ is of dimension $2g$, and is spanned by $g$ pairs $(A_{i},B_{i})$ of so-called $a$- and $b$-cycles, which have the following intersection matrix

\begin{equation}
    \lab{A_{i},B_{j}} = \delta_{ij}
\end{equation}
with all other intersections vanishing.

In terms of the branch cuts comprehended by the hyperelliptic curve, the $a$- and $b$-cycles can be constructed as follows: $A_{i}$ is the contour encircling the branch spanned between $\var{z_i}$ and $\var{z_{i+1}}$, while the $B_i$ encircle $\var{z_i}$ and $\var{z_{2g+1}}$. Schematically, the picture is given below\footnote{It is a pleasure to thank Cristian Vergu for teaching me how to prepare the cycles.}.

\begin{equation}
    \includegraphics[valign = c,scale=0.3]{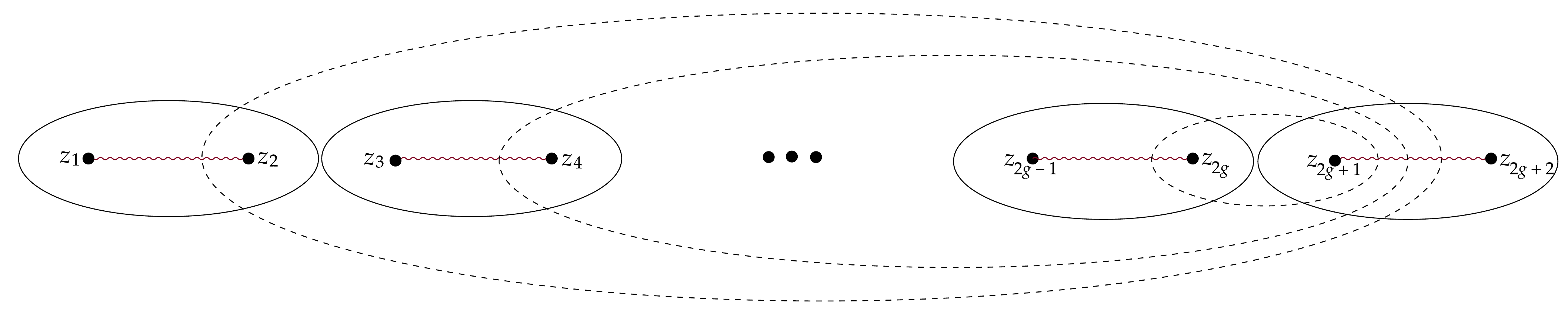}
\end{equation}

Since the elliptic case is given by the restriction to $g=1$, we have for it a representation in terms of a general quartic

\begin{equation}\label{eq:B4}
    y(\var{z})^2 = (\var{z}-\bur{r_1})(\var{z}-\bur{r_2})(\var{z}-\bur{r_3})(\var{z}-\bur{r_4}). 
\end{equation}

Two distinct, but ultimately equivalent, signatures can be extracted for a given elliptic curve. The most robust of these are the so-called periods, which are obtained by computing the $a$- and $b$-cycle integrals of the canonical differential forms

\begin{equation}
    \omega(\var{z}) = \frac{\dbar\var{z}}{y(\var{z})}.
\end{equation}
and
\begin{equation}
    \omega_{\bur{r}}(\var{z}) = \frac{\dbar\var{z}}{(\var{z}-\bur{r})y(\var{z})}.
\end{equation}
The complexity of the functions involved---to which we will shortly arrive---in the calculation of these periods renders the second method of diagnosing differences between elliptic curves, namely the notion of the $j$-invariant. 

The definition of the $j$-invariant relies on the fact that any two elliptic curves related by a birational transformation are equivalent. Specifically, the birational transformation---which is known to exist---that converts the elliptic curve in (\ref{eq:B4}) into the so-called Weierstrass form

\begin{equation}
    y(\var{\alpha})^2 = 4\var{\alpha}^3 -g_2\var{\alpha} - g_{3} 
\end{equation}
where

\begin{equation}
    g_{2} = \bur{f_1} - \frac{\bur{f_2f_{4}}}{4} + \frac{\bur{f_3}^2}{12}
\end{equation}
and

\begin{equation}
    g_3 = -\frac{\bur{f_3}^3}{216}+\frac{\bur{f_1 f_3}}{6}+\frac{\bur{f_2 f_4 f_3}}{48} -\frac{\bur{f_2}^2}{16}-\frac{\bur{f_1 }\bur{f_4}^2}{16} 
\end{equation}
where $\bur{f_i}$ is the ratio of the coefficient of $\var{\alpha}^{5-i}$ and $\var{\alpha}^0$ in the quartic.

The $j$-invariant is defined as the canonical ratio

\begin{equation}
    j = \frac{g_2^3}{g_2^3 - 27 g_3^2}.
\end{equation}
It turns out that this quantity is a genuine invariant of the elliptic curve. In other words, two elliptic curves are equivalent if and only if they have matching $j$-invariants.

\paragraph*{Example B.2. }A simple illustration of this is the elliptic curve of the generic double box. Consider the following choice of reference kinematics

\begin{center}
\begin{tabular}{ |cccccccccccc| } 
 \hline
 $\bur{a}$ & $\bur{A}$ & $\bur{b}$ &$\bur{B}$ & $\bur{c}$ & $\bur{C}$ &$\bur{d}$ & $\bur{D}$ & $\bur{e}$&$\bur{E}$&$\bur{f}$&$\bur{F}$ \\ 
 \hline
 0& 1& 0& 2& 1& 0& 1& 1& 0& 3& 3& 0\\
 4& 0& 2& 2& 5& 3& 2& 0& 2& 3& 4& 4\\
 0& 2& 3& 0& 1& 1& 5& 3& 3& 3& 0& 0\\
 0& 0& 2& 0& 4& 0& 4& 3& 2& 3& 1& 2\\
 \hline
\end{tabular}
\end{center}
and depending on whether we eliminate $\var{\alpha_1}$ versus $\var{\alpha_2}$ we obtain the following two---allegedly dissimilar---elliptic curves:

\begin{equation}
   y(\var{z}) =  705600 + 3248880 \var{z} + 5497081 \var{z}^2 + 4031950\var{z}^3 + 1078345 \var{z}^4
\end{equation}
and
\begin{equation}
    y(\var{z}) = -30656 + 526624 \var{z} - 984428 \var{z}^2 - 69612 \var{z}^3 + 737001 \var{z}^4.
\end{equation}

The $j$-invariants however both turn out to be

\begin{equation}
    j = -\frac{128765871783947280410631964986481}{43587981692158343567022489600000}
\end{equation}
showing that they are actually the same elliptic curve.

\hfill
\vspace{0.7cm}

Despite the invariance properties conveniently manifested by the Weierstrass form, the period integrals turn out to be more easily calculated using the quartic representation of the elliptic curve. Indeed, as expressed in the formulae (\ref{eq:3.40}), (\ref{ee:3.41}) and their $b$-cycle counterparts, the period integrals are captured by two elliptic functions, namely the the complete elliptic integrals of the first and second kind. We'll now move to an overview of the rudiments of these functions, followed by some comments on numerical and hypergeometric aspects thereof.

The elliptic integral of the first kind, denoted by $K[k]$, is defined by the following canonical integral

\begin{equation}
    K[k] = \int_{0}^{1}\frac{dx}{\sqrt{(1-x^2)(1-k^2x^2)}}.
\end{equation}
The number $k$ is referred to as the \emph{modulus}. The complete elliptic curve of the third kind is expressed as a similar integral

\begin{equation}
    \Pi[n;k] = \int_0^1\frac{dx}{(1-n^2x^2)\sqrt{(1-x^2)(1-k^2x^2)}}.
\end{equation}

These functions are conveniently implemented in most commonly used numerical software packages, including in \textsc{Mathematica}. Parenthetically, Monte Carlo evaluations of these integrals turn out to be extremely stable, and can be evaluated to arbitrary precision.

The use of hypergeometric representations is sometimes valuable during numerical implementation, as series solutions are well known in the literature on this subject. Consider first the following definition of the so-called Lauricella hypergeometric function

\begin{equation}
    F^{n}\bigg(a,b_1,\dots,b_n,c\bigg|x_1,\dots,x_n\bigg) = \int_0^1 \bl{x}^{a-1}(1-\bl{x})^{c-a-1}(1-x_1\bl{x})^{-b_1}\dots(1-x_n\bl{x})^{-b_n}.
\end{equation}
Starting now with the elliptic periods

\begin{equation}
    \underset{\text{a-cycle}}{\oint}\frac{\dbar\var{z}}{y(\var{z})} = \frac{2}{2\pi i}\int_{\bur{r_1}}^{\bur{r_2}}\frac{d\var{z}}{\sqrt{(\var{z}-\bur{r_1})(\var{z}-\bur{r_2})(\var{z}-\bur{r_3})(\var{z}-\bur{r_4})}}.
\end{equation}
and

\begin{equation}
    \underset{\text{a-cycle}}{\oint}\frac{\dbar\var{z}}{(\var{z}-\bur{r})y(\var{z})} = \frac{2}{2\pi i}\int_{\bur{r_1}}^{\bur{r_2}}\frac{d\var{z}}{(\var{z}-\bur{r})\sqrt{(\var{z}-\bur{r_1})(\var{z}-\bur{r_2})(\var{z}-\bur{r_3})(\var{z}-\bur{r_4})}}.
\end{equation}
we will show that they can be rewritten as hypergeometric functions. Indeed, making the transformation

\begin{equation}
    \var{\alpha} = \frac{\var{z}-\bur{r_1}}{\bur{r_2}-\bur{r_1}}
\end{equation}
the integrals are transformed into

\begin{equation}
    \underset{\text{a-cycle}}{\oint}\frac{\dbar\var{z}}{y(\var{z})} = \frac{i}{\pi \sqrt{\bur{r_{31}r_{41}}}}\int_{0}^{1}\frac{d\var{\alpha}}{\sqrt{\var{\alpha}(1-\var{\alpha})(1-(\bur{21;31})\var{\alpha})(1-(\bur{21;41})\var{\alpha})}}.
\end{equation}
and

\begin{equation}
\begin{aligned}
     &\underset{\text{a-cycle}}{\oint}\frac{\dbar\var{z}}{(\var{z}-\bur{r})y(\var{z})} =\\& \frac{-i}{\pi (\bur{r}-\bur{r_1})\sqrt{\bur{r_{31}r_{41}}}}\int_{0}^{1}\frac{d\var{\alpha}}{(1-\var{\alpha}(\bur{21;r1}))\sqrt{\var{\alpha}(1-\var{\alpha})(1-(\bur{21;31})\var{\alpha})(1-(\bur{21;41})\var{\alpha})}}.
\end{aligned}
\end{equation}
where

\begin{equation}
    (\bur{ij;k\ell}) = \frac{\bur{r_i}-\bur{r_j}}{\bur{r_k}-\bur{r_{\ell}}}
\end{equation}
Both are clearly of hypergeometric form; the first has the identification

\begin{equation}
    (a,b_1,b_2,b_3,c) = \left(\frac{1}{2},\frac{1}{2},\frac{1}{2},\frac{1}{2},1\right)
\end{equation}
and the second

\begin{equation}
    (a,b_1,b_2,b_3,b_4,c) = \left(\frac{1}{2},\frac{1}{2},\frac{1}{2},\frac{1}{2},1,1\right).
\end{equation}

\clearpage
\section{Symbolic Implementation of Integrands and Cuts}\label{app:C}
To make the formalism described in the work, especially that of section \ref{sec:2}, less obscure, we have prepared---and attached with the present work ---a \textsc{Mathematica} package called `\texttt{one\_loop\_integrands.wl}' that implements integrands parametrized in twistor coordinates in a manner that makes the evaluation of cuts and the computation of quantities like Jacobians analytically and numerically streamlined. In this appendix, we will provide a short user manual for the functions (and functionality) included in \texttt{one\_loop\_integrands.wl}.

The package itself can be loaded by setting the notebook directory to the one with the package using the \texttt{SetDirectory[]} instruction, followed by the instruction 

\mathematica{0.9}{\texttt{<<one\_loop\_integrands.wl}}{\texttt{One Loop Integrands}}

The package, once loaded, makes use of a number of abstract symbols, which may be listed as follows. 

\begin{itemize}
    \item \texttt{\fun{ab}[\msymb{a},\msymb{b},\msymb{c},\msymb{d}\texttt{]}} is a symbolic representation of the four-bracket notation for the twistor determinant. The package supports the use of three kinds of geometric quantities as arguments of \texttt{\fun{ab}[]}: kinematic twistors labelled by integers $1, 2, \dots$, loop twistors labelled by \texttt{\fun{loop}[\fun{A}]} and \texttt{\fun{loop}[\fun{B}]}, and the shifted twistor labelled by \texttt{\fun{shift}[\{\msymb{a},\msymb{b},\msymb{c}\},\{\msymb{x},\msymb{y}\}\texttt{]}}. However, unlike in the case of the \texttt{two\_loop\_amplitudes.m} package due to Bourjaily and Trnka \cite{Bourjaily:2015jna}, numerical evaluation is only supported for arguments of the first kind, and all abstract arguments must be ultimately replaced by their expansions in terms of external twistors for numerics.
    
    \item \texttt{\fun{loop}[\fun{A}]} and \texttt{\fun{loop}[\fun{B}]} are symbols demarcating the twistor variables defining the loop momentum (since we restrict our attention in the attendant package to one loop) according to

    \begin{equation}
        \ell \Longleftrightarrow (AB).
    \end{equation}

    \item \texttt{\fun{shift}[\{\msymb{a},\msymb{b},\msymb{c}\},\{\msymb{x},\msymb{y}\}\texttt{]}} defines a twistor according to a two parameter shift spanned by $\varc{x}$ and $\varc{y}$:

    \begin{equation}
        \texttt{\fun{shift}[\{\msymb{a},\msymb{b},\msymb{c}\},\{\msymb{x},\msymb{y}\}\texttt{]}}\Longleftrightarrow Z_{\varc{a}} + \varc{x}Z_{\varc{b}} + \varc{y}Z_{\varc{b}}.
    \end{equation}
    This is geometrically equivalent to picking a point spanned by the twistors labelled by $\varc{a}$, $\varc{b}$ and $\varc{c}$.
\end{itemize}

The package furnishes two auxiliary functions that perform operations on angle brackets. Specifically, we have the following two replacement rules.

\begin{itemize}
    \item \texttt{\fun{shiftRule}} expands all instances of \texttt{\fun{ab}[]} where one has a \texttt{\fun{shift}[]} in the argument. For example, we have 

    \mathematica{0.9}{\texttt{\fun{ab}[1, \fun{shift}[\{2, 3, 4\}, \{\fun{x}, \fun{y}\}], 5, 6]/.shiftRule}}{\texttt{ab[1, 2, 5, 6] + x ab[1, 3, 5, 6] + y ab[1, 4, 5, 6]}}.

    \item \texttt{\fun{deleteZeros}} enforces the rudimentary fact that angle brackets vanish when any two of the twistors match. For example, $\langle{1213\rangle}$ is zero as two of the columns of the attendant determinant match. Indeed, we have the following illustration of its functionality:

    \mathematica{0.9}{\texttt{\fun{ab}[1,2,3,4]}\texttt{\fun{ab}[3,4,5,6]} + \texttt{\fun{ab}[1,2,1,4]}\texttt{\fun{ab}[1,3,5,6]}/.\texttt{deleteZeros}}{\texttt{ab[1,2,3,4]}\texttt{ab[3,4,5,6]}}
\end{itemize}

The main content of the package revolves around a set of functions useful for setting up one loop cuts, generalized readily to higher loops. The main purpose of these functions is to make easy the calculation of auxiliary quantities such as Jacobians; we will go over a few examples and the instructions in the following. 

\begin{itemize}
    \item \texttt{\fun{loopRules}[\msymb{a},\msymb{b},\msymb{c},\msymb{d}\texttt{]}} takes as input four twistors, labelled by $a$, $b$, $c$ and $d$, and expands the loop momentum twistors in this basis by setting

    \begin{equation}
        A = Z_{\varc{b}} + \var{\alpha_1}Z_{\varc{a}} + \var{\alpha_2} Z_{\varc{c}}
    \end{equation}
    and

    \begin{equation}
        B = Z_{\varc{d}} + \var{\beta_1}Z_{\varc{a}} + \var{\beta_2}Z_{\varc{c}}.
    \end{equation}
    An example of its use can be provided by the following.

    \mathematica{0.9}{\texttt{\fun{loopRules}[1,2,3,4]}}{\{\texttt{loop[A]}$\to$\texttt{shift[\{2,1,3\},$\lbrace{\alpha_1,\alpha_2\rbrace}$]},\texttt{loop[B]}$\to$\texttt{shift[\{4,1,3\},$\lbrace{\beta_1,\beta_2\rbrace}$]}\}}

    \item Two functions \texttt{\fun{propListOneLoop}[\msymb{vars}\texttt{]}} and \texttt{\fun{propList}[\msymb{vars},\msymb{list}\texttt{]}} are provided, which assist in preparing propagators for a single loop. The former is an auxiliary of the latter, which takes as input a list \textit{\varc{vars}} of length $8$ and a list \textit{\varc{list}} of length $4$. The task it performs is to create a list of four propagators of the form 

    $$
    \text{\texttt{\fun{ab}[\fun{loop}[A],\fun{loop}[B],list$[\![$i$]\!]$,list$[\![$i+1$]\!]$\texttt{]}}}
    $$
    followed by a replacement of the loop variables according to the rules defined by \texttt{\fun{loopRules}[vars\texttt{]}}, expanded and simplified according to the rules \texttt{\fun{shiftRule}} and \texttt{\fun{deleteZeros}}. An instance of the use of this function can be seen by the following snippet.

    \mathematica{0.9}{\texttt{\fun{propList}[\{1,2,3,4,5,6,7,8\},\{1,2,3,4\}]}$[\![$1$]\!]$}{\texttt{ab[3,4,1,2]}$\alpha_1$}

    \item Three functions named \texttt{\fun{bubble}[\normalfont{\textit{\msymb{list}}}]}, \texttt{\fun{triangle}[\normalfont{\textit{\msymb{list}}}]} and \texttt{\fun{box}[\normalfont{\textit{\msymb{list}}}]} implement integrands---matching the name of the function---given \texttt{list}, a list of four propagators, which we will usually take to be determined by the function \texttt{\fun{propList}[]}, although the functions themselves support any list of four propagators. Due to the nature of the numerics of this package, it is recommended that all \texttt{\fun{shift}[]} functions be fully expanded before the use of these functions. The convention here will be to pick the first $n$ of the elements of \texttt{list} to prepare the integrand, where $n$ is $2$, $3$ or $4$ for \texttt{\fun{bubble}[]}, \texttt{\fun{triangle}[]} and \texttt{\fun{box}[]} respectively. For example,

    \mathematica{0.9}{\texttt{\fun{bubble}[\texttt{\fun{propList}[\{1,2,3,4,5,6,7,8\},\{1,2,3,4\}]}]}}{$\frac{\texttt{1}}{\texttt{ab[2,1,3,4]ab[3,4,1,2]}\alpha_2\beta_1}$}
    
    The broader utility of the integrand functions is to make it easy to compute Jacobians of non-maximal cuts in terms of angle brackets. An example of this is for the case of the triangle cut, which involves one final variable. We set this up by the following three instructions.

    \mathematica{0.9}{\texttt{props = }\texttt{\fun{propList}[\{1,2,3,4,5,6,7,8\},\{1,2,3,4\}];}\newline \texttt{triangleIntegrand = \fun{triangle}[\texttt{\fun{propList}[\{1,2,3,4,5,6,7,8\},\{1,2,3,4\}]}];}\newline\texttt{poles = Solve[Table[props$[\![i]\!]$==0,\{i,1,3\}], \{$\alpha_2,\beta_1,\alpha_1$\}]//Flatten}\newline}{$\left\{\alpha_2\to \texttt{0},\beta_1\to \texttt{0},\alpha_1\to \frac{-\beta_2 \texttt{ab[2,3,5,6]-ab[2,4,5,6]}}{\beta_2\texttt{ab[1,3,5,6]+ab[1,4,5,6]}}\right\}$}

    Computing the triple cut is straightforward; we simply evaluate the residues around the three poles given by the latter expression:

    \mathematica{0.9}{\texttt{Residue[triangleIntegrand,\{$\alpha_2$,$\alpha_2$\symbol{92}.poles$[\![1]\!]$}\}]\newline \texttt{Residue[\%,\{$\beta_1$,$\beta_1$\symbol{92}.poles$[\![2]\!]$}\}] \newline \texttt{Residue[\%,\{$\alpha_1$,$\alpha_1$\symbol{92}.poles$[\![3]\!]$}\}]\newline}{$\frac{\texttt{1}}{\texttt{ab[2,1,3,4]ab[3,4,1,2](ab[1,4,5,6]+ab[1,3,5,6]$\beta_2$)}}$}

    which is the correct form (up to a redefinition of $\beta_2$) of the Jacobian on the triangle cut at one loop.
\end{itemize}

Finally, there are three functions provided to make numerical evaluation easy for the user. 

\begin{itemize}
    \item \texttt{\fun{randomKinematics}[\msymb{n}\texttt{]}} generates a set of $n$ momentum twistors, where the entries are (pseudo)random rational numbers. The entries are printed as a table upon calling the function. For $n=4$, we may find the result given below.

    \mathematica{0.9}{\texttt{\fun{randomKinematics}[4]}}{$\begin{array}{cccc}
 \mathtt{Z_1} & \mathtt{Z_2} & \mathtt{Z_3} & \mathtt{Z_4} \\
 \mathtt{2} & -\fract{2}{5} & \mathtt{-1} & -\fract{1}{5} \\
 \fract{5}{2} & -2 & -\fract{1}{4} & \fract{3}{4} \\
 \mathtt{1} & \mathtt{-2} & -\fract{3}{5} & \mathtt{0} \\
 \fract{5}{2} & -\fract{1}{5} & \fract{3}{5} & \fract{1}{5} \\
\end{array}$}

\item \texttt{\fun{numeric}} is a replacement rule; it acts on \texttt{\fun{ab}[]} symbols, and converts them into their corresponding numerical values when a specific set of random kinematics is initialized. Importantly, it only supports the evaluation of \texttt{\fun{ab}[]} symbols when the arguments are $1, 2, \dots$ etc. Accordingly, the user is required to perform the expansion of any other such symbols with symbolic arguments using \texttt{\fun{shiftReplace}} and a choice of \texttt{\fun{loopRules}[]}. As an example, for the latter set of four twistors,  we find the following.

\mathematica{0.9}{\texttt{\fun{ab}[1,\fun{shift}[\{1,2,3\},\{x,y\}],3,4]/.\fun{shiftReplace}/.\fun{numeric}}}{$-\fract{423 x}{100}$}

\item Finally, in homage to \cite{Bourjaily:2010wh}, we have defined a replacement rule \texttt{\fun{nice}}, albeit with poorer functionality; it simply replaces all angle brackets (which the user is again recommended to expand fully) into human-readable form. See the example below.

\mathematica{0.9}{\texttt{\fun{ab}[1,2,3,4]\fun{ab}[1,2,3,5]/.\fun{nice}}}{$\langle 1,2,3,4\rangle  \langle 1,2,3,5\rangle$}
\end{itemize}
\clearpage
\addcontentsline{toc}{section}{References}
\bibliographystyle{JHEP}
\bibliography{main}

\providecommand{\noopsort}[1]{}\providecommand{\singleletter}[1]{#1}%

\providecommand{\href}[2]{#2}\begingroup\raggedright\begin{thebibliography}{100}

\bibitem{Bern:1987tw}
Z.~Bern and D.~A. Kosower, \emph{{A New Approach to One Loop Calculations in
  Gauge Theories}},
  \href{http://dx.doi.org/10.1103/PhysRevD.38.1888}{\emph{Phys. Rev. D}
  {\bfseries 38} (1988) 1888}.

\bibitem{Bern:1990cu}
Z.~Bern and D.~A. Kosower, \emph{{Efficient calculation of one loop QCD
  amplitudes}},
  \href{http://dx.doi.org/10.1103/PhysRevLett.66.1669}{\emph{Phys. Rev. Lett.}
  {\bfseries 66} (1991) 1669--1672}.

\bibitem{Bern:1990ux}
Z.~Bern and D.~A. Kosower, \emph{{Color decomposition of one loop amplitudes in
  gauge theories}},
  \href{http://dx.doi.org/10.1016/0550-3213(91)90567-H}{\emph{Nucl. Phys. B}
  {\bfseries 362} (1991) 389--448}.

\bibitem{Bern:1990qr}
Z.~Bern and D.~A. Kosower, \emph{{Efficient calculation of one-loop polarized
  QCD amplitudes}}, \href{http://dx.doi.org/10.1063/1.40513}{\emph{AIP Conf.
  Proc.} {\bfseries 223} (1991) 358--363}.

\bibitem{Bern:1991aq}
Z.~Bern and D.~A. Kosower, \emph{{The Computation of loop amplitudes in gauge
  theories}}, \href{http://dx.doi.org/10.1016/0550-3213(92)90134-W}{\emph{Nucl.
  Phys. B} {\bfseries 379} (1992) 451--561}.

\bibitem{Bern:1992ee}
Z.~Bern, L.~J. Dixon and D.~A. Kosower, \emph{{The Five gluon amplitude and one
  loop integrals}},  in \emph{{7th Meeting of the APS Division of Particles
  Fields}}, pp.~901--905, 11, 1992.

\bibitem{Bern:1992em}
Z.~Bern, L.~J. Dixon and D.~A. Kosower, \emph{{Dimensionally regulated one loop
  integrals}},
  \href{http://dx.doi.org/10.1016/0370-2693(93)90400-C}{\emph{Phys. Lett. B}
  {\bfseries 302} (1993) 299--308},
  [\href{https://arxiv.org/abs/hep-ph/9212308}{{\ttfamily hep-ph/9212308}}].

\bibitem{Bern:1993mq}
Z.~Bern, L.~J. Dixon and D.~A. Kosower, \emph{{One loop corrections to five
  gluon amplitudes}},
  \href{http://dx.doi.org/10.1103/PhysRevLett.70.2677}{\emph{Phys. Rev. Lett.}
  {\bfseries 70} (1993) 2677--2680},
  [\href{https://arxiv.org/abs/hep-ph/9302280}{{\ttfamily hep-ph/9302280}}].

\bibitem{Bern:1993qk}
Z.~Bern, G.~Chalmers, L.~J. Dixon and D.~A. Kosower, \emph{{One loop N gluon
  amplitudes with maximal helicity violation via collinear limits}},
  \href{http://dx.doi.org/10.1103/PhysRevLett.72.2134}{\emph{Phys. Rev. Lett.}
  {\bfseries 72} (1994) 2134--2137},
  [\href{https://arxiv.org/abs/hep-ph/9312333}{{\ttfamily hep-ph/9312333}}].

\bibitem{Bern:1994ju}
Z.~Bern, L.~J. Dixon, D.~C. Dunbar and D.~A. Kosower, \emph{{One loop gauge
  theory amplitudes with an arbitrary number of external legs}},  in
  \emph{{Workshop on Continuous Advances in QCD}}, 2, 1994,
  \href{https://arxiv.org/abs/hep-ph/9405248}{{\ttfamily hep-ph/9405248}}.

\bibitem{Bern:1994zx}
Z.~Bern, L.~J. Dixon, D.~C. Dunbar and D.~A. Kosower, \emph{{One loop n point
  gauge theory amplitudes, unitarity and collinear limits}},
  \href{http://dx.doi.org/10.1016/0550-3213(94)90179-1}{\emph{Nucl. Phys. B}
  {\bfseries 425} (1994) 217--260},
  [\href{https://arxiv.org/abs/hep-ph/9403226}{{\ttfamily hep-ph/9403226}}].

\bibitem{Bern:1994fz}
Z.~Bern, L.~J. Dixon and D.~A. Kosower, \emph{{One loop corrections to two
  quark three gluon amplitudes}},
  \href{http://dx.doi.org/10.1016/0550-3213(94)00542-M}{\emph{Nucl. Phys. B}
  {\bfseries 437} (1995) 259--304},
  [\href{https://arxiv.org/abs/hep-ph/9409393}{{\ttfamily hep-ph/9409393}}].

\bibitem{Bern:1994cg}
Z.~Bern, L.~J. Dixon, D.~C. Dunbar and D.~A. Kosower, \emph{{Fusing gauge
  theory tree amplitudes into loop amplitudes}},
  \href{http://dx.doi.org/10.1016/0550-3213(94)00488-Z}{\emph{Nucl. Phys. B}
  {\bfseries 435} (1995) 59--101},
  [\href{https://arxiv.org/abs/hep-ph/9409265}{{\ttfamily hep-ph/9409265}}].

\bibitem{Bern:1995ix}
Z.~Bern and G.~Chalmers, \emph{{Factorization in one loop gauge theory}},
  \href{http://dx.doi.org/10.1016/0550-3213(95)00226-I}{\emph{Nucl. Phys. B}
  {\bfseries 447} (1995) 465--518},
  [\href{https://arxiv.org/abs/hep-ph/9503236}{{\ttfamily hep-ph/9503236}}].

\bibitem{Bern:1995db}
Z.~Bern and A.~G. Morgan, \emph{{Massive loop amplitudes from unitarity}},
  \href{http://dx.doi.org/10.1016/0550-3213(96)00078-8}{\emph{Nucl. Phys. B}
  {\bfseries 467} (1996) 479--509},
  [\href{https://arxiv.org/abs/hep-ph/9511336}{{\ttfamily hep-ph/9511336}}].

\bibitem{Bern:1996ka}
Z.~Bern, L.~J. Dixon, D.~A. Kosower and S.~Weinzierl, \emph{{One loop
  amplitudes for e+ e- ---\ensuremath{>} anti-q q anti-Q Q}},
  \href{http://dx.doi.org/10.1016/S0550-3213(96)00703-1}{\emph{Nucl. Phys. B}
  {\bfseries 489} (1997) 3--23},
  [\href{https://arxiv.org/abs/hep-ph/9610370}{{\ttfamily hep-ph/9610370}}].

\bibitem{Bern:1997nh}
Z.~Bern, J.~S. Rozowsky and B.~Yan, \emph{{Two loop four gluon amplitudes in
  N=4 superYang-Mills}},
  \href{http://dx.doi.org/10.1016/S0370-2693(97)00413-9}{\emph{Phys. Lett. B}
  {\bfseries 401} (1997) 273--282},
  [\href{https://arxiv.org/abs/hep-ph/9702424}{{\ttfamily hep-ph/9702424}}].

\bibitem{Bern:1997sc}
Z.~Bern, L.~J. Dixon and D.~A. Kosower, \emph{{One loop amplitudes for e+ e- to
  four partons}},
  \href{http://dx.doi.org/10.1016/S0550-3213(97)00703-7}{\emph{Nucl. Phys. B}
  {\bfseries 513} (1998) 3--86},
  [\href{https://arxiv.org/abs/hep-ph/9708239}{{\ttfamily hep-ph/9708239}}].

\bibitem{Ossola:2006us}
G.~Ossola, C.~G. Papadopoulos and R.~Pittau, \emph{{Reducing full one-loop
  amplitudes to scalar integrals at the integrand level}},
  \href{http://dx.doi.org/10.1016/j.nuclphysb.2006.11.012}{\emph{Nucl. Phys. B}
  {\bfseries 763} (2007) 147--169},
  [\href{https://arxiv.org/abs/hep-ph/0609007}{{\ttfamily hep-ph/0609007}}].

\bibitem{Britto:2007tt}
R.~Britto and B.~Feng, \emph{{Integral coefficients for one-loop amplitudes}},
  \href{http://dx.doi.org/10.1088/1126-6708/2008/02/095}{\emph{JHEP} {\bfseries
  02} (2008) 095}, [\href{https://arxiv.org/abs/0711.4284}{{\ttfamily
  0711.4284}}].

\bibitem{Forde:2007mi}
D.~Forde, \emph{{Direct extraction of one-loop integral coefficients}},
  \href{http://dx.doi.org/10.1103/PhysRevD.75.125019}{\emph{Phys. Rev. D}
  {\bfseries 75} (2007) 125019},
  [\href{https://arxiv.org/abs/0704.1835}{{\ttfamily 0704.1835}}].

\bibitem{Badger:2008cm}
S.~D. Badger, \emph{{Direct Extraction Of One Loop Rational Terms}},
  \href{http://dx.doi.org/10.1088/1126-6708/2009/01/049}{\emph{JHEP} {\bfseries
  01} (2009) 049}, [\href{https://arxiv.org/abs/0806.4600}{{\ttfamily
  0806.4600}}].

\bibitem{Arkani-Hamed:2010pyv}
N.~Arkani-Hamed, J.~L. Bourjaily, F.~Cachazo and J.~Trnka, \emph{{Local
  Integrals for Planar Scattering Amplitudes}},
  \href{http://dx.doi.org/10.1007/JHEP06(2012)125}{\emph{JHEP} {\bfseries 06}
  (2012) 125}, [\href{https://arxiv.org/abs/1012.6032}{{\ttfamily 1012.6032}}].

\bibitem{Bourjaily:2011hi}
J.~L. Bourjaily, A.~DiRe, A.~Shaikh, M.~Spradlin and A.~Volovich, \emph{{The
  Soft-Collinear Bootstrap: N=4 Yang-Mills Amplitudes at Six and Seven Loops}},
  \href{http://dx.doi.org/10.1007/JHEP03(2012)032}{\emph{JHEP} {\bfseries 03}
  (2012) 032}, [\href{https://arxiv.org/abs/1112.6432}{{\ttfamily 1112.6432}}].

\bibitem{Bourjaily:2013mma}
J.~L. Bourjaily, S.~Caron-Huot and J.~Trnka, \emph{{Dual-Conformal
  Regularization of Infrared Loop Divergences and the Chiral Box Expansion}},
  \href{http://dx.doi.org/10.1007/JHEP01(2015)001}{\emph{JHEP} {\bfseries 01}
  (2015) 001}, [\href{https://arxiv.org/abs/1303.4734}{{\ttfamily 1303.4734}}].

\bibitem{Bourjaily:2015jna}
J.~L. Bourjaily and J.~Trnka, \emph{{Local Integrand Representations of All
  Two-Loop Amplitudes in Planar SYM}},
  \href{http://dx.doi.org/10.1007/JHEP08(2015)119}{\emph{JHEP} {\bfseries 08}
  (2015) 119}, [\href{https://arxiv.org/abs/1505.05886}{{\ttfamily
  1505.05886}}].

\bibitem{Bourjaily:2016evz}
J.~L. Bourjaily, P.~Heslop and V.-V. Tran, \emph{{Amplitudes and Correlators to
  Ten Loops Using Simple, Graphical Bootstraps}},
  \href{http://dx.doi.org/10.1007/JHEP11(2016)125}{\emph{JHEP} {\bfseries 11}
  (2016) 125}, [\href{https://arxiv.org/abs/1609.00007}{{\ttfamily
  1609.00007}}].

\bibitem{Bourjaily:2019iqr}
J.~L. Bourjaily, E.~Herrmann, C.~Langer, A.~J. McLeod and J.~Trnka,
  \emph{{Prescriptive Unitarity for Non-Planar Six-Particle Amplitudes at Two
  Loops}}, \href{http://dx.doi.org/10.1007/JHEP12(2019)073}{\emph{JHEP}
  {\bfseries 12} (2019) 073},
  [\href{https://arxiv.org/abs/1909.09131}{{\ttfamily 1909.09131}}].

\bibitem{Bourjaily:2019gqu}
J.~L. Bourjaily, E.~Herrmann, C.~Langer, A.~J. McLeod and J.~Trnka,
  \emph{{All-Multiplicity Nonplanar Amplitude Integrands in Maximally
  Supersymmetric Yang-Mills Theory at Two Loops}},
  \href{http://dx.doi.org/10.1103/PhysRevLett.124.111603}{\emph{Phys. Rev.
  Lett.} {\bfseries 124} (2020) 111603},
  [\href{https://arxiv.org/abs/1911.09106}{{\ttfamily 1911.09106}}].

\bibitem{Britto:2004ap}
R.~Britto, F.~Cachazo and B.~Feng, \emph{{New recursion relations for tree
  amplitudes of gluons}},
  \href{http://dx.doi.org/10.1016/j.nuclphysb.2005.02.030}{\emph{Nucl. Phys. B}
  {\bfseries 715} (2005) 499--522},
  [\href{https://arxiv.org/abs/hep-th/0412308}{{\ttfamily hep-th/0412308}}].

\bibitem{Britto:2005fq}
R.~Britto, F.~Cachazo, B.~Feng and E.~Witten, \emph{{Direct proof of tree-level
  recursion relation in Yang-Mills theory}},
  \href{http://dx.doi.org/10.1103/PhysRevLett.94.181602}{\emph{Phys. Rev.
  Lett.} {\bfseries 94} (2005) 181602},
  [\href{https://arxiv.org/abs/hep-th/0501052}{{\ttfamily hep-th/0501052}}].

\bibitem{ArkaniHamed:2010kv}
N.~Arkani-Hamed, J.~L. Bourjaily, F.~Cachazo, S.~Caron-Huot and J.~Trnka,
  \emph{{The All-Loop Integrand For Scattering Amplitudes in Planar N=4 SYM}},
  \href{http://dx.doi.org/10.1007/JHEP01(2011)041}{\emph{JHEP} {\bfseries 01}
  (2011) 041}, [\href{https://arxiv.org/abs/1008.2958}{{\ttfamily 1008.2958}}].

\bibitem{Arkani-Hamed:2013jha}
N.~Arkani-Hamed and J.~Trnka, \emph{{The Amplituhedron}},
  \href{http://dx.doi.org/10.1007/JHEP10(2014)030}{\emph{JHEP} {\bfseries 10}
  (2014) 030}, [\href{https://arxiv.org/abs/1312.2007}{{\ttfamily 1312.2007}}].

\bibitem{Trnka:2013eth}
J.~Trnka, \emph{{Grassmannian origin of scattering amplitudes}}, Ph.D. thesis,
  Princeton U., 2013.

\bibitem{Franco:2013nwa}
S.~Franco, D.~Galloni and A.~Mariotti, \emph{{The Geometry of On-Shell
  Diagrams}}, \href{http://dx.doi.org/10.1007/JHEP08(2014)038}{\emph{JHEP}
  {\bfseries 08} (2014) 038},
  [\href{https://arxiv.org/abs/1310.3820}{{\ttfamily 1310.3820}}].

\bibitem{Arkani-Hamed:2014bca}
N.~Arkani-Hamed, J.~L. Bourjaily, F.~Cachazo, A.~Postnikov and J.~Trnka,
  \emph{{On-Shell Structures of MHV Amplitudes Beyond the Planar Limit}},
  \href{http://dx.doi.org/10.1007/JHEP06(2015)179}{\emph{JHEP} {\bfseries 06}
  (2015) 179}, [\href{https://arxiv.org/abs/1412.8475}{{\ttfamily 1412.8475}}].

\bibitem{Franco:2015rma}
S.~Franco, D.~Galloni, B.~Penante and C.~Wen, \emph{{Non-Planar On-Shell
  Diagrams}}, \href{http://dx.doi.org/10.1007/JHEP06(2015)199}{\emph{JHEP}
  {\bfseries 06} (2015) 199},
  [\href{https://arxiv.org/abs/1502.02034}{{\ttfamily 1502.02034}}].

\bibitem{Frassek:2015rka}
R.~Frassek, D.~Meidinger, D.~Nandan and M.~Wilhelm, \emph{{On-shell diagrams,
  Gra\ss{}mannians and integrability for form factors}},
  \href{http://dx.doi.org/10.1007/JHEP01(2016)182}{\emph{JHEP} {\bfseries 01}
  (2016) 182}, [\href{https://arxiv.org/abs/1506.08192}{{\ttfamily
  1506.08192}}].

\bibitem{Jin:2015pua}
Q.~Jin and B.~Feng, \emph{{Boundary Operators of BCFW Recursion Relation}},
  \href{http://dx.doi.org/10.1007/JHEP04(2016)123}{\emph{JHEP} {\bfseries 04}
  (2016) 123}, [\href{https://arxiv.org/abs/1507.00463}{{\ttfamily
  1507.00463}}].

\bibitem{Benincasa:2015zna}
P.~Benincasa, \emph{{On-shell diagrammatics and the perturbative structure of
  planar gauge theories}},  \href{https://arxiv.org/abs/1510.03642}{{\ttfamily
  1510.03642}}.

\bibitem{Heslop:2016plj}
P.~Heslop and A.~E. Lipstein, \emph{{On-shell diagrams for $ \mathcal{N} $ = 8
  supergravity amplitudes}},
  \href{http://dx.doi.org/10.1007/JHEP06(2016)069}{\emph{JHEP} {\bfseries 06}
  (2016) 069}, [\href{https://arxiv.org/abs/1604.03046}{{\ttfamily
  1604.03046}}].

\bibitem{Herrmann:2016qea}
E.~Herrmann and J.~Trnka, \emph{{Gravity On-shell Diagrams}},
  \href{http://dx.doi.org/10.1007/JHEP11(2016)136}{\emph{JHEP} {\bfseries 11}
  (2016) 136}, [\href{https://arxiv.org/abs/1604.03479}{{\ttfamily
  1604.03479}}].

\bibitem{Bourjaily:2016mnp}
J.~L. Bourjaily, S.~Franco, D.~Galloni and C.~Wen, \emph{{Stratifying On-Shell
  Cluster Varieties: the Geometry of Non-Planar On-Shell Diagrams}},
  \href{http://dx.doi.org/10.1007/JHEP10(2016)003}{\emph{JHEP} {\bfseries 10}
  (2016) 003}, [\href{https://arxiv.org/abs/1607.01781}{{\ttfamily
  1607.01781}}].

\bibitem{Bork:2016xfn}
L.~V. Bork and A.~I. Onishchenko, \emph{{Wilson lines, Grassmannians and gauge
  invariant off-shell amplitudes in $ \mathcal{N}=4 $ SYM}},
  \href{http://dx.doi.org/10.1007/JHEP04(2017)019}{\emph{JHEP} {\bfseries 04}
  (2017) 019}, [\href{https://arxiv.org/abs/1607.02320}{{\ttfamily
  1607.02320}}].

\bibitem{Benincasa:2016awv}
P.~Benincasa and D.~Gordo, \emph{{On-shell diagrams and the geometry of planar
  $ \mathcal{N}<4 $ SYM theories}},
  \href{http://dx.doi.org/10.1007/JHEP11(2017)192}{\emph{JHEP} {\bfseries 11}
  (2017) 192}, [\href{https://arxiv.org/abs/1609.01923}{{\ttfamily
  1609.01923}}].

\bibitem{Boels:2016jmi}
R.~H. Boels and H.~Luo, \emph{{On-shell recursion relations for generic
  integrands}},  \href{https://arxiv.org/abs/1610.05283}{{\ttfamily
  1610.05283}}.

\bibitem{Herderschee:2019ofc}
A.~Herderschee, S.~Koren and T.~Trott, \emph{{Massive On-Shell Supersymmetric
  Scattering Amplitudes}},
  \href{http://dx.doi.org/10.1007/JHEP10(2019)092}{\emph{JHEP} {\bfseries 10}
  (2019) 092}, [\href{https://arxiv.org/abs/1902.07204}{{\ttfamily
  1902.07204}}].

\bibitem{Armstrong:2020ljm}
C.~Armstrong, J.~A. Farrow and A.~E. Lipstein, \emph{{$ \mathcal{N} $ = 7
  On-shell diagrams and supergravity amplitudes in momentum twistor space}},
  \href{http://dx.doi.org/10.1007/JHEP01(2021)181}{\emph{JHEP} {\bfseries 01}
  (2021) 181}, [\href{https://arxiv.org/abs/2010.11813}{{\ttfamily
  2010.11813}}].

\bibitem{Bartsch:2022pyi}
C.~Bartsch, K.~Kampf and J.~Trnka, \emph{{Recursion relations for one-loop
  Goldstone boson amplitudes}},
  \href{http://dx.doi.org/10.1103/PhysRevD.106.076008}{\emph{Phys. Rev. D}
  {\bfseries 106} (2022) 076008},
  [\href{https://arxiv.org/abs/2206.04694}{{\ttfamily 2206.04694}}].

\bibitem{ArkaniHamed:2008yf}
N.~Arkani-Hamed and J.~Kaplan, \emph{{On Tree Amplitudes in Gauge Theory and
  Gravity}}, \href{http://dx.doi.org/10.1088/1126-6708/2008/04/076}{\emph{JHEP}
  {\bfseries 04} (2008) 076},
  [\href{https://arxiv.org/abs/0801.2385}{{\ttfamily 0801.2385}}].

\bibitem{ArkaniHamed:2010gh}
N.~Arkani-Hamed, J.~L. Bourjaily, F.~Cachazo and J.~Trnka, \emph{{Local
  Integrals for Planar Scattering Amplitudes}},
  \href{http://dx.doi.org/10.1007/JHEP06(2012)125}{\emph{JHEP} {\bfseries 06}
  (2012) 125}, [\href{https://arxiv.org/abs/1012.6032}{{\ttfamily 1012.6032}}].

\bibitem{ArkaniHamed:2012nw}
N.~Arkani-Hamed, J.~L. Bourjaily, F.~Cachazo, A.~B. Goncharov, A.~Postnikov and
  J.~Trnka, \emph{{Grassmannian Geometry of Scattering Amplitudes}}.
\newblock Cambridge University Press, 2016,
  \href{http://dx.doi.org/10.1017/CBO9781316091548}{10.1017/CBO9781316091548}.

\bibitem{Cachazo:2013hca}
F.~Cachazo, S.~He and E.~Y. Yuan, \emph{{Scattering of Massless Particles in
  Arbitrary Dimensions}},
  \href{http://dx.doi.org/10.1103/PhysRevLett.113.171601}{\emph{Phys. Rev.
  Lett.} {\bfseries 113} (2014) 171601},
  [\href{https://arxiv.org/abs/1307.2199}{{\ttfamily 1307.2199}}].

\bibitem{Cachazo:2013iaa}
F.~Cachazo, S.~He and E.~Y. Yuan, \emph{{Scattering in Three Dimensions from
  Rational Maps}}, \href{http://dx.doi.org/10.1007/JHEP10(2013)141}{\emph{JHEP}
  {\bfseries 10} (2013) 141},
  [\href{https://arxiv.org/abs/1306.2962}{{\ttfamily 1306.2962}}].

\bibitem{Cachazo:2014fwa}
F.~Cachazo and A.~Strominger, \emph{{Evidence for a New Soft Graviton
  Theorem}},  \href{https://arxiv.org/abs/1404.4091}{{\ttfamily 1404.4091}}.

\bibitem{Cachazo:2014nsa}
F.~Cachazo, S.~He and E.~Y. Yuan, \emph{{Einstein-Yang-Mills Scattering
  Amplitudes From Scattering Equations}},
  \href{http://dx.doi.org/10.1007/JHEP01(2015)121}{\emph{JHEP} {\bfseries 01}
  (2015) 121}, [\href{https://arxiv.org/abs/1409.8256}{{\ttfamily 1409.8256}}].

\bibitem{Cachazo:2014xea}
F.~Cachazo, S.~He and E.~Y. Yuan, \emph{{Scattering Equations and Matrices:
  From Einstein To Yang-Mills, DBI and NLSM}},
  \href{http://dx.doi.org/10.1007/JHEP07(2015)149}{\emph{JHEP} {\bfseries 07}
  (2015) 149}, [\href{https://arxiv.org/abs/1412.3479}{{\ttfamily 1412.3479}}].

\bibitem{Mizera:2017rqa}
S.~Mizera, \emph{{Scattering Amplitudes from Intersection Theory}},
  \href{http://dx.doi.org/10.1103/PhysRevLett.120.141602}{\emph{Phys. Rev.
  Lett.} {\bfseries 120} (2018) 141602},
  [\href{https://arxiv.org/abs/1711.00469}{{\ttfamily 1711.00469}}].

\bibitem{Mizera:2019gea}
S.~Mizera, \emph{{Aspects of Scattering Amplitudes and Moduli Space
  Localization}}, Ph.D. thesis, Perimeter Inst. Theor. Phys., 2019.
\newblock \href{https://arxiv.org/abs/1906.02099}{{\ttfamily 1906.02099}}.

\bibitem{Mizera:2019blq}
S.~Mizera, \emph{{Kinematic Jacobi Identity is a Residue Theorem: Geometry of
  Color-Kinematics Duality for Gauge and Gravity Amplitudes}},
  \href{http://dx.doi.org/10.1103/PhysRevLett.124.141601}{\emph{Phys. Rev.
  Lett.} {\bfseries 124} (2020) 141601},
  [\href{https://arxiv.org/abs/1912.03397}{{\ttfamily 1912.03397}}].

\bibitem{Britto:2010xq}
R.~Britto, \emph{{Loop Amplitudes in Gauge Theories: Modern Analytic
  Approaches}},
  \href{http://dx.doi.org/10.1088/1751-8113/44/45/454006}{\emph{J. Phys. A}
  {\bfseries 44} (2011) 454006},
  [\href{https://arxiv.org/abs/1012.4493}{{\ttfamily 1012.4493}}].

\bibitem{Chetyrkin:1981qh}
K.~G. Chetyrkin and F.~V. Tkachov, \emph{{Integration by Parts: The Algorithm
  to Calculate beta Functions in 4 Loops}},
  \href{http://dx.doi.org/10.1016/0550-3213(81)90199-1}{\emph{Nucl. Phys. B}
  {\bfseries 192} (1981) 159--204}.

\bibitem{Kotikov:1990kg}
A.~V. Kotikov, \emph{{Differential equations method: New technique for massive
  Feynman diagrams calculation}},
  \href{http://dx.doi.org/10.1016/0370-2693(91)90413-K}{\emph{Phys. Lett. B}
  {\bfseries 254} (1991) 158--164}.

\bibitem{Kotikov:1991pm}
A.~V. Kotikov, \emph{{Differential equation method: The Calculation of N point
  Feynman diagrams}},
  \href{http://dx.doi.org/10.1016/0370-2693(91)90536-Y}{\emph{Phys. Lett. B}
  {\bfseries 267} (1991) 123--127}.

\bibitem{Tarasov:1996br}
O.~V. Tarasov, \emph{{Connection between Feynman integrals having different
  values of the space-time dimension}},
  \href{http://dx.doi.org/10.1103/PhysRevD.54.6479}{\emph{Phys. Rev. D}
  {\bfseries 54} (1996) 6479--6490},
  [\href{https://arxiv.org/abs/hep-th/9606018}{{\ttfamily hep-th/9606018}}].

\bibitem{Remiddi:1997ny}
E.~Remiddi, \emph{{Differential equations for Feynman graph amplitudes}},
  \href{http://dx.doi.org/10.1007/BF03185566}{\emph{Nuovo Cim. A} {\bfseries
  110} (1997) 1435--1452},
  [\href{https://arxiv.org/abs/hep-th/9711188}{{\ttfamily hep-th/9711188}}].

\bibitem{Gehrmann:1999as}
T.~Gehrmann and E.~Remiddi, \emph{{Differential equations for two loop four
  point functions}},
  \href{http://dx.doi.org/10.1016/S0550-3213(00)00223-6}{\emph{Nucl. Phys. B}
  {\bfseries 580} (2000) 485--518},
  [\href{https://arxiv.org/abs/hep-ph/9912329}{{\ttfamily hep-ph/9912329}}].

\bibitem{Laporta:2000dsw}
S.~Laporta, \emph{{High precision calculation of multiloop Feynman integrals by
  difference equations}},
  \href{http://dx.doi.org/10.1142/S0217751X00002159}{\emph{Int. J. Mod. Phys.
  A} {\bfseries 15} (2000) 5087--5159},
  [\href{https://arxiv.org/abs/hep-ph/0102033}{{\ttfamily hep-ph/0102033}}].

\bibitem{Laporta:2003jz}
S.~Laporta, \emph{{Calculation of Feynman integrals by difference equations}},
  {\emph{Acta Phys. Polon. B} {\bfseries 34} (2003) 5323--5334},
  [\href{https://arxiv.org/abs/hep-ph/0311065}{{\ttfamily hep-ph/0311065}}].

\bibitem{Lee:2009dh}
R.~N. Lee, \emph{{Space-time dimensionality D as complex variable: Calculating
  loop integrals using dimensional recurrence relation and analytical
  properties with respect to D}},
  \href{http://dx.doi.org/10.1016/j.nuclphysb.2009.12.025}{\emph{Nucl. Phys. B}
  {\bfseries 830} (2010) 474--492},
  [\href{https://arxiv.org/abs/0911.0252}{{\ttfamily 0911.0252}}].

\bibitem{Grozin:2011mt}
A.~G. Grozin, \emph{{Integration by parts: An Introduction}},
  \href{http://dx.doi.org/10.1142/S0217751X11053687}{\emph{Int. J. Mod. Phys.
  A} {\bfseries 26} (2011) 2807--2854},
  [\href{https://arxiv.org/abs/1104.3993}{{\ttfamily 1104.3993}}].

\bibitem{Henn:2013pwa}
J.~M. Henn, \emph{{Multiloop integrals in dimensional regularization made
  simple}}, \href{http://dx.doi.org/10.1103/PhysRevLett.110.251601}{\emph{Phys.
  Rev. Lett.} {\bfseries 110} (2013) 251601},
  [\href{https://arxiv.org/abs/1304.1806}{{\ttfamily 1304.1806}}].

\bibitem{Henn:2014qga}
J.~M. Henn, \emph{{Lectures on differential equations for Feynman integrals}},
  \href{http://dx.doi.org/10.1088/1751-8113/48/15/153001}{\emph{J. Phys. A}
  {\bfseries 48} (2015) 153001},
  [\href{https://arxiv.org/abs/1412.2296}{{\ttfamily 1412.2296}}].

\bibitem{vonManteuffel:2014ixa}
A.~von Manteuffel and R.~M. Schabinger, \emph{{A novel approach to integration
  by parts reduction}},
  \href{http://dx.doi.org/10.1016/j.physletb.2015.03.029}{\emph{Phys. Lett. B}
  {\bfseries 744} (2015) 101--104},
  [\href{https://arxiv.org/abs/1406.4513}{{\ttfamily 1406.4513}}].

\bibitem{Zhang:2016kfo}
Y.~Zhang, \emph{{Lecture Notes on Multi-loop Integral Reduction and Applied
  Algebraic Geometry}},  12, 2016,
  \href{https://arxiv.org/abs/1612.02249}{{\ttfamily 1612.02249}}.

\bibitem{Kosower:2018obg}
D.~A. Kosower, \emph{{Direct Solution of Integration-by-Parts Systems}},
  \href{http://dx.doi.org/10.1103/PhysRevD.98.025008}{\emph{Phys. Rev. D}
  {\bfseries 98} (2018) 025008},
  [\href{https://arxiv.org/abs/1804.00131}{{\ttfamily 1804.00131}}].

\bibitem{Smirnov:2008iw}
A.~V. Smirnov, \emph{{Algorithm FIRE -- Feynman Integral REduction}},
  \href{http://dx.doi.org/10.1088/1126-6708/2008/10/107}{\emph{JHEP} {\bfseries
  10} (2008) 107}, [\href{https://arxiv.org/abs/0807.3243}{{\ttfamily
  0807.3243}}].

\bibitem{vonManteuffel:2012np}
A.~von Manteuffel and C.~Studerus, \emph{{Reduze 2 - Distributed Feynman
  Integral Reduction}},  \href{https://arxiv.org/abs/1201.4330}{{\ttfamily
  1201.4330}}.

\bibitem{Lee:2012cn}
R.~N. Lee, \emph{{Presenting LiteRed: a tool for the Loop InTEgrals
  REDuction}},  \href{https://arxiv.org/abs/1212.2685}{{\ttfamily 1212.2685}}.

\bibitem{Maierhofer:2017gsa}
P.~Maierh\"ofer, J.~Usovitsch and P.~Uwer, \emph{{Kira\textemdash{}A Feynman
  integral reduction program}},
  \href{http://dx.doi.org/10.1016/j.cpc.2018.04.012}{\emph{Comput. Phys.
  Commun.} {\bfseries 230} (2018) 99--112},
  [\href{https://arxiv.org/abs/1705.05610}{{\ttfamily 1705.05610}}].

\bibitem{Bourjaily:2021hcp}
J.~L. Bourjaily, C.~Langer and Y.~Zhang, \emph{{Illustrations of
  Integrand-Basis Building at Two Loops}},
  \href{https://arxiv.org/abs/2112.05157}{{\ttfamily 2112.05157}}.

\bibitem{Bourjaily:2017wjl}
J.~L. Bourjaily, E.~Herrmann and J.~Trnka, \emph{{Prescriptive Unitarity}},
  \href{http://dx.doi.org/10.1007/JHEP06(2017)059}{\emph{JHEP} {\bfseries 06}
  (2017) 059}, [\href{https://arxiv.org/abs/1704.05460}{{\ttfamily
  1704.05460}}].

\bibitem{Mastrolia:2018uzb}
P.~Mastrolia and S.~Mizera, \emph{{Feynman Integrals and Intersection Theory}},
  \href{http://dx.doi.org/10.1007/JHEP02(2019)139}{\emph{JHEP} {\bfseries 02}
  (2019) 139}, [\href{https://arxiv.org/abs/1810.03818}{{\ttfamily
  1810.03818}}].

\bibitem{Frellesvig:2019kgj}
H.~Frellesvig, F.~Gasparotto, S.~Laporta, M.~K. Mandal, P.~Mastrolia,
  L.~Mattiazzi and S.~Mizera, \emph{{Decomposition of Feynman Integrals on the
  Maximal Cut by Intersection Numbers}},
  \href{http://dx.doi.org/10.1007/JHEP05(2019)153}{\emph{JHEP} {\bfseries 05}
  (2019) 153}, [\href{https://arxiv.org/abs/1901.11510}{{\ttfamily
  1901.11510}}].

\bibitem{Frellesvig:2019uqt}
H.~Frellesvig, F.~Gasparotto, M.~K. Mandal, P.~Mastrolia, L.~Mattiazzi and
  S.~Mizera, \emph{{Vector Space of Feynman Integrals and Multivariate
  Intersection Numbers}},
  \href{http://dx.doi.org/10.1103/PhysRevLett.123.201602}{\emph{Phys. Rev.
  Lett.} {\bfseries 123} (2019) 201602},
  [\href{https://arxiv.org/abs/1907.02000}{{\ttfamily 1907.02000}}].

\bibitem{Mizera:2019vvs}
S.~Mizera and A.~Pokraka, \emph{{From Infinity to Four Dimensions: Higher
  Residue Pairings and Feynman Integrals}},
  \href{http://dx.doi.org/10.1007/JHEP02(2020)159}{\emph{JHEP} {\bfseries 02}
  (2020) 159}, [\href{https://arxiv.org/abs/1910.11852}{{\ttfamily
  1910.11852}}].

\bibitem{Frellesvig:2020qot}
H.~Frellesvig, F.~Gasparotto, S.~Laporta, M.~K. Mandal, P.~Mastrolia,
  L.~Mattiazzi and S.~Mizera, \emph{{Decomposition of Feynman Integrals by
  Multivariate Intersection Numbers}},
  \href{http://dx.doi.org/10.1007/JHEP03(2021)027}{\emph{JHEP} {\bfseries 03}
  (2021) 027}, [\href{https://arxiv.org/abs/2008.04823}{{\ttfamily
  2008.04823}}].

\bibitem{Caron-Huot:2021xqj}
S.~Caron-Huot and A.~Pokraka, \emph{{Duals of Feynman integrals. Part I.
  Differential equations}},
  \href{http://dx.doi.org/10.1007/JHEP12(2021)045}{\emph{JHEP} {\bfseries 12}
  (2021) 045}, [\href{https://arxiv.org/abs/2104.06898}{{\ttfamily
  2104.06898}}].

\bibitem{Chestnov:2022alh}
V.~Chestnov, F.~Gasparotto, M.~K. Mandal, P.~Mastrolia, S.~J. Matsubara-Heo,
  H.~J. Munch and N.~Takayama, \emph{{Macaulay matrix for Feynman integrals:
  linear relations and intersection numbers}},
  \href{http://dx.doi.org/10.1007/JHEP09(2022)187}{\emph{JHEP} {\bfseries 09}
  (2022) 187}, [\href{https://arxiv.org/abs/2204.12983}{{\ttfamily
  2204.12983}}].

\bibitem{Chestnov:2022xsy}
V.~Chestnov, H.~Frellesvig, F.~Gasparotto, M.~K. Mandal and P.~Mastrolia,
  \emph{{Intersection Numbers from Higher-order Partial Differential
  Equations}},  \href{https://arxiv.org/abs/2209.01997}{{\ttfamily
  2209.01997}}.

\bibitem{Giroux:2022wav}
M.~Giroux and A.~Pokraka, \emph{{Loop-by-loop Differential Equations for Dual
  (Elliptic) Feynman Integrals}},
  \href{https://arxiv.org/abs/2210.09898}{{\ttfamily 2210.09898}}.

\bibitem{Bourjaily:2020qca}
J.~L. Bourjaily, E.~Herrmann, C.~Langer and J.~Trnka, \emph{{Building bases of
  loop integrands}},
  \href{http://dx.doi.org/10.1007/JHEP11(2020)116}{\emph{JHEP} {\bfseries 11}
  (2020) 116}, [\href{https://arxiv.org/abs/2007.13905}{{\ttfamily
  2007.13905}}].

\bibitem{Bourjaily:2017bsb}
J.~L. Bourjaily, A.~J. McLeod, M.~Spradlin, M.~von Hippel and M.~Wilhelm,
  \emph{{Elliptic Double-Box Integrals: Massless Scattering Amplitudes beyond
  Polylogarithms}},
  \href{http://dx.doi.org/10.1103/PhysRevLett.120.121603}{\emph{Phys. Rev.
  Lett.} {\bfseries 120} (2018) 121603},
  [\href{https://arxiv.org/abs/1712.02785}{{\ttfamily 1712.02785}}].

\bibitem{Laporta:2004rb}
S.~Laporta and E.~Remiddi, \emph{{Analytic treatment of the two loop equal mass
  sunrise graph}},
  \href{http://dx.doi.org/10.1016/j.nuclphysb.2004.10.044}{\emph{Nucl. Phys. B}
  {\bfseries 704} (2005) 349--386},
  [\href{https://arxiv.org/abs/hep-ph/0406160}{{\ttfamily hep-ph/0406160}}].

\bibitem{Adams:2013nia}
L.~Adams, C.~Bogner and S.~Weinzierl, \emph{{The two-loop sunrise graph with
  arbitrary masses}}, \href{http://dx.doi.org/10.1063/1.4804996}{\emph{J. Math.
  Phys.} {\bfseries 54} (2013) 052303},
  [\href{https://arxiv.org/abs/1302.7004}{{\ttfamily 1302.7004}}].

\bibitem{Bloch:2013tra}
S.~Bloch and P.~Vanhove, \emph{{The elliptic dilogarithm for the sunset
  graph}}, \href{http://dx.doi.org/10.1016/j.jnt.2014.09.032}{\emph{J. Number
  Theor.} {\bfseries 148} (2015) 328--364},
  [\href{https://arxiv.org/abs/1309.5865}{{\ttfamily 1309.5865}}].

\bibitem{Adams:2014vja}
L.~Adams, C.~Bogner and S.~Weinzierl, \emph{{The two-loop sunrise graph in two
  space-time dimensions with arbitrary masses in terms of elliptic
  dilogarithms}}, \href{http://dx.doi.org/10.1063/1.4896563}{\emph{J. Math.
  Phys.} {\bfseries 55} (2014) 102301},
  [\href{https://arxiv.org/abs/1405.5640}{{\ttfamily 1405.5640}}].

\bibitem{Adams:2015gva}
L.~Adams, C.~Bogner and S.~Weinzierl, \emph{{The two-loop sunrise integral
  around four space-time dimensions and generalisations of the Clausen and
  Glaisher functions towards the elliptic case}},
  \href{http://dx.doi.org/10.1063/1.4926985}{\emph{J. Math. Phys.} {\bfseries
  56} (2015) 072303}, [\href{https://arxiv.org/abs/1504.03255}{{\ttfamily
  1504.03255}}].

\bibitem{Adams:2015ydq}
L.~Adams, C.~Bogner and S.~Weinzierl, \emph{{The iterated structure of the
  all-order result for the two-loop sunrise integral}},
  \href{http://dx.doi.org/10.1063/1.4944722}{\emph{J. Math. Phys.} {\bfseries
  57} (2016) 032304}, [\href{https://arxiv.org/abs/1512.05630}{{\ttfamily
  1512.05630}}].

\bibitem{Adams:2016xah}
L.~Adams, C.~Bogner, A.~Schweitzer and S.~Weinzierl, \emph{{The kite integral
  to all orders in terms of elliptic polylogarithms}},
  \href{http://dx.doi.org/10.1063/1.4969060}{\emph{J. Math. Phys.} {\bfseries
  57} (2016) 122302}, [\href{https://arxiv.org/abs/1607.01571}{{\ttfamily
  1607.01571}}].

\bibitem{vonManteuffel:2017hms}
A.~von Manteuffel and L.~Tancredi, \emph{{A non-planar two-loop three-point
  function beyond multiple polylogarithms}},
  \href{http://dx.doi.org/10.1007/JHEP06(2017)127}{\emph{JHEP} {\bfseries 06}
  (2017) 127}, [\href{https://arxiv.org/abs/1701.05905}{{\ttfamily
  1701.05905}}].

\bibitem{Bogner:2017vim}
C.~Bogner, A.~Schweitzer and S.~Weinzierl, \emph{{Analytic continuation and
  numerical evaluation of the kite integral and the equal mass sunrise
  integral}},
  \href{http://dx.doi.org/10.1016/j.nuclphysb.2017.07.008}{\emph{Nucl. Phys. B}
  {\bfseries 922} (2017) 528--550},
  [\href{https://arxiv.org/abs/1705.08952}{{\ttfamily 1705.08952}}].

\bibitem{Muller:2022gec}
H.~M\"uller and S.~Weinzierl, \emph{{A Feynman integral depending on two
  elliptic curves}},
  \href{http://dx.doi.org/10.1007/JHEP07(2022)101}{\emph{JHEP} {\bfseries 07}
  (2022) 101}, [\href{https://arxiv.org/abs/2205.04818}{{\ttfamily
  2205.04818}}].

\bibitem{Broedel:2019kmn}
J.~Broedel, C.~Duhr, F.~Dulat, R.~Marzucca, B.~Penante and L.~Tancredi,
  \emph{{An analytic solution for the equal-mass banana graph}},
  \href{http://dx.doi.org/10.1007/JHEP09(2019)112}{\emph{JHEP} {\bfseries 09}
  (2019) 112}, [\href{https://arxiv.org/abs/1907.03787}{{\ttfamily
  1907.03787}}].

\bibitem{Broadhurst:1993mw}
D.~J. Broadhurst, J.~Fleischer and O.~V. Tarasov, \emph{{Two loop two point
  functions with masses: Asymptotic expansions and Taylor series, in any
  dimension}}, \href{http://dx.doi.org/10.1007/BF01474625}{\emph{Z. Phys. C}
  {\bfseries 60} (1993) 287--302},
  [\href{https://arxiv.org/abs/hep-ph/9304303}{{\ttfamily hep-ph/9304303}}].

\bibitem{Muller-Stach:2011qkg}
S.~M\"uller-Stach, S.~Weinzierl and R.~Zayadeh, \emph{{A Second-Order
  Differential Equation for the Two-Loop Sunrise Graph with Arbitrary Masses}},
  \href{http://dx.doi.org/10.4310/CNTP.2012.v6.n1.a5}{\emph{Commun. Num. Theor.
  Phys.} {\bfseries 6} (2012) 203--222},
  [\href{https://arxiv.org/abs/1112.4360}{{\ttfamily 1112.4360}}].

\bibitem{Remiddi:2013joa}
E.~Remiddi and L.~Tancredi, \emph{{Schouten identities for Feynman graph
  amplitudes; The Master Integrals for the two-loop massive sunrise graph}},
  \href{http://dx.doi.org/10.1016/j.nuclphysb.2014.01.009}{\emph{Nucl. Phys. B}
  {\bfseries 880} (2014) 343--377},
  [\href{https://arxiv.org/abs/1311.3342}{{\ttfamily 1311.3342}}].

\bibitem{Adams:2018yfj}
L.~Adams and S.~Weinzierl, \emph{{The $\varepsilon$-form of the differential
  equations for Feynman integrals in the elliptic case}},
  \href{http://dx.doi.org/10.1016/j.physletb.2018.04.002}{\emph{Phys. Lett. B}
  {\bfseries 781} (2018) 270--278},
  [\href{https://arxiv.org/abs/1802.05020}{{\ttfamily 1802.05020}}].

\bibitem{Bloch:2014qca}
S.~Bloch, M.~Kerr and P.~Vanhove, \emph{{A Feynman integral via higher normal
  functions}}, \href{http://dx.doi.org/10.1112/S0010437X15007472}{\emph{Compos.
  Math.} {\bfseries 151} (2015) 2329--2375},
  [\href{https://arxiv.org/abs/1406.2664}{{\ttfamily 1406.2664}}].

\bibitem{Bloch:2016izu}
S.~Bloch, M.~Kerr and P.~Vanhove, \emph{{Local mirror symmetry and the sunset
  Feynman integral}},
  \href{http://dx.doi.org/10.4310/ATMP.2017.v21.n6.a1}{\emph{Adv. Theor. Math.
  Phys.} {\bfseries 21} (2017) 1373--1453},
  [\href{https://arxiv.org/abs/1601.08181}{{\ttfamily 1601.08181}}].

\bibitem{Adams:2017ejb}
L.~Adams and S.~Weinzierl, \emph{{Feynman integrals and iterated integrals of
  modular forms}},
  \href{http://dx.doi.org/10.4310/CNTP.2018.v12.n2.a1}{\emph{Commun. Num.
  Theor. Phys.} {\bfseries 12} (2018) 193--251},
  [\href{https://arxiv.org/abs/1704.08895}{{\ttfamily 1704.08895}}].

\bibitem{Ablinger:2017bjx}
J.~Ablinger, J.~Bl\"umlein, A.~De~Freitas, M.~van Hoeij, E.~Imamoglu, C.~G.
  Raab, C.~S. Radu et~al., \emph{{Iterated Elliptic and Hypergeometric
  Integrals for Feynman Diagrams}},
  \href{http://dx.doi.org/10.1063/1.4986417}{\emph{J. Math. Phys.} {\bfseries
  59} (2018) 062305}, [\href{https://arxiv.org/abs/1706.01299}{{\ttfamily
  1706.01299}}].

\bibitem{Primo:2017ipr}
A.~Primo and L.~Tancredi, \emph{{Maximal cuts and differential equations for
  Feynman integrals. An application to the three-loop massive banana graph}},
  \href{http://dx.doi.org/10.1016/j.nuclphysb.2017.05.018}{\emph{Nucl. Phys. B}
  {\bfseries 921} (2017) 316--356},
  [\href{https://arxiv.org/abs/1704.05465}{{\ttfamily 1704.05465}}].

\bibitem{Broedel:2017kkb}
J.~Broedel, C.~Duhr, F.~Dulat and L.~Tancredi, \emph{{Elliptic polylogarithms
  and iterated integrals on elliptic curves. Part I: general formalism}},
  \href{http://dx.doi.org/10.1007/JHEP05(2018)093}{\emph{JHEP} {\bfseries 05}
  (2018) 093}, [\href{https://arxiv.org/abs/1712.07089}{{\ttfamily
  1712.07089}}].

\bibitem{Hidding:2017jkk}
M.~Hidding and F.~Moriello, \emph{{All orders structure and efficient
  computation of linearly reducible elliptic Feynman integrals}},
  \href{http://dx.doi.org/10.1007/JHEP01(2019)169}{\emph{JHEP} {\bfseries 01}
  (2019) 169}, [\href{https://arxiv.org/abs/1712.04441}{{\ttfamily
  1712.04441}}].

\bibitem{Lee:2017qql}
R.~N. Lee, A.~V. Smirnov and V.~A. Smirnov, \emph{{Solving differential
  equations for Feynman integrals by expansions near singular points}},
  \href{http://dx.doi.org/10.1007/JHEP03(2018)008}{\emph{JHEP} {\bfseries 03}
  (2018) 008}, [\href{https://arxiv.org/abs/1709.07525}{{\ttfamily
  1709.07525}}].

\bibitem{Lee:2018ojn}
R.~N. Lee, A.~V. Smirnov and V.~A. Smirnov, \emph{{Evaluating
  \textquoteleft{}elliptic\textquoteright{} master integrals at special
  kinematic values: using differential equations and their solutions via
  expansions near singular points}},
  \href{http://dx.doi.org/10.1007/JHEP07(2018)102}{\emph{JHEP} {\bfseries 07}
  (2018) 102}, [\href{https://arxiv.org/abs/1805.00227}{{\ttfamily
  1805.00227}}].

\bibitem{Broedel:2018qkq}
J.~Broedel, C.~Duhr, F.~Dulat, B.~Penante and L.~Tancredi, \emph{{Elliptic
  Feynman integrals and pure functions}},
  \href{http://dx.doi.org/10.1007/JHEP01(2019)023}{\emph{JHEP} {\bfseries 01}
  (2019) 023}, [\href{https://arxiv.org/abs/1809.10698}{{\ttfamily
  1809.10698}}].

\bibitem{Weinzierl:2020fyx}
S.~Weinzierl, \emph{{Modular transformations of elliptic Feynman integrals}},
  \href{http://dx.doi.org/10.1016/j.nuclphysb.2021.115309}{\emph{Nucl. Phys. B}
  {\bfseries 964} (2021) 115309},
  [\href{https://arxiv.org/abs/2011.07311}{{\ttfamily 2011.07311}}].

\bibitem{Walden:2020odh}
M.~Walden and S.~Weinzierl, \emph{{Numerical evaluation of iterated integrals
  related to elliptic Feynman integrals}},
  \href{http://dx.doi.org/10.1016/j.cpc.2021.108020}{\emph{Comput. Phys.
  Commun.} {\bfseries 265} (2021) 108020},
  [\href{https://arxiv.org/abs/2010.05271}{{\ttfamily 2010.05271}}].

\bibitem{Frellesvig:2021hkr}
H.~Frellesvig, \emph{{On epsilon factorized differential equations for elliptic
  Feynman integrals}},
  \href{http://dx.doi.org/10.1007/JHEP03(2022)079}{\emph{JHEP} {\bfseries 03}
  (2022) 079}, [\href{https://arxiv.org/abs/2110.07968}{{\ttfamily
  2110.07968}}].

\bibitem{Bourjaily:2020hjv}
J.~L. Bourjaily, N.~Kalyanapuram, C.~Langer, K.~Patatoukos and M.~Spradlin,
  \emph{{Elliptic, Yangian-Invariant \textquotedblleft{}Leading
  Singularity\textquotedblright{}}},
  \href{http://dx.doi.org/10.1103/PhysRevLett.126.201601}{\emph{Phys. Rev.
  Lett.} {\bfseries 126} (2021) 201601},
  [\href{https://arxiv.org/abs/2012.14438}{{\ttfamily 2012.14438}}].

\bibitem{Bourjaily:2021vyj}
J.~L. Bourjaily, N.~Kalyanapuram, C.~Langer and K.~Patatoukos,
  \emph{{Prescriptive unitarity with elliptic leading singularities}},
  \href{http://dx.doi.org/10.1103/PhysRevD.104.125009}{\emph{Phys. Rev. D}
  {\bfseries 104} (2021) 125009},
  [\href{https://arxiv.org/abs/2102.02210}{{\ttfamily 2102.02210}}].

\bibitem{Bourjaily:2022tep}
J.~L. Bourjaily and N.~Kalyanapuram, \emph{{The Stratification of Rigidity}},
  \href{https://arxiv.org/abs/2207.00596}{{\ttfamily 2207.00596}}.

\bibitem{Bourjaily:2021ujs}
J.~L. Bourjaily, E.~Herrmann, C.~Langer, K.~Patatoukos, J.~Trnka and M.~Zheng,
  \emph{{Integrands of less-supersymmetric Yang-Mills at one loop}},
  \href{http://dx.doi.org/10.1007/JHEP03(2022)126}{\emph{JHEP} {\bfseries 03}
  (2022) 126}, [\href{https://arxiv.org/abs/2112.06901}{{\ttfamily
  2112.06901}}].

\bibitem{Bourjaily:2018yfy}
J.~L. Bourjaily, A.~J. McLeod, M.~von Hippel and M.~Wilhelm, \emph{{Bounded
  Collection of Feynman Integral Calabi-Yau Geometries}},
  \href{http://dx.doi.org/10.1103/PhysRevLett.122.031601}{\emph{Phys. Rev.
  Lett.} {\bfseries 122} (2019) 031601},
  [\href{https://arxiv.org/abs/1810.07689}{{\ttfamily 1810.07689}}].

\bibitem{Paulos:2012nu}
M.~F. Paulos, M.~Spradlin and A.~Volovich, \emph{{Mellin Amplitudes for Dual
  Conformal Integrals}},
  \href{http://dx.doi.org/10.1007/JHEP08(2012)072}{\emph{JHEP} {\bfseries 08}
  (2012) 072}, [\href{https://arxiv.org/abs/1203.6362}{{\ttfamily 1203.6362}}].

\bibitem{Caron-Huot:2012awx}
S.~Caron-Huot and K.~J. Larsen, \emph{{Uniqueness of two-loop master
  contours}}, \href{http://dx.doi.org/10.1007/JHEP10(2012)026}{\emph{JHEP}
  {\bfseries 10} (2012) 026},
  [\href{https://arxiv.org/abs/1205.0801}{{\ttfamily 1205.0801}}].

\bibitem{Nandan:2013ip}
D.~Nandan, M.~F. Paulos, M.~Spradlin and A.~Volovich, \emph{{Star Integrals,
  Convolutions and Simplices}},
  \href{http://dx.doi.org/10.1007/JHEP05(2013)105}{\emph{JHEP} {\bfseries 05}
  (2013) 105}, [\href{https://arxiv.org/abs/1301.2500}{{\ttfamily 1301.2500}}].

\bibitem{Goncharov:1998kja}
A.~B. Goncharov, \emph{{Multiple polylogarithms, cyclotomy and modular
  complexes}}, \href{http://dx.doi.org/10.4310/MRL.1998.v5.n4.a7}{\emph{Math.
  Res. Lett.} {\bfseries 5} (1998) 497--516},
  [\href{https://arxiv.org/abs/1105.2076}{{\ttfamily 1105.2076}}].

\bibitem{Panzer:2014caa}
E.~Panzer, \emph{{Algorithms for the symbolic integration of hyperlogarithms
  with applications to Feynman integrals}},
  \href{http://dx.doi.org/10.1016/j.cpc.2014.10.019}{\emph{Comput. Phys.
  Commun.} {\bfseries 188} (2015) 148--166},
  [\href{https://arxiv.org/abs/1403.3385}{{\ttfamily 1403.3385}}].

\bibitem{chiPersonal}
C.~Zhang, \emph{private comunication}, .

\bibitem{Bourjaily:2021iyq}
J.~L. Bourjaily, C.~Langer and Y.~Zhang, \emph{{All two-loop, color-dressed,
  six-point amplitude integrands in supersymmetric Yang-Mills theory}},
  \href{http://dx.doi.org/10.1103/PhysRevD.105.105015}{\emph{Phys. Rev. D}
  {\bfseries 105} (2022) 105015},
  [\href{https://arxiv.org/abs/2112.06934}{{\ttfamily 2112.06934}}].

\bibitem{Drummond:2006rz}
J.~M. Drummond, J.~Henn, V.~A. Smirnov and E.~Sokatchev, \emph{{Magic
  identities for conformal four-point integrals}},
  \href{http://dx.doi.org/10.1088/1126-6708/2007/01/064}{\emph{JHEP} {\bfseries
  01} (2007) 064}, [\href{https://arxiv.org/abs/hep-th/0607160}{{\ttfamily
  hep-th/0607160}}].

\bibitem{Bern:2008ap}
Z.~Bern, L.~J. Dixon, D.~A. Kosower, R.~Roiban, M.~Spradlin, C.~Vergu and
  A.~Volovich, \emph{{The Two-Loop Six-Gluon MHV Amplitude in Maximally
  Supersymmetric Yang-Mills Theory}},
  \href{http://dx.doi.org/10.1103/PhysRevD.78.045007}{\emph{Phys. Rev. D}
  {\bfseries 78} (2008) 045007},
  [\href{https://arxiv.org/abs/0803.1465}{{\ttfamily 0803.1465}}].

\bibitem{Drummond:2008aq}
J.~M. Drummond, J.~Henn, G.~P. Korchemsky and E.~Sokatchev, \emph{{Hexagon
  Wilson loop = six-gluon MHV amplitude}},
  \href{http://dx.doi.org/10.1016/j.nuclphysb.2009.02.015}{\emph{Nucl. Phys. B}
  {\bfseries 815} (2009) 142--173},
  [\href{https://arxiv.org/abs/0803.1466}{{\ttfamily 0803.1466}}].

\bibitem{bourjailyCode}
J.~Bourjaily, \emph{private comunication}, .

\bibitem{Bourjaily:2019jrk}
J.~L. Bourjaily, F.~Dulat and E.~Panzer, \emph{{Manifestly Dual-Conformal Loop
  Integration}},
  \href{http://dx.doi.org/10.1016/j.nuclphysb.2019.03.022}{\emph{Nucl. Phys. B}
  {\bfseries 942} (2019) 251--302},
  [\href{https://arxiv.org/abs/1901.02887}{{\ttfamily 1901.02887}}].

\bibitem{Hodges:2009hk}
A.~Hodges, \emph{{Eliminating spurious poles from gauge-theoretic amplitudes}},
  \href{http://dx.doi.org/10.1007/JHEP05(2013)135}{\emph{JHEP} {\bfseries 05}
  (2013) 135}, [\href{https://arxiv.org/abs/0905.1473}{{\ttfamily 0905.1473}}].

\bibitem{Witten:2003nn}
E.~Witten, \emph{{Perturbative gauge theory as a string theory in twistor
  space}}, \href{http://dx.doi.org/10.1007/s00220-004-1187-3}{\emph{Commun.
  Math. Phys.} {\bfseries 252} (2004) 189--258},
  [\href{https://arxiv.org/abs/hep-th/0312171}{{\ttfamily hep-th/0312171}}].

\bibitem{Bourjaily:2010wh}
J.~L. Bourjaily, \emph{{Efficient Tree-Amplitudes in N=4: Automatic BCFW
  Recursion in Mathematica}},
  \href{https://arxiv.org/abs/1011.2447}{{\ttfamily 1011.2447}}.

\end{thebibliography}\endgroup


\providecommand{\noopsort}[1]{}\providecommand{\singleletter}[1]{#1}%

\providecommand{\href}[2]{#2}\begingroup\raggedright\endgroup
\end{document}